%% file: fh3january.tex
\documentclass[12pt]{article}
\usepackage{amsfonts,amssymb,graphicx,fullpage}
\usepackage[tight]{subfigure}
\usepackage{rotating}
\pdfoutput=1
\usepackage{ifpdf}
\ifpdf
  \usepackage{color}
  \definecolor{darkblue}{rgb}{0.3,0.3,0.6}
  \usepackage[pdftitle={Rigid D-branes on T6/Z(2)xZ(2M)},pdfauthor={Stefan F\"orste and Gabriele Honecker},pdfsubject={String theory},pdfkeywords={string theory, rigid branes},colorlinks=true,linkcolor=black,urlcolor=darkblue,filecolor=darkblue,citecolor=darkblue,menucolor=darkblue,breaklinks=true,bookmarks=true,anchorcolor=black,unicode=true]{hyperref}
\else
  
\fi
\usepackage[numbers,compress]{natbib}
\usepackage{a4wide}
\usepackage{longtable}
\usepackage{graphicx}
\usepackage{subfigure}
\usepackage[centertags]{amsmath}

\newcommand{\bCentering}{\centering}
\newcommand{\bCaption}{\caption}
\newcommand{\sgn}{{\rm sgn}}
\newcommand{\unity}{{\footnotesize\mbox{1\!\!I}}}
\newcommand{\bC}{\mathbb{C}}
\newcommand{\bcross}{\textcolor{blue}{\times}}
\newcommand{\rcircle}{\textcolor{red}{\bigcirc}}

\def\Z{\mathbb{Z}}
\def\unity{1\!\!{\rm I}}
\def\ov{\overline}
\def\N{\mathbf{N}}
\def\Sym{\mathbf{Sym}}
\def\Anti{\mathbf{Anti}}
\def\Adj{\mathbf{Adj}}

\def\ov{\overline}
\def\1{{\bf 1}}
\def\2{{\bf 2}}
\def\3{{\bf 3}}
\def\4{{\bf 4}}
\def\6{{\bf 6}}
\def\OR{\Omega\mathcal{R}}

\def\pp{\uparrow\uparrow}

\newcommand{\bCaptionfonts}{\small}
\makeatletter
\long\def\@makecaption#1#2{%
  \vskip\abovecaptionskip
  \sbox\@tempboxa{{\bCaptionfonts #1: #2}}%
  \ifdim \wd\@tempboxa >\hsize
    {\bCaptionfonts #1: #2\par}
  \else
    \hbox to\hsize{\hfil\box\@tempboxa\hfil}%
  \fi
  \vskip\belowcaptionskip}
\makeatother

\makeatletter
\let\ORIGINALlatex@openbib@code=\@openbib@code
\renewcommand{\@openbib@code}{\ORIGINALlatex@openbib@code\setlength{\itemsep}{1ex plus.5ex minus.5ex}\setlength{\parsep}{0pt}}
\makeatother

\def\mathsidetabfix#1#2#3{\begin{sidewaystable}[H]\bCentering\resizebox{\linewidth}{!}{$#1$}\bCaption{#3}\label{tab:#2}\end{sidewaystable}}

\renewcommand{\arraystretch}{1.3}

\begin{document}
\begin{center}
\begin{flushright}
{\small MZ-TH/10-39\\NSF-KITP-10-133\\October 2010}
\end{flushright}

\vspace{30mm}
{\Large\bf Rigid D6-branes on {\boldmath $T^6/(\Z_2 \times \Z_{2M}
    \times \OR)$} with discrete torsion} 

\vspace{15mm}
{\large Stefan F\"orste$^1$ and Gabriele Honecker$^{2,3}$
}

\vspace{10mm}
{~$^1$\it Bethe Center for Theoretical Physics and Physikalisches
  Institut der Universit\"at Bonn, Nussallee 12, D - 53115 Bonn,
  Germany  \; 
{\tt forste@th.physik.uni-bonn.de}}\\[1ex]
{~$^2$\it Institut f\"ur Physik  (WA THEP), Johannes-Gutenberg-Universit\"at, D - 55099 Mainz, Germany
\; {\tt Gabriele.Honecker@uni-mainz.de}}\\[1ex]
{~$^3$\it  Institute of Theoretical Physics, K.U.Leuven, Celestijnenlaan 200D, B - 3001 Leuven, Belgium}

\vspace{20mm}{\bf Abstract}\\[2ex]\parbox{140mm}{
We give a complete classification of $T^6/(\Z_2 \times \Z_{2M} \times \OR)$ orientifolds on factorisable tori
and rigid D6-branes on them. The analysis includes the supersymmetry, RR tadpole cancellation and K-theory 
conditions and complete massless open and closed string spectrum
(i.e.\ non-chiral as well as chiral)  
for fractional
or rigid D6-branes for all inequivalent compactification lattices, without and with discrete torsion.
We give examples for each orbifold background, which show that on $\Z_2 \times \Z_6$ and $\Z_2 \times \Z_6'$ 
there exist completely rigid D6-branes despite the self-intersections of orbifold image cycles.
This opens up a new avenue for improved Standard Model building.
On the other hand, we show that Standard and GUT model building on the $\Z_2 \times \Z_4$ background 
is ruled out by simple arguments.}
\end{center}

\thispagestyle{empty}
\clearpage

\tableofcontents
\newpage
\setlength{\parskip}{1em plus1ex minus.5ex}
\section{Introduction}\label{S:intro}

To date the Standard Model of particle physics (SM) is in very good
agreement with experiment. Still, there are several theoretical
reasons to believe that it is an effective theory valid only up to
some scale\footnote{There are also some experimental and observational
  hints such as e.g.\ dark matter.}. The SM has about 17 parameters
determined only by fitting with data. In an underlying more
fundamental theory these parameters might be related via symmetry
and/or determined dynamically due to expectation values of extra SM
singlets. In particular the electro-weak symmetry breaking scale is
not explained within the SM. Therefore, it is natural to believe that
the SM is an effective description of nature up to that scale. The,
perhaps, most prominent theory beyond the SM (BSM) is low energy
supersymmetry. It explains neatly many open questions of the SM (for
reviews see e.g.\ \cite{Nilles:1983ge,Martin:1997ns,Luty:2005sn}). In
the supersymmetric extension of the SM some parameters are indeed
related by symmetry, e.g.\ the quartic Higgs coupling is expressed in
terms of the $SU(2)_L$ and $U(1)_Y$ gauge couplings. However, it is
not known which mechanism of supersymmetry breaking is realised in
nature. Pragmatically, all possible schemes are parameterised by the
so called soft breaking terms introducing 105 new parameters. So, the
number of parameters actually grows but just because details of the
BSM are not known yet. Conceptually the number is reduced by choosing
a particular breaking scheme.

So far, our discussion focused on the so called bottom up approach:
The BSM consists of the SM plus extensions. In a top down approach, on
the other hand, one starts with more fundamental questions, for
instance: How gravity is quantised? In an ideal world the answer to
such a question would ultimately lead to the SM with the correct
values for its 17 parameters. Indeed, string theory consistently
includes quantum gravity and has only one parameter, the string
tension $1/\alpha^\prime$. However, there is a huge landscape of
consistent string vacua and no conceptual way of choosing one is
known. Parameters are indeed replaced by expectation values of
moduli. There are, however, many ways to stabilise these moduli. 
Again, one can postpone the question of vacuum selection
and pragmatically look for string vacua reproducing correctly the SM
at low energies. There are several approaches to identify promising
string vacua each of which is well motivated, see 
e.g.~\cite{Buchmuller:2005jr,Lebedev:2006kn,Buchmuller:2006ik,Lebedev:2008un}
for heterotic orbifolds,~\cite{Braun:2005nv,Bouchard:2005ag} for heterotic Calabi-Yau compactifications
with $SU(N)$ bundles and~\cite{Blumenhagen:2005ga,Blumenhagen:2006ux}  with $U(N)$ bundles,
\cite{Aldazabal:2000sa,Verlinde:2005jr,Conlon:2008wa,Krippendorf:2010hj} for local IIB models, 
\cite{Beasley:2008dc,Donagi:2008ca,Weigand:2010wm} for F-theory and~\cite{Dijkstra:2004ym,Dijkstra:2004cc} 
for Gepner models. 
While the techniques for identifying the gauge group and chiral matter
spectrum are rather straightforward in all approaches, an investigation
of the exact field theory is typically constrained to vacua, where e.g.
conformal field theory methods can be used.

In the present paper, we take intersecting D6-branes of type IIA string
theory as our starting point. There are several good reviews on this
kind of model building containing also references to the original
literature (including the T-dual constructions of magnetised branes
in IIB string theory), e.g.\ 
\cite{Uranga:2003pz,Blumenhagen:2005mu,Blumenhagen:2006ci,Dudas:2006bj,Marchesano:2007de,Lust:2007kw,Angelantonj:2002ct}. 
Geometrically,
intersecting D6-brane constructions are quite intuitive. Closed strings
move through the bulk. Their excitations provide the gravitational
sector of the low energy theory. Open strings ending on the same stack
of D6-branes give rise to the gauge sector, while strings stretched
between different stacks of D6-branes provide potentially chiral
matter. Type IIA string theory will be compactified on an orbifold of
$T^6$ such that the amount of supersymmetry is reduced to ${\cal N}
=2$. This simple geometry of the compact space has the advantage that
conformal field theory techniques can be applied. Explicit computations 
of the complete spectrum, interaction terms and instanton corrections
are possible. Gauging further an orientifold symmetry
reduces the amount of supersymmetry to ${\cal N} = 1$. In type IIA,
orientifold symmetries include a reflection of an odd number of
directions. We choose this number equal to three and add thus
orientifold-six-planes (O6-planes). RR-charges are finally cancelled
by adding D6-branes. 
Throughout this article, we focus on {\it ``globally consistent string 
compactifications''} in the sense of cancellation of {\it all} untwisted and 
twisted RR tadpoles and fulfillment of the K-theory constraint, where we also know the 
{\it full} closed and open string spectrum and can in principle compute the exact 
moduli dependence of couplings
to all orders. This is in contrast to a recent trend of calling
{\it ``locally consistent''} or anomaly-free gauge quivers ``globally consistent'',
see e.g.~\cite{Cvetic:2009yh,Anastasopoulos:2010hu}.

In the present paper, we will take ${\mathbb Z}_2 \times {\mathbb
  Z}_{2M}$ with $2M = 2,4,6$ and $6'$ as orbifold group. These belong
to a set 
for which discrete torsion can be turned on \cite{Vafa:1986wx,Font:1988mk}. There
the authors consider ${\mathbb Z}_M \times {\mathbb Z}_N$  orbifolds
and discuss possible phase factors in front of twisted sector
contributions to the torus amplitude. Non-trivial phase factors
correspond to discrete torsion. For our subset discrete torsion is
related to a non-trivial second root of unity, i.e.\ a sign
choice. Our motivation is that with discrete torsion one can have {\it
  rigid D6-branes}. These are D6-branes wrapping a fractional bulk three-cycle
plus an exceptional three-cycle which is collapsed to a lower
dimensional fixed cycle in the orbifold limit of the Calabi-Yau space
\cite{Blumenhagen:2002wn}. Since D6-branes wrapping such cycles cannot
change position (at least in some direction) the corresponding adjoint
moduli are absent from the open string spectrum. (However, for $2M>2$
adjoints can ``reappear through the back-door'' from open strings
stretching to intersecting orbifold images
\cite{Blumenhagen:2005tn,Blumenhagen:2002gw}. In that case adjoint
moduli are associated with brane recombination. Whether these moduli
are really present has to be seen in a case by case study.)
In the absence of adjoint matter, the arbitraryness of breaking the 
gauge group along a flat direction is removed, and the values of beta
function coefficients are improved in view of phenomenology. 
Rigid branes also admit non-vanishing instanton contributions to the
superpotential, K\"ahler potential~\cite{Billo:2007sw,Billo:2007py} and gauge kinetic function
due to their minimal number of zero modes (see e.g.
the review~\cite{Blumenhagen:2009qh} and references therein), which can generate non-perturbative
neutrino masses~\cite{Blumenhagen:2006xt,Ibanez:2006da}, $\mu$-terms~\cite{Ibanez:2008my}, 
perturvatively forbidden Yukawa 
couplings~\cite{Abel:2006yk,Blumenhagen:2007zk}
or might even trigger supersymmetry breaking~\cite{Cvetic:2008mh}.

Orbifolds with discrete torsion are also of phenomenological interest,
since they have a reduced number of twisted moduli compared to their
counterparts without torsion. This is due to the fact that discrete torsion
does not only exchange K\"ahler for complex structure moduli in some twist sectors,
but also projects out states at points fixed under all orbifold generators.

While twisted moduli are frozen at the orbifold point and
untwisted complex structure moduli are stabilised by the supersymmetry conditions
on D6-branes, the stabilisation of the dilaton and untwisted K\"ahler moduli 
 requires the introduction of closed string background fluxes.
The impact on D6-brane model building has to date been mainly discussed on the torus 
background, see e.g.~\cite{Marchesano:2006ns}
with some partial first results for orbifolds in~\cite{Ihl:2006pp,Ihl:2007ah,Cvetic:2007ju}.
A non-trivial $H$-flux is closely connected to Scherk-Schwarz compactifications
(see e.g.~\cite{Angelantonj:2005hs} and references therein) and 
freeley-acting orbifolds (e.g.~\cite{Serone:2003sv}).

In \cite{Larosa:2003mz,Dudas:2005jx,Blumenhagen:2005tn} the case $2M=2$ (${\mathbb Z}_2 \times
{\mathbb Z}_2$) was studied and toy models with removed adjoint moduli
were presented.\footnote{For D6-branes on the $T^6/\Z_2 \times \Z_2$ orbifold 
without discrete torsion and three adjoints per stack, 
see e.g. the first non-chiral~\cite{Forste:2000hx} and 
chiral models in~\cite{Cvetic:2001tj,Cvetic:2001nr},
a statistical treatment in~\cite{Blumenhagen:2004xx,Gmeiner:2005vz} 
and the further references in the review articles~\cite{Uranga:2003pz,Blumenhagen:2005mu,Blumenhagen:2006ci,Dudas:2006bj,Marchesano:2007de,Lust:2007kw}. } 
Another, empirical, observation favouring the discrete
torsion orbifolds has been made in \cite{Forste:2008ex}: In cases
where there are no three family models without discrete torsion
({\it viz.} ${\mathbb Z}_2\times {\mathbb Z}_2$ on the AAA lattice
\cite{Cvetic:2001tj,Cvetic:2001nr} or on non-factorisable
$T^6$-orbifolds \cite{Forste:2007zb}) it has been demonstrated that
with discrete torsion and rigid D6-branes there are three family models.

The other orbifolds for $2M \in \{4,6,6'\}$ with discrete torsion 
have to our knowledge not been studied before in view of their potential for 
D6-brane model building. 
Note in particular, that for $2M > 2$, the IIA models on $T^6/\Z_2 \times \Z_{2M}$,
both without and with discrete torsion, in this article are not
T-dual to any of the IIB models on $T^6/\Z_N \times \Z_M$, see 
e.g.~\cite{Klein:2000tf,Klein:2000qw,Rabadan:2000ma,Karp:2000hp,Lust:2006zg}, 
since T-duality maps a symmetric orbifold
to an asymmetric one.

The first case $2M=4$ is closely related to 
the same orbifold background without torsion, which has been studied 
before in~\cite{Honecker:2003vq,Honecker:2003vw,Cvetic:2006by}, and for which a  
no-go theorem for three supersymmetric SM generations exists~\cite{Honecker:2004np}. 
The second case $2M=6$ has $T^6/\Z_6'$ as a subsector, which 
has proven to be able to accommodate the SM gauge group and 
chiral spectrum~\cite{Gmeiner:2007zz,Gmeiner:2008xq}, however, 
always with some adjoint matter.
The last case $2M=6'$ has $T^6/\Z_6$ as a subsector, for which also 
three generation models with some adjoint matter content
are known~\cite{Honecker:2004kb,Gmeiner:2007we}.

In the context of fractional D6-branes on $T^6/\Z_{2N}$ backgrounds, 
a method of deriving the complete matter spectrum from the beta function coefficients,
which are computed along with the gauge thresholds,\footnote{For earlier computations of 
gauge thresholds for intersecting D6-branes, see~\cite{Lust:2003ky,Akerblom:2007np} 
on the torus background and~\cite{Blumenhagen:2007ip} on the $T^6/\Z_2 \times \Z_2$ 
background with discrete torsion and~\cite{Angelantonj:2009yj} 
for the dual description with D9-branes in IIB. For gauge thresholds in local models
see~\cite{Conlon:2009xf,Conlon:2009kt,Conlon:2009qa}. } 
has been derived~\cite{Gmeiner:2009fb},
on which we will heavily rely after appropriate modifications to $T^6/\Z_2 \times \Z_{2M}$.
This method will in particular be used to confirm the absence of adjoint matter, i.e.\
the complete rigidity of the D6-branes, for explicit examples.

\subsection*{Outline}
This paper is organised as follows:
in section~\ref{S:orbifolds}, we discuss the general set-up of D6-branes on orbifolds with discrete torsion.
This includes the various twist sectors, three cycles, supersymmetry and RR tadpole cancellation and 
K-theory conditions and chiral spectrum, as well as exotic O6-planes.

In section~\ref{S:Z2Z2orient}, we review the known $T^6/\Z_2 \times \Z_2'$ example and generalise it to 
arbitrary backgrounds with tilted tori. The K-theory constraints for the case with tilted tori 
are discussed in great detail, since the argumentation carries over to the 
$T^6/\Z_2 \times \Z_6$ and $T^6/\Z_2 \times \Z_6'$ backgrounds.
The $T^6/\Z_2 \times \Z_4$ orbifold without and with discrete torsion is presented in 
section~\ref{S:Z2Z4orient}, and the no-go theorem on three generation models
is generalised from the known case without torsion to the one with discrete torsion.
The $T^6/\Z_2 \times \Z_6$ and $T^6/\Z_2 \times \Z_6'$ cases are discussed in sections~\ref{Ss:Z2Z6orient}
 and~\ref{S:Z2Z6porient}, respectively, where the full lattices of three-cycles are worked out.
We give some globaly consistent, i.e. RR tadpole cancelling and K-theory constraint fulfilling,
 supersymmetric example for each orbifold background at the end of the corresponding section,
for $T^6/\Z_2 \times \Z_2$ and $T^6/\Z_2 \times \Z_4$ only in the presence of discrete torsion,
but for $T^6/\Z_2 \times \Z_6$ and $T^6/\Z_2 \times \Z_6'$ for both choices of torsion.

Sections~\ref{S:Z2Z2orient} to~\ref{S:Z2Z6porient} can basically be read independently of each other.

Section~\ref{S:Conclusions} contains our conclusions and outlook.
Finally, in appendices~\ref{App:A} to~\ref{App:ZN}, we collect technical details on the computation 
of the complete massless open and closed string spectrum, a complete list of assignments of 
exceptional three-cycles to a given bulk three-cycle for $T^6/\Z_2 \times \Z_2$, 
$T^6/\Z_2 \times \Z_6$ and $T^6/\Z_2 \times \Z_6'$, 
and last but not least we present the relations of product orbifolds $T^6/\Z_2 \times \Z_{2M}$ 
without and with discrete torsion to  orbifolds with only one generator, $T^4/\Z_N$ and $T^6/\Z_N$.

{\boldmath
\section{IIA string theory with D6-branes on $T^6/\Z_2 \times \Z_{2M}$ orbi- and orientifolds without and with torsion}\label{S:orbifolds}}

{\boldmath
\subsection{Generalities on $T^6/\Z_2 \times \Z_{2M}$ orbifolds}\label{Ss:general-orb}
}

\subsubsection{Geometric set-up}

For the six extra dimensions of type IIA string theory we choose complex
coordinates $z_1, z_2, z_3$ with $z_k \equiv x_{2k-1} + i \, x_{2k}$.  
These are compactified on a factorisable
six-torus $T^6=(T^2)^3$. This $T^6$ is constructed by dividing each of the
complex planes by a two dimensional lattice to be specified
shortly. From $T^6$ we obtain an orbifold via identification 
of points related by a discrete subgroup of $SU(3)$ which is our
orbifold group ${\mathbb Z}_2 \times {\mathbb Z}_{2M}$. Its generators
$\theta$ and $\omega$, respectively, act on the coordinates as
\begin{equation}
\theta:\quad z_i \, \to \, \mbox{e}^{2 \pi \mbox{\scriptsize i} v_i}\,
z_i\, , \quad \quad \omega:\quad z_i \, \to \, \mbox{e}^{2 \pi
  \mbox{\scriptsize i} w_i}\, z_i,
\end{equation}
with $\sum_{i=1}^3 v_i = \sum_{i=1}^3 w_i =0$. The $v_i$'s and $w_i$'s
are integer multiples of $1/2$ and $1/2M$, respectively. For the list
of orbifolds to be considered, they are explicitly taken as
\begin{equation}
\overrightarrow{v} = \frac{1}{2}\left( 1 , -1 ,0\right) , \quad
\left\{ \begin{array}{l} \overrightarrow{w} = \frac{1}{2M}\left(
      0,1,-1\right)\quad 
  \mbox{with} \quad 2M \in \left\{2,4,6\right\}\\
\overrightarrow{w} = \frac{1}{6}\left( -2,1,1\right) \quad
\mbox{for ${\mathbb Z}_2 \times {\mathbb Z}_6
  ^\prime$}\end{array}\right. .
\end{equation}
Finally, we obtain an orientifold by identifying strings related by
$\Omega {\cal R}$ where $\Omega$ changes the orientation of the string worldsheet
and ${\cal R}$ acts as complex conjugation on the $z_i$. The lattices
have to be taken such that ${\mathbb Z}_2 \times {\mathbb Z}_{2M}$ as
well as ${\cal R}$ are lattice automorphisms.

\begin{figure}[ht]
\begin{center}
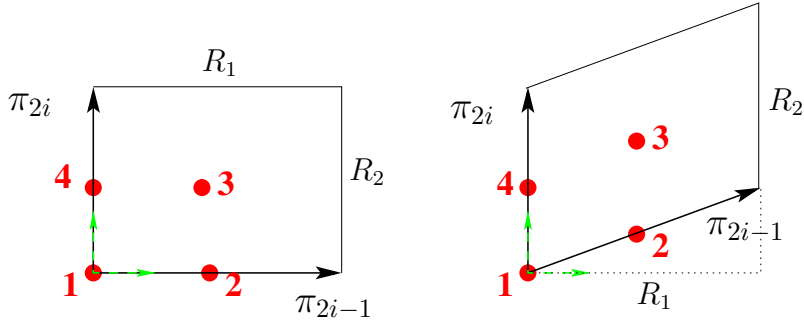
\end{center}
\caption{The $\Z_2$ invariant {\bf a}-type (left) and {\bf b}-type (right) lattices. The ${\cal R}$ invariant $x_{2i-1}$ axis is along 
the 1-cycle $\pi_{2i-1} - b \, \pi_{2i}$ with $b=0$ for the {\bf a}-type lattice and $b=\frac{1}{2}$ for the {\bf b}-type lattice.
${\cal R}$ acts as reflection along the $x_{2i}$ axis, which is spanned by the 1-cycle $\pi_{2i}$.
$\Z_2$ fixed points are depicted in red. For the {\bf a}-type lattice, all $\Z_2$ fixed points are invariant under ${\cal R}$, whereas
for the {\bf b}-type lattice, only 1 and 4 are invariant while $2 \stackrel{\cal R}{\leftrightarrow} 3$.
The $\Z_4$ and $\Z_6$ invariant lattices displayed in figure~\protect\ref{Fig:Z4-Z6lattice} can be mapped onto these two lattices
for specific values of the radii as follows: the {\bf A}-type $\Z_4$ invariant lattice is obtained by setting $R_1=R_2$ on the $\Z_2$ invariant
{\bf a}-type lattice. The {\bf b}-type lattice with $R_2/R_1=2, 2\sqrt{3}, 2/\sqrt{3}$ is a reparameterisation of the $\Z_4$ invariant {\bf B}-type 
lattice and the $\Z_6$ invariant {\bf A}- and {\bf B}-type lattices, respectively. 
}
\label{Fig:Z2-lattice}
\end{figure}

\begin{figure}[ht]
\begin{center}
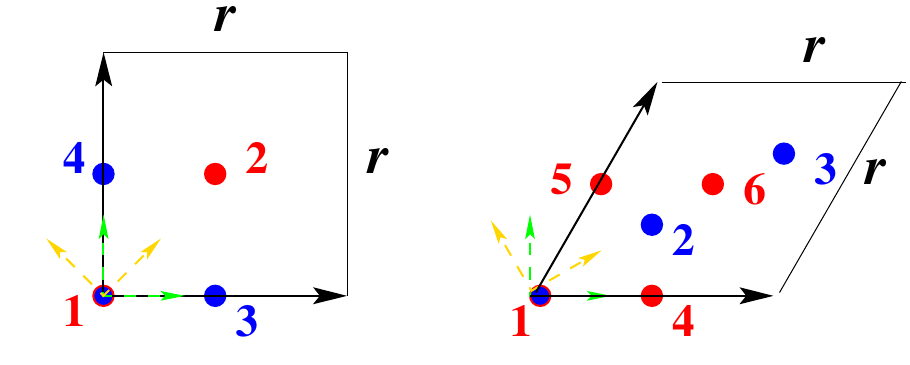
\end{center}
\caption{The $\Z_4$ (left) and $\Z_3$ (right) invariant lattices. The $\Z_4$ invariant lattice is the root lattice of $B_2=SO(5)$ and   
has two fixed points (1,2) under $\Z_4$ and two further points that are fixed under the $\Z_2$ subsymmetry, but interchanged by the $\Z_4$ action.
The $\Z_3$ (and $\Z_6$) invariant lattice is the root lattice of $A_2=SU(3)$ or equivalently $G_2$ (observe that the lattices coincide even though the simple roots differ).
It has fixed points under the $\Z_3$ subsymmetry (1,2,3) and under the $\Z_2$ subsymmetry (1,4,5,6), where $2 \stackrel{\Z_6}{\leftrightarrow} 3$ and  $4 \stackrel{\Z_6}{\rightarrow} 5 \stackrel{\Z_6}{\rightarrow} 6$. The {\bf A}-type lattices have the coordinate
orientation $x_{2i-1}$ along the 1-cycle $\pi_{2i-1}$ (depicted in green); for the {\bf B}-type lattices  $x_{2i-1}$ is along 
 $\pi_{2i-1}+\pi_{2i}$ (depicted in yellow).  
}
\label{Fig:Z4-Z6lattice}
\end{figure}

For any two dimensional lattice, ${\mathbb Z}_2$ is always an
automorphism. Non-trivial restrictions come from imposing ${\mathbb
  Z}_3$ or ${\mathbb Z}_4$, and ${\cal R}$ invariance. Figure
\ref{Fig:Z2-lattice} displays the two choices imposed by ${\cal R}$
invariance,   
%
%
whereas figure \ref{Fig:Z4-Z6lattice} shows ${\mathbb Z}_4$ and
${\mathbb Z}_3$ invariant lattices (the latter one is also ${\mathbb
  Z}_2 \times {\mathbb Z}_3 = {\mathbb Z}_6$ invariant). Imposing
${\mathbb Z}_4$ or ${\mathbb Z}_3$ invariance on the ${\cal R}$
invariant lattices fixes the ratio of the lengths of the lattice basis,
or radii of the corresponding one-cycles as explained in the caption
of figure \ref{Fig:Z2-lattice}. Alternatively, one can impose ${\cal
  R}$ invariance on the orbifold invariant lattice. This fixes the
orientation of the real axis ($x_{2i -1}$) within the plane leaving
the green or yellow coordinate systems of the $z_i$ in figure
\ref{Fig:Z4-Z6lattice}. 
%

Of particular importance are fixed points under the orbifold
group. States of twisted sector closed strings are localised at such
points. More importantly for us, also exceptional cycles to be
discussed below are situated at fixed points. Figures
\ref{Fig:Z2-lattice} and \ref{Fig:Z4-Z6lattice} show the location of a
fixed point (or torus) within one complex direction. More details are
discussed in the captions. 

Orientifold planes extend along the non-compact directions and an
${\cal R}$ invariant three cycle on the orbifold (on $T^6$ this
corresponds to an ${\cal R}$ invariant cycle plus its orbifold
images). To cancel RR charges of O6-planes, D6-branes are added. They
as well extend along the non-compact directions and wrap three-cycles
on the underlying $T^6$.  Bulk cycles correspond to such a three-cycle
plus all its orbifold images. Further, there are exceptional and fractional cycles 
stuck at the orbifold fixed points to be discussed in section \ref{sec:3cyc}.

\subsubsection{Discrete Torsion and Exotic O-Planes}\label{Sss:DT+EO}

Here, we consider string theory compactified on an orbifold. Imposing
modular invariance on torus amplitudes shows the importance of twisted
sector contributions. This can be easily seen as follows. The torus
amplitude contains a trace over orbifold invariant states. This is
ensured by placing a projection operator onto orbifold invariant states
inside the trace. The amplitude splits into several terms containing
an insertion of an element in the orbifold group. Modular
transformation mixes twist sectors and insertions. So, the contribution
of twisted sectors is essential. If there are subsets within all
contributions to the torus amplitude which are mapped onto themselves
by modular transformations, there can be an arbitrary number in front
of the contribution of these subsets. This corresponds to discrete
torsion. A simple argument restricting the possible numbers can be
found in \cite{Vafa:1995fj}. Denote the factor not fixed by imposing
modular invariance by
\begin{equation} \label{eq:epsilon}
\epsilon\left( x,y\right) , 
\end{equation}
where $x$ and $y$ are in the orbifold group. The expression
(\ref{eq:epsilon}) appears in front of a contribution from the
$x$--twisted sector trace with a $y$ insertion. In the direct
computation of the torus amplitude as a trace such ambiguities can be
viewed as an unfixed $y$-eigenvalue of the $x$-twisted vacuum. As
such it should form a representation of the orbifold group, i.e.\ ($z$
is also an orbifold group element)
\begin{equation} \label{eq:epsrep}
\epsilon\left( x,y\right) \epsilon\left( x,z\right) = \epsilon\left(
  x, yz\right) .
\end{equation}
Such a relation leaves a discrete set of possibilities. Let us
illustrate that at the ${\mathbb Z}_2\times {\mathbb Z}_2$ example. 
Any element in the orbifold group leaves the untwisted vacuum
invariant and hence $\epsilon\left( 1, \cdots\right) = 1$. Since
modular transformations exchange twist and insertion also
$\epsilon\left( \cdots, 1\right) =1$. So, for the ${\mathbb Z}_2\times
{\mathbb Z}_2$ orbifold we have
\begin{equation}\label{eq:untweps}
\epsilon\left( 1, \theta\right) = \epsilon\left( 1, \omega\right) =
\epsilon\left( \theta , 1\right) = \epsilon\left( \omega, 1\right) = 1
.
\end{equation}
However, the $\theta$ eigenvalue of the $\omega$ twisted vacuum is not
fixed by modular invariance but just related to the $\omega$
eigenvalue of the $\theta$ twisted vacuum, i.e.
\begin{equation}
\epsilon\left( \theta, \omega\right) = \epsilon\left( \omega,
  \theta\right) = \eta
\end{equation}
with a so far undetermined $\eta$. Equations (\ref{eq:epsrep}) and
({\ref{eq:untweps}) lead finally to
\begin{equation}
\eta = \pm 1
\end{equation}
corresponding to the orbifold without ($\eta = 1$) and with ($\eta =
-1$) discrete torsion. 
The situation is tabulated in table~\ref{Tab:Orbits-Z2Z2}.
\begin{table}[h!]
\renewcommand{\arraystretch}{1.3}
  \begin{center}
\begin{equation*}
    \begin{array}{|c||c|c|c|c|} \hline
        \multicolumn{5}{|c|}{\rule[-3mm]{0mm}{8mm}
\text{\bf Torus orbits of $T^6/\Z_2 \times \Z_2$}
}\\ \hline\hline
\begin{array}{c} \text{twist sector} \\ \text{insertion} \end{array} \theta^k \omega^l & 00 & 01 & 10 & 11 
\\\hline\hline
00 & \bcross & \bcross & \bcross & \bcross
\\\hline
01 & \bcross & \bcross & \bullet & \bullet
\\\hline
10 & \bcross &  \bullet   & \bcross &  \bullet 
\\\hline
11 & \bcross & \bullet  & \bullet    & \bcross 
\\\hline
     \end{array}
    \end{equation*}
\end{center}
\caption{The two orbits of the $T^6/\Z_2 \times \Z_2$ orbifold without ($\eta=1$) and with ($\eta=-1$) discrete torsion. 
The twisted sectors (rows) and 
insertions (columns) are labeled by the powers of the two $\Z_2$ orbifold generators, i.e. $kl$ corresponds to
$\theta^k\omega^l$. $\bcross$ labels the untwisted orbit on which the discrete torsion does not act. $\bullet$ denotes the orbit with
discrete torsion $\eta=\pm 1$.}
\label{Tab:Orbits-Z2Z2}
\end{table}

Now, for the ${\mathbb Z}_2 \times {\mathbb Z}_4$ case there are three
orbits under modular invariance depicted by crosses, dots and circles
in table \ref{Tab:Orbits-Z2Z4}. 
\begin{table}[h!]
\renewcommand{\arraystretch}{1.3}
  \begin{center}
$\begin{array}{|c||c|c|c|c|c|c|c|c|} \hline
        \multicolumn{9}{|c|}{\rule[-3mm]{0mm}{8mm}
\text{\bf The three torus orbits of $T^6/\Z_2 \times \Z_4$}
}\\ \hline\hline
\begin{array}{c} \text{twist sector} \\ \text{insertion} \end{array} & 00 & 01 & 02 & 03 & 10 & 11 & 12 & 13 
\\\hline\hline
00 & \bcross & \bcross & \bcross & \bcross & \bcross & \bcross & \bcross & \bcross
\\\hline
01 & \bcross & \bcross & \bcross & \bcross & \bullet & \bullet & \bullet & \bullet
\\\hline
02 & \bcross & \bcross & \bcross & \bcross & \rcircle & \bcross & \rcircle & \bcross
\\\hline
03 & \bcross & \bcross & \bcross & \bcross & \bullet & \bullet & \bullet & \bullet 
\\\hline
10 & \bcross & \bullet & \rcircle & \bullet & \bcross & \bullet & \rcircle & \bullet  
\\\hline
11 & \bcross & \bullet & \bcross & \bullet & \bullet & \bcross & \bullet & \bcross 
\\\hline
12 & \bcross & \bullet & \rcircle & \bullet & \rcircle & \bullet & \bcross & \bullet   
\\\hline
13 & \bcross & \bullet & \bcross & \bullet & \bullet & \bcross & \bullet & \bcross
\\ \hline
     \end{array}$

  \end{center}
\caption{The three orbits of the $T^6/\Z_2 \times \Z_4$ orbifold without ($\eta=1$) and with ($\eta=-1$) discrete torsion. The twisted sectors (rows) and 
insertions (columns) are labeled by the powers of the $\Z_2$ and $\Z_4$ orbifold generators, i.e. $kl$ corresponds to
$\theta^k\omega^l$. $\bcross$ labels the untwisted orbit on which the torsion does not act. $\bullet$ denotes the orbit with
discrete torsion $\eta=\pm 1$. The orbit $\rcircle$ transforms as $\eta^2=1$.}
\label{Tab:Orbits-Z2Z4}
\end{table}
The orbit with the crosses contains
the untwisted sector and hence the corresponding epsilons are one. The
orbit with the dot contains ${\mathbb Z}_2$ generators acting on a
twisted state. By the same argument as for the ${\mathbb Z}_2
\times {\mathbb Z}_2$ case the corresponding epsilons are either all
plus or all minus one,  corresponding to no discrete torsion or
discrete torsion, respectively. To fix the orbit depicted by  empty circles
we can again use (\ref{eq:epsrep}), e.g.
\begin{equation}\label{eq:notwist}
\epsilon\left(\theta ,\omega^2\right) = \epsilon\left(\theta,
  \omega\right)^2 = \left( \pm 1\right)^2 =1 .
\end{equation}
So, again there are only two discrete choices.

Finally, the situation for ${\mathbb Z}_2 \times {\mathbb Z}_6$ (and
${\mathbb Z}_2 \times {\mathbb Z}_6^\prime$)  is summarised in table
\ref{Tab:Orbits-Z2Z6} . Again there are three orbits of modular
transformations denoted by crosses, dots and circles. In the orbit
denoted by a cross epsilon is again 1, and for the dotted one we
obtain as before $\epsilon = \pm 1$ by, for instance, imposing
\begin{equation}
\epsilon\left( \omega, \theta\right)^2 = \epsilon\left(  \omega, 1\right) =
  1 .
\end{equation}
The epsilon in the orbit denoted by circles is the same as in the one
with dots as can be seen from, e.g.
\begin{equation}
\epsilon\left( \theta, \omega^3\right) = \epsilon\left(\theta,
  \omega\right)^3 .
\end{equation}
%
\begin{table}[h!]
\renewcommand{\arraystretch}{1.3}
  \begin{center}
    \begin{equation*}
      \begin{array}{|c||c|c|c|c|c|c|c|c|c|c|c|c|} \hline
        \multicolumn{13}{|c|}{\rule[-3mm]{0mm}{8mm}
\text{\bf The three torus orbits of $T^6/\Z_2 \times \Z_6$ and $T^6/\Z_2 \times \Z_6'$}
}\\ \hline\hline
\begin{array}{c} \text{twist sector} \\ \text{insertion} \end{array} & 00 & 01 & 02 & 03 & 04 & 05 & 10 & 11 & 12 & 13 & 14 & 15
\\\hline\hline
00 & \bcross & \bcross & \bcross & \bcross & \bcross & \bcross & \bcross & \bcross & \bcross & \bcross & \bcross & \bcross
\\\hline
01 & \bcross & \bcross & \bcross & \bcross & \bcross & \bcross & \bullet & \bullet & \bullet & \bullet & \bullet & \bullet
\\\hline
02 & \bcross & \bcross & \bcross & \bcross & \bcross & \bcross & \bcross & \bcross & \bcross & \bcross & \bcross & \bcross 
\\\hline
03 & \bcross & \bcross & \bcross & \bcross & \bcross & \bcross & \rcircle & \bullet & \bullet & \rcircle & \bullet & \bullet 
\\\hline
04 & \bcross & \bcross & \bcross & \bcross & \bcross & \bcross & \bcross & \bcross & \bcross & \bcross & \bcross & \bcross 
\\\hline
05 & \bcross & \bcross & \bcross & \bcross & \bcross & \bcross & \bullet & \bullet & \bullet & \bullet & \bullet & \bullet 
\\\hline
10 & \bcross & \bullet & \bcross & \rcircle & \bcross & \bullet & \bcross & \bullet & \bcross & \rcircle & \bcross & \bullet    
\\\hline
11 & \bcross & \bullet & \bcross & \bullet & \bcross & \bullet & \bullet & \bcross & \bullet & \bcross & \bullet & \bcross     
\\\hline
12 & \bcross & \bullet & \bcross & \bullet & \bcross & \bullet & \bcross & \bullet & \bcross & \bullet & \bcross & \bullet   
\\\hline
13 & \bcross & \bullet & \bcross & \rcircle & \bcross & \bullet & \rcircle & \bcross & \bullet & \bcross & \bullet & \bcross       
\\ \hline
14 & \bcross & \bullet & \bcross & \bullet & \bcross & \bullet & \bcross & \bullet & \bcross & \bullet & \bcross & \bullet    
\\ \hline
15 & \bcross & \bullet & \bcross & \bullet & \bcross & \bullet & \bullet & \bcross & \bullet & \bcross & \bullet & \bcross       
\\ \hline
     \end{array}
    \end{equation*}
  \end{center}
\caption{The three orbits of $T^6/\Z_2 \times \Z_6$ and  $T^6/\Z_2 \times \Z_6'$ 
without ($\eta=1$) and with ($\eta=-1$) discrete torsion. The twisted sectors (rows) and 
insertions (columns) are labeled by the powers of the $\Z_2$ and $\Z_6$ (or $\Z_6'$) orbifold generators, i.e. $kl$ corresponds to
$\theta^k\omega^l$. $\bcross$ labels the untwisted orbit on which the torsion does not act.  $\bullet$ denotes the orbit with
discrete torsion $\eta=\pm 1$. The orbit $\rcircle$ transforms as $\eta^3=\eta$.}
\label{Tab:Orbits-Z2Z6}
\end{table}

The assignment of discrete torsion on the various twist sectors fixes the Hodge numbers of $T^6/\Z_2 \times \Z_{2M}$ orbifolds
on a given lattice completely. Those for factorisable lattices are listed in table~\ref{Tab:Hodge-T6ZNxZM-torsion}.
\begin{table}[ht]
\renewcommand{\arraystretch}{1.3}
  \begin{center}
    \begin{equation*}
\mbox{\resizebox{\textwidth}{!}{%
      $\begin{array}{|c|c||c||c|c|c|c|c|c|c||c|} \hline
        \multicolumn{11}{|c|}{\rule[-3mm]{0mm}{8mm}
\text{\bf Hodge numbers per twist sector on $T^6/\Z_2 \times \Z_{2M}$ without and with discrete torsion}
}\\ \hline\hline
\begin{array}{c} T^6/ \\
{\rm torsion} \end{array}&\!\!\!\begin{array}{c} {\rm lattice} \\ \text{Hodge numbers} \end{array}\!\!\!& \begin{sideways}\!\!\!\!\!\!\textcolor{blue}{Untwisted}\; \end{sideways}&  \vec{w} 
&  2\vec{w}  & 3\vec{w}
& \vec{v}   & (\vec{v}+\vec{w})
& (\vec{v}+2\vec{w}) &  (\vec{v}+3\vec{w})
& \text{total}
\\\hline\hline
\Z_2 \times \Z_2 & SU(2)^6 & & {\color{blue} (0,\frac{1}{2},-\frac{1}{2})} & \multicolumn{2}{|c|}{} &  {\color{blue} (\frac{1}{2},-\frac{1}{2},0)} &  {\color{blue} (\frac{1}{2},0,-\frac{1}{2})}
 & \multicolumn{2}{|c||}{} & 
\\\hline
\eta =1& h_{11} & 3 & 16 & \multicolumn{2}{|c|}{} & 16 & 16 & \multicolumn{2}{|c||}{} & 51
\\
 & h_{21} & {\color{blue} 3} & {\color{blue} 0} &  \multicolumn{2}{|c|}{} & {\color{blue} 0} & {\color{blue} 0} & \multicolumn{2}{|c||}{} & {\color{blue} 3}
\\\hline
\eta =-1 & h_{11} & 3 & 0 &  \multicolumn{2}{|c|}{} & 0 & 0 & \multicolumn{2}{|c||}{} & 3 
\\
& h_{21} & {\color{blue} 3} & {\color{blue} 16} & \multicolumn{2}{|c|}{} & {\color{blue} 16} & {\color{blue} 16} & \multicolumn{2}{|c||}{} & {\color{blue} 51}
\\\hline\hline
\Z_2 \times \Z_4 &\!\!\!SU(2)^2 \times SO(5)^2\!\!\!& & (0,\frac{1}{4},-\frac{1}{4}) & {\color{blue} (0,\frac{1}{2},-\frac{1}{2})} & & {\color{blue} (\frac{1}{2},-\frac{1}{2},0)} & 
(\frac{1}{2},-\frac{1}{4},-\frac{1}{4}) & {\color{blue} (\frac{1}{2},0,-\frac{1}{2})} &  &
\\\hline
\eta =1 & h_{11} & 3 & 8 & 10 & & 12 & 16 & 12 &  & 61  
 \\
& h_{21} & {\color{blue} 1} & 0 & {\color{blue} 0} & & {\color{blue} 0} & 0 & {\color{blue} 0} &  & {\color{blue} 1}
\\\hline
\eta =-1 & h_{11} & 3 & 0 & 10 &  & 4 & 0 & 4 & & 21
 \\
& h_{21} & {\color{blue} 1} & 8 & {\color{blue} 0} & & {\color{blue} 0} & 0 & {\color{blue} 0} &  & {\color{blue} 1+ } 8
\\\hline\hline
\Z_2 \times \Z_6 &\!\!\!SU(2)^2 \times SU(3)^2\!\!\!& & (0,\frac{1}{6},-\frac{1}{6}) & (0,\frac{1}{3},-\frac{1}{3}) & {\color{blue} (0,\frac{1}{2},-\frac{1}{2})} & {\color{blue} (\frac{1}{2},-\frac{1}{2},0)} & (\frac{1}{2},-\frac{1}{3},-\frac{1}{6}) & (\frac{1}{2},-\frac{1}{6},-\frac{1}{3}) &  {\color{blue} (\frac{1}{2},0,-\frac{1}{2})} &  
\\\hline
\eta =1 & h_{11} & 3 & 2 & 8 & 6 & 8 & 8 & 8 & 8 & 51
\\
& h_{21} & {\color{blue} 1} & 0 & 2 & {\color{blue} 0} & {\color{blue} 0} & 0 & 0 & {\color{blue} 0} & {\color{blue} 1} + 2
\\\hline
\eta =-1 &  h_{11} & 3 & 0 & 8 & 0 & 0 & 4 & 4 & 0 & 19
\\
& h_{21} & {\color{blue} 1} & 2 & 2 & {\color{blue} 6} & {\color{blue} 4} & 0 & 0 & {\color{blue} 4} & {\color{blue} 15} + 4
\\ \hline\hline
\Z_2 \times \Z_6' & SU(3)^3 & &  (-\frac{1}{3},\frac{1}{6},\frac{1}{6}) & (-\frac{2}{3},\frac{1}{3},\frac{1}{3}) & {\color{blue} (0,\frac{1}{2},-\frac{1}{2})} & {\color{blue} (\frac{1}{2},-\frac{1}{2},0)} & (\frac{1}{6},-\frac{1}{3},\frac{1}{6}) & (-\frac{1}{6},-\frac{1}{6},\frac{1}{3})  &  {\color{blue} (\frac{1}{2},0,-\frac{1}{2})} &  
\\\hline
\eta =1 & h_{11} & 3 & 2 & 9 & 6 & 6 & 2 & 2 & 6 & 36
\\
 & h_{21} & {\color{blue} 0} & 0 & 0 & {\color{blue} 0} & {\color{blue} 0} & 0 & 0 & {\color{blue} 0} & {\color{blue} 0}
\\\hline
\eta =-1 & h_{11} & 3 & 1 & 9 & 0 & 0 & 1 & 1 & 0 & 15 
\\
& h_{21} & {\color{blue} 0} & 0 & 0 & {\color{blue} 5} & {\color{blue} 5} & 0 & 0 & {\color{blue} 5} & {\color{blue} 15}
\\ \hline
     \end{array}$}}
    \end{equation*}
  \end{center}
\caption{Hodge numbers per twist-sector for orbifolds without and with discrete torsion. The three-cycles, which can be wrapped by intersecting D6-branes, stem from the untwisted and various $\Z_2$ twisted sectors and are highlighted in blue. As for the $T^6/\Z_N$ orbifold limits, the total number of three cycles is $b_3=2 h_{21}+2$ with the two additional three cycles 
$(h_{30}=h_{03}=1)$ arising in the untwisted sector.}
\label{Tab:Hodge-T6ZNxZM-torsion}
\end{table}

The computation of the ${\cal N}=2$ supersymmetric type IIA closed
string spectrum on these orbifolds is discussed in
appendix~\ref{App:A}. The orientifolded ${\cal N}=1$ supersymmetric
closed string spectrum, in particular the decomposition of the
$h_{11}$ ${\cal N}=2$ vector multiplets into ${\cal N}=1$ chiral
(K\"ahler moduli containing) and vector multiplets,
$(h_{11}^-,h_{11}^+)$, depends on the choice of an exotic O6-plane as
discussed below. The untwisted and twisted orientifolded closed string spectrum on
$T^6/(\Z_2 \times \Z_{2M} \times \OR)$ is tabulated
in table~\ref{Tab:N2-N1-IIAuntwistedOrb}   and
table~\ref{tab:Closed-string-T6ZNxZM-torsion}, respectively, in appendix~\ref{App:A}. 

The three-cycles will be discussed in detail in
section~\ref{S:Z2Z2orient} for $T^6/\Z_2 \times \Z_2$,
section~\ref{S:Z2Z4orient}
for $T^6/\Z_2 \times \Z_4$, and sections~\ref{Ss:Z2Z6orient}
and~\ref{S:Z2Z6porient} for 
$T^6/\Z_2 \times \Z_6$ and $T^6/\Z_2 \times \Z_6'$, respectively.  

The twist sectors of $T^6/\Z_2 \times \Z_{2M}$ are inherited from
various underlying $T^4/\Z_N$ and $T^6/\Z_N$ cases as discussed in
appendix~\ref{App:ZN}. This is on the one hand of interest in view of the
fact that intersecting D6-branes on $T^6/\Z_6$ and $T^6/\Z_6'$ have
been discussed before  
in~\cite{Honecker:2004kb,Gmeiner:2007we}
and~\cite{Gmeiner:2007zz,Gmeiner:2008xq}, respectively, and on the
other hand when one considers the blow-up of orbifold singularities as
discussed e.g. in~\cite{Reffert:2007im}.

So far, we did not discuss orientifolds. There arises another
ambiguity, {\it viz.} eigenvalues of $\Omega {\cal R}$. 
In the following we briefly review what such eigenvalues mean and how
they relate to discrete torsion. Our arguments follow
\cite{Gimon:1996rq, Blumenhagen:2005tn}.
An $\Omega {\cal R}$ eigenvalue appears in the loop channel and can be
$\pm 1$ since 
$\left( \Omega {\cal R}\right)^2 =1$. To understand what 
that means it is useful to relate the loop amplitude to the tree
amplitude drawn in figure \ref{Fig:KBtree}. 
%
\begin{figure}[h!]
\begin{center}
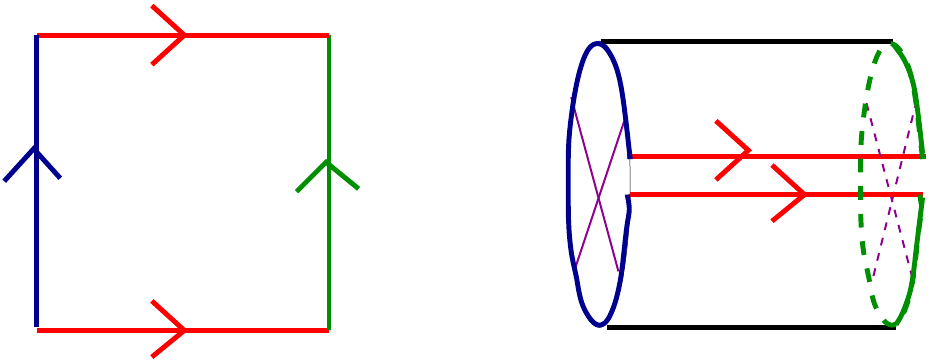
\end{center}
\caption{The figure on the left hand side shows the `triangulated'
  version of a tree channel Klein Bottle amplitude. In difference to a
  torus the left and right edges are not glued together but to
  themselves to form crosscaps as indicated in the figure on the right
  hand side by purple lines.  
}
\label{Fig:KBtree}
\end{figure}
%
Figure \ref{Fig:KBloop} reviews how this relates to the loop amplitude. 
\begin{figure}[h!]
\begin{center}
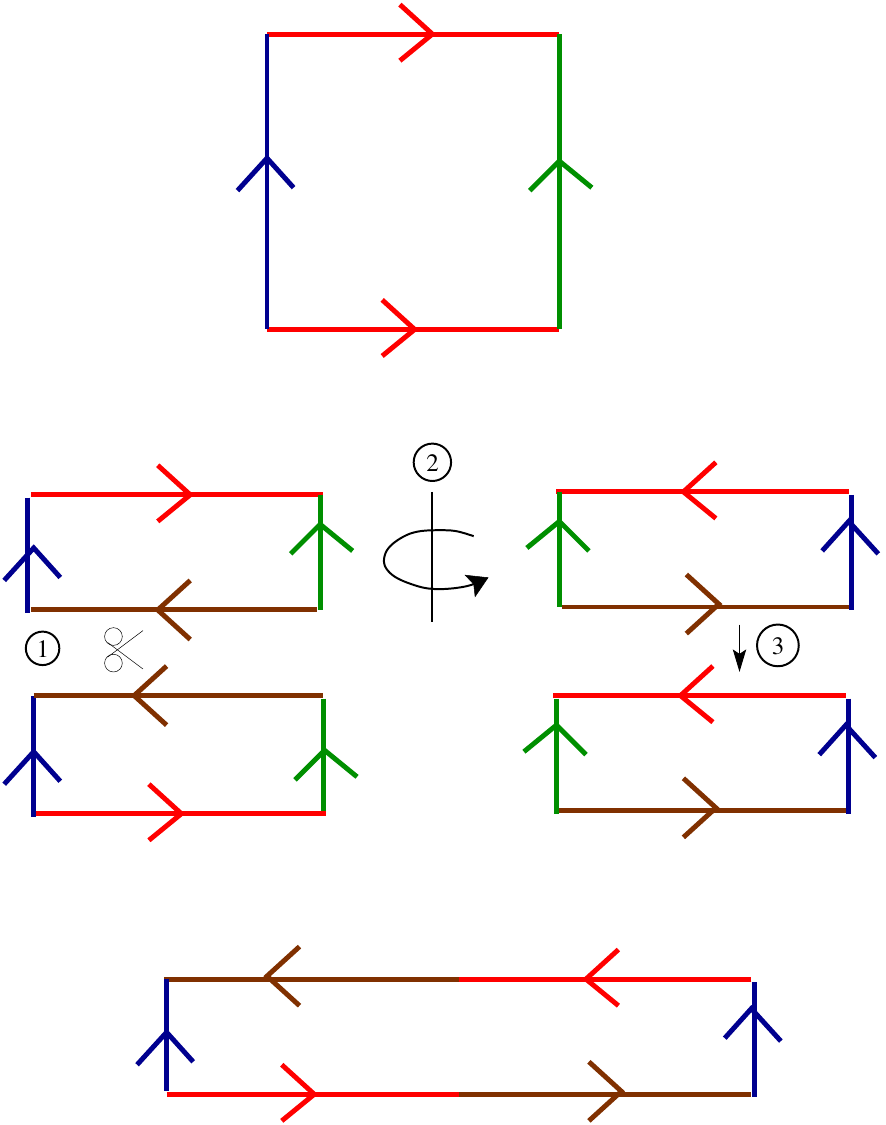
\end{center}
\caption{The figure on the top shows the 
  tree channel Klein Bottle amplitude (see figure \ref{Fig:KBtree}). 
Topology preserving operations mapping this to the loop channel
amplitude are illustrated in the middle: \textcircled{1}
cut along the 
brown line with indicated orientation, \textcircled{2} flip upper
rectangle over its right edge, \textcircled{3} push flipped rectangle
down and glue along the green line. Finally, the loop channel diagram
is displayed at the bottom. The red and brown line merge into a string
closed upon a twist (visible from its tree channel origin). The
crosscap map of the blue O6-plane appears as a trace insertion.} 
\label{Fig:KBloop}
\end{figure}
To be specific, consider first a particular tree channel amplitude
\begin{equation}\label{eq:treeloopa}
\left< \Omega {\cal R}\right| e^{-\ell H_{cl}}\left| \Omega {\cal R}
    \theta^n \omega^m\right> = \mbox{Tr}_{\theta^n\omega^m}\left(
    \Omega {\cal R} e^{-2\pi t H}\right) ,
\end{equation}
where on the rhs the loop channel form is displayed.
First, note that to the tree channel amplitude there is no twisted
sector closed string contribution since \cite {Blumenhagen:1999ev}
\begin{equation}\label{eq:untwi}
\left(\Omega {\cal R}\right)^2 = \left(\Omega {\cal
    R}\theta^n\omega^m\right)^2 =1 .
\end{equation}
The twist sector of the loop channel can be inferred by closing the
path containing a red and brown line in the central drawing in figure
\ref{Fig:KBloop},
$$
\left( \Omega {\cal R}\right)^{-1} \Omega {\cal R}\theta^n\omega^m =
\theta^n\omega^m .
$$
The insertion is given by the crosscap map of the left (blue in figure
\ref{Fig:KBloop}) O6-plane, {\it viz.} $\Omega {\cal R}$. The
eigenvalue of $\Omega {\cal R}$ can be $\pm 1$. Changing it changes
signs on both sides of (\ref{eq:treeloopa}). Hence, changing the
eigenvalue of $\Omega {\cal R}$ in the loop channel corresponds to
changing one of the O6-planes in the tree channel to an exotic O6-plane
(of opposite charge).

To see how this relates to discrete torsion let $\theta^b \omega^a$ be
an orbifold group element acting on the $\theta^n\omega^m$  twisted
sector with an additional sign upon turning on discrete torsion ($\eta
= -1$). Consider
\begin{equation}\label{eq:treeloopb}
\left< \Omega {\cal R}\theta^b\omega^a\right| e^{-\ell H_{cl}}\left|
    \Omega {\cal R} 
    \theta^{b+n} \omega^{a+m}\right> =\eta \; \mbox{Tr}_{\theta^n\omega^m}\left(
    \Omega {\cal R}\theta^b\omega^a e^{-2\pi t H}\right) , 
\end{equation}
where the identification of twist sector and insertion works in the
same way as before. The effect of discrete torsion has been extracted into
$\eta$  on the rhs of (\ref{eq:treeloopb}). So, turning on discrete
torsion changes sign of one of the O6-plane charges on the lhs of
(\ref{eq:treeloopb}). Denoting the extra sign of an O6-plane charge by eta
with its crosscap map as a subscript we can summarise the situation as
\cite{Blumenhagen:2005tn} 
\begin{equation}\label{eq:ocharge}
\eta_{\Omega {\cal R}}\,\eta_{\Omega {\cal R} \theta^n \omega^m}\,
\eta_{\Omega {\cal  R} \theta^b \omega^a}\,\eta_{\Omega {\cal R}
  \theta^{b+n}\omega^{a +m}} = \eta ,
\end{equation}
where $\eta$ is the discrete torsion sign appearing in the torus
amplitude element with $\theta^n \omega^m$ twist and
$\theta^b\omega^a$ insertion. These considerations generalise the
discussion from the ${\mathbb Z}_2 \times {\mathbb Z}_2$ case considered in
\cite{Angelantonj:1999ms,Blumenhagen:2005tn} to any $\Z_2 \times \Z_{2M}$. For all orbifolds
considered in this article, ~(\ref{eq:ocharge}) boils down to the fact that, in the 
presence of discrete torsion, one (or three) of the four O6-plane orbits under the 
orbifold action has (have) to be exotic.

\subsection{Three-Cycles \label{sec:3cyc}}

Three-cycles within the compact space will be wrapped by D6-branes and
O6-planes. Here, we discuss the cycles of interest more explicitly.  
First, we consider three-cycles on the underlying $T^6$.  These can be
written as a direct product of one-cycles $\pi_i$ (see figures
\ref{Fig:Z2-lattice} and \ref{Fig:Z4-Z6lattice})
\begin{equation}\label{eq:toruscyc}
\Pi^{\mbox{\scriptsize torus}}= \bigotimes_{j=1}^3 \left( n^j \pi_{2j -1}
  + m^j \pi_{2j} \right),
\end{equation}
where $n^j$, $m^j$ are co-prime integers denoting wrapping numbers. A bulk
cycle on the orbifold is obtained by adding all its orbifold images to
the torus cycle. The ${\mathbb
  Z}_2 \times {\mathbb Z}_2$ 
subgroup of ${\mathbb Z}_2 \times {\mathbb Z}_{2M}$ maps torus cycles
passing through the origin
to themselves. Therefore, for such cycles, adding the corresponding
images yields an 
overall factor of four times a sum over images under coset ${\mathbb
  Z}_2 \times {\mathbb Z}_{2M}/\left( {\mathbb
  Z}_2 \times {\mathbb Z}_2\right)$ representatives. Explicitly one obtains for
bulk cycles
\begin{equation}\label{eq:inher}
\Pi^{\mbox{\scriptsize bulk}}= 4 \sum_{m=0}^{M-1} \omega^m
\left[ \bigotimes_{j=1}^3 \left( n^j \pi_{2j -1} + m^j \pi_{2j} \right) \right].
\end{equation}
Let us discuss first the case of vanishing torsion. From table
\ref{Tab:Hodge-T6ZNxZM-torsion} (and caption) one can see that in this
case the third Betti number, $b_3$, receives contributions only from
untwisted sectors (except for $T^6/\Z_2 \times \Z_6$, which also has a contribution
from the $\Z_3$ twisted sector). That means that the dimensionality of the lattice
(\ref{eq:inher}) equals $b_3$. However,
on the orbifold the cycles $\Pi^{bulk}$ form only a sublattice of the
integral homology lattice $H_3\left( T^6/{\mathbb Z}_2\times {\mathbb
    Z}_{2M}, {\mathbb Z}\right)$. This can be seen by picking a basis
of these bulk cycles and computing the determinant of the intersection
form to differ from $\pm 1$. A general lattice vector in $H_3\left(
  T^6/{\mathbb Z}_2\times {\mathbb 
    Z}_{2M}, {\mathbb Z}\right)$ without discrete torsion, i.e. $\eta=1$,
is instead
\begin{equation}\label{eq:fracwo}
\Pi^{\mbox{\tiny frac}} = \frac{1}{2}\Pi^{\mbox{\scriptsize bulk}}
\end{equation}
with integer wrapping numbers $n^j$, $m^j$
\cite{Blumenhagen:2005mu}.  D6-branes wrapping such a fractional cycle
are not completely rigid. Only part of the adjoint moduli is
stabilised due to a superpotential \cite{Blumenhagen:2005tn}. 

A stack of $N$ D6-branes wrapping four times a torus cycle supports the
gauge group $U\left( 4 N\right)$. If the torus cycle passes through
the origin, then the ${\mathbb Z}_2 \times {\mathbb Z}_2$ action breaks
this gauge group: one ${\mathbb Z}_2$ to $U\left( 2N\right) \times
U\left( 2N\right)$ which is broken by the second ${\mathbb Z}_2$ to
$U\left( 2N\right)$ \cite{Berkooz:1996dw}. Representatives in the
coset ${\mathbb 
  Z}_2 \times {\mathbb Z}_{2M}/\left( {\mathbb
  Z}_2 \times {\mathbb Z}_2\right)$ fix the gauge group on the
corresponding orbifold image stacks. Hence, $N$ D6-branes wrapping a bulk cycle
$\Pi^{\mbox{\scriptsize bulk}}$ support the gauge group $U\left(
    2N\right)$, $N$ D-branes wrapping a fractional cycle
  (\ref{eq:fracwo}) carry gauge group $U\left( N\right)$.

For the case with discrete torsion there are also twisted contributions
to $b_3$ (see table \ref{Tab:Hodge-T6ZNxZM-torsion}). These
additional contributions come from exceptional three-cycles. Given an
element of the ${\mathbb Z}_2 \times {\mathbb Z}_2$ subgroup, a
corresponding exceptional cycle wraps an $S^1$ inherited from the $T^2$
left invariant under that element and is localised at a fixed point
within the other two complex directions. Such a fixed point becomes an
$S^2$ in the blown up version of the orbifold. Explicitly the relevant
exceptional three-cycles are
\begin{equation}\label{eq:def-Z2ex-cycle}
\Pi^{{\mathbb Z}_2^{(i)}} = 2 \, (-1)^{\tau_{0}^{(i)}} \,
 \sum_{m=0}^{M-1}  \omega^m \left[
  \sum_{(\alpha,\beta) \in T_j \times T_k}  
(-1)^{\tau_j^{\alpha} +\tau_k^{\beta}}
e_{\alpha\beta}^{(i)}
 \bigotimes \left(  n^i \pi_{2i-1} + m^i \pi_{2i} \right) 
\right] ,
\end{equation}
where $i$ labels a $T^2$ on which ${\mathbb Z}_2^{(i)}$ acts trivially
and $\left( i,j,k\right)$ are cyclic permutations of $\left(
  1,2,3\right)$. The factor two arises since we take the cycle to pass
through the origin of the invariant $T^2$. Hence, the other ${\mathbb
  Z}_2 = \left({\mathbb Z}_2 \times {\mathbb Z}_2\right)/{\mathbb
  Z}_2^{(i)}$ maps the exceptional cycle to itself. The phase factors $\tau_j^{\alpha}$
geometrically correspond to the orientation of the collapsed $S^2$ and
physically are associated to discrete Wilson lines, whereas $(-1)^{\tau_0^{(i)}}$ are the $\Z_2^{(i)}$
eigenvalues with $(-1)^{\tau_0^{(i)} + \tau_0^{(j)}} = (-1)^{\tau_0^{(k)}}$. 
Including such cycles
together with the bulk cycles (\ref{eq:inher}), the dimensionality of
the resulting lattice matches $b_3$ at least for $T^6/\Z_2 \times \Z_2$ and $T^6/\Z_2 \times \Z_6'$. 
We will briefly comment on the $\Z_4$ exceptional three-cycles for $T^6/\Z_2 \times \Z_4$ and 
the $\Z_6$ and $\Z_3$ exceptional three-cycles for $T^6/\Z_2 \times \Z_6$ in section~\ref{S:Z2Z4orient} and~\ref{Ss:Z2Z6orient},
respectively, none of which can be used for our D6-brane model building purposes.

Following~\cite{Blumenhagen:2005tn}, we take a fractional cycle as a superposition
of a fractional bulk cycle and fractional exceptional cycles at $\Z_2$ fixed points,
\begin{equation}\label{eq:fracw}
\Pi^{\mbox{\tiny frac}} = \frac{1}{4}\left( \Pi^{\mbox{\tiny bulk}} +
  \sum_{i=1}^3 \Pi^{{\mathbb Z}_2^{(i)}}\right) ,
\end{equation}
where only exceptional cycles through which the bulk cycle passes
contribute. 

The cycles for ${\mathbb Z}_2 \times {\mathbb Z}_2$ invariant
fractional D6-branes can be summarised as
\begin{equation}\label{eq:fracgen}
\Pi^{\rm frac} = \frac{3+\eta}{8} \, 
\Pi^{\rm bulk}
+\frac{1-\eta}{8} \, \sum_{i=1}^{3} \Pi^{\Z_2^{(i)}},
\end{equation}
where, as previously,  $\eta = \pm 1$ labels the case without and with
discrete torsion. For $T^6/\Z_2 \times \Z_2$ and $T^6/\Z_2 \times \Z_6'$,
cycles of the form~(\ref{eq:fracgen}) generate an unimodular lattice; for
more details see section~\ref{S:Z2Z2orient} and~\ref{S:Z2Z6porient}, respectively.
For $T^6/\Z_2 \times \Z_6$, the cycles~(\ref{eq:fracgen}) generate a sub-lattice
of the unimodular lattice of three-cycles, which also contains contributions
from $\Z_3^{(1)}$ twisted sectors as discussed in section~\ref{Ss:Z2Z6orient}.
Finally, $T^6/\Z_2 \times \Z_4$ does not have any three-cycles of the form~(\ref{eq:fracgen})
for $\eta=-1$ since all $\Z_2$ exceptional three-cycles are projected out for any choice of discrete torsion
as discussed in section~\ref{S:Z2Z4orient}, see also
table~\ref{Tab:Hodge-T6ZNxZM-torsion}. Instead, also for $\eta=-1$,
fractional three-cycles 
are of the form~(\ref{eq:fracwo}) for $T^6/\Z_2 \times \Z_4$.

To construct orientifold planes one can start with a set of $\OR$
fixed points on $T^6$ and add orbifold images. Here, however,
identical cycles count only once due to the fact that the  O6-plane is
a non-dynamical object. Schematically we write
\begin{equation}\label{eq:length-O6planes}
\Pi_{O6} = \frac{1}{4}\Pi^{\mbox{\tiny bulk}}_{O6} .
\end{equation}
From identity (\ref{eq:untwi}) one learns that there are no
twisted sector contributions to the Klein-Bottle. That means that
O6-planes do not wrap exceptional cycles. There are no 
contributions from twisted cycles to the M\"obius strip either
since O-planes do not carry twisted RR charges. 
To the annulus diagram bulk cycles as
well as ${\mathbb Z}_2$ twisted cycles can contribute. That other
twisted cycles cannot contribute can be seen as follows. 
The twist sector of
the tree-level cylinder diagram becomes an insertion in the annulus
amplitude. This insertion should leave the boundary conditions of the
open string invariant (otherwise the trace vanishes). But that means
that it should leave the D6-branes invariant which is possible only for
the identity and a ${\mathbb Z}_2$ generator. To summarise, only
fractional bulk cycles and ${\mathbb Z}_2$  twisted cycles enter
tadpole cancellation conditions.  

In the following section, we discuss how to use the fractional three-cycles~(\ref{eq:fracgen})
and O6-planes~(\ref{eq:length-O6planes}) for model building.

\subsection{Model building with D6-branes in IIA orientifolds}\label{Ss:ModelBuilding}

Stacks of $N_a$ identical D$6_a$-branes generically support the gauge groups $U(N_a)$.
We discuss  the following model building rules for generic Calabi-Yau backgrounds (cf.\ the
review~\cite{Blumenhagen:2006ci}) in terms  
of the three-cycles $\Pi_a$ and $\Pi_{O6}$ wrapped by D$6_a$-branes and O6-planes, respectively,
and comment on their explicit form on $T^6/(\Z_2 \times \Z_{2M} \times \OR)$ backgrounds:
\begin{itemize}
\item
The {\bf RR tadpole cancellation conditions} read
\begin{equation}\label{eq:RRtad}
 \sum_a N_a \left( \Pi_a + \Pi_{a'} \right) - 4 \, \Pi_{O6} =0. 
\end{equation}
For $T^6/(\Z_2 \times \Z_{2M} \times \OR)$ backgrounds, the RR tadpole cancellation conditions
can be split into bulk and twisted ones,
\begin{equation}
\begin{aligned}
& \frac{1}{2^{(3-\eta)/2}} \sum_a N_a \left( \Pi_a^{\rm bulk} + \Pi_{a'}^{\rm bulk} \right) - 4 \, \Pi_{O6} =0,
\\
& \sum_a N_a \left( \Pi_a^{\Z_2^{(i)}}  + \Pi_{a'}^{\Z_2^{(i)}} \right)= 0
\quad {\rm for} \quad i \in \{1,2,3\}
\text{ and } 2M \in \{2,6,6'\}.
\end{aligned}
\end{equation}
For $T^6/\Z_2 \times \Z_4$, only bulk cycles exist, and the prefactor $\frac{1}{2^{(3-\eta)/2}}$ has
to be replaced by $\frac{1}{2}$ also for the choice of discrete torsion $\eta=-1$.
\item
The {\bf chiral spectrum} is given by table~\ref{Tab:ChiralSpectrum}.
\begin{table}[h!]
\renewcommand{\arraystretch}{1.3}
  \begin{center}
\begin{equation*}
\begin{array}{|c|c|}\hline
\multicolumn{2}{|c|}{\text{\bf Chiral spectrum}}
\\\hline\hline
\text{representation} & \text{net chirality}
\\\hline\hline
(\N_a,\ov{\N}_b) & \Pi_a \circ \Pi_b 
\\
(\N_a,\N_b) & \Pi_a \circ \Pi_{b'} 
\\
{\bf Anti}_a & \frac{1}{2} \left( \Pi_a \circ \Pi_{a'} + \Pi_a \circ \Pi_{O6} \right) 
\\
{\bf Sym}_a &   \frac{1}{2} \left( \Pi_a \circ \Pi_{a'} - \Pi_a \circ \Pi_{O6} \right)
\\\hline
\end{array}
\end{equation*}
\end{center}
\caption{The chiral spectrum on intersecting D6-branes in terms of intersection numbers of
the wrapped three-cycles.}
\label{Tab:ChiralSpectrum}
\end{table}
For $T^6/\Z_2 \times \Z_{2M}$ orbifolds, the intersection numbers are given in terms of 
the toroidal wrapping numbers, $\Z_2^{(i)}$ eigenvalues, discrete displacements and Wilson lines,
\begin{equation}
\begin{aligned}
\left(\frac{3+\eta}{8} \Pi^{\rm bulk}_a \right) \circ \left(\frac{3+\eta}{8} \Pi^{\rm bulk}_b \right)
=& - 2^{\eta-1}  \sum_{m=0}^{M-1} I_{a(\omega^m b)}
,
\\
\left(\frac{1}{4} \Pi^{\Z_2^{(i)}}_a \right) \circ \left(\frac{1}{4} \Pi^{\Z_2^{(i)}}_b \right)
=&  - 2^{\eta-1}  \sum_{m=0}^{M-1} I_{a(\omega^m b)}^{\Z_2^{(i)}}
,
\end{aligned}
\end{equation}
with  $I_{ab}^{(j)} \equiv n_a^j m_b^j - m_a^j n_b^j$ and $I_{ab} \equiv \prod_{j=1}^3 I_{ab}^{(j)}$
and the $\Z_2^{(i)}$ invariant intersection number weighted with
relative discrete Wilson lines and $\Z_2^{(i)}$  
eigenvalue,
\begin{equation}
I^{\Z_2^{(i)}}_{a(\theta^n\omega^m b)} = (-1)^{\tau_{0,a}^{(i)} +
  \tau_{0,b}^{(i)}}
\hspace*{-0.4in}
\sum_{\footnotesize 
\begin{array}{c}(\alpha_a,\beta_a)\\[-1ex] (\alpha_b,\beta_b)\end{array}
  \in T_j \times T_k}  
\hspace*{-0.4in}
(-1)^{\tau_j^{\alpha_a} +\tau_k^{\beta_a}+\tau_j^{\alpha_b} +\tau_k^{\beta_b}} 
\delta_{\alpha_a(\theta^n \omega^m  \alpha_b)} \delta_{\beta_a(\theta^n \omega^m  \beta_b)} 
I^{(i)}_{a(\theta^n \omega^m b)}
.
\end{equation}
More details on this expression can be found in appendix A.1 of~\cite{Gmeiner:2009fb} in the context of $T^6/\Z_{2N}$ models.
In order to compute the multiplicites of chiral antisymmetric and symmetric representations, one also needs the intersection number
with the orientifold planes,
\begin{equation}
\Pi_a^{\rm frac} \circ \Pi_{O6} =  - 2^{\frac{\eta-3}{2}} \sum_{n=0}^1 \sum_{m=0}^{2M-1}
\eta_{\OR\theta^n\omega^{m}}\,  
,
\end{equation}
where $\eta_{\OR\theta^n\omega^{m}} \equiv \eta_{\OR\theta^n\omega^{m+2}}$ and $\tilde{I}_{a, \OR\theta^n\omega^{m}} \equiv N_{\OR\theta^n\omega^{m}}  \, I_{a, \OR\theta^n\omega^{m}}$.
\\
For {\bf orbifolds}, one can go further and compute the {\bf non-chiral spectrum} 
either by means of Chan-Paton labels of open string states or by using the beta function coefficients
which arise in the CFT computation of the gauge thresholds. Details for $T^6/(\Z_2 \times \Z_{2M} \times \OR)$ without and with discrete 
torsion are given in appendix~\ref{S:App-OpenStrings}, where again the discussion follows the one for $T^6/\Z_{2N}$ in~\cite{Gmeiner:2009fb}. 
\item
The {\bf K-theory constraint} states that for any probe D6-brane with $Sp(2)$ gauge group,
one has to impose~\cite{Uranga:2000xp}
\begin{equation}\label{Eq:general-Ktheory}
\sum_a N_a \Pi_a \circ \Pi_{Sp \text{-probe}} = 0 \text{ mod } 2.
\end{equation}
This requires the complete classification of $\OR$ invariant three-cycles on which such probe 
D6-branes can be wrapped. For each  $T^6/(\Z_2 \times \Z_{2M} \times \OR)$ with $2M\in \{2,4,6,6'\}$ on factorisable lattices,
 we perform the full classification in the corresponding section,
 and we verify that the three-cycles support $Sp(2M)$ gauge factors, not $SO(2M)$, in appendix~\ref{sec:thres}. 
\item
{\bf Massless U(1) factors}  are linear combinations of various Abelian factors,
$U(1)_X = \sum_a c_a U(1)_a$, with $U(1)_a \subset U(N_a)$ and constants $c_a$.
These U(1)s are massless if their effective three-cycle is orientifold invariant, 
\begin{equation}
\Pi_{X'}\stackrel{!}{=} \Pi_X 
\qquad
\text{ with }
\qquad
\Pi_X = \sum_a N_a c_a \Pi_a.
\end{equation}
\item
The {\bf supersymmetry conditions} state that the 
D6-branes have to wrap special Lagrangian three-cycles with the same calibration as those wrapped by the O6-planes.
In terms of the complex number ${\mathcal Z}_a \equiv \int_{\Pi_a} \Omega_3$ with the holomorphic volume form $\Omega_3$, 
the supersymmetry conditions can be written as  
\begin{equation}\label{Eq:SUSYgeneral}
{\rm Im} ({\mathcal Z}_a) = 0,
\qquad 
{\rm Re} ({\mathcal Z}_a) > 0.
\end{equation}
For $T^6/\Z_N$ and $T^6/\Z_2 \times \Z_{2M}$, the bulk supersymmetry
condition can be obtained from
\begin{equation}\label{Eq:Z-Orb}
\mathcal{Z}_a =\prod_{k=1}^3 \left(  e^{- i \pi \tilde{\phi}_k}   x^k_a \right)  
\end{equation}
with $\tilde{\phi}_k$ encoding the orientation of a given two
dimensional lattice and $x_k^a$ the 1-cycle wrapped by the D$6_a$-brane on the same lattice,  
\begin{align}\label{Eq:Z-contribOrb}
\tilde{\phi}_k&= \left\{\begin{array}{cc} 0 & \Z_2: {\bf a/b}; \,
    \Z_4,\Z_3,\Z_6: {\bf A} 
\\
1/4 & \Z_4: {\bf B}
\\
1/6 & \Z_3,\Z_6: {\bf B}
\end{array}\right.
,\nonumber \\
x^k_a &=\left\{\begin{array}{cc}
n^k_a + i \frac{R_2^{(k)}}{R_1^{(k)}} \left(m^k_a + b_k n^k_a \right) & \Z_2
\\
n^k_a + i \, m^k_a & \Z_4
\\
n^k_a + e^{\pi i/3} m^k_a & \Z_3,\Z_6
\end{array}\right.
.
\end{align}
The supersymmetry conditions can be verified using the angles of a toroidal three-cycle with respect to
the $\OR$ plane expressed in terms of the wrapping numbers $(n_a,m_a)$ per two-torus,
\begin{equation}\label{Eq:angles}
\tan (\pi \phi_a) = \left\{\begin{array}{cc}
\frac{m_a + b n_a}{n_a} \frac{R_2}{R_1} & \Z_2: {\bf a},{\bf b}
\\
\frac{m_a}{n_a} & \Z_4: {\bf A}
\\
\frac{m_a - n_a}{m_a+n_a} & \Z_4: {\bf B}
\\
\sqrt{3}\frac{m_a}{2n_a + m_a} & \Z_3: {\bf A}
\\
\frac{1}{\sqrt{3}} \frac{m_a - n_a}{m_a+n_a} & \Z_3: {\bf B}
\end{array}\right.
.
\end{equation}
In terms of the angles in~(\ref{Eq:angles}), the first condition in~(\ref{Eq:SUSYgeneral}) 
reads \linebreak $\sum_{i=1}^3 \tan (\pi \phi_a^{(i)}) - \prod_{i=1}^3 \tan (\pi \phi_a^{(i)})=0$.
\\
A fractional three-cycle is supersymmetric if the bulk part is supersymmetric
and only those exceptional cycles are wrapped, whose $\Z_2^{(i)}$ fixed points are
trasversed by the bulk cycle.
For $T^6/\Z_{2N}$, three of four signs from the fixed point
contributions are independent 
and correspond to the choice of two discrete Wilson lines and the $\Z_2$ eigenvalue.
Similarly, on $T^6/\Z_2 \times \Z_{2M}$ with discrete torsion (and $2M \neq 4$), five of the eight signs 
have the physical interpretation as the choice of three discrete Wilson lines and two independent $\Z_2^{(i)}$
eigenvalues. 
\end{itemize}

{\boldmath
\section{The $T^6/(\Z_2 \times \Z_2 \times \OR)$ orientifolds on
  tilted tori}\label{S:Z2Z2orient} 
}

The $T^6/\Z_2 \times \Z_2$ orbifold is generated by the two shift vectors
\begin{equation}
\vec{v} = \frac{1}{2}(1,-1,0),
\qquad
\vec{w} = \frac{1}{2}(0,1,-1),
\end{equation}
which correspond to $\Z_2^{(3)}$ and $\Z_2^{(1)}$, respectively. The third
orbifold twist $\Z_2^{(2)}$ is generated by the sum, $\vec{v}+\vec{w}=\frac{1}{2}(1,0,-1)$.

On the factorisable lattice $SU(2)^6$, the relevant Hodge numbers for the 
three-cycles are 
\begin{equation}
\eta=1: \; h_{21} = 3_{\rm bulk}, 
\qquad
\eta=-1: \; h_{21} = 3_{\rm bulk} + 48_{\Z_2},
\end{equation}
where in the case with discrete torsion, the exceptional 3-cycles are evenly distributed over the three $\Z_2^{(i)}$ twisted sectors.
The complete list of Hodge numbers per untwisted and twisted sector is given in table~\ref{Tab:Hodge-T6ZNxZM-torsion} for 
both choices of discrete torsion.

\subsection{The bulk three-cycles}\label{Ss:Z2Z2bulk}

A fractional cycle for the case without discrete torsion is (cf.\ (\ref{eq:toruscyc})) 
\begin{equation}\label{eq:z2z2cyc}
\begin{aligned}
\frac{\Pi_a^{\rm bulk}}{2} &= 2\bigotimes_{j=1}^3 \left( n^j
  \pi_{2j-1} + m^j \pi_{2j}\right)\\ &= 2 \, \left(n^1_a n^2_a n^3_a
  \Pi_{135} + \sum_{i=1}^3 n^i_a m^j_a m^k_a \Pi_{2i-1;2j;2k} + m^1_a
  m^2_a m^3_a \Pi_{246}\right. \\&\hspace*{.5in}\left. +
  \sum_{i=1}^3 m^i_a n^j_a n^k_a 
  \Pi_{2i;2j-1;2k-1} \right) .
\end{aligned}
\end{equation}
Again, $(i,j,k)$ is a cyclic permutation of $(1,2,3)$. The appearing
three-cycles are direct products of three one-cycles as encoded in the
triple index. The index $a$ labels a stack of fractional D6-branes
wrapping the three-cycle. For the case with discrete torsion (half) the
above cycle contributes as in (\ref{eq:fracgen}).   

A D6-brane wrapping the cycle (\ref{eq:z2z2cyc}) has to be accompanied
by its orientifold image wrapping the cycle
\begin{equation}
\begin{aligned}
\frac{1}{2}\Pi_{a'}^{\rm bulk} &=  2 \left( \left(\prod_{i=1}^3 n^i_a\right) \Pi_{135} 
+ \sum_{i=1}^3 n^i_a\left( m^j_a+2b_jn^j_a\right)\left( m^k_a +2b_kn^k_a\right) \Pi_{2i-1;2j;2k} 
\right.
\\
& \qquad \left. -\left( \prod_{i=1}^3 \left( m^i_a + 2 b_i n^i_a \right) \right)\Pi_{246} 
- \sum_{i=1}^3 \left( m^i_a  + 2 b_i n^i_a\right) n^j_a n^k_a \Pi_{2i;2j-1;2k-1} \right)
 .
\end{aligned}
\end{equation}
Here, it matters whether an underlying $T^2$ is of {\bf a} or {\bf
  b} orientation (see figure \ref{Fig:Z2-lattice}). If the
$i^{\mbox{\tiny th}}$ two-torus is of the {\bf 
  a} type then $b_i =0$, otherwise $b_i =1/2$. 

The supersymmetry conditions~(\ref{Eq:SUSYgeneral}) can be written as follows
\begin{equation}
\begin{aligned}
\sum_{k=1}^3 \frac{1}{\varrho_k} n^i_a n^j_a (m^k_a +b_kn^k_a ) - \prod_{i=1}^3 (m^i_a+b_i n^i_a) &= 0,
\\
n^1_a n^2_a n^3_a - \sum_{i=1}^3 \varrho_i \, n^i_a (m^j_a +b_j n^j_a)(m^k_a +b_kn^k_a) &>0 ,
\end{aligned}
\end{equation}
with $\varrho_k \equiv \frac{R^{(i)}_2}{R^{(i)}_1} \frac{R^{(j)}_2}{R^{(j)}_1}$ and $(i,j,k)$ 
cyclic permutations of $(1,2,3)$ as before.

For the orientifold fixed cycles there is again a difference between
{\bf a} and {\bf b} lattices. Consider figure
\ref{Fig:Z2-lattice}. For the left torus horizontal lines passing
through the origin as well as through points shifted vertically by
half a lattice vector are invariant under complex conjugation, whereas
for the right torus only the horizontal line passing through the
origin consists of fixed points under complex conjugation. Therefore,
it is useful to introduce the numerical factor
\begin{equation}
N_{O6} = 2^3 \prod_{i=1}^3 \left( 1 - b_i\right) .
\end{equation}
The cycle fixed under complex conjugations, i.e.\ ${\cal R}$, is
\begin{equation}\label{eq:def-ORplane-Z2Z2}
 \Pi_{\OR} = \left(\prod_{i=1}^3 \frac{1}{1-b_i}
 \right)\left(\Pi_{135} - \sum_{i=1}^3 b_i \Pi_{2i;2j-1;2k-1} 
+ \sum_{k=1}^3 b_i b_j \Pi_{2i;2j;2k-1} -b_1b_2b_3 \, \Pi_{246}\right) .
\end{equation}
Next, we specify the set fixed under ${\cal R}{\mathbb Z}_2^{(k)}$,
where ${\mathbb Z}_2 ^{(k)}$ is an element in the ${\mathbb Z}_2
\times {\mathbb Z}_2$ orbifold group leaving the
$k^{\mbox{\tiny th}}$ two-torus invariant and $(i,j,k)$  are cyclic permutations
of $(1,2,3)$,
\begin{equation}\label{eq:def-ORZ2planes-Z2Z2}
\Pi_{\OR\Z_2^{(k)}} =\frac{1}{1-b_k} \left(-\Pi_{2i;2j;2k-1} + b_k
  \Pi_{246}   \right) . 
\end{equation}
With these ingredients, we can write down the overall cycle wrapped by the
O6-plane in the ${\mathbb Z}_2 \times {\mathbb Z}_2$ orientifold,
\begin{equation}
\begin{aligned}
\Pi_{O6} &\equiv N_{O6} \left(\eta_{\OR} \Pi_{\OR} + \sum_{i=1}^3
  \eta_{\OR{\mathbb Z}_2^{(i)}} \Pi_{\OR\Z_2^{(i)}} \right) 
\\
&= 8 \, \left( \eta_{\OR} \,  \Pi_{135} -  \eta_{\OR} \,  \sum_{i=1}^3
  b_i \,   \Pi_{2i;2j-1;2k-1}\right. \\ 
  &\qquad \left. + \sum_{k=1}^3
  \left\{ b_i b_j 
    \eta_{\OR} -(1-b_i)(1- b_j) \eta_{\OR\Z_2^{(k)}}  \right\}
  \Pi_{2i;2j;2k-1} \right. 
\\
&\qquad  \left. + \left\{ -b_1b_2b_3 \, \eta_{\OR} + \sum_{k=1}^3
    (1-b_i)(1- b_j)b_k \,   \eta_{\OR\Z_2^{(k)}} \right\}  \,
  \Pi_{246} 
\right) .
\end{aligned}
\end{equation}
If there is no discrete torsion, a consistent option is to take none of
the O6-planes to be exotic. 
With discrete torsion, one needs
an odd number of exotic O6-planes, 
due to the relation~(\ref{eq:ocharge}) which can be rewritten in this case as 
\begin{equation}
\eta_{\OR} \prod_{k=1}^3 \eta_{\OR\Z_2^{(k)}} = \eta.
\end{equation}

The O6-planes are affected by discrete
torsion only by these signs. Because of (\ref{eq:untwi}) they
    do not wrap exceptional cycles \cite{Blumenhagen:2002wn}.

The untwisted RR tadpole cancellation conditions (\ref{eq:RRtad}) on 
$T^6/\Z_2 \times \Z_2$ are explicitly
\begin{equation}\label{Eq:Z2Z2-bulkRRtcc}
\begin{aligned}
0= & \left[\sum_a N_a  n^1_a n^2_a n^3_a - 2^{\frac{7-\eta}{2}} \,
  \eta_{\OR} \right] \\
  &\qquad \times 
\left( \Pi_{1,3,5}+ \sum_{\{i,j,k\} \text{ cyclic}} (b_jb_k
  \Pi_{2i-1,2j,2k} - b_i \Pi_{2i,2j-1,2k-1})  
-b_1b_2b_3 \Pi_{2,4,6}
\right)
\\
& + \sum_{\{i,j,k\} \text{ cyclic}} \left[\sum_a N_a  n^i_a
  \left(m^j_a  + b_j n^j_a \right) \left( m^k_a  + b_k n^k_a \right)  
+ 2^{\frac{7-\eta}{2}} \, (1-b_j)(1-b_k) \, \eta_{\OR\Z_2^{(i)}}
\right] \\
& \qquad \times 
\left(\Pi_{2i-1,2j,2k} - b_i \Pi_{2,4,6}
\right) .
\end{aligned}
\end{equation}

\subsection{The exceptional three-cycles}\label{Ss:Z2Z2ex}

The discrete torsion factor $\eta$ acts by adding a phase to the $\Z_2^{(i)}$ projection on the $\Z_2^{(k)}$ twisted
sectors according to table~\ref{Tab:Orbits-Z2Z2}. This can be summarised as
\begin{equation}\label{eq:Z2Z2-ex-2cycles}
e_{\alpha\beta}^{(k)} \stackrel{\Z_2^{(i)}}{\longrightarrow} \eta \, e_{\alpha\beta}^{(k)} ,
\qquad 
\left\{\begin{array}{c}
\pi_{2k-1}\stackrel{\Z_2^{(i)}}{\longrightarrow} -\pi_{2k-1}
\\
\pi_{2k} \stackrel{\Z_2^{(i)}}{\longrightarrow} -\pi_{2k}
\end{array}\right.
\text{ for }
i \neq k
,
\end{equation}
where $\alpha\beta$ labels a $\Z_2^{(k)}$ fixed point on $T^2_{(i)} \times T^2_{(j)}$.
The orbifold invariant three-cycles are obtained by tensoring the exceptional two-cycles with
toroidal one-cycles,\footnote{In~\cite{Blumenhagen:2005tn}, these cycles were called $\alpha_{,n}$ and $\alpha_{,m}$.
However, throughout this article, we stick to the notation that $\Z_2$ exceptional three-cycles are labeled by 
$\varepsilon$ and $\tilde{\varepsilon}$ with intersection numbers $\varepsilon \circ \varepsilon =\tilde{\varepsilon} \circ \tilde{\varepsilon}=0$.} 
\begin{equation}\label{eq:Z2Z2-ex-3cycles}
\begin{aligned}
\varepsilon_{\alpha\beta}^{(k)} \equiv 2 \, e_{\alpha\beta}^{(k)} \otimes \pi_{2k-1} 
&\stackrel{\Z_2^{(i)}}{\longrightarrow} - \eta \, \varepsilon_{\alpha\beta}^{(k)}
,
\\
\tilde{\varepsilon}_{\alpha\beta}^{(k)} \equiv  2 \, e_{\alpha\beta}^{(k)} \otimes \pi_{2k} &\stackrel{\Z_2^{(i)}}{\longrightarrow}  
- \eta \, 
\tilde{\varepsilon}_{\alpha\beta}^{(k)} 
.
\end{aligned}
\end{equation}
In the absence of discrete torsion, $\eta=1$, the two-cycles~(\ref{eq:Z2Z2-ex-2cycles}) survive
the orbifold projections from the other sectors, whereas with discrete torsion, $\eta=-1$, the 
three-cycles~(\ref{eq:Z2Z2-ex-3cycles}) survive. In the following, we will concentrate on this 
latter case.

The intersection number among exceptional three-cycles reads
\begin{equation}
\varepsilon_{\alpha\beta}^{(k)} \circ \tilde{\varepsilon}_{\alpha'\beta'}^{(k')}
= -4 \, 
\delta^{kk'} \delta_{\alpha\alpha'}\delta_{\beta\beta'} 
,
\end{equation}
and all others vanish.

The orientifold images of the exceptional three-cycles are derived by combining the action of ${\cal R}$ on
the $\Z_2^{(i)}$ fixed points and toroidal one-cycles with the signs from $\Omega$ and the 
exotic O6-plane,
\begin{equation}\label{eq:def-eta-i}
\eta_{(k)} \equiv \eta_{\OR}  \cdot \eta_{\OR\Z_2^{(k)}}
,
\end{equation}
on the exceptional cycles as discussed around equation~(\ref{eq:treeloopb}).
The results are given in table~\ref{Tab:Ex-OR-Z2Z2}.
\begin{table}[h!]
\renewcommand{\arraystretch}{1.3}
  \begin{center}
\begin{equation*}
\begin{array}{|c||c|} \hline
        \multicolumn{2}{|c|}{\rule[-3mm]{0mm}{8mm}
\OR \; \text{\bf projection on exceptional 3-cycles on } T^6/\Z_2 \times \Z_2
}\\ \hline\hline
\varepsilon^{(k)}_{\alpha\beta} &  \eta_{(k)}\,  \left(- \varepsilon^{(k)}_{{\cal R}(\alpha\beta)} + 2 \,  b_k \,   \tilde{\varepsilon}^{(k)}_{{\cal R}(\alpha\beta)}    \right)
\\\hline
\tilde{\varepsilon}^{(k)}_{\alpha\beta} &  \eta_{(k)} \, \tilde{\varepsilon}^{(k)}_{{\cal R}(\alpha\beta)} 
\\\hline
     \end{array}
    \end{equation*}
\end{center}
\caption{The orientifold projection on exceptional three-cycles on  $T^6/(\Z_2 \times \Z_2 \times \OR)$ with 
discrete torsion preserves the twist 
sector. The $\Z_2^{(k)}$ fixed points $(\alpha\beta)$ on $T_i \times T_j$ transform under ${\cal R}$ as described in the
caption of figure~\protect\ref{Fig:Z2-lattice}. The signs $\eta_{(k)}$ are as defined in~(\protect\ref{eq:def-eta-i}).}
\label{Tab:Ex-OR-Z2Z2}
\end{table}

It is often convenient to work with the sign factors $\eta_{(k)}$ of the orientifold projection on 
$\Z_2^{(k)}$ twisted sectors, which
are in one-to-one correspondence with the signs of the O6-planes, see table~\ref{Tab:correspondence-etas}.
\begin{table}[h!]
\renewcommand{\arraystretch}{1.3}
  \begin{center}
\begin{equation*}
\begin{array}{|c|c|}\hline
\multicolumn{2}{|c|}{\text{\bf Relation between the sets of signs}}
\\\hline\hline
(\eta_{\OR},\eta_{\OR\Z_2^{(1)}},\eta_{\OR\Z_2^{(2)}},\eta_{\OR\Z_2^{(3)}}) & (\eta_{(1)}, \eta_{(2)},\eta_{(3)})
\\\hline\hline
(-1,1,1,1) & (-1.-1.-1)
\\
(1,-1,1,1) & (-1,1,1)
\\
(1,1,-1,1) & (1,-1,1)
\\
(1,1,1,-1) & (1,1,-1)
\\\hline
\end{array}
\end{equation*}
\end{center}
\caption{Relation between the signs of exotic O6-planes and the orientifold projection on $\Z_2$ twisted sectors for $T^6/\Z_2 \times \Z_2$,
$T^6/\Z_2 \times \Z_6$ and $T^6/\Z_2 \times \Z_6'$ with discrete torsion.
Since supersymmetric model building only admits one exotic O6-plane, there is a 1-1 correspondence between the sets 
$(\eta_{\OR},\eta_{\OR\Z_2^{(1)}},\eta_{\OR\Z_2^{(2)}},\eta_{\OR\Z_2^{(3)}})$ and $ (\eta_{(1)}, \eta_{(2)},\eta_{(3)})$.}
\label{Tab:correspondence-etas}
\end{table}

On $T^6/\Z_2 \times \Z_2$ with discrete torsion, the exceptional cycles~(\ref{eq:def-Z2ex-cycle})
can be written as
\begin{equation}
\Pi^{{\mathbb Z}_2^{(k)}} = \sum_{(\alpha,\beta) \in T_i \times T_j} \left(
x_{\alpha\beta}^{(k)} \, \varepsilon^{(k)}_{\alpha\beta}
+ y_{\alpha\beta}^{(k)} \, \tilde{\varepsilon}^{(k)}_{\alpha\beta}
\right),
\end{equation}
where 
\begin{equation}
\begin{array}{ll}
x_{\alpha_1\beta_1}^{(k)}= (-1)^{\tau_0^{(k)}} \, n^k , 
& x_{\alpha_2\beta_1}^{(k)}= (-1)^{\tau_0^{(k)} + \tau_i} \, n^k ,\\ 
x_{\alpha_1\beta_2}^{(k)}= (-1)^{\tau_0^{(k)} + \tau_j} \, n^k ,
& x_{\alpha_2\beta_2}^{(k)}= (-1)^{\tau_0^{(k)} + \tau_i + \tau_j} \, n^k , 
\\
y_{\alpha_1\beta_1}^{(k)}=(-1)^{\tau_0^{(k)}} \,m^k , 
& y_{\alpha_2\beta_1}^{(k)}=(-1)^{\tau_0^{(k)} + \tau_i} \,m^k , \\
y_{\alpha_1\beta_2}^{(k)}=(-1)^{\tau_0^{(k)}+ \tau_j} \,m^k , 
& y_{\alpha_2\beta_2}^{(k)}=(-1)^{\tau_0^{(k)} + \tau_i + \tau_j} \,m^k , 
\end{array}
\end{equation}
and $\tau_0^{(k)}\in \{0,1\}$ parameterises the $\Z_2^{(k)}$ eigenvalue, whereas 
$\tau_i,\tau_j \in \{0,1\}$ correspond to the choice of discrete Wilson lines on $T^2_{(i)} \times T^2_{(j)}$.

The twisted RR tadpole cancellation conditions take the form
\begin{equation*}
\begin{aligned}
0= & \sum_{k=1}^3 \sum_{(\alpha,\beta) \in T_i \times T_j} \left\{ x_{\alpha\beta}^{(k)}
\left( \varepsilon^{(k)}_{\alpha\beta} - \eta_{(k)}\,  \varepsilon^{(k)}_{{\cal R}(\alpha\beta)} + 2\, b_k \,\eta_{(k)} \,\tilde{\varepsilon}^{(k)}_{{\cal R}(\alpha\beta)}    \right)  
+ y_{\alpha\beta}^{(k)} \left( \tilde{\varepsilon}^{(k)}_{\alpha\beta} +  \eta_{(k)} \, \tilde{\varepsilon}^{(k)}_{{\cal R}(\alpha\beta)}  \right)  
\right\},
\end{aligned}
\end{equation*}
where ${\cal R}(\alpha\beta)$ is the orientifold image of the fixed point $\alpha\beta$.

\subsection{The K-theory constraint}\label{Ss:Z2Z2Ktheory}

For discrete torsion and the {\bf aaa} torus, the K-theory constraint has been derived before in~\cite{Blumenhagen:2005tn}. We will generalise their result to 
an arbitrary choice of tilted tori.

As a first step, we classify all cycles which are topologically $\OR$ invariant and whose bulk part is parallel to some $\OR\Z_2^{(i)}$ invariant plane, where we shorten the notation by setting $\OR \equiv \OR\Z_2^{(0)}$. The bulk part of these cycles is the same as one of the O6-planes in~(\ref{eq:def-ORplane-Z2Z2}) and~(\ref{eq:def-ORZ2planes-Z2Z2}). 
In table~\ref{Tab:bulk-Ktheory-onZ2Z2}, we compute the bulk contributions to the K-theory constraints 
explicitly, where the last column shows the simplified result after the untwisted RR tadpole cancellation 
conditions~(\ref{Eq:Z2Z2-bulkRRtcc}) have been inserted.
One can read off that e.g. for the case without discrete torsion and some tilted torus $b_k=1/2$, 
the two K-theory constraints associated to the probe branes $Sp(2)_0$
and   $Sp(2)_k$ become trivial. Note, that in the case without
discrete torsion ($\eta =1$) at least one of the two-tori should be
tilted for interesting models with three generations
\cite{Cvetic:2001tj,Cvetic:2001nr}.  
\begin{table}[h]
  \renewcommand{\arraystretch}{1.3}
  \begin{center}
  \begin{equation*}
    \begin{array}{|c|c|c|c|}   \hline
\multicolumn{4}{|c|}{\text{\bf Bulk parts of K-theory constraints on } T^6/\Z_2 \times \Z_2}
\\\hline\hline
 & & & \text{after RR tcc - mod 2}
\\\hline\hline
\OR & \frac{\sum_a N_a \Pi_a^{\rm bulk} \circ \Pi^{\rm bulk}_{Sp(2)_0}}{2^{3-\eta}} 
& \frac{\sum_a N_a \prod_{i=1}^3 (m^i_a + b_i n^i_a)}{2^{1-\eta} \, \prod_{i=1}^3 (1-b_i)} 
& \frac{\sum_a N_a \prod_{i=1}^3 m^i_a - \sum_a N_a \, b_jb_k \, m^i_an^j_an^k_a
}{2^{1-\eta} \, \prod_{i=1}^3 (1-b_i)} 
\\\hline
\begin{array}{c} \OR\Z_2^{(k)}\\ k=1,2,3 \end{array}
& \frac{\sum_a N_a \Pi_a^{\rm bulk} \circ \Pi^{\rm bulk}_{Sp(2)_k}}{2^{3-\eta}} 
&  \frac{- \sum_a N_a n^i_an^j_a (m^k_a + b_k n^k_a)}{2^{1-\eta} \, (1-b_k)} 
& \frac{- \sum_a N_a n^i_an^j_a m^k_a}{2^{1-\eta} \, (1-b_k)} 
\\\hline
   \end{array}
   \end{equation*}
    \caption{Bulk contribution to the K-theory constraint on $T^6/\Z_2
      \times \Z_2$ without 
$(\eta=1)$ and with $(\eta=-1)$ discrete torsion. In the last column,
the result has been simplified by  
applying the untwisted RR tadpole cancellation
conditions~(\ref{Eq:Z2Z2-bulkRRtcc}). } 
\label{Tab:bulk-Ktheory-onZ2Z2}
  \end{center}
\end{table}

The invariant exceptional three-cycles have the toroidal one-cycle either parallel to the $\OR$ invariant plane ({\it `horizontal'}),
\begin{equation}
\begin{aligned}
\Pi^{\Z_2^{(k)}}_{h} &=\Pi^{\Z_2^{(k)}}_{h'}= (-1)^{\tau_0^{(k)}} \Bigl(
\varepsilon_{\alpha_1\beta_1}^{(k) \; \prime} 
+(-1)^{\tau_i}\varepsilon_{\alpha_2\beta_1}^{(k) \; \prime} 
+ (-1)^{\tau_j}\varepsilon_{\alpha_1\beta_2}^{(k) \; \prime} 
+(-1)^{\tau_i + \tau_j}\varepsilon_{\alpha_2\beta_2}^{(k) \; \prime}
\Bigr)
\\
&\stackrel{!}{=} -\eta_{(k)} (-1)^{\tau_0^{(k)}} 
 \Bigl(
\!\varepsilon_{{\cal R}(\alpha_1\beta_1)}^{(k) \; \prime}\! 
+(-1)^{\tau_i}\varepsilon_{{\cal R}(\alpha_2\beta_1)}^{(k) \; \prime}
+ (-1)^{\tau_j}\varepsilon_{{\cal R}(\alpha_1\beta_2)}^{(k) \; \prime}
+(-1)^{\tau_i + \tau_j}\varepsilon_{{\cal R}(\alpha_2\beta_2)}^{(k) \; \prime}
\!\Bigr) ,
\end{aligned}
\end{equation}
where for brevity we defined $\varepsilon_{\alpha\beta}^{(k) \;
  \prime} \equiv  \frac{1}{1-b_k}(\varepsilon_{\alpha\beta}^{(k)} -
b_k \, \tilde{\varepsilon}_{\alpha\beta}^{(k)})$, 
or are at angle $\pi/2$ to the $\OR$ invariant plane ({\it `vertical'}),
\begin{equation}
\begin{aligned}
\Pi^{\Z_2^{(k)}}_{v} &=\Pi^{\Z_2^{(k)}}_{v'}=(-1)^{\tau_0^{(k)}}  \left(
  \tilde{\varepsilon}_{\alpha_1\beta_1}^{(k)} 
+(-1)^{\tau_i}\tilde{\varepsilon}_{\alpha_2\beta_1}^{(k)}
+ (-1)^{\tau_j}\tilde{\varepsilon}_{\alpha_1\beta_2}^{(k)}
+(-1)^{\tau_i + \tau_j}\tilde{\varepsilon}_{\alpha_2\beta_2}^{(k)}
\right)
\\
&\stackrel{!}{=} \eta_{(k)} (-1)^{\tau_0^{(k)}}  \left(
\tilde{\varepsilon}_{{\cal R}(\alpha_1\beta_1)}^{(k)}
+(-1)^{\tau_i}\tilde{\varepsilon}_{{\cal R}(\alpha_2\beta_1)}^{(k)}
+ (-1)^{\tau_j}\tilde{\varepsilon}_{{\cal R}(\alpha_1\beta_2)}^{(k)}
+(-1)^{\tau_i + \tau_j}\tilde{\varepsilon}_{{\cal R}(\alpha_2\beta_2)}^{(k)}
\right) .
\end{aligned}
\end{equation}
In order to classify those fractional three-cycles,
\begin{equation}\label{Eq:define-probes}
\begin{aligned}
\Pi_{{\rm probe},0} &\equiv \frac{1}{4} \Pi_{\OR}^{\rm bulk} +\frac{1}{4} \sum_{k=1}^3  \Pi^{\Z_2^{(k)}}_h 
,
\\
\Pi_{{\rm probe},i} & \equiv \frac{1}{4} \Pi_{\OR\Z_2^{(i)}}^{\rm bulk} +\frac{1}{4} \left(
 \Pi^{\Z_2^{(i)}}_h +  \Pi^{\Z_2^{(j)}}_v -  \Pi^{\Z_2^{(k)}}_v
\right)
,
\end{aligned}
\end{equation}
which are $\OR$ invariant, one needs to evaluate the orientifold images of the orbifold fixed points $\alpha\beta$ at which the  exceptional three-cycles are stuck, see table~\ref{Tab:Z2Z2-fixedpoints-SOSp}.
\begin{table}[h!]
\renewcommand{\arraystretch}{1.3}
  \begin{center}
\begin{equation*}
\begin{array}{|c|cc|}\hline
\multicolumn{3}{|c|}{\text{\bf $\Z_2$ fixed points for $\OR\Z_2^{(i)}$ invariant branes on } T^6/\Z_2 \times \Z_2}
\\\hline\hline
(\sigma) & \alpha_1,\beta_1 & \alpha_2,\beta_2
\\\hline\hline
h(0) & 1 & \frac{2}{1-b}
\\
h(1) & 4(1-b) & 3
\\\hline
v(0) & 1 & 4
\\
v(1) & 2 & 3
\\\hline 
\end{array}
\end{equation*}
\end{center}
\caption{$\Z_2$ fixed points which are trasversed by a one-cycle parallel to the $\OR$ invariant plane ($h$) or 
perpendicular to it ($v$). The argument $\sigma$ parameterises the discrete displacement of the one-cycle
on the same two-torus.}
\label{Tab:Z2Z2-fixedpoints-SOSp}
\end{table}

The $\Z_2$ fixed points per two-torus transform under the orientifold projection as follows,
\begin{equation}
\begin{aligned} 
1 &\stackrel{\OR}{\longrightarrow} 1,
\\
\frac{2}{1-b} &\stackrel{\OR}{\longrightarrow} \frac{2}{1-b},
\\
3  &\stackrel{\OR}{\longrightarrow} 3-2b,
\\
4(1-b)  &\stackrel{\OR}{\longrightarrow} 4-2b.
\end{aligned} 
\end{equation}
A careful analysis of all combinations of untilted and tilted tori with discrete displacements from the origin 
in table~\ref{tab:Z2fixed_Z2Z2}
leads to the complete list of invariance conditions in table~\ref{Tab:Class-OR-inv-cycles} in the appendix, which can in short be summarised as the relations among displacements $\sigma_i$, Wilson lines $\tau_i$, the shape of the tori $b_i$ and the orientifold projection $\eta_{(k)}$ on twisted sectors for $i \neq k$ in table~\ref{Tab:Conditions-on_b+t+s-SOSp}.
\begin{table}[h!]
\renewcommand{\arraystretch}{1.3}
  \begin{center}
\begin{equation*}
\begin{array}{|c|c|}\hline
\multicolumn{2}{|c|}{\text{\bf Existence of $\OR$ invariant three-cycles on } T^6/\Z_2 \times \Z_{2M}}
\\\hline\hline
\begin{array}{c}
\text{parallel to}
\\ \text{O6-plane}  
\end{array}
& (\eta_{(1)},\eta_{(2)},\eta_{(3)}) \stackrel{!}{=}
\\\hline\hline
\OR  & \left( -(-1)^{2(b_2\sigma_2\tau_2 + b_3\sigma_3 \tau_3)} , - (-1)^{2(b_1\sigma_1\tau_1 + b_3\sigma_3 \tau_3)} ,  -(-1)^{2(b_1\sigma_1\tau_1 + b_2\sigma_2 \tau_2)} \right)
\\
\OR\Z_2^{(1)}  & \left( -(-1)^{2(b_2\sigma_2\tau_2 + b_3\sigma_3 \tau_3)} ,  (-1)^{2(b_1\sigma_1\tau_1 + b_3\sigma_3 \tau_3)} ,  (-1)^{2(b_1\sigma_1\tau_1 + b_2\sigma_2 \tau_2)} \right)
\\
\OR\Z_2^{(2)}  &  \left( (-1)^{2(b_2\sigma_2\tau_2 + b_3\sigma_3 \tau_3)} , - (-1)^{2(b_1\sigma_1\tau_1 + b_3\sigma_3 \tau_3)} ,  (-1)^{2(b_1\sigma_1\tau_1 + b_2\sigma_2 \tau_2)} \right)
\\
\OR\Z_2^{(3)}  &  \left( (-1)^{2(b_2\sigma_2\tau_2 + b_3\sigma_3 \tau_3)} ,  (-1)^{2(b_1\sigma_1\tau_1 + b_3\sigma_3 \tau_3)} ,  -(-1)^{2(b_1\sigma_1\tau_1 + b_2\sigma_2 \tau_2)} \right)
\\\hline
\end{array}
\end{equation*}
\end{center}
\caption{Conditions on the existence of $\OR$ invariant fractional three-cycles on $T^6/(\Z_2 \times \Z_{2M} \times \OR)$ with
discrete torsion for $2M \in \{2,6,6'\}$ using the relations in table~\protect\ref{Tab:Class-OR-inv-cycles}.}
\label{Tab:Conditions-on_b+t+s-SOSp}
\end{table}

The $\OR$ invariant branes can either carry $SO(2M)$ or $Sp(2M)$ gauge factors. The analysis in appendix~\ref{sec:thres}
reveals that the complete classification of $\OR$ invariant branes provides $Sp(2M)$ gauge factors.

{\boldmath
\subsection{A $T^6/\Z_2 \times \Z_2$ example with discrete torsion on
  a tilted torus}\label{Ss:Z2Z2-example} 
}

Here, we present an example with discrete torsion inspired by
\cite{Blumenhagen:2005tn}. We take the horizontal O-plane to be exotic
and all the other ones non-exotic. In difference to
\cite{Blumenhagen:2005tn} we choose the {\bf aab} lattice for
compactification, i.e.\ we tilt the third two-torus. In our brane
setup we aim for simplicity and not to get a phenomenologically
relevant model. The brane configuration  to be discussed is given in
table \ref{tab:Z2Z2mod}.

\begin{table}[h]
  \renewcommand{\arraystretch}{1.5}
  \begin{center}
    \begin{tabular}{|c||c|c|c|c|}\hline
\multicolumn{5}{|c|}{\bf Bulk cycles for a $T^6/\Z_2 \times \Z_2$ example with discrete torsion}
      \\\hline\hline
      $N_\alpha $  &  $(n_\alpha ^{1},m_\alpha ^{1})$  &  $(n_\alpha
      ^{2},m_\alpha ^{2})$ 
      &  $(n_\alpha ^{3},\tilde{m}_\alpha ^{3})$ & $m_\alpha ^3
      =\tilde{m}_\alpha ^{3} -\frac{1}{2}n_\alpha ^{3}  $\\
      \hline
      \hline $N_{a_1} = 2$ & $(-1,1)$ & $(-2,1)$ & $(-1,\frac{1}{2})$ & $1$ \\
      \hline $N_{a_2} = 2$ & $(1,-1)$ & $(2,-1)$ & $(-1,\frac{1}{2})$ & $1$ \\
      \hline $N_{a_3} = 2$ & $(-1,1)$ & $(2,-1)$ & $(1,-\frac{1}{2})$ & $-1$ \\
      \hline $N_{a_4}= 2$ & $(1,-1)$ &  $ (-2,1)$  &
      $(1,-\frac{1}{2})$ & $-1$ \\
      \hline
      \hline $N_{b_1}= 2$ & $(1,0)$ &  $ (0,1)$  & $(0,-1)$
      & $-1$ \\
\hline $N_{b_2}= 2$ & $(-1,0)$ &  $ (0,1)$  & $(0,1)$
      & $1$ \\
\hline
\hline $N_{c_1}= 2$ & $(0,1)$ &  $ (0,-1)$  & $(2,0)$ &$-1$\\
      \hline $N_{c_2}= 2$ & $(0,1)$ &  $ (0,1)$  & $(-2,0)$ & $1$\\
           \hline
    \end{tabular}
    \caption{Solution to tadpole conditions for tilted $T^2_{(3)}$
      (i.e. $b_3=\frac{1}{2}$), 
      exceptional $\Omega {\cal R}$ plane, the $\OR\Z_2^{(i)}$ planes are 
regular.\label{tab:Z2Z2mod}}
  \end{center}
\end{table}
All branes pass through the origin and have trivial discrete Wilson
lines. To specify the $\tau_0^{(i)}$'s we adopted the compressed notation of
\cite{Blumenhagen:2005tn}. For instance, all branes labeled by $a_i$,
$i=1,\ldots, 4$ wrap the same fractional bulk cycle but differ in
their $\tau_0^{(i)}$'s. The $\tilde{m}^i$'s are useful numbers for
  computing tadpole cancellation and torus intersection numbers
  whereas the numbers denoted by $m^i$ should be integers. The latter
  ones should be used when determining through which fixed points the
  corresponding brane passes. Note, for instance, that for the
  $\Omega{\cal R}$ image of the branes labeled by $a_i$,
  $i=1,\ldots,4$, the wrapping number $m^3$ vanishes. This reflects
  the earlier observation that on a tilted torus the second and third
  fixed points are interchanged by $\Omega {\cal R}$. Naively, the 
  gauge group is $U(4)^4 
  \times U(2)^2 \times U(2)^2$. Anomalous $U(1)$'s become massive via
  the Green--Schwarz mechanism leaving as a non-anomalous gauge group
  $SU(2)^4 \times SU(2)^2\times SU(2)^2$, as we will see shortly. 
Supersymmetry conditions on the torus
  moduli arise from the $a$-type branes and are
\begin{equation}
\varrho_1 = 4 + \frac{2 \varrho_1}{\varrho_3 + \varrho_2},
\end{equation}
from Im$\left( {\cal Z}\right) = 0$, and then Re$\left({\cal Z}\right)
> 0$ for any positive radii.  

The model gives rise to a spectrum of chiral multiplets transforming
under $U(2)^8$ as follows  (the ordering of the group factors follows
the ordering of the branes in table \ref{tab:Z2Z2mod})

\begin{align*}
& 
2 \left( \mbox{\boldmath $\overline{2}$},\mbox{\boldmath $1$},
  \mbox{\boldmath   
    $1$},  \mbox{\boldmath $\overline{2}$},\mbox{\boldmath
    $1$},\mbox{\boldmath 
    $1$},\mbox{\boldmath $1$},\mbox{\boldmath $1$}\right)
+2 \left( \mbox{\boldmath $1$},\mbox{\boldmath $\overline{2}$},
  \mbox{\boldmath $\overline{2}$},  \mbox{\boldmath
    $1$},\mbox{\boldmath $1$},\mbox{\boldmath $1$},\mbox{\boldmath
    $1$},\mbox{\boldmath $1$}\right) 
+ 2 \left( \mbox{\boldmath $\overline{2}$},\mbox{\boldmath $1$},
  \mbox{\boldmath $1$},  \mbox{\boldmath $1$},\mbox{\boldmath
  $1$},\mbox{\boldmath $\overline{2}$},\mbox{\boldmath $1$},\mbox{\boldmath
  $1$} \right) \\ 
&+ 2 \left( 
  \mbox{\boldmath $1$},\mbox{\boldmath $\overline{2}$},  \mbox{\boldmath 
    $1$},  \mbox{\boldmath $1$},\mbox{\boldmath
    $2$},\mbox{\boldmath $1$},\mbox{\boldmath
    $1$},\mbox{\boldmath $1$}\right)
+  
2 \left( \mbox{\boldmath $1$},\mbox{\boldmath $1$},  \mbox{\boldmath
    $\overline{2}$},  \mbox{\boldmath $1$},\mbox{\boldmath
    $1$},\mbox{\boldmath $2$},\mbox{\boldmath $1$},\mbox{\boldmath
    $1$}\right)
+ 
2 \left( \mbox{\boldmath $1$},\mbox{\boldmath $1$},  \mbox{\boldmath
    $1$},  \mbox{\boldmath $\overline{2}$},\mbox{\boldmath
    $\overline{2}$},\mbox{\boldmath $1$},\mbox{\boldmath
    $1$},\mbox{\boldmath $1$}\right) \\
& +  2 \left( 
  \mbox{\boldmath $\overline{2}$},\mbox{\boldmath $1$},  \mbox{\boldmath 
    $1$},  \mbox{\boldmath $1$},\mbox{\boldmath $1$},\mbox{\boldmath
$1$},\mbox{\boldmath $1$},\mbox{\boldmath $\overline{2}$}\right) 
+ 
2 \left( 
  \mbox{\boldmath $1$},\mbox{\boldmath $\overline{2}$},  \mbox{\boldmath 
    $1$},  \mbox{\boldmath $1$}, \mbox{\boldmath $1$},\mbox{\boldmath
    $1$},\mbox{\boldmath $1$},\mbox{\boldmath
    $2$}\right) 
+ 
2 \left( \mbox{\boldmath $1$},\mbox{\boldmath $1$},  \mbox{\boldmath 
    $\overline{2}$},  \mbox{\boldmath $1$},\mbox{\boldmath
    $1$},\mbox{\boldmath $1$},\mbox{\boldmath $\overline{2}$},
  \mbox{\boldmath $1$}\right) \\
& + 
2 \left( \mbox{\boldmath $1$},\mbox{\boldmath $1$},  \mbox{\boldmath
    $1$},  \mbox{\boldmath $\overline{2}$},\mbox{\boldmath
    $1$},\mbox{\boldmath $1$},\mbox{\boldmath $2$},\mbox{\boldmath
    $1$}\right) 
+
2  \left( \mbox{\boldmath $1$},\mbox{\boldmath $1$},  \mbox{\boldmath
    $1$},  \mbox{\boldmath $1$},\mbox{\boldmath
    $\overline{2}$},\mbox{\boldmath $1$},\mbox{\boldmath $2$},\mbox{\boldmath
    $1$}\right) 
+
2 \left( \mbox{\boldmath $1$},\mbox{\boldmath $1$},  \mbox{\boldmath
    $1$},  \mbox{\boldmath $1$},\mbox{\boldmath
    $2$},\mbox{\boldmath $1$},\mbox{\boldmath $1$},\mbox{\boldmath
    $2$}\right) \\
& +
2 \left( \mbox{\boldmath $1$},\mbox{\boldmath $1$},  \mbox{\boldmath
    $1$},  \mbox{\boldmath $1$},\mbox{\boldmath
    $1$},\mbox{\boldmath $2$},\mbox{\boldmath
    $\overline{2}$},\mbox{\boldmath $1$}\right)
+
2 \left( \mbox{\boldmath $1$},\mbox{\boldmath $1$},  \mbox{\boldmath
    $1$},  \mbox{\boldmath $1$},\mbox{\boldmath
    $1$},\mbox{\boldmath $\ov{2}$},\mbox{\boldmath
    $1$},\mbox{\boldmath $\ov{2}$}\right)
+
6\left( \mbox{\boldmath $\overline{3}$}_S,\mbox{\boldmath $1$},
  \mbox{\boldmath  
    $1$},  \mbox{\boldmath $1$},\mbox{\boldmath $1$},\mbox{\boldmath
    $1$},\mbox{\boldmath $1$},\mbox{\boldmath $1$}\right) \\
& + 
6\left( \mbox{\boldmath $1$},\mbox{\boldmath $\overline{3}$}_S,
  \mbox{\boldmath  
    $1$},  \mbox{\boldmath $1$},\mbox{\boldmath $1$},\mbox{\boldmath
    $1$},\mbox{\boldmath $1$},\mbox{\boldmath $1$}\right)
+
6\left( \mbox{\boldmath $1$},
  \mbox{\boldmath  
    $1$}, \mbox{\boldmath $\overline{3}$}_S, \mbox{\boldmath
    $1$},\mbox{\boldmath $1$},\mbox{\boldmath 
    $1$},\mbox{\boldmath $1$},\mbox{\boldmath $1$}\right)
+
6\left( \mbox{\boldmath $1$},
  \mbox{\boldmath  
    $1$},  \mbox{\boldmath $1$},\mbox{\boldmath
    $\overline{3}$}_S,\mbox{\boldmath $1$},\mbox{\boldmath 
    $1$},\mbox{\boldmath $1$},\mbox{\boldmath $1$}\right) .
\end{align*}
Here, the subscript $S$ denotes a symmetric representation of $U(2)$. 
So far, we have
listed the chiral spectrum. It can be seen that the normal $U(1)$
subgroups within each $U(2)$ are anomalous. The corresponding $U(1)$
gauge fields become massive via the Green--Schwarz mechanism.  In
addition to the chiral spectrum there can be non-chiral matter as
discussed in appendix \ref{S:App-OpenStrings}: For each $a_i a_j$,
$b_i b_j$, $c_i c_j$ pair with $i\not= j$ there is an additional
non-chiral pair transforming in the bifundamental of the corresponding
$U(2)^2$, and for each stack $b_i$, $c_i$ there is a non-chiral
pair of antisymmetric representations of the corresponding $U(2)$ factor.
%

{\boldmath
\section{The $T^6/(\Z_2 \times \Z_4 \times \OR)$
  orientifolds}\label{S:Z2Z4orient} 
}

The $T^6/\Z_2 \times \Z_4$ orbifold is generated by the two shift vectors
\begin{equation}
\vec{v}=\frac{1}{2}(1,-1,0) ,
\quad
\vec{w}=\frac{1}{4}(0,1,-1).
\end{equation}
On the factorisable lattice $SU(2)^2 \times SO(5)^2$, the number of three-cycles 
is given by
\begin{equation}
\eta=1: \; h_{21}= 1_{\rm bulk},
\qquad\qquad
\eta=-1:\; h_{21}= 1_{\rm bulk} + 8_{\Z_4}.
\end{equation}
The Hodge number $h_{11}$ can be found in table~\ref{Tab:Hodge-T6ZNxZM-torsion}.

{\boldmath
\subsection{Model building on $T^6/\Z_2 \times \Z_4$ without and with discrete torsion}\label{Ss:Z2Z4-model-building}
}

For both cases, without and with torsion, only the four independent
bulk three-cycles 
\begin{equation}
\rho_1= 4  \left( \pi_{135} - \pi_{146} \right) , \,
\rho_2= 4  \left( \pi_{236} + \pi_{245}  \right), \,
\rho_3= 4  \left( \pi_{246} - \pi_{235} \right) , \,
\rho_4= 4  \left( \pi_{145} + \pi_{136} \right) ,
\end{equation}
with non-vanishing intersection numbers
\begin{equation}
\rho_1 \circ \rho_3 = \rho_2 \circ \rho_4 = -4
\end{equation}
can be used for model building.
This orbifold differs from the other ones discussed in this article 
in that D6-branes wrap the same half-bulk three-cycles 
independently of the choice of discrete torsion. The case without discrete torsion was first investigated 
in~\cite{Honecker:2003vq,Honecker:2003vw}, and in~\cite{Honecker:2004np} it was shown that 
three Standard Model generations cannot be obtained. The number theoretic proof used the minimal
requirements of (a) three intersections of the QCD stack with a second stack, 
(b) no symmetric representations on the QCD stack and (c) supersymmetry. These constraints are common to all
Standard Model, Pati-Salam and $SU(5)$ GUT models on any Calabi-Yau or orbifold background. 
For $T^6/\Z_2 \times \Z_4$, the three-cycles on which D6-branes can be wrapped do not depend on the 
choice of torsion, and therefore conditions (a) and (c) are the same for both choices of $\eta=\pm 1$.
The discrete torsion enters only in condition (b) in terms of the sign of one of the O6-planes.
In~\cite{Cvetic:2006by}, the no-go theorem for the $T^6/\Z_2 \times \Z_4$ orbifold without discrete torsion
was confirmed for some examples.

For the sake of completeness of discussing $T^6/\Z_2 \times \Z_{2M}$ orbifolds on factorisable tori,
we give in the following all model building ingredients.

The $\Z_4$ orbifold permutes wrapping numbers on the second and third torus,
\begin{equation}\label{Eq:Z4-1cycle-Orb}
\left(\begin{array}{cc}
n^1 & m^1 \\ n^2 & m^2 \\ n^3 & m^3 
\end{array}\right) \stackrel{\Z_4^{(1)}}{\rightarrow}
\left(\begin{array}{cc}
n^1 & m^1 \\   -m^2 & n^2 \\ m^3 & - n^3 
\end{array}\right) ,
\end{equation}
and a generic orbifold invariant three-cycle is parameterised by
\begin{equation}\label{Eq:Z2Z4-bulkWrappings}
\begin{aligned}
\Pi^{\rm bulk} =& P \, \rho_1 + Q \, \rho_2 + U \, \rho_3 + V \, \rho_4 ,
\\
P &= n^1 \, \left( n^2 n^3 - m^2 m^3 \right),
\\
Q &= m^1 \, \left( n^2  m^3 + m^2 n^3 \right), 
\\
U &= m^1 \, \left( m^2 m^3 - n^2 n^3 \right) ,
\\
V &= n^1 \, \left( n^2 m^3 + m^2 n^3 \right) .
\end{aligned}
\end{equation}
The supersymmetry conditions read
\begin{equation}
\begin{aligned}
{\bf a/bAA:} \qquad & V + \varrho (-U+b \, P) =0 ,\qquad P-  \varrho(Q +b \, V) > 0 
,
\\
{\bf a/bAB:} \qquad & [V-P]+ \varrho(-U+Q+b \, [P+V]  )  =0 ,\qquad  P +V - \varrho (  Q+U+b \, [V-P]  ) > 0 
,
\\
{\bf a/bBB:} \qquad & P- \varrho (Q+b \, V)=0 ,\qquad   V+  \varrho (-U+b \, P)> 0 
,
\end{aligned}
\end{equation}
where $\varrho \equiv \frac{R_2}{R_1}$ is the complex structure modulus on the first two-torus.

The orientifold projection for all six inequivalent choices of lattice orientations is given in table~\ref{Tab:Z2Z4-ORprojection}.
\begin{table}[ht]
\renewcommand{\arraystretch}{1.3}
  \begin{center}
\begin{equation*}
\begin{array}{|c||c|c|c|c|}\hline
\multicolumn{5}{|c|}{\text{\bf Orientifold projection on bulk 3-cycles for } T^6/\Z_2 \times \Z_4}
\\\hline\hline
{\rm 3-cycle} & \rho_1 & \rho_2 & \rho_3 & \rho_4
\\\hline\hline
{\bf a/bAA} & \rho_1 + (2b) \rho_3 & \rho_2 & -\rho_3 & -\rho_4 + (2b) \rho_2
\\\hline
{\bf a/bAB} & \rho_4 - (2b) \rho_2 & \rho_3 & \rho_2 & \rho_1 + (2b) \rho_3
\\\hline
{\bf a/bBB} & -\rho_1 -(2b) \rho_3 & -\rho_2 & \rho_3 & \rho_4 -(2b) \rho_2
\\\hline
\end{array}
\end{equation*}
\end{center}
\caption{The orientifold projection on bulk three-cycles on  $T^6/(\Z_2 \times \Z_4 \times \OR)$ for the six inequivalent
lattices.}
\label{Tab:Z2Z4-ORprojection}
\end{table}

Representants of the four O6-plane orbits are given in table~\ref{Tab:Z2Z4-O6planes-1}, and the resulting bulk 
wrapping numbers of the O6-planes are listed in table~\ref{Tab:Z2Z4-O6planes-2}.
\begin{table}[ht]
\renewcommand{\arraystretch}{1.3}
  \begin{center}
\begin{equation*}
\begin{array}{|c|c||c|c|c|}\hline
\multicolumn{5}{|c|}{\text{\bf Torus wrapping numbers of O6-planes on } T^6/\Z_2 \times \Z_4}
\\\hline\hline
\text{O6-plane orbit} & \frac{\rm angle}{\pi} \text{ w.r.t. } \OR & {\bf a/bAA} & {\bf a/bAB} & {\bf a/bBB}
\\\hline
& & \multicolumn{3}{|c|}{(n^1,m^1;n^2,m^2;n^3,m^3) \text{ for one representant} }
\\\hline\hline
\OR & (0,0,0) & (\frac{1}{1-b},\frac{-b}{1-b};1,0;1,0) &  (\frac{1}{1-b},\frac{-b}{1-b};1,0;1,1)  &  (\frac{1}{1-b},\frac{-b}{1-b};1,1;1,1)
\\
\OR\Z_4^{(1)} & (0,-\frac{1}{4},\frac{1}{4}) &  (\frac{1}{1-b},\frac{-b}{1-b};1,-1;1,1) &  (\frac{1}{1-b},\frac{-b}{1-b};1,-1;0,1) &   (\frac{1}{1-b},\frac{-b}{1-b};1,0;0,1)
\\
\OR\Z_2^{(3)} & (\frac{1}{2},-\frac{1}{2},0)  & ( 0,1;0,-1;1,0) &  ( 0,1;0,-1;1,1) &  ( 0,1;1,-1;1,1) 
\\
\!\!\!\!\!\OR\Z_4^{(1)}\Z_2^{(3)}\!\!\!\!\!& (-\frac{1}{2}, \frac{1}{4} ,\frac{1}{4}) & (0,-1;1,1;1,1) & (0,-1;1,1;0,1) & (0,-1;0,1;0,1)
\\\hline
\end{array}
\end{equation*}
\end{center}
\caption{Torus wrapping numbers for one representant per O6-plane orbit on $T^6/(\Z_2 \times \Z_4 \times \OR)$. The corresponding bulk wrapping numbers 
per orbit are given in table~\protect\ref{Tab:Z2Z4-O6planes-2}. }
\label{Tab:Z2Z4-O6planes-1}
\end{table}

\begin{table}[ht]
\renewcommand{\arraystretch}{1.3}
  \begin{center}
\begin{equation*}
\begin{array}{|c||c|c|c|c|c||c|c|c|c|c||c|c|c|c|c|}\hline
\multicolumn{16}{|c|}{\text{\bf Bulk wrapping numbers for the O6-planes on } T^6/\Z_2 \times \Z_4}
\\\hline\hline
 & \multicolumn{5}{|c||}{\bf a/bAA}   & \multicolumn{5}{|c||}{\bf a/bAB} & \multicolumn{5}{|c|}{\bf a/bBB}   
\\
{\rm orbit} & P & Q & U & V & N_{O6} 
&  P & Q & U & V  & N_{O6} 
&  P & Q & U & V  & N_{O6} 
\\\hline\hline
\OR & \frac{1}{1-b} & 0 & \frac{b}{1-b} & 0  & 8(1-b) 
& \frac{1}{1-b} & \frac{-b}{1-b} & \frac{b}{1-b} & \frac{1}{1-b} & 4(1-b) 
& 0 & \frac{-2b}{1-b} & 0 & \frac{2}{1-b} & 2(1-b) 
\\
\OR\Z_4^{(1)} &  \frac{2}{1-b} & 0 & \frac{2 \, b}{1-b} & 0 & 2(1-b) 
& \frac{1}{1-b} & \frac{-b}{1-b} & \frac{b}{1-b} & \frac{1}{1-b} & 4(1-b) 
& 0 & \frac{-b}{1-b} & 0 & \frac{1}{1-b} & 8(1-b) 
\\
\OR\Z_2^{(3)} & 0 & -1 & 0 & 0 & 8(1-b)
& 0 & -1 & -1 & 0 & 4(1-b)  
& 0 & 0 & -2 & 0 & 2(1-b) 
\\
\!\!\!\OR\Z_4^{(1)}\Z_2^{(3)}\!\!\! & 0 & -2 & 0 & 0 & 2(1-b) 
& 0 & -1 & -1 & 0 & 4(1-b) 
& 0 & 0 & -1 & 0 & 8(1-b) 
\\\hline
\end{array}
\end{equation*}
  \end{center}
\caption{Bulk wrapping numbers of the O6-plane orbits on $T^6/(\Z_2 \times \Z_4 \times \OR)$ computed from the torus wrapping numbers in table~\protect\ref{Tab:Z2Z4-O6planes-1}
using equation~(\protect\ref{Eq:Z2Z4-bulkWrappings}).}
\label{Tab:Z2Z4-O6planes-2}
\end{table}
The charge assignment condition on O6-planes~(\ref{eq:ocharge}) can be written as 
\begin{equation}
\eta_{\OR} \eta_{\OR\Z_4^{(1)}}  \eta_{\OR\Z_2^{(3)}}  \eta_{\OR\Z_4^{(1)}\Z_2^{(3)}} = \eta,
\end{equation}
where, e.g., $ \eta_{\OR\Z_4^{(1)}}$ is the charge of the orbit of O6-planes consisting of $\OR\omega$ and $\OR\omega^3$.

As stated in section~\ref{Sss:DT+EO}, in the absence of torsion
the number of exotic O6-plane orbits must be even, whereas in the presence
of discrete torsion, 
an odd number needs to be exotic.

Using the orientifold images of three-cycles wrapped by D6-branes in table~\ref{Tab:Z2Z4-ORprojection} and the O6-planes in table~\ref{Tab:Z2Z4-O6planes-2},
the RR tadpole cancellation conditions on $T^6/\Z_2 \times \Z_4$ read 
\begin{equation}\label{Eq:Z2Z4-RRtcc}
\begin{aligned}
{\bf a/bAA} \quad  & \left[ \sum_a N_a \, P_a  - \left( 8 \, \eta_{\OR} + 4 \, \eta_{\OR\Z_4^{(1)}} \right) \right] \left(\rho_1 + b \rho_3 \right)
\\
 & \quad +  \left[\sum_a N_a \left(Q_a + b V_a  \right) + (1-b) \left(
     8 \, \eta_{\OR\Z_2^{(3)}} + 4 \, \eta_{\OR\Z_4^{(1)}\Z_2^{(3)}}
   \right) \right] \rho_2  =0
,
\\
\\
{\bf a/bAB}  \quad  & \left[\sum_a N_a \left(P_a + V_a  \right)
  -\left( 8 \, \eta_{\OR} + 8 \, \eta_{\OR\Z_4^{(1)}} \right)  \right]
\left(\rho_1 + \rho_4  + b (\rho_3 - \rho_2) \right)  
\\
 & \hspace*{-1in} +  \left[\sum_a N_a \left(Q_a + U_a -b(P_a-V_a)  \right)  +
   (1-b) \left( 8 \, \eta_{\OR\Z_2^{(3)}} + 8 \,
     \eta_{\OR\Z_4^{(1)}\Z_2^{(3)}} \right) \right] \left(\rho_2 +
   \rho_3  \right)  = 0
,
\\
\\
{\bf a/bBB}  \quad  & \left[\sum_a N_a \, V_a  - \left( 4 \, \eta_{\OR} + 8 \, \eta_{\OR\Z_4^{(1)}} \right) \right] \left(\rho_4 - b \rho_2 \right) 
\\
 &\quad + \left[\sum_a N_a \left( U_a - b P_a \right) + (1-b)
   \left( 4 \, \eta_{\OR\Z_2^{(3)}} + 8 \,
     \eta_{\OR\Z_4^{(1)}\Z_2^{(3)}} \right)   \right] \rho_3  = 0
,
\end{aligned}
\end{equation}
for both choices of discrete torsion $\eta = \pm 1$.
These are per lattice $2=1 + h_{21}^U$ independent conditions.

Next, we list how redundancies due to symmetires are removed in the counting of inequivalent models:
\begin{itemize}
\item
The $\Z_4$ projection~(\ref{Eq:Z4-1cycle-Orb}) on $T^2_{(3)}$ exchanges the torus wrapping numbers $(n^3,m^3)=$
(odd,even) $\leftrightarrow$ (even,odd) and preserves (odd,odd).
On can therefore choose \framebox{$n^3 = {\rm odd}$}. This singles out one orbifold image for $m^3$ even, 
but the redundancy for $(n^3,m^3)$=(odd,odd) survives. 
\item
The orientation of the one-cycle on $T^2_{(3)}$ is fixed by requiring \framebox{$n^3 >0$}.
\item
Avoiding simultaneous sign flips on $T^2_{(1)} \times T^2_{(2)}$ and no double-counting of the orientifold image is achieved by requiring
$0 \leq \pi\phi^{(1)} \leq \frac{\pi}{2}$ in equation~(\ref{Eq:angles}): 
\framebox{$m^1 + b \, n^1 \geq 0$  and $n^1 \geq 0$}.
These conditions are sufficient for non-trivial angles.
\item
For $\pi\phi^{(1)} \in \{0, \frac{\pi}{2}\}$ one has to impose an additional condition on the third two-torus, e.g. by demanding 
\framebox{$m^3 \leq 0$ on {\bf A} and $|m^3| \leq n^3$ on {\bf B} for $T^2_{(3)}$} .\\
The limiting cases $\pi\phi^{(3)} \in \{0, -\frac{\pi}{2}\}$ correspond to three-cycles parallel to the O6-planes.
\end{itemize}
With these constraints on torus wrapping numbers, a classification of supersymmetric three-cycles can be performed 
in an economic way.

{\boldmath
\subsection{A discussion on three generations on $T^6/(\Z_2 \times \Z_4 \times \OR)$ without and with discrete torsion}\label{Ss:Z2Z4-3generations}
}

Since the bulk three-cycles are identical for both choices of torsion, the no-go theorem on three generations found in~\cite{Honecker:2004np}
is expected to still be valid, as we discuss below for several scenarios.
\begin{enumerate}
\item
The number of chiral bifundamental representations of the QCD stack
with a second stack is given by 
\begin{equation}\label{Eq:Z2Z4-bifund}
\Pi_a \circ \left( \Pi_b + \Pi_{b'} \right) = \left\{\begin{array}{cc}
2 U_a P_b +2 V_a Q_b + 2b (V_aV_b - P_aP_b) & {\bf a/bAA} \\
\begin{array}{ll}(V_a - P_a) (Q_b + U_b) + (U_a -Q_a) (P_b + V_b)\\[-0.5ex]
-2b (P_aV_b + V_aP_b) \end{array}& {\bf a/bAB} \\ 
-2Q_a V_b -2 P_a U_b + 2 b (P_a P_b - V_a V_b) & {\bf a/bBB}
\end{array} \right.
\end{equation}
On the {\bf aAA} and {\bf aBB} lattices, it is obvious that only an even number of generations can occur.
This leaves at first glance the {\bf bAA}, {\bf bBB} and {\bf a/bAB} lattices as potentially interesting backgrounds for model building.
\item
The number of chiral symmetrics on the QCD stack is given by
\begin{equation}
\chi^{\Sym_a} =\frac{1}{2} \left(\Pi_a \circ \Pi_{a'} - \Pi_a \circ \Pi_{O6}\right)
\end{equation}
with
\begin{equation}\label{Eq:Z2Z4-anti+sym}
\begin{aligned}
\Pi_a \circ \Pi_{a'} &= \left\{\begin{array}{cc}
2P_a U_a +2 V_a Q_a+ 2b (V_a^2 - P_a^2) & {\bf a/bAA} \\
2 U_aV_a - 2 P_a Q_a -4b P_a V_a & {\bf a/bAB} \\
-2P_a U_a - 2 V_a Q_a - 2b (V_a^2 - P_a^2)  & {\bf a/bBB}
\end{array} \right.
,
\\
\Pi_a \circ \Pi_{O6} &= \left\{\!\!\!\begin{array}{cc}
\begin{array}{l}(4 \eta_{\OR} + 2\eta_{\OR\Z_4^{(1)}}) ( U_a - b P_a)\\[-0.5ex]
- (4 \eta_{\OR\Z_2^{(3)}} + 2\eta_{\OR\Z_2^{(3)}\Z_4^{(1)}})
(1-b)V_a\end{array}  
& {\bf a/bAA} \\
\begin{array}{l}
(4 \eta_{\OR} + 4\eta_{\OR\Z_4^{(1)}}) ( U_a - Q_a - b ( P_a + V_a) )\\[-0.5ex]
 + (4 \eta_{\OR\Z_2^{(3)}} + 4\eta_{\OR\Z_2^{(3)}\Z_4^{(1)}}) (1-b)(P_a - V_a)
\end{array}
 & {\bf a/bAB} \\
\begin{array}{l}-(2 \eta_{\OR} + 4\eta_{\OR\Z_4^{(1)}}) (Q_a + b V_a)\\[-0.5ex]
+ (2 \eta_{\OR\Z_2^{(3)}} + 4\eta_{\OR\Z_2^{(3)}\Z_4^{(1)}}) (1-b) P_a\end{array}
 & {\bf a/bBB}
\end{array} \right.
,
\end{aligned}
\end{equation}
and the number of chiral antisymmetrics is computed from
\begin{equation}
\chi^{\Anti_a} =\frac{1}{2} \left(\Pi_a \circ \Pi_{a'} + \Pi_a \circ \Pi_{O6}\right)
\stackrel{\chi^{\Sym_a} =0} {=} \Pi_a \circ \Pi_{a'} 
,
\end{equation}
where in the second step we inserted the requirement of no chiral symmetric representation for phenomenological reasons.
Furthermore, 
\begin{equation}
0 \leq |\chi^{\Anti_a}| \leq 3
 \end{equation}
 ensures the absence of an excess of right-handed quarks, where $\chi^{\Anti_a}=0$ is required for left-right symmetric and Pati-Salam  models and 
 $ |\chi^{\Anti_a}| = 3$ for $SU(5)$ GUTs. Formula~(\ref{Eq:Z2Z4-anti+sym}) reveals, that an odd number of antisymmetric generations
 is only possible on the {\bf bAA} and {\bf bBB} lattices. This means that $SU(5)$ GUTs with three generations and no chiral exotics
  are excluded on all other lattices.
\item
Supersymmetry ensures that there are only two choices, $(n^1_a,m^1_a) \in \{(\frac{1}{1-b},\frac{-b}{1-b}),(0,1)\}$,
for which 
\begin{equation}
\chi^{\Anti_a} = \chi^{\Sym_a} =0.
 \end{equation}
Supersymmetry on the D$6_a$-brane with these special values for  $(n^1_a,m^1_a)$ also  renders the number of bifundamental chiral
generations with an arbitrary (supersymmetric or non-supersymmetric) D$6_b$-brane  zero for any lattice, see equation~(\ref{Eq:Z2Z4-bifund}). 
This implies that supersymmetric left-right symmetric or Pati-Salam models are excluded on any lattice.
\end{enumerate}
The discussion above is independent of the choice of discrete torsion or an exotic O6-plane. We expect that a thorough case by case study of 
the remaining possibilities with some right-handed quarks in the antisymmetric representation 
will rule out model building on $T^6/(\Z_2 \times \Z_4 \times \OR)$ completey.

{\boldmath
\subsection{The K-theory constraint on $T^6/\Z_2 \times \Z_4$}\label{Ss:Z2Z4-Ktheory}
}

The K-theory constraint is for  $T^6/(\Z_2 \times \Z_4 \times \OR)$ independent of the choice of discrete torsion.
There exist four possible probe D6-branes per lattice, namely those parallel to the O6-planes with bulk wrapping numbers given in table~\ref{Tab:Z2Z4-O6planes-2}, 
which provide for at most two independent K-theory conditions. 
Computing the intersection numbers~(\ref{Eq:general-Ktheory}) and using the RR tadpole cancellation 
conditions~(\ref{Eq:Z2Z4-RRtcc}) to simplify the K-theory constraint, one ends up with
\begin{equation}
\begin{array}{crcl}
{\bf aAA:} & \qquad
\sum_a N_a U_a
& = 0 \text{ mod } 2 = &
\sum_a N_a V_a
,
\\
{\bf aAB:} & \qquad
\sum_a N_a (Q_a - U_a)
& = 0 \text{ mod } 2 =  & 
\sum_a N_a (P_a - V_a)
,
\\
{\bf aBB:} & \qquad
\sum_a N_a Q_a
& = 0 \text{ mod } 2 = & 
\sum_a N_a P_a
,
\end{array}
\end{equation}
and all constraints trivially fulfilled on the {\bf b} type torus $T^2_{(1)}$.
These conditions may be too restrictive, if some of the probe branes had $SO(2M)$ instead of $Sp(2M)$ gauge groups. 
We will discuss this subtlety further in appendix~\ref{sec:thres}, and in section~\ref{Ss:Z2Z4-example-torsion} we give an 
example where an orientifold invariant D6-brane indeed supports an $SO(2M)$ gauge factor.

{\boldmath
\subsection{Exeptional three-cycles for $T^6/\Z_2 \times \Z_4$ with discrete torsion}\label{Ss:Z2Z4-exceptional}
}

The exceptional three-cycles in the $\Z_4$ twisted sectors for the case of discrete torsion arise  as follows 
\begin{equation}
\begin{aligned}
& 
\delta^{(k)}_l = 2 \, \tilde{ d}_{ij}^{(k)} \otimes \pi_1,
\qquad
\tilde{\delta}^{(k)}_l =2 \, \tilde{ d}_{ij}^{(k)} \otimes \pi_2
& l \in \{1 \ldots 4 \} \; \Leftrightarrow \;  (ij) \in \{(11),(12),(21),(22)\},
\qquad
 k \in \{1,2\}
\\
& \delta^{(k)}_l \circ \tilde{\delta}^{(k')}_{l'} \sim  \; \delta^{kk'} \delta_{ll'}
\end{aligned}
\end{equation}
with the intersection form of two-cycles inherited from $T^4/\Z_4$,
\begin{equation}
d_{ij}^{(k)} \circ d_{i'j'}^{(k')} =
\delta_{ii'} \delta_{jj'} \left(\begin{array}{ccc}
-2 & 1 & 0 
\\ 1 & -2 & 1
\\ 0 & 1 & -2
\end{array}\right)_{kk'}
,
\end{equation}
where $d^{(k)}$ runs over the two exceptional cycles  $\tilde{d}^{(k)}$ associated to $\Z_4$ twisted sectors and on exceptional cycle
belonging to the $\Z_2$ twisted sector at the same fixed point.
The orientifold projection on the twisted three-cycles is on this orbifold independent of the lattice orientation 
as displayed in table~\ref{Tab:Z2Z4twisted-Orient}. 
\begin{table}[h!]
\renewcommand{\arraystretch}{1.3}
  \begin{center}
\begin{equation*}
\begin{array}{|c||c|c|}\hline
 \multicolumn{3}{|c|}{\text{\bf $\OR$  projection on exceptional 3-cycles on } T^6/\Z_2 \times \Z_4}
\\\hline\hline
{\rm 3-cycle} & \delta^{(k)}_l & \tilde{\delta}^{(k)}_l
\\\hline\hline
\left.\begin{array}{c}
{\bf a/bAA} \\ {\bf a/bAB} \\ {\bf a/bBB} 
\end{array}\right\}
& - \eta_{\Z_4^{(1)}} \, \delta^{(k)}_l 
& \eta_{\Z_4^{(1)}}  \; \tilde{\delta}^{(k)}_l
\\\hline
\end{array}
\end{equation*}
\end{center}
\caption{The orientifold projection on twisted three-cycles on $T^6/(\Z_2 \times \Z_4 \times \OR)$ with discrete torsion. The prefactor $\eta_{\Z_4^{(1)}} \equiv \eta_{\OR} \cdot \eta_{\OR\Z_4^{(1)}}$ depends on the choice of the 
exotic O6-plane, cf. eq.~(\ref{eq:ocharge}).
}
\label{Tab:Z2Z4twisted-Orient}
\end{table}

{\boldmath
\subsection{A $T^6/\Z_2 \times \Z_4$ example with discrete torsion}\label{Ss:Z2Z4-example-torsion}
}

In~\cite{Honecker:2004np}, a supersymmetric model on the {\bf bBB} lattice on $T^6/\Z_2 \times \Z_4$ without discrete  torsion was 
presented. We list the torus and bulk wrapping numbers of the D6-branes in that model in table~\ref{Tab:Z2Z4-example-cycles}
after choosing each orbifold and orientifold representant as discussed at the end of section~\ref{Ss:Z2Z4-model-building}.
\begin{table}[h]
  \renewcommand{\arraystretch}{1.3}
  \begin{center}
  \begin{equation}
    \begin{array}{|c|c|c||c|c|c|c|}   \hline
\multicolumn{7}{|c|}{\text{\bf Supersymmetric bulk 3-cycles on the bBB lattice on } T^6/\Z_2 \times \Z_4}
\\\hline\hline
\text{brane/orbit} & \frac{\text{angle}}{\pi} & (n^1,m^1;n^2,m^2;n^3,m^3) & P & Q & U & V 
\\\hline\hline
a_1 & (\frac{1}{4},-\frac{1}{4},0) & (1,0;1,0;1,1) & 1 & 0 & 0 & 1
\\\hline
a_2 & (\frac{1}{4},0,-\frac{1}{4}) & (1,0;1,1;1,0) & 1 & 0 & 0 & 1
\\\hline
b_1/\OR\Z_4^{(1)} &  (0,\frac{1}{4},-\frac{1}{4}) & (2,-1;0,1;1,0) & 0 & -1 & 0 & 2
\\\hline
b_2/\OR & (0,0,0) & (2,-1;1,1;1,1) & 0 & -2 & 0 & 4
\\\hline
c_1/\OR\Z_4^{(1)}\Z_2^{(3)} & (\frac{1}{2},-\frac{1}{4},-\frac{1}{4}) & (0,1;1,0;1,0) & 0 & 0 & -1 & 0 
\\\hline
c_2/\OR\Z_2^{(3)} & (\frac{1}{2},0,-\frac{1}{2}) & (0,1;1,1;1,-1) & 0 & 0 & -2 & 0
\\\hline
   \end{array}
   \end{equation}
    \caption{Some examples of supersymmetric three-cycles on the  {\bf bBB} lattice on $T^6/(\Z_2 \times \Z_4 \times \OR)$
with complex structure modulus $\varrho =2$.
In~\protect\cite{Honecker:2004np}, RR tadpole cancellation was achieved by setting $N_{a_i}=3$ and $N_{b_i}=N_{c_i}=1$ for $i=1,2$. }
\label{Tab:Z2Z4-example-cycles}
  \end{center}
\end{table}

The RR tadpole cancellation conditions~(\ref{Eq:Z2Z4-RRtcc}) on {\bf bBB} with the choice of an exotic O6-plane
$\eta_{\OR} =-1$ read 
\begin{equation}
\begin{aligned}
\sum_a N_a \, V_a=&4,
\\
\sum_a N_a \left( U_a - \frac{1}{2} \, P_a \right)=&  - 6.
\end{aligned}
\end{equation}
A supersymmetric solution is given for $N_{a_1}=4$ and $N_{c_2}=2$ with the cycles as defined in table~\ref{Tab:Z2Z4-example-cycles}, i.e. 
\begin{equation}
\begin{aligned}
\Pi_{a_1} &= \frac{1}{2} \left( \rho_1 + \rho_4 \right)
,
\\
\Pi_{c_2} &=- \rho_2
.
\end{aligned}
\end{equation}
The gauge group is $U(4)_{a_1} \times SO(4)_{c_2}$, where the diagonal $U(1) \subset U(4)_{a_1}$ is anomalous and massive.
The relevant intersection numbers are 
\begin{equation}
\begin{aligned}
\Pi_{a_1}  \circ \Pi_{c_2} &=-2,
\\
\Pi_{a_1}  \circ \Pi_{a_1'} &=0, 
\\
\Pi_{a_1}  \circ \Pi_{O6} &=2,
\end{aligned}
\end{equation}
from which the chiral part of the spectrum is derived,
\begin{equation}
({\bf 6}_{a_1},\1) + (\ov{\bf 10}_{a_1},\1) + 2 \, (\ov{\bf 4}_{a_1},{\bf 4}_{c_2}) .
\end{equation}
Using the computational methods described in appendix~\ref{S:App-OpenStrings}, 
the massless spectrum contains also the following non-chiral matter,
\begin{itemize}
\item
three multiplets in the adjoint representation $({\bf 16}_{a_1},\1)$ of  $U(4)_{a_1}$ from the $a_1a_1$ sector plus 
two further adjoints at orbifold intersections $a_1(\omega a_1)$,
\item
one non-chiral pair of bifundamentals $[ ({\bf 4}_{a_1},{\bf 4}_{c_2})+c.c.]$ under $U(4)_{a_1} \times SO(4)_{c_2}$,
\item
two non-chiral pairs of antisymmetrics $2 \times [({\bf 6}_{a_1},\1) + c.c.]$ of $U(4)_{a_1}$, 
\item
two multiplets in the antisymmetric and one in the symmetric representation of $SO(4)_{c_2}$ from
the $c_2c_2$ sector plus four more antisymmetrics from the $c_2 (\omega c_2)$ sector.
\end{itemize}
The K-theory constraint is trivially fulfilled on the {\bf bBB} lattice, and also due to the construction with gauge groups of even
rank only.

{\boldmath
\section{The $T^6/(\Z_2 \times \Z_6 \times \OR)$
  orientifolds}\label{Ss:Z2Z6orient} 
}

The orbifold shifts for $T^6/\Z_2 \times \Z_6$ are given by
\begin{equation}
\vec{v}=\frac{1}{2}(1,-1,0),
\quad
\vec{w}=\frac{1}{6}(0,1,-1).
\end{equation}
There are two sub-sectors, $\vec{v} + \vec{w} = \frac{1}{6}(3,-2,-1)$ and $\vec{v} + 2\vec{w} = \frac{1}{6}(3,-1,-2)$, which
 are the generators of two different $T^6/\Z_6'$ orbifolds. As will be discussed in detail below, the compactification
lattices and bulk cycles of $T^6/\Z_2 \times \Z_6$ and $T^6/\Z_6'$ are identical up to normalisation.
Also two of the three $\Z_2^{(i)}$ twisted sectors of $T^6/\Z_2 \times \Z_6$ with discrete torsion are inherited 
from the two $T^6/\Z_6'$ subsectors, as discussed in detail in table~\ref{Tab:T6Z2Z6} in appendix~\ref{App:ZN}.
This correspondence allows for using partial results, e.g. the classification of supersymmetric bulk three-cycles, 
from the well-studied $T^6/\Z_6'$ orbifold~\cite{Gmeiner:2007zz,Gmeiner:2008xq} in order to construct 
D6-brane models on the $T^6/\Z_2 \times \Z_6$ orbifold with discrete torsion.
It also gives reason to hope that on this product orbifold, new phenomenologically appealing spectra with three
Standard Model generations will be found.

Choosing the factorisable torus lattice $SU(2)^2 \times SU(3)^2$ leads to the Hodge numbers for $T^6/\Z_2 \times \Z_6$ 
without and with discrete torsion listed in table~\ref{Tab:Hodge-T6ZNxZM-torsion} and the decomposition into $h_{11}^{\pm}$ upon orientifolding
in table~\ref{tab:DecompositionHodge-T6ZNxZM-torsion}.
The three-cycles, on which D6-branes can be wrapped, are counted by the following Hodge numbers 
\begin{equation}
\eta=1: \; h_{21} = 1_{\rm bulk} + 2_{\Z_3} ,
\qquad\qquad
\eta=-1: \; h_{21} = 1_{\rm bulk} + 14_{\Z_2} + 2_{\Z_6} + 2_{\Z_3},
\end{equation}
where only the bulk and $\Z_2$ twisted cycles have an interpretation in terms of open string loop amplitudes
as discussed in section~\ref{S:orbifolds}.

In the absence of discrete torsion, four independent bulk three-cycles can be used for model building, where supersymmetry projects out a one-dimensional sub-space, but leaves enough freedom for engineering chirality.

\subsection{The bulk part}\label{Ss:Z2Z6-bulk-part}

A basis of bulk three-cycles is given by
\begin{equation}
\begin{aligned}
\rho_1 \equiv 4\,\sum_{i=0}^2 \theta^i (\pi_{135}) &= 4 \, \left(\pi_{135} + \pi_{145} - 2 \, \pi_{146} + \pi_{136} \right) ,
\\
\rho_2 \equiv 4\,\sum_{i=0}^2 \theta^i (\pi_{136}) &=4 \, \left(2\, \pi_{136} + 2 \,  \pi_{145} -\pi_{146} -\pi_{135}\right) ,
\\
\rho_3\equiv 4\,\sum_{i=0}^2 \theta^i (\pi_{235})  &=4 \, \left(\pi_{235} + \pi_{245} - 2 \, \pi_{246} +\pi_{236}\right) ,
\\
\rho_4\equiv 4\,\sum_{i=0}^2  \theta^i (\pi_{236})  &=4 \, \left(2 \, \pi_{236} + 2 \, \pi_{245} - \pi_{246} - \pi_{235} \right) ,
\end{aligned}
\end{equation}
with non-vanishing intersection numbers
\begin{equation}
\begin{aligned}
\rho_1 \circ \rho_3 = \rho_2 \circ \rho_4 &= 8 ,
\\
\rho_1 \circ \rho_4 = \rho_2 \circ \rho_3 &= 4 .
\end{aligned}
\end{equation}

The wrapping numbers of a factorisable three-cycle on the underlying torus lattice transform as
\begin{equation}\label{Eq:Z6-1cycle-Orb}
\left(\begin{array}{cc}
n^1 & m^1 \\ n^2 & m^2 \\ n^3 & m^3 
\end{array}\right) \stackrel{\Z_6^{(1)}}{\rightarrow}
\left(\begin{array}{cc}
n^1 & m^1 \\ -m^2 & n^2 +m^2 \\ n^3 + m^3 & -n^3 
\end{array}\right) \stackrel{\Z_6^{(1)}}{\rightarrow}
\left(\begin{array}{cc}
n^1 & m^1 \\-(n^2+m^2) & n^2 \\ m^3 & -(n^3 + m^3)
\end{array}\right) 
\end{equation}
under the $\Z_6$ symmetry.
A bulk three-cycle on the $T^6/\Z_2 \times \Z_6$ orbifold takes thus the form
\begin{equation}
\Pi^{\rm bulk}= P\; \rho_1 + Q \; \rho_2 + U \; \rho_3 + V \; \rho_4
\end{equation}
with the bulk wrapping numbers 
\begin{equation}\label{Eq:Def-PQUV-Z2Z6}
\begin{aligned}
 P \equiv&  n^1 \, X ,
\quad 
Q\equiv n^1 \, Y ,
\quad
U \equiv m^1 \, X ,
\quad
V \equiv m^1 \, Y ,
\\
\text{with }  & \quad X \equiv n^2 n^3 - m^2 m^3 ,
\quad
Y \equiv n^2 m^3  + m^2 n^3 + m^2 m^3 , 
\end{aligned}
\end{equation}
and the intersection number of two bulk cycles  is given by
\begin{equation}
\Pi^{\rm bulk}_a \circ \Pi^{\rm bulk}_b = 4 \left\{ 2 \left( P_aU_b -P_bU_a + Q_a V_b - Q_b V_a \right)
+ P_a V_b - P_b V_a + Q_a U_b - Q_b U_a
\right\} .
\end{equation}
The cycles $\frac{1}{2}\Pi^{\rm bulk}$ thus all have integer intersection numbers among each other. 
They do, however, not form an unimodular basis even in the absence of discrete torsion since there are 
always three-cycles from the $\Z_3^{(1)}$-twisted sector present as briefly discussed below in section~\ref{Ss:Z2Z6-Z3twist}.

The orientifold projection on the bulk cycles is independent of the choice of discrete torsion and the exotic O6-plane and given
for all six inequivalent choices of lattice orientations in table~\ref{Tab:Z2Z6bulk-Orient}.
\begin{table}[h!]
\renewcommand{\arraystretch}{1.3}
  \begin{center}
\begin{equation*}
\begin{array}{|c||c|c|c|c|}\hline
 \multicolumn{5}{|c|}{\text{\bf The orientifold projection on bulk 3-cycles for } T^6/\Z_2 \times \Z_6}
\\\hline\hline
{\rm 3-cycle} & \rho_1 & \rho_2 & \rho_3 & \rho_4
\\\hline\hline
{\bf a/bAA} & \rho_1-(2b)\rho_3 & \rho_1 - \rho_2-(2b)[\rho_3-\rho_4] & -\rho_3 & \rho_4 - \rho_3 
\\\hline
{\bf a/bAB} & \rho_2-(2b)\rho_4 & \rho_1-(2b)\rho_3 & -\rho_4 & -\rho_3
\\\hline
{\bf a/bBB} & \rho_2 -\rho_1 - (2b)[\rho_4-\rho_3] & \rho_2 - (2b) \rho_4 & \rho_3 - \rho_4 & -\rho_4
\\\hline
\end{array}
\end{equation*}
\end{center}
\caption{The orientifold projection on bulk three-cycles on  $T^6/(\Z_2 \times \Z_6 \times \OR)$ without and with discrete torsion.
$b=0, 1/2$ labels the untilted and tilted two-torus $T^2_{(1)}$.}
\label{Tab:Z2Z6bulk-Orient}
\end{table}

The supersymmetry conditions~(\ref{Eq:SUSYgeneral}) on bulk three-cycles with the explicit orbifold expressions~(\ref{Eq:Z-Orb}) and~(\ref{Eq:Z-contribOrb})
read  
\begin{equation}\label{Eq:SUSY-Z2Z6}
\begin{array}{lll}
{\bf a/bAA:}\quad &  3 \, Q + \varrho [2U +V +b \, (2P +Q )] =0 ,\quad & 2 \, P +Q -\varrho[V +b \, Q ]>0 ,
\\
{\bf a/bAB:} \quad&  Q -P + \varrho[U +V  + b \, (P +Q )]=0 ,\quad & 3 \, (P +Q ) + \varrho [U -V +b \, (P -Q )]>0  , 
\\
{\bf a/bBB:} \quad & -3 \, P + \varrho [U +2 \, V  + b \, (P +2 \, Q )] =0  ,\quad &  P + 2 \, Q + \varrho (U +b \, P )>0 ,
\end{array}
\end{equation}
in terms of the bulk wrapping numbers and the complex structure modulus $\varrho \equiv \sqrt{3} \frac{R_2}{R_1}$ on the first 
two-torus.\footnote{The supersymmetry conditions agree with those on $T^6/\Z_6'$ in~\cite{Gmeiner:2007zz,Gmeiner:2008xq,Gmeiner:2009fb} up to permutation of the two-tori and 
with the change in the definition of the complex structure modulus by a factor of two.}

The orientifold planes lie in four orbits, $\OR\omega^{2l}$, $\OR\theta$, $\OR\omega^{2l+1}$ and   $\OR\theta\omega^{2l+1}$ under the 
$\Z_6^{(1)}$ orbifold generator;
we label these orbits in the following by one of their representants namely, $\OR$, $\OR\Z_2^{(3)}$, $\OR\Z_2^{(1)}$ and $\OR\Z_2^{(2)}$,
respectively. The torus wrapping numbers for one representant of each orbit are given in table~\ref{Tab:Z2Z6-Oplanes-torus}
for all inequivalent lattices, and in table~\ref{Tab:Z2Z6-Oplanes-bulk} the corresponding bulk wrapping numbers using
the definition~(\ref{Eq:Def-PQUV-Z2Z6}) are listed.
\begin{table}[h!]
\renewcommand{\arraystretch}{1.3}
  \begin{center}
\begin{equation*}
\begin{array}{|c|c||c|c|c|}\hline
\multicolumn{5}{|c|}{\text{\bf Torus wrapping numbers for the four O6-plane orbits on  } T^6/\Z_2 \times \Z_6}
\\\hline\hline
\text{O6-plane} & \frac{\rm angle}{\pi} & {\bf a/bAA} & {\bf a/bAB} & {\bf a/bBB}
\\\hline
& & \multicolumn{3}{|c|}{(n^1,m^1;n^2,m^2;n^3,m^3)}
\\\hline\hline
\OR & (0,0,0) & (\frac{1}{1-b},\frac{-b}{1-b};1,0;1,0) &  (\frac{1}{1-b},\frac{-b}{1-b};1,0;1,1) 
&  (\frac{1}{1-b},\frac{-b}{1-b};1,1;1,1)
\\
\OR\Z_2^{(1)} & (0,\frac{1}{2},-\frac{1}{2}) &  (\frac{1}{1-b},\frac{-b}{1-b};-1,2;1,-2) &   (\frac{1}{1-b},\frac{-b}{1-b};-1,2;1,-1) 
&   (\frac{1}{1-b},\frac{-b}{1-b};-1,1;1,-1)
\\
\OR\Z_2^{(3)} & (\frac{1}{2},-\frac{1}{2},0) & (0,1;1,-2;1,0) & (0,1;1,-2;1,1) & (0,1;1,-1;1,1)
\\
\OR\Z_2^{(2)}  &  (\frac{1}{2},0,-\frac{1}{2}) &  (0,1;1,0;1,-2) &  (0,1;1,0;1,-1) & (0,1;1,1;1,-1)
\\\hline
\end{array}
\end{equation*}
\end{center}
\caption{The torus wrapping numbers for one representant of each O6-plane orbit on $T^6/(\Z_2 \times \Z_6 \times \OR)$. The angle w.r.t. the $\OR$ invariant 
plane is listed in the second column. The other two torus cycles per orbit can be computed by using~(\protect\ref{Eq:Z6-1cycle-Orb}).}
\label{Tab:Z2Z6-Oplanes-torus}
\end{table}
%
\begin{table}[h!]
\renewcommand{\arraystretch}{1.3}
  \begin{center}
\begin{equation*}
\begin{array}{|c||c|c|c|c||c|c|c|c||c|c|c|c|}\hline
\multicolumn{13}{|c|}{\text{\bf Bulk wrapping numbers for the four O6-plane orbits on  } T^6/\Z_2 \times \Z_6}
\\\hline\hline
 & \multicolumn{4}{|c|}{\bf a/bAA}   & \multicolumn{4}{|c|}{\bf a/bAB} & \multicolumn{4}{|c|}{\bf a/bBB}   
\\
{\rm orbit} & P & Q & U & V &  P & Q & U & V &  P & Q & U & V 
\\\hline\hline
\OR & \frac{1}{1-b} & 0  & \frac{-b}{1-b} & 0  & \frac{1}{1-b}  & \frac{1}{1-b}  & \frac{-b}{1-b} & \frac{-b}{1-b}
& 0  & \frac{3}{1-b} & 0  & \frac{-3\,b}{1-b}
\\
\OR\Z_2^{(1)} & \frac{3}{1-b} & 0  & \frac{-3\,b}{1-b} & 0  & \frac{1}{1-b}  & \frac{1}{1-b}  & \frac{-b}{1-b} & \frac{-b}{1-b}
& 0  & \frac{1}{1-b} & 0  & \frac{-b}{1-b}
\\
\OR\Z_2^{(3)} & 0 & 0 & 1 & -2 & 0 & 0 & 3 & -3 & 0 & 0 & 2 & -1
\\
\OR\Z_2^{(2)} &  0 & 0 & 1 & -2 & 0 & 0 & 1 & -1 & 0 & 0 & 2 & -1
\\\hline
\end{array}
\end{equation*}
\end{center}
\caption{Bulk wrapping numbers for the O6-plane orbits on $T^6/(\Z_2 \times \Z_6 \times \OR)$. The number of identical O6-planes is 
$N_{O6}=2(1-b)$ with $b=0,1/2$ parameterising the shape on the first two-torus.}
\label{Tab:Z2Z6-Oplanes-bulk}
\end{table}

The charge assignment condition~(\ref{eq:ocharge}) on exotic O6-planes reads the same as for $\Z_2 \times \Z_2$,
\begin{equation}
\eta_{\OR} \prod_{k=1}^3 \eta_{\OR\Z_2^{(k)}} = \eta.
\end{equation}
This is due to the fact that the four O6-plane orbits split exactly into the orbits of $\OR$ and $\OR\Z_2^{(k)}$ with $k=1,2,3$ as discussed above. 

The orientifold projection on bulk 3-cycles in table~\ref{Tab:Z2Z6bulk-Orient} and the O6-plane wrapping numbers in table~\ref{Tab:Z2Z6-Oplanes-bulk}
lead to the bulk RR tadpole cancellation conditions for all six lattices,
\begin{equation}\label{Eq:Z2Z6-bulkRRtcc}
\begin{aligned}
&{\bf a/bAA:}\quad  \left[ \sum_a N_a \left( 2 \, P_a + Q_a \right) -
  2^{\frac{5-\eta}{2}} \, \left(\eta_{\OR} + 3 \,
    \eta_{\OR\Z_2^{(1)}}\right)\right](\rho_1 - b\rho_3) =
\\
 & -  \left[\sum_a N_a \left(V_a + b Q_a  \right) +  2^{\frac{5-\eta}{2}}  \, (1-b) \, \left(\eta_{\OR\Z_2^{(2)}} + \eta_{\OR\Z_2^{(3)}}\right)\right]
(-\rho_3 + 2 \, \rho_4) 
,
\\
\\
& {\bf a/bAB:} \quad   \left[\sum_a N_a \left(P_a + Q_a  \right) -  2^{\frac{5-\eta}{2}} \,  \left(\eta_{\OR} + \eta_{\OR\Z_2^{(1)}}\right)\right]
(\rho_1 + \rho_2 -b(\rho_3 + \rho_4)) =
\\
 & - \left[ \sum_a N_a \left(U_a - V_a +b(P_a - Q_a)  \right) -
   2^{\frac{5-\eta}{2}} \, (1-b) \,  \left( \eta_{\OR\Z_2^{(2)}} + 3
     \, \eta_{\OR\Z_2^{(3)}}\right)\right](\rho_3 - \rho_4) 
,
\\
\\
&{\bf a/bBB:}\quad    \left[\sum_a N_a \left(P_a + 2 \, Q_a  \right) -
  2^{\frac{5-\eta}{2}}  \, \left( 3\, \eta_{\OR} +
    \eta_{\OR\Z_2^{(1)}} \right) \right](\rho_2 - b \rho_4) =
\\
 & - \left[\sum_a N_a \left( U_a + b P_a \right) - 2^{\frac{5-\eta}{2}}  \, (1-b) \,  \left(\eta_{\OR\Z_2^{(2)}} + \eta_{\OR\Z_2^{(3)}}\right)\right]
(2\rho_3 - \rho_4) 
.
\end{aligned}
\end{equation}
Per lattice, two independent bulk RR tadpole cancellation conditions arise which depend on the choice of discrete torsion in two ways: 
$2^{\frac{\eta-1}{2}}$ parameterises the fact that without discrete torsion, the D6-branes wrap cycles $\frac{1}{2} \Pi^{\rm bulk}$,
whereas with discrete torsion they wrap $\frac{1}{4}\left(\Pi^{\rm bulk} + \sum^i \Pi^{\Z_2^{(i)}} \right)$, while in both cases
the non-dynamical O6-planes span $\frac{1}{4}\Pi^{\rm bulk}$. 
Furthermore, in the absence of discrete torsion $\eta_{\OR} = \eta_{\OR\Z_2^{(i)}}=1$ for all $i=1,2,3$, whereas with discrete torsion, one O6-plane has to be chosen exotic.

Before discussing the exceptional three-cycles, let us discuss how a multiple counting of equivalent models is avoided.
The following conditions can be imposed:
\begin{itemize}
\item
\framebox{$(n^3,m^3)=$(odd,odd)} selects the representation through a specific $\Z_6$ orbifold image,
\item
\framebox{$n^3>0$} fixes the orientation on the last two-torus $T^2_{(3)}$,
\item
\framebox{$m^1+b \, n^1 \geq 0$ and $n^1 >  0$} or \framebox{$(n^1,m^1) =(0,1)$}
fixes the orientation on $T^2_{(1)}$ and selects the orientifold image with angle $0 \leq \pi \phi^{(1)} \leq \frac{1}{2}$,
\item
\framebox{for $(n^1,m^1) \in \{(\frac{1}{1-b}, \frac{-b}{1-b}), (0,1)\}$}, the orientifold images are not yet distinguished,
 in this case impose as additional condition on $T^2_{(3)}$ that $-\frac{\pi}{2} < \pi \phi^{(3)} < 0$, which in terms of the wrapping numbers
reads \framebox{$-2n^3 <m^3 < 0$ on {\bf A} and $|m^3| <n^3$ on {\bf B}}.
\item
D6-branes parallel to some $\OR\theta^n\omega^m$ invariant plane are treated separately. 
The torus wrapping numbers of one representant per orbit are listed in table~\ref{Tab:Z2Z6-Oplanes-torus}.
\end{itemize}
These conditions simplify the systematic analysis of fractional supersymmetric three-cycles in the case with discrete torsion significantly, since the assignments of exceptional three-cycles in tables~\ref{Tab:Z2Z6-Z1sector}
and~\ref{Tab:Z2Z6-Z2sector} only need to be considered for $(n^3,m^3)=$ (odd,odd).

{\boldmath
\subsection{The $\Z_2$ twisted parts}\label{Ss:Z2Z6-twisted-part}
}

The $T^6/\Z_2 \times \Z_6$ orbifold  with discrete torsion has three sectors of exceptional three-cycles at $\Z_2^{(i)}$ fixed points
with $i=1,2,3$ labelling the two-torus which is left invariant. These exceptional three-cycles are:
\begin{enumerate}
\item
in the $\Z_2^{(1)}$ sector: twelve independent three-cycles
\begin{equation}
\begin{aligned}
\varepsilon^{(1)}_1 = 6 \, e^{(1)}_{11} \otimes \pi_1, 
\qquad\qquad\quad\quad
&,\quad\quad 
\tilde{\varepsilon}^{(1)}_1 = 6\, e^{(1)}_{11}  \otimes \pi_2,
\\
\varepsilon^{(1)}_1 =2 \left(e^{(1)}_{41} + e^{(1)}_{51} + e^{(1)}_{61}  \right) \otimes \pi_1 
&,\quad\quad 
\tilde{\varepsilon}^{(1)}_1 =2 \left(e^{(1)}_{41} + e^{(1)}_{51} + e^{(1)}_{61}  \right) \otimes \pi_2,
\\
 \varepsilon^{(1)}_2 =2 \left(e^{(1)}_{14} + e^{(1)}_{15} + e^{(1)}_{16}  \right) \otimes \pi_1 
&,\quad\quad 
\tilde{\varepsilon}^{(1)}_2 =2 \left(e^{(1)}_{14} + e^{(1)}_{15} + e^{(1)}_{16}  \right) \otimes \pi_2,
\\
\varepsilon^{(1)}_3 =2 \left(e^{(1)}_{44} + e^{(1)}_{56} + e^{(1)}_{65}  \right) \otimes \pi_1 
&,\quad\quad 
\tilde{\varepsilon}^{(1)}_3 =2 \left(e^{(1)}_{44} + e^{(1)}_{56} + e^{(1)}_{65}  \right) \otimes \pi_2,
\\
 \varepsilon^{(1)}_4 =2 \left(e^{(1)}_{45} + e^{(1)}_{54} + e^{(1)}_{66}  \right) \otimes \pi_1 
&,\quad\quad 
\tilde{\varepsilon}^{(1)}_4 =2 \left(e^{(1)}_{45} + e^{(1)}_{54} + e^{(1)}_{66}  \right) \otimes \pi_2,
\\ 
 \varepsilon^{(1)}_5 =2 \left(e^{(1)}_{46} + e^{(1)}_{55} + e^{(1)}_{64}  \right) \otimes \pi_1 
&,\quad\quad 
\tilde{\varepsilon}^{(1)}_5 =2 \left(e^{(1)}_{46} + e^{(1)}_{55} + e^{(1)}_{64}  \right) \otimes \pi_2,
\end{aligned}
\end{equation}
with intersection form
\begin{equation}
\begin{aligned}
\varepsilon^{(1)}_0 \circ \tilde{\varepsilon}^{(1)}_0 &= -12 , 
\\
\varepsilon^{(1)}_i \circ \tilde{\varepsilon}^{(1)}_j &= -4 \, \delta_{ij} \qquad \text{ for } i,j \in \{1 \ldots 5\}
,
\end{aligned}
\end{equation}
and all other intersections vanishing.
\item
in the $\Z_2^{(\alpha)}$ sector with $\alpha \in \{2,3\}$, we label the fixed points on $T^2_{(1)}$ by $k=1,2,3,4$
and define the following exceptional three-cycles
\begin{equation}
\begin{aligned}
&\varepsilon^{(\alpha)}_k =2 \, \left( e^{(\alpha)}_{k4} \otimes \pi_3 + e^{(\alpha)}_{k6} \otimes \pi_{-4} + e^{(\alpha)}_{k5} \otimes \pi_{4-3}  \right)
,
\\
& \tilde{\varepsilon}^{(\alpha)}_k=2 \, \left( e^{(\alpha)}_{k4} \otimes \pi_4 + e^{(\alpha)}_{k6} \otimes \pi_{3-4} + e^{(\alpha)}_{k5} \otimes \pi_{-3}  \right)
,
\end{aligned}
\end{equation}
with intersection form
\begin{equation}
\varepsilon^{(\alpha)}_k  \circ \tilde{\varepsilon}^{(\alpha)}_l= -4 \,  \delta_{kl} ,
\end{equation}
and all others vanishing.
These two sectors are (up to normalisation and permutation of tori) equivalent to the exceptional three-cycles at $\Z_2$ fixed points on 
the $T^6/\Z_6'$ orbifold discussed in~\cite{Gmeiner:2007zz,Gmeiner:2008xq}.
\end{enumerate}

Supersymmetric fractional three-cycles are composed of a bulk cycle satisfying~(\ref{Eq:SUSY-Z2Z6}) and those
exceptional cycles whose fixed points are traversed by the bulk cycle for a given displacement $\vec{\sigma}$ on 
$T^2_{(1)} \times T^2_{(2)} \times T^2_{(3)}$. The assignemt of a single $\Z_2^{(i)}$ fixed point and its 
exceptional cycle is given in table~\ref{Tab:Z2fixedpoints+cycles}.
\begin{table}[ht]
\renewcommand{\arraystretch}{1.3}
  \begin{minipage}[b]{0.5\linewidth}\centering
   \begin{equation*}
\begin{array}{|c|c|}\hline
\multicolumn{2}{|c|}{\Z_2^{(1)} \;  \text{\bf fixed points and 3-cycles}}
\\\hline\hline
\!\! {\rm f.p.}^{(1)} \otimes (n^1 \pi_{1} + m^1 \pi_{2})\!\! & {\rm orbit} 
\\\hline\hline
11 & n^1 \varepsilon^{(1)}_0 + m^1 \tilde{\varepsilon}^{(1)}_0
\\\hline
41, \; 51, \; 61 
 & n^1 \varepsilon^{(1)}_1 + m^1 \tilde{\varepsilon}^{(1)}_1
\\\hline
14, \; 15, \; 16  & n^1 \varepsilon^{(1)}_2 + m^1 \tilde{\varepsilon}^{(1)}_2
\\\hline
44, \; 56, \; 65 &  n^1 \varepsilon^{(1)}_3 + m^1 \tilde{\varepsilon}^{(1)}_3
\\\hline 
45, \; 54, \; 66 &  n^1 \varepsilon^{(1)}_4 + m^1 \tilde{\varepsilon}^{(1)}_4
\\\hline
46, \; 55,  \; 64 &  n^1 \varepsilon^{(1)}_5 + m^1 \tilde{\varepsilon}^{(1)}_5
\\\hline
\end{array}
\end{equation*}
\end{minipage}
\hspace{0.5cm}
\begin{minipage}[b]{0.5\linewidth}
\centering
\begin{equation*}\!\!\!\!\!\!\!\!\!\!
\begin{array}{|c|c|}\hline
\multicolumn{2}{|c|}{\Z_2^{(2)} \;  \text{\bf and } \Z_2^{(3)} \; \text{\bf fixed points and 3-cycles}}
\\\hline\hline
\!\!{\rm f.p.}^{(\alpha)} \otimes (n^{\alpha} \pi_{2\alpha-1} + m^{\alpha} \pi_{2\alpha})\!\! & {\rm orbit} 
\\\hline\hline
k1 & - 
\\\hline
k4 & n^{\alpha}  \varepsilon^{(\alpha)}_k + m^{\alpha} \tilde{\varepsilon}^{(\alpha)}_k
\\\hline
k5 & m^{\alpha}  \varepsilon^{(\alpha)}_k -(n^{\alpha} + m^{\alpha}) \tilde{\varepsilon}^{(\alpha)}_k
\\\hline
k6 &  -(n^{\alpha} +m^{\alpha}) \varepsilon^{(\alpha)}_k + n^{\alpha} \tilde{\varepsilon}^{(\alpha)}_k
\\\hline
\end{array}
\end{equation*}
 \end{minipage}
\caption{Correspondence between $\Z_2^{(i)}$ fixed points and
  exceptional three-cycles for $i=1$ on the left 
and $\alpha \in \{2,3\}$ on the right on the $T^6/\Z_2 \times \Z_6$ orbifold.}
\label{Tab:Z2fixedpoints+cycles}
\end{table}
We relegate the detailed discussion on the composition of exceptional
three-cycles in dependence of wrapping  
numbers, discrete Wilson lines and displacements to tables~\ref{Tab:Z2Z6-Z1sector},~\ref{Tab:Z2Z6-Z2sector} 
and~\ref{Tab:Z2Z6-Z3sector} in appendix~\ref{App:Tables-Ex-Sectors}, but notice here that only the configurations
with $(n^3,m^3)$ need to be evaluated explicitly.

The orientifold projection on exceptional three-cycles depends on the transformation of the $\Z_2^{(i)}$ fixed 
points under ${\cal R}$ and the sign factor~(\ref{eq:def-eta-i}) under $\Omega$, that is $\eta_{(i)} \equiv \eta_{\OR} \eta_{\OR\Z_2^{(i)}}$, which depends on the choice of 
the exotic O6-plane. The results for the $i^{\rm th}$ exceptional sector are given in 
table~\ref{Tab:Z2Z6-exOrientifoldImages1},~\ref{Tab:Z2Z6-exOrientifoldImages2} 
and~\ref{Tab:Z2Z6-exOrientifoldImages3} for $i=1,2,3$, respectively.
  
\begin{table}[ht]
\renewcommand{\arraystretch}{1.3}
\begin{center}
\begin{equation*}
\begin{array}{|c||c|c||c|c|}\hline
\multicolumn{5}{|c|}{\OR \; \text{\bf  on exceptional 3-cycles for $T^6/\Z_2 \times \Z_6$, Part I}}
\\\hline\hline
\text{3-cycle} & \varepsilon^{(1)}_i & \tilde{\varepsilon}^{(1)}_i & i=i' & i \leftrightarrow i'
\\\hline\hline
\begin{array}{c}
{\bf a/bAA} \\\hline {\bf a/bAB} \\\hline {\bf a/bBB} 
\end{array}
& \eta_{(1)} \left( -\varepsilon^{(1)}_{i'} + (2b) \tilde{\varepsilon}^{(1)}_{i'} \right)
& \eta_{(1)} \, \tilde{\varepsilon}^{(1)}_{i'}
&\begin{array}{c}
0,1,2,3 \\\hline 0,1,2,5 \\\hline 0,1,2,4  
\end{array} 
&\begin{array}{c}
4,5 \\\hline 3,4  \\\hline  3,5 
\end{array} 
\\\hline
\end{array}
\end{equation*}
 \end{center}
\caption{Orientifold projection on exceptional three-cycles in the $\Z_2^{(1)}$ twisted sector on 
$T^6/(\Z_2 \times \Z_6 \times \OR)$ with discrete torsion. The
sign factor $\eta_{(1)} \equiv \eta_{\OR} \eta_{\OR\Z_2^{(1)}}$ depends on the choice of the exotic O6-plane. }
\label{Tab:Z2Z6-exOrientifoldImages1}
\end{table}

\begin{table}[ht]
\renewcommand{\arraystretch}{1.3}
\begin{center}
\begin{equation*}
\begin{array}{|c||c|c||c|c|}\hline
\multicolumn{5}{|c|}{\OR \; \text{\bf  on exceptional 3-cycles for $T^6/\Z_2 \times \Z_6$, Part II}}
\\\hline\hline
\text{3-cycle} & \varepsilon^{(2)}_i & \tilde{\varepsilon}^{(2)}_i & i=i' & i \leftrightarrow i'
\\\hline\hline
\begin{array}{c}
{\bf a/bAA} \\\hline {\bf a/bAB} \\\hline {\bf a/bBB} 
\end{array}
&\begin{array}{c} - \eta_{(2)} \, \varepsilon^{(2)}_{i'} \\\hline
\eta_{(2)} \,  \tilde{\varepsilon}^{(2)}_{i'} \\\hline
\eta_{(2)} \left( \tilde{\varepsilon}^{(2)}_{i'} -\varepsilon^{(2)}_{i'} \right)
\end{array}
&\begin{array}{c} \eta_{(2)} \left( \tilde{\varepsilon}^{(2)}_{i'} -\varepsilon^{(2)}_{i'} \right)\\\hline
\eta_{(2)} \, \varepsilon^{(2)}_{i'} \\\hline
\eta_{(2)} \,  \tilde{\varepsilon}^{(2)}_{i'}
\\\end{array}
& 1,4
& 2+2b, 3-3b
\\\hline
\end{array}
\end{equation*}
 \end{center}
\caption{Orientifold projection on exceptional cycles in the $\Z_2^{(2)}$ twisted sector on 
$T^6/(\Z_2 \times \Z_6 \times \OR)$ with discrete torsion. }
\label{Tab:Z2Z6-exOrientifoldImages2}
\end{table}

\begin{table}[ht]
\renewcommand{\arraystretch}{1.3}
\begin{center}
\begin{equation*}
\begin{array}{|c||c|c||c|c|}\hline
\multicolumn{5}{|c|}{\OR \; \text{\bf  on exceptional 3-cycles for $T^6/\Z_2 \times \Z_6$, Part III}}
\\\hline\hline
\text{3-cycle} & \varepsilon^{(3)}_i & \tilde{\varepsilon}^{(3)}_i & i=i' & i \leftrightarrow i'
\\\hline\hline
\begin{array}{c}
{\bf a/bAA} \\\hline {\bf a/bAB} \\\hline {\bf a/bBB} 
\end{array}
&\begin{array}{c} - \eta_{(3)} \, \varepsilon^{(3)}_{i'} \\\hline
- \eta_{(3)} \,  \tilde{\varepsilon}^{(3)}_{i'} \\\hline
\eta_{(3)} \left( \tilde{\varepsilon}^{(3)}_{i'} -\varepsilon^{(3)}_{i'} \right)
\end{array}
&\begin{array}{c} \eta_{(3)} \left( \tilde{\varepsilon}^{(3)}_{i'} -\varepsilon^{(3)}_{i'} \right)\\\hline
- \eta_{(3)} \, \varepsilon^{(3)}_{i'} \\\hline
\eta_{(3)} \,  \tilde{\varepsilon}^{(3)}_{i'}
\\\end{array}
& 1,4
& 2+2b, 3-3b
\\\hline
\end{array}
\end{equation*}
 \end{center}
\caption{Orientifold projection on exceptional cycles  in the $\Z_2^{(3)}$ twisted sector on 
$T^6/(\Z_2 \times \Z_6 \times \OR)$ with discrete torsion. }
\label{Tab:Z2Z6-exOrientifoldImages3}
\end{table}

A fractional three-cycle can be formally expanded as 
\begin{equation}\label{Eq:Expand-Frac-Z2Z6}
\begin{aligned}
\Pi^{\rm frac} =  \frac{1}{4} \Pi^{\rm bulk}  + \frac{1}{4} \sum_{\alpha=1}^3 \Pi^{\Z_2^{(\alpha)}}
=& \quad\frac{1}{4} \left( P \rho_1 + Q  \rho_2 + U \rho_3 + V \rho_4 \right)\\
& \hspace*{-1in}+ \frac{1}{4} \left[\sum_{i=0}^5 \left( x^{(1)}_i \,
    \varepsilon_i^{(1)} + y ^{(1)}_i \, \tilde{\varepsilon}_i^{(1)}
  \right) 
+ \sum_{\alpha=2,3} \sum_{i=1}^4 \left( x^{(\alpha)}_i \, \varepsilon_i^{(\alpha)} + y ^{(\alpha)}_i \, \tilde{\varepsilon}_i^{(\alpha)} \right)
\right]
,
\end{aligned}
\end{equation}
where the integer coefficients $(x^{(\alpha)}_i,y^{(\alpha)}_i)$ can be read off from tables~\ref{Tab:Z2Z6-Z1sector} to \ref{Tab:Z2Z6-Z3sector} in appendix~\ref{App:Tables-Ex-Sectors}.
The coefficients can be classified as follows: each coefficient can receive contributions from one (I) or two (II) fixed points, and 
\begin{itemize}
\item
for the twist sector $\alpha=1$, exactly three pairs $(x^{(1)}_i,y^{(1)}_i)$ are non-vanishing for 
$(\sigma_2,\sigma_3) \neq (0,0)$, and four pairs for $(\sigma_2,\sigma_3) = (0,0)$. The remaining three (or two) pairs 
have zero entries.
\begin{equation}\nonumber
\begin{array}{|c|c|}\hline
\multicolumn{2}{|c|}{(x^{(1)}_i ,y^{(1)}_i)}
\\\hline\hline
{\rm I.} & {\rm II.}
\\\hline\hline
\pm (n^1,m^1) & \pm z \, (n^1,m^1)
\\\hline
\end{array}
\qquad
\text{with}
\quad
z=\left\{\begin{array}{c}
1+(-1)^{\tau_2}
\\
1+(-1)^{\tau_3}
\\
(-1)^{\tau_2}+(-1)^{\tau_3}
\end{array}\right\}
\in \{0,\pm 2\}
.
\end{equation}
The global signs depend on the choice of the $\Z_2^{(1)}$ eigenvalue and the discrete Wilson lines $(\tau_2,\tau_3)$ as detailed in table~\ref{Tab:Z2Z6-Z1sector}.
\item
per twist sector $\alpha \in \{2,3\}$, exactly two of four pairs
$(x^{(\alpha)}_i,y^{(\alpha)}_i)$  
are non-vanishing.
The remaining two pairs with $i \in \{1 \ldots 4\}$ have zero-entries.
As for the $\Z_2^{(1)}$ twisted sector, there are exactly six different possibilities (up to global signs)
\begin{equation}\nonumber
\begin{array}{|c|c|}\hline
\multicolumn{2}{|c|}{(x^{(\alpha)}_i,y^{(\alpha)}_i)}
\\\hline\hline
{\rm I.} & {\rm II.}
\\\hline\hline
\pm (n^{\alpha},m^{\alpha}) & \pm (-z \, n^{\alpha} + [1-z]\, m^{\alpha}, [z -1] \, n^{\alpha} -m^{\alpha})
\\
\pm (-n^{\alpha}-m^{\alpha},n^{\alpha}) & \pm (n^{\alpha} + z\, m^{\alpha}, -z \, n^{\alpha} +[1-z ] \, m^{\alpha})
\\
\pm (m^{\alpha},-n^{\alpha}-m^{\alpha}) &  \pm ([1-z ] \, n^{\alpha}-z \,  m^{\alpha} , z\, n^{\alpha} + m^{\alpha})
\\\hline
\end{array}
\end{equation}
with $z=(-1)^{\tau_{\beta}}$ and $(\alpha,\beta) \in \{(2,3),(3,2)\}$. The global signs depend on the choice of the $\Z_2^{(\alpha)}$ eigenvalue 
and the discrete Wilson lines $(\tau_1,\tau_{\beta})$ as detailed in tables~\ref{Tab:Z2Z6-Z2sector} and~\ref{Tab:Z2Z6-Z3sector}.
\end{itemize}
The shape of the coefficients $(x^{(\alpha)}_i,y^{(\alpha)}_i)$ is relevant for possible simplifications of 
the \mbox{K-theory} constraint to be discussed below in section~\ref{Ss:Z2Z6-Ktheory}.

In terms of the expansion~(\ref{Eq:Expand-Frac-Z2Z6}), the twisted RR tadpole cancellation conditions on \mbox{$T^6/(\Z_2 \times \Z_6 \times \OR)$} with discrete torsion read on the six different choices of lattice orientations,
\begin{equation}\label{Eq:Z2Z6-twistedRRtcc_AA}
\begin{aligned}
& {\bf a/bAA:}\\  
&\sum_{i=0}^3 \left(\sum_a N_a (1-\eta_{(1)}) x^{(1)}_{i,a}
\right)\varepsilon_{i}^{(1)}  
+ \sum_{i=0}^3 \left(\sum_a N_a [(1+ \eta_{(1)}) y^{(1)}_{i,a} +
  \eta_{(1)} 2b \, x^{(1)}_{i,a} ] \right)
\tilde{\varepsilon}_{i}^{(1)} =
\\
& - \left(\sum_a N_a (x^{(1)}_{4,a} - \eta_{(1)} x^{(1)}_{5,a}) \right) [ \varepsilon_{4}^{(1)} - \eta_{(1)} \varepsilon_{5}^{(1)} - b \, (\tilde{\varepsilon}_{4}^{(1)} - \eta_{(1)} \tilde{\varepsilon}_{5}^{(1)})]\\
& - \left(\sum_a N_a [y^{(1)}_{4,a} + \eta_{(1)}y^{(1)}_{5,a} + b \, (x^{(1)}_{4,a} + \eta_{(1)}x^{(1)}_{5,a})] \right) \,( \tilde{\varepsilon}_{4}^{(1)} + \eta_{(1)}\tilde{\varepsilon}_{5}^{(1)})
,
\\
& \sum_{\alpha=2}^3 \sum_{i=1,4} \left\{ \left(\sum_a N_a  
[(1-\eta_{(\alpha)})x^{(\alpha)}_{i,a} -\eta_{(\alpha)} y^{(\alpha)}_{i,a} ] \right)\varepsilon_{i}^{(\alpha)} 
\!+\! \left(\!\sum_a N_a  (1+\eta_{(\alpha)})y^{(\alpha)}_{i,a}
\right)\tilde{\varepsilon}_{i}^{(\alpha)} \right\} =
\\
& - \sum_{\alpha=2}^3\left( \sum_a N_a [x^{(\alpha)}_{2+2b,a} - \eta_{(\alpha)} x^{(\alpha)}_{3-2b,a} + \frac{1}{2}(y^{(\alpha)}_{2+2b,a} - \eta_{(\alpha)} y^{(\alpha)}_{3-2b,a})]  \right)( \varepsilon_{2+2b}^{(\alpha)} - \eta_{(\alpha)} \varepsilon_{3-2b}^{(\alpha)})
\\
& - \sum_{\alpha=2}^3\left(\sum_a N_a (y^{(\alpha)}_{2+2b,a} + \eta_{(2)} y^{(\alpha)}_{3-2b,a}) \right) [ \tilde{\varepsilon}_{2+2b}^{(\alpha)} - \eta_{(\alpha)} \tilde{\varepsilon}_{3-2b}^{(\alpha)}-\frac{1}{2} (\varepsilon_{2+2b}^{(\alpha)} + \eta_{(\alpha)} \varepsilon_{3-2b}^{(\alpha)} )]
,
\end{aligned}
\end{equation}

\begin{equation}\label{Eq:Z2Z6-twistedRRtcc_AB}
\begin{aligned}
&{\bf a/bAB:} \\
& \!\!\sum_{i=0,1,2,5} \!\left(\sum_a N_a (1-\eta_{(1)}) x^{(1)}_{i,a}  \right)\varepsilon_{i}^{(1)} 
+\!\! \!\!\sum_{i=0,1,2,5}\! \left(\sum_a N_a [(1+ \eta_{(1)}) y^{(1)}_{i,a} +
  \eta_{(1)} 2b \, x^{(1)}_{i,a} ] \right)
\tilde{\varepsilon}_{i}^{(1)} =
\\
& - \left(\sum_a N_a(x^{(1)}_{3,a} - \eta_{(1)} x^{(1)}_{4,a})  \right)
[ \varepsilon_{3}^{(1)} - \eta_{(1)} \varepsilon_{4}^{(1)} - b\, (\tilde{\varepsilon}_{3}^{(1)} - \eta_{(1)} \tilde{\varepsilon}_{4}^{(1)})]
\\
& - \left(\sum_a N_a[y^{(1)}_{3,a} + \eta_{(1)}y^{(1)}_{4,a} + b \, (x^{(1)}_{3,a} + \eta_{(1)}x^{(1)}_{4,a})]  \right)\,( \tilde{\varepsilon}_{3}^{(1)} + \eta_{(1)}\tilde{\varepsilon}_{4}^{(1)})
,
\\
& \sum_{\alpha=2}^3\sum_{i=1,4}   \left(\sum_a N_a 
(x^{(\alpha)}_{i,a} +(-1)^{\alpha} \, \eta_{(\alpha)} \,
y^{(\alpha)}_{i,a} ) \right) [ \varepsilon_{i}^{(\alpha)} +
(-1)^{\alpha} \,\eta_{(\alpha)}  \tilde{\varepsilon}_i^{(\alpha)}] =
\\
& - \sum_{\alpha=2}^3\left(\sum_a N_a (x^{(\alpha)}_{2+2b,a} + (-1)^{\alpha} \,\eta_{(\alpha)}y^{(\alpha)}_{3-2b,a} ) \right) (\varepsilon_{2+2b}^{(\alpha)} + (-1)^{\alpha} \,\eta_{(\alpha)} \tilde{\varepsilon}_{3-2b}^{(\alpha)})
\\
& - \sum_{\alpha=2}^3\left(\sum_a N_a (x^{(\alpha)}_{3-2b,a} + (-1)^{\alpha} \,\eta_{(\alpha)}y^{(\alpha)}_{2+2b,a} ) \right) (\varepsilon_{3-2b}^{(\alpha)} + (-1)^{\alpha} \,\eta_{(\alpha)} \tilde{\varepsilon}_{2+2b}^{(\alpha)}) 
,
\end{aligned}
\end{equation}
\begin{equation}\label{Eq:Z2Z6-twistedRRtcc_BB}
\begin{aligned}
&{\bf a/bBB:} \\&\!\!\! \sum_{i=0,1,2,4}\!\! \left(\sum_a N_a (1-\eta_{(1)}) x^{(1)}_{i,a}  \right)\varepsilon_{i}^{(1)} 
+\!\!\!\! \sum_{i=0,1,2,4}\!\! \left(\sum_a N_a [(1+ \eta_{(1)})
  y^{(1)}_{i,a} +  \eta_{(1)} 2b \, x^{(1)}_{i,a} ] \right)
\tilde{\varepsilon}_{i}^{(1)} =
\\
& -\left(  \sum_a N_a(x^{(1)}_{3,a} - \eta_{(1)} x^{(1)}_{5,a}) \right)
[ \varepsilon_{3}^{(1)} - \eta_{(1)} \varepsilon_{5}^{(1)} - b \, (\tilde{\varepsilon}_{3}^{(1)} - \eta_{(1)} \tilde{\varepsilon}_{5}^{(1)})]
\\
& - \left(\sum_a N_a[y^{(1)}_{3,a} + \eta_{(1)}y^{(1)}_{5,a} + b \,  (x^{(1)}_{3,a} + \eta_{(1)}x^{(1)}_{5,a})] \right) \,( \tilde{\varepsilon}_{3}^{(1)} + \eta_{(1)}\tilde{\varepsilon}_{5}^{(1)})
,
\\
&\! \sum_{\alpha=2}^3 \!\!\sum_{i=1,4}\! \left\{ \left(\! \sum_a N_a
    (1-\eta_{(\alpha)})x^{(\alpha)}_{i,a} \right)
  \varepsilon_{i}^{(\alpha)}  
+ \left( \sum_a N_a [(1+ \eta_{(\alpha)})y^{(\alpha)}_{i,a}  +
  \eta_{(\alpha)} x^{(\alpha)}_{i,a}  ] \right)
\tilde{\varepsilon}_{i}^{(\alpha)}  \right\} =
\\
& -\sum_{\alpha=2}^3\left( \sum_a N_a (x^{(\alpha)}_{2+2b,a} - \eta_{(\alpha)} x^{(\alpha)}_{3-2b,a}) \right)
[\varepsilon_{2+2b}^{(\alpha)} - \eta_{(\alpha)}\varepsilon_{3-2b}^{(\alpha)} - \frac{1}{2}( \tilde{\varepsilon}_{2+2b}^{(\alpha)} - \eta_{(\alpha)} \tilde{\varepsilon}_{3-2b}^{(\alpha)} )]
\\
& - \sum_{\alpha=2}^3 \left( \sum_a N_a [y^{(\alpha)}_{2+2b,a} + \eta_{(\alpha)} y^{(\alpha)}_{3-2b,a} + \frac{1}{2} (x^{(\alpha)}_{2+2b,a} + \eta_{(\alpha)} x^{(\alpha)}_{3-2b,a})] \right)
(\tilde{\varepsilon}_{2+2b}^{(\alpha)} + \eta_{(\eta)} \tilde{\varepsilon}_{3-2b}^{(\alpha)} ) 
.
\end{aligned}
\end{equation}

We will use the bulk and twisted RR tadpole cancellation
conditions~(\ref{Eq:Z2Z6-bulkRRtcc})
and~(\ref{Eq:Z2Z6-twistedRRtcc_AA}}) 
to~(\ref{Eq:Z2Z6-twistedRRtcc_BB})
 in section~\ref{Ss:Z2Z6-example-with-torsion} to construct an example of a globally consistent D6-brane model
on the $T^6/(\Z_2 \times \Z_6 \times \OR)$ orientifold with discrete torsion.

\subsection{The K-theory constraint}\label{Ss:Z2Z6-Ktheory}

The discussion of the K-theory constraint follows closely the one for $T^6/(\Z_2 \times \Z_2 \times \OR)$ in section~\ref{Ss:Z2Z2Ktheory}. 
The classification of $\OR$ invariant three-cycles turns out to be given again by table~\ref{Tab:Conditions-on_b+t+s-SOSp}, 
and the exceptional building blocks of the three-cycles in~(\ref{Eq:define-probes}) are listed in tables~\ref{Tab:Z2Z6-Ktheory_AA} 
to~\ref{Tab:Z2Z6-Ktheory_BB}.

\begin{table}[h!]
\renewcommand{\arraystretch}{1.3}
  \begin{center}
\begin{equation*}
\begin{array}{|c|c|}\hline
\multicolumn{2}{|c|}{\text{\bf Exceptional contributions to $\OR$ invariant 3-cycles for $T^6/\Z_2 \times \Z_6$ on a/bAA} }
\\\hline\hline
\multicolumn{2}{|c|}{{\bf a/bAA}:  \Z_2^{(1)}}
\\\hline\hline
\Pi_{h,(0,0)}^{\Z_2^{(1)}} &  
\frac{1}{1-b} \Bigl\{ [ \varepsilon^{(1)}_0 -b \tilde{\varepsilon}^{(1)}_0] + (-1)^{\tau_2}[ \varepsilon^{(1)}_1 -b \tilde{\varepsilon}^{(1)}_1]
+ (-1)^{\tau_3}[  \varepsilon^{(1)}_2 -b \tilde{\varepsilon}^{(1)}_2] 
+ (-1)^{\tau_2 + \tau_3} [  \varepsilon^{(1)}_3 -b \tilde{\varepsilon}^{(1)}_3]\Bigr\}
\\\hline
\Pi_{h,(1,0)}^{\Z_2^{(1)}} &  \frac{1}{1-b} \Bigl\{[1+(-1)^{\tau_2}][ \varepsilon^{(1)}_1 -b  \tilde{\varepsilon}^{(1)}_1]
 +(-1)^{\tau_2+ \tau_3}[ \varepsilon^{(1)}_4 -b \tilde{\varepsilon}^{(1)}_4]
 +(-1)^{\tau_3}[ \varepsilon^{(1)}_5 -b \tilde{\varepsilon}^{(1)}_5]
 \Bigr\}
\\\hline
\Pi_{h,(0,1)}^{\Z_2^{(1)}} &  \frac{1}{1-b} \Bigl\{ [1+(-1)^{\tau_3}][ \varepsilon^{(1)}_2 -b \tilde{\varepsilon}^{(1)}_2]
 + (-1)^{\tau_2}[ \varepsilon^{(1)}_4 -b \tilde{\varepsilon}^{(1)}_4]
 + (-1)^{\tau_2+ \tau_3}[ \varepsilon^{(1)}_5 -b \tilde{\varepsilon}^{(1)}_5]\Bigr\}
\\\hline
\Pi_{h,(1,1)}^{\Z_2^{(1)}} &  \frac{1}{1-b} \Bigl\{ [1 + (-1)^{\tau_2+ \tau_3}][ \varepsilon^{(1)}_3 -b \tilde{\varepsilon}^{(1)}_3]
 +(-1)^{\tau_3}[ \varepsilon^{(1)}_4 -b  \tilde{\varepsilon}^{(1)}_4] 
+ (-1)^{\tau_2}[ \varepsilon^{(1)}_5 -b  \tilde{\varepsilon}^{(1)}_5]\Bigr\}
\\\hline\hline
\Pi_{v,(0,0)}^{\Z_2^{(1)}} & \tilde{\varepsilon}^{(1)}_0 +  (-1)^{\tau_2} \tilde{\varepsilon}^{(1)}_1
 + (-1)^{\tau_3}  \tilde{\varepsilon}^{(1)}_2 + (-1)^{\tau_2 + \tau_3} \tilde{\varepsilon}^{(1)}_3
\\\hline
\Pi_{v,(1,0)}^{\Z_2^{(1)}} & [1+(-1)^{\tau_2}] \tilde{\varepsilon}^{(1)}_1
+(-1)^{\tau_2+ \tau_3} \tilde{\varepsilon}^{(1)}_4
 +(-1)^{\tau_3} \tilde{\varepsilon}^{(1)}_5 
\\\hline
\Pi_{v,(0,1)}^{\Z_2^{(1)}} &  [1+(-1)^{\tau_3}] \tilde{\varepsilon}^{(1)}_2
 + (-1)^{\tau_2} \tilde{\varepsilon}^{(1)}_4 + (-1)^{\tau_2+ \tau_3}\tilde{\varepsilon}^{(1)}_5
\\\hline
\Pi_{v,(1,1)}^{\Z_2^{(1)}} & [1 + (-1)^{\tau_2+ \tau_3}] \tilde{\varepsilon}^{(1)}_3
 +(-1)^{\tau_3} \tilde{\varepsilon}^{(1)}_4 + (-1)^{\tau_2} \tilde{\varepsilon}^{(1)}_5 
\\\hline\hline\hline 
\multicolumn{2}{|c|}{{\bf a/bAA}: \Z_2^{(i)} \text{ with } (i,j) \in \{(2,3),(3,2)\} }
\\\hline\hline
\Pi_{h,(\sigma_1,0)}^{\Z_2^{(i)}} &  -(-1)^{\tau_j} \left[
\varepsilon^{(i)}_{k_1} + (-1)^{\tau_1} \varepsilon^{(i)}_{k_2} \right] 
\\\hline
\Pi_{h,(\sigma_1,1)}^{\Z_2^{(i)}} &   (-1)^{\tau_j} \varepsilon^{(i)}_{k_1} + [1-(-1)^{\tau_j}] \tilde{\varepsilon}^{(i)}_{k_1} 
+ (-1)^{\tau_1}\left[  (-1)^{\tau_j} \varepsilon^{(i)}_{k_2} 
+ [1-(-1)^{\tau_j}] \tilde{\varepsilon}^{(i)}_{k_2} \right]
\\\hline\hline
\Pi_{v,(\sigma_1,0)}^{\Z_2^{(i)}} &  (-1)^{\tau_j}\left[
- \varepsilon^{(i)}_{k_1} + 2 \tilde{\varepsilon}^{(i)}_{k_1}
+ (-1)^{\tau_1}\left(- \varepsilon^{(i)}_{k_2} + 2 \tilde{\varepsilon}^{(i)}_{k_2}
\right)
\right]
\\\hline
\Pi_{v,(\sigma_1,1)}^{\Z_2^{(i)}} &  [ 2  - (-1)^{\tau_j}  ] \varepsilon^{(i)}_{k_1} 
+[ -1-(-1)^{\tau_j} ] \tilde{\varepsilon}^{(i)}_{k_1}
+ (-1)^{\tau_1}\left[[ 2  - (-1)^{\tau_j}  ] \varepsilon^{(i)}_{k_2} 
+[ -1-(-1)^{\tau_j} ] \tilde{\varepsilon}^{(i)}_{k_2}
\right]
\\\hline
\end{array}
\end{equation*}
\end{center}
\caption{Exceptional three-cycles which enter the K-theory constraint on the {\bf a/bAA} lattice for $T^6/(\Z_2 \times \Z_6 \times \OR)$ with discrete torsion.
All cycles are multiplied by the $\Z_2^{(i)}$ eigenvalue $(-1)^{\tau_0^{(i)}}$. The subscript for the $\Z_2^{(i)}$ sector labels 
if the 1-cycle is parallel to the $\OR$-plane ($h$) or perpendicular to it ($v$) and lists the discrete displacements
$(\sigma_j,\sigma_k)$ on $T_j \times T_k$. 
}
\label{Tab:Z2Z6-Ktheory_AA}
\end{table}

\begin{table}[h!]
\renewcommand{\arraystretch}{1.3}
  \begin{center}
\begin{equation*}
\begin{array}{|c|c|}\hline
\multicolumn{2}{|c|}{\text{\bf Exceptional contributions to $\OR$ invariant 3-cycles for $T^6/\Z_2 \times \Z_6$ on a/bAB} }
\\\hline\hline
\multicolumn{2}{|c|}{{\bf a/bAB}:  \Z_2^{(1)}}
\\\hline\hline
\Pi_{h,(0,0)}^{\Z_2^{(1)}} &  \frac{1}{1-b}\Bigl\{  
[ \varepsilon^{(1)}_0 -b  \tilde{\varepsilon}^{(1)}_0]
+ (-1)^{\tau_2}[ \varepsilon^{(1)}_1 -b  \tilde{\varepsilon}^{(1)}_1]
+ (-1)^{\tau_3}[  \varepsilon^{(1)}_2 -b  \tilde{\varepsilon}^{(1)}_2]
+ (-1)^{\tau_2 + \tau_3} [  \varepsilon^{(1)}_5 -b \tilde{\varepsilon}^{(1)}_5]  \Bigr\}
\\\hline
\Pi_{h,(1,0)}^{\Z_2^{(1)}} &  \frac{1}{1-b}\Bigl\{
[1+(-1)^{\tau_2}][ \varepsilon^{(1)}_1 -b \tilde{\varepsilon}^{(1)}_1]
 +(-1)^{\tau_3} [ \varepsilon^{(1)}_3 -b \tilde{\varepsilon}^{(1)}_3] 
 +(-1)^{\tau_2 + \tau_3}[ \varepsilon^{(1)}_4 -b \tilde{\varepsilon}^{(1)}_4]
\\\hline
\Pi_{h,(0,1)}^{\Z_2^{(1)}} &  \frac{1}{1-b}\Bigl\{
[1+(-1)^{\tau_3}][  \varepsilon^{(1)}_2 -b \tilde{\varepsilon}^{(1)}_2]
 +(-1)^{\tau_2} [ \varepsilon^{(1)}_3 -b  \tilde{\varepsilon}^{(1)}_3]
 +(-1)^{\tau_2 + \tau_3}[ \varepsilon^{(1)}_4 -b  \tilde{\varepsilon}^{(1)}_4]
  \Bigr\}
\\\hline
\Pi_{h,(1,1)}^{\Z_2^{(1)}} & \frac{1}{1-b}\Bigl\{
 (-1)^{\tau_2 + \tau_3}[ \varepsilon^{(1)}_3 -b \tilde{\varepsilon}^{(1)}_3]
+ [ \varepsilon^{(1)}_4 -b \tilde{\varepsilon}^{(1)}_4]
 + [(-1)^{\tau_2} + (-1)^{\tau_3}][ \varepsilon^{(1)}_5 -b \tilde{\varepsilon}^{(1)}_5]
  \Bigr\}
\\\hline\hline
\Pi_{v,(0,0)}^{\Z_2^{(1)}} &  \tilde{\varepsilon}^{(1)}_0 + (-1)^{\tau_2} \tilde{\varepsilon}^{(1)}_1
+ (-1)^{\tau_3} \tilde{\varepsilon}^{(1)}_2 + (-1)^{\tau_2 + \tau_3}  \tilde{\varepsilon}^{(1)}_5
\\\hline
\Pi_{v,(1,0)}^{\Z_2^{(1)}} & [1+(-1)^{\tau_2}] \tilde{\varepsilon}^{(1)}_1
+(-1)^{\tau_3}  \tilde{\varepsilon}^{(1)}_3 +(-1)^{\tau_2 + \tau_3} \tilde{\varepsilon}^{(1)}_4
\\\hline
\Pi_{v,(0,1)}^{\Z_2^{(1)}} & [1+(-1)^{\tau_3}] \tilde{\varepsilon}^{(1)}_2
 +(-1)^{\tau_2}  \tilde{\varepsilon}^{(1)}_3 +(-1)^{\tau_2 + \tau_3} \tilde{\varepsilon}^{(1)}_4
\\\hline
\Pi_{v,(1,1)}^{\Z_2^{(1)}} & (-1)^{\tau_2 + \tau_3} \tilde{\varepsilon}^{(1)}_3
+ \tilde{\varepsilon}^{(1)}_4 + [(-1)^{\tau_2} + (-1)^{\tau_3}] \tilde{\varepsilon}^{(1)}_5
\\\hline\hline\hline 
\multicolumn{2}{|c|}{{\bf a/bAB}: \Z_2^{(2)}}
\\\hline\hline
\Pi_{h,(\sigma_1,0)}^{\Z_2^{(2)}} & (-1)^{\tau_3}\left[
- \varepsilon^{(2)}_{k_1} +  \tilde{\varepsilon}^{(2)}_{k_1}
+ (-1)^{\tau_1} \left(- \varepsilon^{(2)}_{k_2} +  \tilde{\varepsilon}^{(2)}_{k_2}
\right)\right]
\\\hline
\Pi_{h,(\sigma_1,1)}^{\Z_2^{(2)}} &  
 \varepsilon^{(2)}_{k_1}  -(-1)^{\tau_3}  \tilde{\varepsilon}^{(2)}_{k_1}
+ (-1)^{\tau_1} \left[
 \varepsilon^{(2)}_{k_2}  -(-1)^{\tau_3}  \tilde{\varepsilon}^{(2)}_{k_2}\right]
\\\hline\hline
\Pi_{v,(\sigma_1,0)}^{\Z_2^{(2)}} & (-1)^{\tau_3}\left[ 
\varepsilon^{(2)}_{k_1} +  \tilde{\varepsilon}^{(2)}_{k_1}
+ (-1)^{\tau_1} \left(\varepsilon^{(2)}_{k_2} +  \tilde{\varepsilon}^{(2)}_{k_2}
\right)\right]
\\\hline
\Pi_{v,(\sigma_1,1)}^{\Z_2^{(2)}} &  [ 1 -2  (-1)^{\tau_3}  ] \varepsilon^{(2)}_{k_1} 
+[-2 + (-1)^{\tau_3} ] \tilde{\varepsilon}^{(2)}_{k_1}
+ (-1)^{\tau_1} \left[ 
[ 1 -2  (-1)^{\tau_3}  ] \varepsilon^{(2)}_{k_2} 
+[-2 + (-1)^{\tau_3} ] \tilde{\varepsilon}^{(2)}_{k_2}
\right]
\\\hline\hline\hline 
\multicolumn{2}{|c|}{{\bf a/bAB}: \Z_2^{(3)}}
\\\hline\hline
\Pi_{h,(\sigma_1,0)}^{\Z_2^{(3)}} &  (-1)^{\tau_2}   \left[ 
\varepsilon^{(3)}_{k_1} +  \tilde{\varepsilon}^{(3)}_{k_1}
+ (-1)^{\tau_1} \left(\varepsilon^{(3)}_{k_2} +  \tilde{\varepsilon}^{(3)}_{k_2}
\right)\right]
\\\hline
\Pi_{h,(\sigma_1,1)}^{\Z_2^{(3)}} & 
[1 -2(-1)^{\tau_2} ]  \varepsilon^{(3)}_{k_1} + [ -2 + (-1)^{\tau_2}] \tilde{\varepsilon}^{(3)}_{k_1}
+ (-1)^{\tau_1} \left[
[1 -2(-1)^{\tau_2} ]  \varepsilon^{(3)}_{k_2} + [ -2 + (-1)^{\tau_2}] \tilde{\varepsilon}^{(3)}_{k_2}
\right]
\\\hline\hline
\Pi_{v,(\sigma_1,0)}^{\Z_2^{(3)}} &  (-1)^{\tau_2} \left[  
\varepsilon^{(3)}_{k_1} - \tilde{\varepsilon}^{(3)}_{k_1}
+ (-1)^{\tau_1} \left(
\varepsilon^{(3)}_{k_2} - \tilde{\varepsilon}^{(3)}_{k_2}
\right)\right]
\\\hline
\Pi_{v,(\sigma_1,1)}^{\Z_2^{(3)}} & 
- \varepsilon^{(3)}_{k_1} +  (-1)^{\tau_2} \tilde{\varepsilon}^{(3)}_{k_1}
+ (-1)^{\tau_1} \left[
- \varepsilon^{(3)}_{k_2} +  (-1)^{\tau_2} \tilde{\varepsilon}^{(3)}_{k_2}
\right]
\\\hline
\end{array}
\end{equation*}
\end{center}
\caption{Exceptional three-cycles which enter the K-theory constraint on the {\bf a/bAB} lattice for $T^6/(\Z_2 \times \Z_6 \times \OR)$ with discrete torsion.
Observe that the calculations can be simplified by noting that 
$(\Pi_{h,(\sigma_1,\sigma_3)}^{\Z_2^{(2)}} , \Pi_{v,(\sigma_1,\sigma_3)}^{\Z_2^{(2)}}) \Leftrightarrow
( - \Pi_{v,(\sigma_1,\sigma_2)}^{\Z_2^{(3)}} , \Pi_{h,(\sigma_1,\sigma_2)}^{\Z_2^{(3)}})$
when the same values for $\sigma_2$ and $\sigma_3$ are taken. For details on the notation see table~\protect\ref{Tab:Z2Z6-Ktheory_AA}.}
\label{Tab:Z2Z6-Ktheory_AB}
\end{table}
\begin{table}[h!]
\renewcommand{\arraystretch}{1.3}
  \begin{center}
\begin{equation*}
\begin{array}{|c|c|}\hline
\multicolumn{2}{|c|}{\text{\bf Exceptional contributions to $\OR$ invariant 3-cycles for $T^6/\Z_2 \times \Z_6$ on a/bBB} }
\\\hline\hline
\multicolumn{2}{|c|}{{\bf a/bBB}:  \Z_2^{(1)}}
\\\hline\hline
\Pi_{h,(0,0)}^{\Z_2^{(1)}} &  \frac{1}{1-b}\Bigl\{   
[ \varepsilon^{(1)}_0 -b  \tilde{\varepsilon}^{(1)}_0] 
+ (-1)^{\tau_2}[ \varepsilon^{(1)}_1 -b \tilde{\varepsilon}^{(1)}_1]
+ (-1)^{\tau_3}[  \varepsilon^{(1)}_2 -b  \tilde{\varepsilon}^{(1)}_2] 
+ (-1)^{\tau_2 + \tau_3} [  \varepsilon^{(1)}_4 -b  \tilde{\varepsilon}^{(1)}_4]
\Bigr\}
\\\hline
\Pi_{h,(1,0)}^{\Z_2^{(1)}} &  \frac{1}{1-b}\Bigl\{ 
[1+(-1)^{\tau_2}][ \varepsilon^{(1)}_1 -b  \tilde{\varepsilon}^{(1)}_1]
+(-1)^{\tau_2 + \tau_3}[ \varepsilon^{(1)}_3 -b  \tilde{\varepsilon}^{(1)}_3]
+(-1)^{\tau_3}[ \varepsilon^{(1)}_5 -b  \tilde{\varepsilon}^{(1)}_5]
\Bigr\}
\\\hline
\Pi_{h,(0,1)}^{\Z_2^{(1)}} &  \frac{1}{1-b}\Bigl\{ 
[1+(-1)^{\tau_3}][ \varepsilon^{(1)}_2 -b  \tilde{\varepsilon}^{(1)}_2]
 +(-1)^{\tau_2 + \tau_3}[ \varepsilon^{(1)}_3 -b \tilde{\varepsilon}^{(1)}_3]
 +(-1)^{\tau_2}[ \varepsilon^{(1)}_5 -b \tilde{\varepsilon}^{(1)}_5]
\Bigr\}
\\\hline
\Pi_{h,(1,1)}^{\Z_2^{(1)}} &  \frac{1}{1-b}\Bigl\{  
[\varepsilon^{(1)}_3 -b \tilde{\varepsilon}^{(1)}_3]
  +[(-1)^{\tau_2} + (-1)^{\tau_3}][ \varepsilon^{(1)}_4 -b \tilde{\varepsilon}^{(1)}_4]
 + (-1)^{\tau_2 + \tau_3}[ \varepsilon^{(1)}_5 -b \tilde{\varepsilon}^{(1)}_5]
\Bigr\}
\\\hline\hline
\Pi_{v,(0,0)}^{\Z_2^{(1)}} &  \tilde{\varepsilon}^{(1)}_0 + (-1)^{\tau_2} \tilde{\varepsilon}^{(1)}_1
+ (-1)^{\tau_3} \tilde{\varepsilon}^{(1)}_2 + (-1)^{\tau_2 + \tau_3}  \tilde{\varepsilon}^{(1)}_4
\\\hline
\Pi_{v,(1,0)}^{\Z_2^{(1)}} & [1+(-1)^{\tau_2}]  \tilde{\varepsilon}^{(1)}_1
 +(-1)^{\tau_2 + \tau_3}  \tilde{\varepsilon}^{(1)}_3
+(-1)^{\tau_3}  \tilde{\varepsilon}^{(1)}_5
\\\hline
\Pi_{v,(0,1)}^{\Z_2^{(1)}} & [1+(-1)^{\tau_3}]  \tilde{\varepsilon}^{(1)}_2
 +(-1)^{\tau_2 + \tau_3} \tilde{\varepsilon}^{(1)}_3
 +(-1)^{\tau_2}  \tilde{\varepsilon}^{(1)}_5
\\\hline
\Pi_{v,(1,1)}^{\Z_2^{(1)}} &  \tilde{\varepsilon}^{(1)}_3 +[(-1)^{\tau_2} 
+ (-1)^{\tau_3}] \tilde{\varepsilon}^{(1)}_4
 + (-1)^{\tau_2 + \tau_3} \tilde{\varepsilon}^{(1)}_5
\\\hline\hline\hline 
\multicolumn{2}{|c|}{{\bf a/bBB}: \Z_2^{(i)} \text{ with } (i,j) \in \{(2,3),(3,2)\} }
\\\hline\hline
\Pi_{h,(\sigma_1,0)}^{\Z_2^{(i)}} & (-1)^{\tau_j} 
\left[  -2  \varepsilon^{(i)}_{k_1} +  \tilde{\varepsilon}^{(i)}_{k_1}
+ (-1)^{\tau_1} \left(   -2  \varepsilon^{(i)}_{k_2} +  \tilde{\varepsilon}^{(i)}_{k_2}
\right)
\right]
\\\hline
\Pi_{h,(\sigma_1,1)}^{\Z_2^{(i)}} & 
[1+  (-1)^{\tau_j} ]   \varepsilon^{(i)}_{k_1} +[1-2(-1)^{\tau_j} ]\tilde{\varepsilon}^{(i)}_{k_1}
+(-1)^{\tau_1} \left[[1+  (-1)^{\tau_j} ]   \varepsilon^{(i)}_{k_2} +[1-2(-1)^{\tau_j} ]\tilde{\varepsilon}^{(i)}_{k_2}
\right]
\\\hline\hline
\Pi_{v,(\sigma_1,0)}^{\Z_2^{(i)}} & (-1)^{\tau_j}\left[ 
\tilde{\varepsilon}^{(i)}_{k_1} +(-1)^{\tau_1} \tilde{\varepsilon}^{(i)}_{k_2}
\right]
\\\hline
\Pi_{v,(\sigma_1,1)}^{\Z_2^{(i)}} &  [1 - (-1)^{\tau_j} ]   \varepsilon^{(i)}_{k_1} -  
\tilde{\varepsilon}^{(i)}_{k_1}
+(-1)^{\tau_1} \left[  [1 - (-1)^{\tau_j} ]   \varepsilon^{(i)}_{k_2} -  
\tilde{\varepsilon}^{(i)}_{k_2}
\right]
\\\hline
\end{array}
\end{equation*}
\end{center}
\caption{Exceptional three-cycles which enter the K-theory constraint on the {\bf a/bBB} lattice for $T^6/(\Z_2 \times \Z_6 \times \OR)$ with discrete torsion.
 For details on the notation see table~\protect\ref{Tab:Z2Z6-Ktheory_AA}.}
\label{Tab:Z2Z6-Ktheory_BB}
\end{table}

The bulk parts of the cycles are, as for $T^6/(\Z_2 \times \Z_2 \times \OR)$, parallel to the O6-planes with bulk wrapping numbers 
given in table~\ref{Tab:Z2Z6-Oplanes-bulk}.

In terms of the expansion~(\ref{Eq:Expand-Frac-Z2Z6}) of a fractional three-cycle on $T^6/\Z_2 \times \Z_6$, the 
intersection numbers read 
\begin{equation}\label{Eq:Z2Z6-Expand-Intersection}
\begin{aligned}
\Pi^{\rm frac}_a \circ \Pi^{\rm frac}_b =& \frac{1}{4}\left( 
2 \left[ P_aU_b -P_bU_a + Q_a V_b - Q_b V_a \right]
+ P_a V_b - P_b V_a + Q_a U_b - Q_b U_a
\right)
\\
&-  \frac{1}{4}
 \sum_{\alpha=1}^3 \left(\vec{x}^{(\alpha)}_a \cdot \vec{y}^{(\alpha)}_b - \vec{x}^{(\alpha)}_b \cdot \vec{y}^{(\alpha)}_a \right)
\\
\text{with}& \quad
\vec{x}^{(1)}_a \cdot \vec{y}^{(1)}_b \equiv  3 \, x^{(1)}_{0,a} y^{(1)}_{0,b} + \sum_{m=1}^5 x^{(1)}_{m,a} y^{(1)}_{m,b}
\\
\text{and} & \quad
\vec{x}^{(\alpha)}_a \cdot \vec{y}^{(\alpha)}_b \equiv \sum_{m=1}^4 x^{(\alpha)}_{m,a} y^{(\alpha)}_{m,b}
\qquad
\alpha \in \{2,3\} 
.
\end{aligned}
\end{equation}
The bulk parts of the intersection numbers, both without and with discrete torsion, are explicitly given in 
table~\ref{Tab:Z2Z4-example-cycles6-Ktheory-bulk-v1}.
\begin{table}[h]
  \renewcommand{\arraystretch}{1.3}
  \begin{center}
\begin{equation*}\!\!\!\!\!\!\!\!\!\!
\begin{array}{|c||c||c|c|c|}\hline
\multicolumn{5}{|c|}{\text{\bf Bulk parts of K-theory constraints on } T^6/\Z_2 \times \Z_6}
\\\hline\hline
& {\rm lattice} & {\bf a/bAA} & {\bf a/bAB} & {\bf a/bBB} 
\\\hline\hline
\OR & \frac{\sum_a N_a \Pi^{\rm bulk}_a \circ \Pi^{\rm bulk}_{Sp(2)_0}}{2^{3-\eta}} 
& \frac{-\sum_a N_a \left(2U_a + V_a + b(2P_a+Q_a)  \right) }{2^{1-\eta}(1-b)} 
& \frac{-3 \sum_a N_a \left(U_a + V_a + b (P_a + Q_a)  \right) }{2^{1-\eta}(1-b)}
& \frac{-3 \sum_a N_a \left(U_a + 2 V_a + b (P_a + 2 Q_a)  \right)}{2^{1-\eta}(1-b)} 
\\\hline
\!\!\!\OR\Z_2^{(1)}\!\!\!& \frac{ \sum_a N_a \Pi^{\rm bulk}_a \circ \Pi^{\rm bulk}_{Sp(2)_1}}{2^{3-\eta}} 
& \frac{-3 \sum_a N_a \left(2U_a + V_a + b(2P_a+Q_a)   \right) }{2^{1-\eta}(1-b)}
& \frac{-3 \sum_a N_a \left(U_a + V_a + b (P_a + Q_a)  \right) }{2^{1-\eta}(1-b)}
& \frac{-\sum_a N_a \left(U_a + 2 V_a + b (P_a + 2 Q_a)  \right)}{2^{1-\eta}(1-b)} 
\\\hline
\!\!\!\OR\Z_2^{(2)}\!\!\!&\frac{\sum_a N_a \Pi^{\rm bulk}_a \circ \Pi^{\rm bulk}_{Sp(2)_2} }{2^{3-\eta}} 
&  - \frac{3}{2^{1-\eta}} \sum_a N_a Q_a & \frac{3}{2^{1-\eta}} \sum_a N_a \left(P_a -Q_a  \right) 
& \frac{3 \sum_a N_a P_a}{2^{1-\eta}}
\\\hline
\!\!\!\OR\Z_2^{(3)}\!\!\!&\frac{ \sum_a N_a \Pi^{\rm bulk}_a \circ \Pi^{\rm bulk}_{Sp(2)_3}}{2^{3-\eta}} 
& - \frac{3}{2^{1-\eta}} \sum_a N_a Q_a & \frac{1}{2^{1-\eta}} \sum_a N_a \left(P_a - Q_a  \right) 
& \frac{3 \sum_a N_a P_a}{2^{1-\eta}}
\\\hline
\end{array}
\end{equation*}
    \caption{Bulk part of the K-theory constraints on $T^6/(\Z_2 \times \Z_6 \times \OR)$ without ($\eta=1$) and with
discrete torsion ($\eta=-1$). By inserting the bulk RR tadpole cancellation conditions~(\protect\ref{Eq:Z2Z6-bulkRRtcc}), the
simpler form in table~\protect\ref{Tab:Z2Z4-example-cycles6-Ktheory-bulk-v2} appears.}
\label{Tab:Z2Z4-example-cycles6-Ktheory-bulk-v1}
  \end{center}
\end{table}
They can be simplified by imposing the bulk RR tadpole cancellation
conditions~(\ref{Eq:Z2Z6-bulkRRtcc}). This leads to the expressions in table~\ref{Tab:Z2Z4-example-cycles6-Ktheory-bulk-v2}.
\begin{table}[h]
  \renewcommand{\arraystretch}{1.3}
  \begin{center}
\begin{equation*}
\begin{array}{|c|c||c|c||c|c|}\hline
\multicolumn{6}{|c|}{\text{\bf Simplification of the bulk K-theory constraints on } T^6/\Z_2 \times \Z_6}
\\\hline\hline
 {\bf a/bAA} &\!\!\!\begin{array}{c} \text{after RR tcc} \\ \text{mod  2} \end{array}\!\!\!\!& {\bf a/bAB}  &\!\!\!\begin{array}{c}\text{after RR tcc} \\ \text{mod  2} \end{array}\!\!\!\!& {\bf a/bBB}  &\!\!\!\begin{array}{c} \text{after RR tcc} \\ \text{mod  2} \end{array}\!\!\!\!
 \\\hline\hline
\OR &\frac{-\sum_a N_a \left(2U_a + V_a  \right) }{2^{1-\eta}(1-b)}
& \OR & \frac{-3 \sum_a N_a \left(U_a + V_a  \right) }{2^{1-\eta}(1-b)}
&\OR\Z_2^{(1)} &\frac{- \sum_a N_a \left(U_a + 2 V_a \right)}{2^{1-\eta}(1-b)}
\\\hline
\OR\Z_2^{(2)}
& 3 \cdot 2^{\eta} \sum_a N_a P_a
 &\OR\Z_2^{(3)} & 2^{\eta} \sum_a N_a P_a
& \OR\Z_2^{(2)} & -3 \cdot 2^{\eta} \sum_a N_a Q_a 
\\\hline
\end{array}
\end{equation*}
    \caption{Simplification of the K-theory constraints on $T^6/(\Z_2 \times \Z_6 \times \OR)$
 upon the bulk RR tadpole cancellation conditions~(\protect\ref{Eq:Z2Z6-bulkRRtcc}).
The remaining two kinds of probe branes are proportional to the constraints in this table, where the 
integer multiplicities can be read off from table~\protect\ref{Tab:Z2Z4-example-cycles6-Ktheory-bulk-v1}. }
\label{Tab:Z2Z4-example-cycles6-Ktheory-bulk-v2}
  \end{center}
\end{table}
On can immediately see that in the case without torsion ($\eta=1$) and a tilted two-torus $T_{(1)}^2$, the
K-theory constraints are trivially fulfilled for all models that satisfy RR tadpole cancellation.
For trivial torsion  ($\eta=1$) and an untilted  $T_{(1)}^2$, still half of the K-theory constraints are trivial.
In the case with discrete torsion ($\eta=-1$), the exceptional contributions of the intersection numbers according 
to equation~(\ref{Eq:Z2Z6-Expand-Intersection}) have to be added, which can again be simplified using the 
twisted RR tadpole cancellation
conditions~(\ref{Eq:Z2Z6-twistedRRtcc_AA})
to~(\ref{Eq:Z2Z6-twistedRRtcc_BB}) and  
the knowledge of the shape of the coefficients $(x^{(\alpha)}_i,y^{(\alpha)}_i)$ as discussed in 
section~\ref{Ss:Z2Z6-twisted-part}.
Due to the large number of combinations of the exceptional cycles in 
tables~\ref{Tab:Z2Z6-Ktheory_AA} to~\ref{Tab:Z2Z6-Ktheory_BB} with $\OR$ and $\OR\Z_2^{(i)}$ invariant bulk three-cycles,
we do not write out explicitly the K-theory constraints on $T^6/(\Z_2 \times \Z_6 \times \OR)$ here.

Instead, we give examples of consistent models where the K-theory constraints are fulfilled trivially by only
having even ranks of the gauge factors.

{\boldmath
\subsection{Exceptional three-cycles at $\Z_3$ singularities}\label{Ss:Z2Z6-Z3twist}
}

The discrete torsion factor acts trivially in the $\Z_3$ twisted sector, as displayed in table~\ref{Tab:Orbits-Z2Z6}.
The exceptional three-cycles in this sector, for both choices of torsion, are given by
\begin{equation}
\begin{aligned}
\delta^{(k)} &= \left( d_{22} + d_{33} - d_{23} - d_{32} \right) \otimes \pi_1,
\\
\tilde{\delta}^{(k)} &= \left( d_{22} + d_{33} - d_{23} - d_{32} \right) \otimes \pi_2,
\\
\text{with } & \quad  \delta^{(k)} \circ \tilde{\delta}^{(l)} = \left(\begin{array}{cc} -2 & 1 \\ 1 & -2 \end{array} \right)_{kl},
\end{aligned}
\end{equation}
and all others vanishing.
These exceptional three-cycles are relevant for determining an unimodular basis of the lattice of three-cycles, but cannot be used
for model building with D6-branes as mentioned in section~\ref{S:orbifolds}.
We will therefore not discuss them further in this article.

{\boldmath
\subsection{A $T^6/(\Z_2 \times \Z_6 \times \OR)$ example without torsion}\label{Ss:Z2Z6-example-without-torsion}
}

In~\cite{Gmeiner:2007zz,Gmeiner:2008xq,Gmeiner:2009fb}, two classes of supersymmetric models
on {\bf aAB} on $T^6/\Z_6'$ were analysed with complex structure parameters 
$\varrho\equiv \sqrt{3} \frac{R_2}{R_1} = \frac{1}{\omega} $ and $\omega=1$ for the models with hidden sectors
and $\omega=2$ for the model without hidden sector.  
The representants per orbifold and orientifold orbit, as discussed at the end of section~\ref{Ss:Z2Z6-bulk-part},
are listed in table~\ref{Tab:Z2Z6-example-bulk-cycles}.
\begin{table}[h]
  \renewcommand{\arraystretch}{1.3}
  \begin{center}
  \begin{equation}
    \begin{array}{|c|c|c||c|c|c|c|}   \hline
\multicolumn{7}{|c|}{\text{\bf Supersymmetric bulk 3-cycles on aAB on } T^6/\Z_2 \times \Z_6}
\\\hline\hline
\text{brane/orbit} & \frac{\text{angle}}{\pi} & (n^1,m^1;n^2,m^2;n^3,m^3) & P & Q & U & V 
\\\hline\hline
a, h_i/\OR\Z_2^{(2)} &  (\frac{1}{2}, 0,-\frac{1}{2})  &  (0,1;1,0;1,-1) & 0 & 0 & 1 & -1 
\\\hline
b & (\frac{1}{6}, -\frac{1}{6},0) & (1,\omega;2,-1;1,1) & 3 & 0 & 3 \, \omega & 0 
\\\hline
c/\OR & (0,0,0) & (1,0;1,0;1,1) &  1 & 1 & 0 & 0 
\\\hline
d,\hat{h}_1 /\OR\Z_2^{(3)} & (\frac{1}{2},-\frac{1}{2},0) & (0,1;1,-2;1,1) & 0 & 0 & 3 & -3
\\\hline\hline
\OR\Z_2^{(1)} & (0,\frac{1}{2},-\frac{1}{2}) & (1,0;-1,2;1,-1) &  1 & 1 & 0 & 0 
\\\hline
\tilde{b} &  (\frac{1}{6},\frac{1}{3},-\frac{1}{2}) & (1,\omega;0,1;1,-1) & 1 & 0 & \omega & 0
\\\hline
   \end{array}
   \end{equation}
    \caption{Examples of supersymmetric bulk three-cycles on the {\bf aAB} lattice on $T^6/(\Z_2 \times \Z_6 \times \OR)$.
The labels $a, \ldots, d, h_i$ and $\hat{h}_1$ correspond to the cycles used for model building on $T^6/\Z_6'$ 
in~\cite{Gmeiner:2009fb} up to permutation of tori. $\OR\Z_2^{(2)}$ denotes the fact that cycle $a$ is parallel to this O6-plane orbit. 
The complex structures, for which $b$ and $\tilde{b}$ are supersymmetric, are $\varrho=\frac{1}{\omega}$.}
\label{Tab:Z2Z6-example-bulk-cycles}
  \end{center}
\end{table}
On the {\bf aAB} lattice without discrete torsion, the untwisted RR tadpole cancellation conditions~(\ref{Eq:Z2Z6-bulkRRtcc}) read 
\begin{equation}
\begin{aligned}
\sum_a N_a \left(P_a + Q_a  \right) &=  8
,
\\
 \sum_a N_a \left(U_a - V_a  \right) &=  16
,
\end{aligned}
\end{equation}
which can be solved by 
\begin{equation}
\Pi_{\tilde{b}}^{\rm frac} = \frac{1}{2}\left(\rho_1 + 2 \, \rho_3 \right)
\qquad \text{with} \qquad
N_{\tilde{b}}=8
\end{equation}
for $\omega=2$ and the complex structure $\varrho=\frac{1}{2}$.
The resulting gauge group is $U(8)$ with the Abelian subgroup $U(1) \subset U(8)$ anomalous and massive, 
and the chiral spectrum is derived from the intersection numbers
 \begin{equation}
\begin{aligned}
\Pi_{\tilde{b}}^{\rm frac} \circ \Pi_{\tilde{b}'}^{\rm frac} &=-4,
\\
\Pi_{\tilde{b}}^{\rm frac} \circ \Pi_{O6} &=-8,
\end{aligned}
\end{equation}
as 
 \begin{equation}
6\, (\ov{\bf 28}) + 2 \, ({\bf 36}).
\end{equation}
The non-chiral massless spectrum is computed along the lines describe in appendix~\ref{S:App-OpenStrings} and gives
\begin{itemize}
\item
three multiplets in the adjoint representation $({\bf 64})$ of $U(8)$ from the $\tilde{b}\tilde{b}$ sector and 
two further multiplets in the adjoint at $\tilde{b}(\omega^k \, \tilde{b})_{k=1,2}$ intersections,
\item
six non-chiral pairs of antisymmetrics $6 \times [ ({\bf 28}) + c.c.]$ of $U(8)$ at $\tilde{b}(\omega^k \tilde{b}')_{k=0,1,2}$ intersections.
\end{itemize}
The K-theory constraint is trivially fulfilled for a single gauge factor $U(8)$ due to its even rank.

{\boldmath
\subsection{A $T^6/(\Z_2 \times \Z_6 \times \OR)$ example with discrete torsion and D6-branes}\label{Ss:Z2Z6-example-with-torsion}
}

On the {\bf aAB} lattice with discrete torsion and the choice $\eta_{\OR\Z_2^{(2)}}=-1$ of the exotic O6-plane, 
the untwisted RR tadpole cancellation conditions~(\ref{Eq:Z2Z6-bulkRRtcc}) read 
\begin{equation}
\begin{aligned}
\sum_a N_a \left(P_a + Q_a  \right) &= 16
,
\\
 \sum_a N_a \left(U_a - V_a  \right) &= 16
,
\end{aligned}
\end{equation}
which can be solved by four kinds of fractional D6-branes adding up to the supersymmetric bulk cycle $\tilde{b}$ with $\omega=1$ 
in table~\ref{Tab:Z2Z6-example-bulk-cycles} with complex structure $\varrho=1$ and $N_{\tilde{b}_m}=4$ for $m=0 \ldots 3$,
\begin{equation}
\begin{aligned}
\Pi_{\tilde{b}_m}^{\rm frac} =& \frac{1}{4}\left(\rho_1 +  \rho_3 \right)
+ \frac{(-1)^{\tau_{0,m}^{(1)}}}{4}\left(2 \, \varepsilon_1^{(1)} + 2  \,\tilde{ \varepsilon}_1^{(1)} 
+ \varepsilon_4^{(1)} + \tilde{ \varepsilon}_4^{(1)} + \varepsilon_5^{(1)} + \tilde{ \varepsilon}_5^{(1)} 
 \right)
\\
& + \frac{(-1)^{\tau_{0,m}^{(2)}}}{4}\left(-\varepsilon_1^{(2)} -\varepsilon_3^{(2)}    \right)
 + \frac{(-1)^{\tau_{0,m}^{(3)}}}{4}\left(\varepsilon_1^{(3)} +\varepsilon_3^{(3)}  \right)
,
\end{aligned}
\end{equation}
where we choose the discrete displacement $\vec{\sigma}=(0,1,0)$ and no Wilson line, $\vec{\tau}=\vec{0}$, for all four D6-branes.

The orientifold image branes are 
\begin{equation}
\begin{aligned}
\Pi_{\tilde{b}_m'}^{\rm frac} =& \frac{1}{4}\left(\rho_2 -  \rho_4 \right)
+ \frac{(-1)^{\tau_{0,m}^{(1)}}}{4}\left(-2 \, \varepsilon_1^{(1)} + 2  \,\tilde{ \varepsilon}_1^{(1)}
- \varepsilon_3^{(1)} + \tilde{ \varepsilon}_3^{(1)} - \varepsilon_5^{(1)} + \tilde{ \varepsilon}_5^{(1)} 
 \right)
\\
& - \frac{(-1)^{\tau_{0,m}^{(2)}}}{4}\left(-\tilde{\varepsilon}_1^{(2)} -\tilde{\varepsilon}_3^{(2)}    \right)
 +\frac{(-1)^{\tau_{0,m}^{(3)}}}{4}\left(-\tilde{\varepsilon}_1^{(3)} -\tilde{\varepsilon}_3^{(3)}  \right)
,
\end{aligned}
\end{equation}
where we used $\eta_{(2)}=-1$ and $\eta_{(1)}=\eta_{(3)}=1$.

The relevant intersection numbers are thus
\begin{equation}
\begin{aligned}
\Pi_{\tilde{b}_m}^{\rm frac} \circ \Pi_{\tilde{b}_n}^{\rm frac} =&0,
\\
\Pi_{\tilde{b}_m}^{\rm frac} \circ \Pi_{\tilde{b}_n'}^{\rm frac} =&-\frac{1}{2} -\frac{5}{2} (-1)^{\tau_{0,m}^{(1)} + \tau_{0,n}^{(1)}}
+\frac{1}{2}(-1)^{\tau_{0,m}^{(2)} + \tau_{0,n}^{(2)}}
+\frac{1}{2} (-1)^{\tau_{0,m}^{(3)} + \tau_{0,n}^{(3)}}
\\
&=\left\{\begin{array}{cc}
-2 & (m,m)\\
-4 & (0,1),(2,3) \\
2 & (0,2),(1,3),(0,3),(1,2)
\end{array}\right.
,
\\
\Pi_{\tilde{b}_m}^{\rm frac} \circ \Pi_{O6} =&-2,
\end{aligned}
\end{equation}
where we choose the following assignment of $\Z_2^{(i)}$ eigenvalues.
\begin{equation*}
\begin{array}{|c|ccc|}\hline
\text{brane} & \tau_{0,m}^{(1)} & \tau_{0,m}^{(2)} & \tau_{0,m}^{(3)}
\\\hline
\tilde{b}_0 & 0 & 0 & 0
\\
\tilde{b}_1 & 0 & 1 & 1 
\\
\tilde{b}_2 & 1 & 0 & 1
\\
\tilde{b}_3 & 1 & 1 & 0
\\\hline
\end{array}
\end{equation*}

This means that there is no net-chirality in the $(\N_m,\ov{\N}_n)_{m \neq n}$ sectors, but in the  $(\N_m,\N_n)_{m \neq n}$
sectors, as well as in $(\Anti_m)$. The chiral spectrum of $U(4)^4$ is in detail
\begin{equation}
\begin{aligned}
& 2 \, (\ov{\bf 6},\1,\1,\1) + 2 \, (\1,\ov{\bf 6},\1,\1) + 2 \, (,\1,\1,\ov{\bf 6},\1) + 2 \, (\1,\1,\1,\ov{\bf 6})
\\
& + 4 \, (\ov{\bf 4},\ov{\bf 4},\1,\1) + 4 \, (,\1,\1,\ov{\bf 4},\ov{\bf 4})
\\
&+ 2 \, ({\bf 4},\1,{\bf 4},\1) +  2 \, (\1,{\bf 4},\1,{\bf 4})
+2 \, ({\bf 4},\1,\1,{\bf 4}) +  2 \, (\1,{\bf 4},{\bf 4},\1) 
.
\end{aligned}
\end{equation}
The complete spectrum is computed via the intersection numbers as described in appendix~\ref{sec:thres}, 
leading to the addition of the non-chiral matter states
\begin{itemize}
\item
${\cal N}=1$ multiplets in bifundamental representations on parallel branes $\tilde{b}_m\tilde{b}_n$,
\begin{equation*}
\bigl[ (\4,\ov{\4},\1,\1) + (\1,\1,\4,\ov{\4}) 
+ (\4,\1,\ov{\4},\1) + (\1,\4,\1,\ov{\4}) 
+ (\4,\1,\1,\ov{\4}) + (\1,\4,\ov{\4},\1) 
+c.c. \bigr]
\end{equation*}
\item
more bifundamentals at intersections $\tilde{b}_m(\omega^k \, \tilde{b}_n)_{k=1,2}$ for $m \neq n$,
\begin{equation*}
\bigl[ 
 (\4,\1,\ov{\4},\1) + (\1,\4,\1,\ov{\4}) 
+ (\4,\1,\1,\ov{\4}) + (\1,\4,\ov{\4},\1) 
+c.c. \bigr]
\end{equation*}
\item
one non-chiral pair of symmetrics per stack of fractional D6-branes, 
\begin{equation*}
[({\bf 10},\1,\1,\1)+ (\1,{\bf 10},\1,\1)+ (\1,\1,{\bf 10},\1)+ (\1,\1,\1,{\bf 10})+ c.c.]
.
\end{equation*}
\end{itemize}
The $U(1)_m \subset U(4)_m$ gauge factors are all anomalous and acquire a mass by the 
generalised Green-Schwarz mechanism.

The K-theory constraint is again trivially fulfilled due to the even rank of each gauge group.

Note that there is no adjoint representation in the massless spectrum, which means that the D6-branes in this example
are completely rigid.

{\boldmath
\section{The $T^6/(\Z_2 \times \Z_6' \times \OR)$
  orientifolds}\label{S:Z2Z6porient} 
}

The  $T^6/\Z_2 \times \Z_6'$ orbifold is generated by the shift vectors
\begin{equation}
\vec{v}=\frac{1}{2}(1,-1,0),
\quad
\vec{w}=\frac{1}{6}(-2,1,1),
\end{equation}
and on the compactification lattice $SU(3)^3$, the Hodge numbers relevant for D6-brane model building
read 
\begin{equation}
\eta=1: \; h_{21} = 0,
\qquad\qquad
\eta=-1: \; h_{21} = 15_{\Z_2}. 
\end{equation}
The complete list of Hodge numbers per twist sector is given in table~\ref{Tab:Hodge-T6ZNxZM-torsion}, 
and the decomposition after the orientifold projection is displayed in table~\ref{tab:DecompositionHodge-T6ZNxZM-torsion}.

In the case without discrete torsion, there exist only two three-cycles. Since the supersymmetry
condition projects onto a one-dimensional sub-space, on $T^6/\Z_2 \times \Z_6'$ without discrete torsion, no
chiral supersymmetric models can be found.
On the other hand, the $T^6/\Z_2 \times \Z_6'$ orbifold with discrete torsion contains a $T^6/\Z_6$ sub-sector,
and in analogy to the known three-generation models there, we expect to find phenomenologically appealing spectra 
in the future~\cite{Forste:201xxx}.
In the following, we concentrate on the latter case with discrete torsion.

\subsection{The bulk part}\label{Ss:Z2Z6p-bulk}

The lattice of three-cycles consists solely of bulk and exceptional cycles from the three $\Z_2^{(i)}$ twisted sectors.

The bulk cycles are (up to normalisation) identical to the $T^6/\Z_6$ case of~\cite{Honecker:2004kb,Gmeiner:2007we}.
The two linearly independent cycles are 
\begin{equation}
\begin{aligned}
\rho_1 &= 4 \left( 1 + \theta + \theta^2 \right) \pi_{135} 
= 4 \left(\pi_{136} + \pi_{145} + \pi_{235} -\pi_{146} - \pi_{245} - \pi_{236}  \right)
,
\\
\rho_2 &= 4 \left( 1 + \theta + \theta^2 \right) \pi_{136} 
= 4 \left(\pi_{136} + \pi_{145} + \pi_{235} -\pi_{246} - \pi_{135}  \right)
,
\end{aligned}
\end{equation}
with intersection form
\begin{equation}
  \rho_1 \circ \rho_2 = 4
.
\end{equation}

The 1-cycle wrapping numbers transform as
\begin{equation}\label{Eq:Z6p-1cycle-Orb}
\left(\begin{array}{cc}
n^1 & m^1 \\ n^2 & m^2 \\ n^3 & m^3 
\end{array}\right) \stackrel{\Z_6'}{\rightarrow}
\left(\begin{array}{cc}
m^1 & -(n^1+m^1) \\ -m^2 & n^2 +m^2 \\-m^3 & n^3 +m^3
\end{array}\right) \stackrel{\Z_6'}{\rightarrow}
\left(\begin{array}{cc}
-(n^1+m^1) & n^1\\-(n^2+m^2) & n^2 \\-(n^3+m^3) & n^3
\end{array}\right) 
\end{equation}
under the $\Z_6'$ symmetry,
and the bulk wrapping numbers $(X,Y)$ along $\rho_1,\rho_2$ are 
obtained by taking the orbifold invariant orbit, 
\begin{equation}
\Pi^{\rm bulk} = 
X \; \rho_1 + Y \; \rho_2.
\end{equation}
This leads to the definition
\begin{equation}\label{Eq:Z2Z6p-Def-XY}
\begin{aligned}
X &\equiv  n^1 n^2 n^3  - m^1 m^2 m^3 -\sum_{i\neq j\neq k\neq i} n^i m^j m^k
,
\\
Y &\equiv \sum_{i\neq j\neq k\neq i} \left(n^i n^j m^k + n^i m^j m^k  \right)
.
\end{aligned}
\end{equation}
The bulk supersymmetry conditions read
\begin{equation}\label{Eq:Z2Z6p-SUSY}
\begin{aligned}
{\bf AAA:} \qquad & Y_a = 0 , \qquad\qquad 2X_a + Y_a > 0,
\\
{\bf AAB:} \qquad & Y_a-X_a =0 , \qquad\qquad X_a + Y_a > 0,
\\
{\bf ABB:} \qquad &  X_a=0, \qquad\qquad X_a + 2 Y_a > 0,
\\
{\bf BBB:}  \qquad&  2 X_a + Y_a =0, \qquad\qquad Y_a >0, 
\end{aligned}
\end{equation}
on the four inequivalent lattice orientations.

The orientifold projection on the bulk three-cycles is listed in table~\ref{Tab:Z2Z6p-ORonBulk}.
\begin{table}[h!]
\renewcommand{\arraystretch}{1.3}
  \begin{center}
\begin{equation*}
\begin{array}{|c||c|c|}\hline
 \multicolumn{3}{|c|}{\OR \; \text{\bf projection on bulk 3-cycles on } T^6/\Z_2 \times \Z_6'}
\\\hline\hline
{\rm 3-cycle} & \rho_1 & \rho_2 
\\\hline\hline
{\bf AAA} & \rho_1 & \rho_1 - \rho_2 
\\\hline
{\bf AAB} & \rho_2 & \rho_1
\\\hline
{\bf ABB} & \rho_2 - \rho_1 & \rho_2 
\\\hline
{\bf BBB} & -\rho_1 & \rho_2 - \rho_1
\\\hline
\end{array}
\end{equation*}
\end{center}
\caption{Orientifold projection on bulk three-cycles on the $T^6/(\Z_2 \times \Z_6' \times \OR)$ orientifold 
without and with discrete torsion.}
\label{Tab:Z2Z6p-ORonBulk}
\end{table}


The torus wrapping numbers of  the O6-planes are listed in table~\ref{Tab:Z2Z6p-ORonBulkO6plane-torus}, and the corresponding bulk wrapping numbers 
are given in table~\ref{Tab:Z2Z6p-ORonBulkO6plane-bulk}.
\begin{table}[h!]
\renewcommand{\arraystretch}{1.3}
  \begin{center}
\begin{equation*}
\begin{array}{|c|c||c|c|c|c|}\hline
 \multicolumn{6}{|c|}{\text{\bf Torus wrapping numbers of O6-planes on } T^6/\Z_2 \times \Z_6'}
\\\hline\hline
{\rm O-plane} & \frac{\rm angle}{\pi} & {\bf AAA} & {\bf AAB} & {\bf ABB} & {\bf BBB} 
\\\hline\hline
& & \multicolumn{4}{|c|}{(n^1,m^1;n^2,m^2;n^3,m^3)}
\\\hline\hline
\OR & (0,0,0) & (1,0;1,0;1,0) & (1,0;1,0;1,1) & (1,0;1,1;1,1) & (1,1;1,1;1,1)
\\
\OR\Z_2^{(1)} & (0,\frac{1}{2},-\frac{1}{2}) & (1,0;-1,2;1,-2) & (1,0;-1,2;1,-1) & (1,0;-1,1;1,-1) & (1,1;-1,1;1,-1)
\\
\OR\Z_2^{(3)} & (\frac{1}{2},-\frac{1}{2},0) & (-1,2;1,-2;1,0) & (-1,2;1,-2;1,1) & (-1,2;1,-1;1,1) & (-1,1;1,-1;1,1)
\\
\OR\Z_2^{(2)} &  (\frac{1}{2},0,-\frac{1}{2}) & (-1,2;1,0;1,-2) & (-1,2;1,0;1,-1) & (-1,2;1,1;1,-1) & (-1,1;1,1;1,-1)
\\\hline
\end{array}
\end{equation*}
\end{center}
\caption{Torus wrapping numbers of one representant per O6-plane orbit on $T^6/(\Z_2 \times \Z_6' \times \OR)$ without and
with discrete torsion. In the second column, the angles w.r.t. the $\OR$ invariant axis are given in units of $\pi$.}
\label{Tab:Z2Z6p-ORonBulkO6plane-torus}
\end{table}

\begin{table}[h!]
\renewcommand{\arraystretch}{1.3}
  \begin{center}
\begin{equation*}
\begin{array}{|c||c|c||c|c||c|c||c|c|}\hline
 \multicolumn{9}{|c|}{\text{\bf Bulk wrapping numbers of O6-planes on } T^6/\Z_2 \times \Z_6'}
\\\hline\hline
 & \multicolumn{2}{|c|}{\bf AAA} & \multicolumn{2}{|c|}{\bf AAB} & \multicolumn{2}{|c|}{\bf ABB}& \multicolumn{2}{|c|}{\bf BBB}   
\\
{\rm orbit} & X & Y & X & Y & X & Y & X & Y 
\\\hline\hline
\OR & 1 & 0 & 1 & 1 & 0 & 3 & -3 & 6
\\
\OR\Z_2^{(1)} & 3 & 0 & 1 & 1 & 0 & 1 & -1 & 2 
\\
\OR\Z_2^{(3)} &  3 & 0 & 3 & 3 & 0 & 3 & -1 & 2 
\\
\OR\Z_2^{(2)} &  3 & 0 &  1 & 1 & 0 & 3 &  -1 & 2 
\\\hline
\end{array}
\end{equation*}
\end{center}
\caption{Bulk wrapping numbers for the four O6-plane orbits on $T^6/(\Z_2 \times \Z_6' \times \OR)$ without and with discrete torsion computed from the torus wrapping numbers in table~\protect\ref{Tab:Z2Z6p-ORonBulkO6plane-torus}
using~(\ref{Eq:Z2Z6p-Def-XY}).}
\label{Tab:Z2Z6p-ORonBulkO6plane-bulk}
\end{table}

The bulk RR tadpole cancellation conditions can now be written as 
\begin{equation}\label{Eq:Z2Z6p-bulkRRtcc}
\begin{aligned}
{\bf AAA:} \quad 0=& \left[\sum_a N_a  (2X_a + Y_a)  - 2^{\frac{3-\eta}{2}}
 \left( \eta_{\OR} + 3\sum_{i=1}^3 \eta_{\OR\Z_2^{(i)}}  \right) \right]
\rho_1
, 
\\
{\bf AAB:} \quad 0=&  \left[\sum_a N_a (X_a+Y_a) -  2^{\frac{3-\eta}{2}}
 \left( \sum_{i=0}^2  \eta_{\OR\Z_2^{(i)}}  + 3\, \eta_{\OR\Z_2^{(3)}}  \right)
 \right] (\rho_1 + \rho_2)
,
\\
{\bf ABB:} \quad 0=&  \left[\sum_a N_a (X_a+2Y_a) - 2^{\frac{3-\eta}{2}}
 \left( 3\sum_{i=0,2,3} \eta_{\OR\Z_2^{(i)}} +  \eta_{\OR\Z_2^{(1)}}  \right)\right]
\rho_2
,
\\
{\bf BBB:} \quad 0=& \left[\sum_a N_a   Y_a - 2^{\frac{3-\eta}{2}}
 \left( 3 \,  \eta_{\OR} +  \sum_{i=1}^3 \eta_{\OR\Z_2^{(i)}}  \right) \right] (-\rho_1 + 2 \rho_2) 
,
\end{aligned}
\end{equation}
with $\eta = \pm 1$ for the case without and with discrete torsion.
This is as expected $1=1+h_{21}^U$ condition per lattice.

Double counting of models is on this orbifold background avoided as follows:
\begin{itemize}
\item
\framebox{$(n^3,m^3)=$ (odd,odd)} selects one $\Z_6'$ orbifold image,
\item
\framebox{$n^3 >0$} fixes the orientation on $T^2_{(3)}$,
\item
\framebox{$0 \leq m^1 \leq 2 |n^1|$ on {\bf A} and $|n^1| \leq m^1$ on {\bf B}} 
fixes the orientation on $T^2_{(1)}$ and singles out a D6-brane compared to its orientifold image,
\item
if the D6-brane is at \framebox{angle $0,\pi/2$ on  $T^2_{(1)}$}, the orientifold image
can be excluded by demanding that the angle $\pi\phi^{(3)}$ w.r.t. the $\OR$ plane is in the range $(-\frac{\pi}{2},0)$;
using the relation~(\ref{Eq:angles}) leads to 
 \framebox{$-2n^3 < m^3 < 0$ on {\bf A} and $-n^3 < m^3 < n^3$ on {\bf B}} 
on  $T^2_{(3)}$,
\item
D6-branes parallel to some $\OR\theta^n\omega^m$ plane are treated separately; their torus wrapping numbers are given in 
table~\ref{Tab:Z2Z6p-ORonBulkO6plane-torus}.
\end{itemize}

{\boldmath
\subsection{The $\Z_2$ twisted parts}\label{Ss:Z2Z6p-twisted}
}

Each $\Z_2^{(\alpha)}$ twisted sector with $\alpha=1,2,3$  is similar to the $\Z_2$ twisted sector of $T^6/\Z_6$~\cite{Honecker:2004kb,Gmeiner:2007we}
 with a different normalisation factor of the exceptional three-cycles,
\begin{equation}
\mbox{\resizebox{0.9\textwidth}{!}{%
$\begin{aligned}
\varepsilon^{(\alpha)}_1 =2 \, \left( e^{(\alpha)}_{41} -  e^{(\alpha)}_{61} \right) \otimes \pi_1 +2 \, \left( e^{(\alpha)}_{61} -  e^{(\alpha)}_{51} \right) \otimes \pi_2,
&\quad
\tilde{\varepsilon}^{(\alpha)}_1 =2 \, \left( e^{(\alpha)}_{51} -  e^{(\alpha)}_{61} \right) \otimes \pi_1 +2 \, \left( e^{(\alpha)}_{41} -  e^{(\alpha)}_{51} \right) \otimes \pi_2,
\\
\varepsilon^{(\alpha)}_2 =2 \, \left( e^{(\alpha)}_{14} -  e^{(\alpha)}_{16} \right) \otimes \pi_1 +2 \, \left( e^{(\alpha)}_{16} -  e^{(\alpha)}_{15} \right) \otimes \pi_2,
&\quad 
\tilde{\varepsilon}^{(\alpha)}_2 =2 \, \left( e^{(\alpha)}_{15} -  e^{(\alpha)}_{16} \right) \otimes \pi_1 +2 \, \left( e^{(\alpha)}_{14} -  e^{(\alpha)}_{15} \right) \otimes \pi_2,
\\
\varepsilon^{(\alpha)}_3 =2 \, \left( e^{(\alpha)}_{44} -  e^{(\alpha)}_{66} \right) \otimes \pi_1 +2 \, \left( e^{(\alpha)}_{66} -  e^{(\alpha)}_{55} \right) \otimes \pi_2,
&\quad
\tilde{\varepsilon}^{(\alpha)}_3 =2 \, \left( e^{(\alpha)}_{55} -  e^{(\alpha)}_{66} \right) \otimes \pi_1 +2 \, \left( e^{(\alpha)}_{44} -  e^{(\alpha)}_{55} \right) \otimes \pi_2,
\\
\varepsilon^{(\alpha)}_4 =2 \, \left( e^{(\alpha)}_{45} -  e^{(\alpha)}_{64} \right) \otimes \pi_1 +2 \, \left( e^{(\alpha)}_{64} -  e^{(\alpha)}_{56} \right) \otimes \pi_2,
&\quad
\tilde{\varepsilon}^{(\alpha)}_4 =2 \, \left( e^{(\alpha)}_{56} -  e^{(\alpha)}_{64} \right) \otimes \pi_1 +2 \, \left( e^{(\alpha)}_{45} -  e^{(\alpha)}_{56} \right) \otimes \pi_2,
\\
\varepsilon^{(\alpha)}_5 =2 \, \left( e^{(\alpha)}_{46} -  e^{(\alpha)}_{65} \right) \otimes \pi_1 +2 \, \left( e^{(\alpha)}_{65} -  e^{(\alpha)}_{54} \right) \otimes \pi_2,
&\quad
\tilde{\varepsilon}^{(\alpha)}_5 =2 \, \left( e^{(\alpha)}_{54} -  e^{(\alpha)}_{65} \right) \otimes \pi_1 +2 \, \left( e^{(\alpha)}_{46} -  e^{(\alpha)}_{54} \right) \otimes \pi_2.
\end{aligned}$}}
\end{equation}
which leads to the intersection form
\begin{equation}
\varepsilon^{(\alpha)}_i \circ \tilde{\varepsilon}^{(\beta)}_j = -4 \, \delta_{ij} \, \delta^{\alpha\beta}
,
\end{equation}
and all other intersections vanish.

The relation between a single $\Z_2^{(\alpha)}$ fixed point and the exceptional three-cycles is given in table~\ref{Tab:Z2Z6p-fps+excycles},
and the complete assignment for a given bulk cycle, displacement $\vec{\sigma}$ and Wilson line
$\vec{\tau}$ is relegated to tables~\ref{Tab:Z2Z6p-Part1} ande~\ref{Tab:Z2Z6p-Part2} in appendix~\ref{App:Tables-Ex-Sectors}.
\begin{table}[h!]
\renewcommand{\arraystretch}{1.3}
  \begin{center}
\begin{equation*}
\begin{array}{|c|c|}\hline
\multicolumn{2}{|c|}{\Z_2^{(\alpha)} \; \text{\bf fixed points and exceptional 3-cycles on } T^6/\Z_2 \times \Z_6'} 
\\\hline\hline
{\rm f.p.}^{(\alpha)} \otimes (n^{\alpha} \pi_{2\alpha-1} + m^{\alpha} \pi_{2\alpha}) & {\rm orbit} 
\\\hline\hline
11 & - 
\\\hline
41 &   n^{\alpha} \varepsilon^{(\alpha)}_1 + m^{\alpha} \tilde{\varepsilon}^{(\alpha)}_1
\\\hline
51 & -(n^{\alpha} + m^{\alpha})  \varepsilon^{(\alpha)}_1 + n^{\alpha} \tilde{\varepsilon}^{(\alpha)}_1
\\\hline
61 & m^{\alpha} \varepsilon^{(\alpha)}_1 - (n^{\alpha} + m^{\alpha})  \tilde{\varepsilon}^{(\alpha)}_1
\\\hline\hline
14 &   n^{\alpha} \varepsilon^{(\alpha)}_2 + m^{\alpha} \tilde{\varepsilon}^{(\alpha)}_2
\\\hline
15 & -(n^{\alpha} + m^{\alpha})  \varepsilon^{(\alpha)}_2 + n^{\alpha} \tilde{\varepsilon}^{(\alpha)}_2
\\\hline
16 &  m^{\alpha} \varepsilon^{(\alpha)}_2 - (n^{\alpha} + m^{\alpha})  \tilde{\varepsilon}^{(\alpha)}_2
\\\hline\hline
44 &   n^{\alpha} \varepsilon^{(\alpha)}_3 + m^{\alpha} \tilde{\varepsilon}^{(\alpha)}_3
\\\hline
45 &    n^{\alpha} \varepsilon^{(\alpha)}_4 + m^{\alpha} \tilde{\varepsilon}^{(\alpha)}_4
\\\hline
46 &    n^{\alpha} \varepsilon^{(\alpha)}_5 + m^{\alpha} \tilde{\varepsilon}^{(\alpha)}_5
\\\hline
54 &  -(n^{\alpha} + m^{\alpha})  \varepsilon^{(\alpha)}_5 + n^{\alpha} \tilde{\varepsilon}^{(\alpha)}_5
\\\hline
55 &  -(n^{\alpha} + m^{\alpha})  \varepsilon^{(\alpha)}_3 + n^{\alpha} \tilde{\varepsilon}^{(\alpha)}_3
\\\hline
56 &  -(n^{\alpha} + m^{\alpha})  \varepsilon^{(\alpha)}_4 + n^{\alpha} \tilde{\varepsilon}^{(\alpha)}_4
\\\hline
64 &  m^{\alpha} \varepsilon^{(\alpha)}_4 - (n^{\alpha} + m^{\alpha})  \tilde{\varepsilon}^{(\alpha)}_4
\\\hline
65 &  m^{\alpha} \varepsilon^{(\alpha)}_5 - (n^{\alpha} + m^{\alpha})  \tilde{\varepsilon}^{(\alpha)}_5
\\\hline
66 &  m^{\alpha} \varepsilon^{(\alpha)}_3 - (n^{\alpha} + m^{\alpha})  \tilde{\varepsilon}^{(\alpha)}_3
\\\hline
\end{array}
\end{equation*}
\end{center}
\caption{Relation between a $\Z_2^{(\alpha)}$ fixed point times a 1-cycle on $T^2_{(\alpha)}$ and the exceptional three-cycles on
$T^6/\Z_2 \times \Z_6'$.}
\label{Tab:Z2Z6p-fps+excycles}
\end{table}

The exceptional three-cycles pick up a sign $-\eta_{(\alpha)}$ under the orientifold projection $\Omega$, while ${\cal R}$ permutes 
the $\Z_2^{(\alpha)}$ fixed points and one-cycles, see figure~\ref{Fig:Z4-Z6lattice}. 
The result of the orientifold projection on exceptional three-cycles from the $\Z_2^{(\alpha)}$ sector
is listed in tables~\ref{Tab:Z2Z6p-OR-ex-1},~\ref{Tab:Z2Z6p-OR-ex-2} and~\ref{Tab:Z2Z6p-OR-ex-3} for $\alpha=1,2,3$.
\begin{table}[h!]
\renewcommand{\arraystretch}{1.3}
  \begin{center}
\begin{equation*}
\begin{array}{|c||c|c||c|c|}\hline
\multicolumn{5}{|c|}{\OR \; \text{\bf  on exceptional 3-cycles on $T^6/\Z_2 \times \Z_6'$, Part I}}
\\\hline\hline
\text{3-cycle} & \varepsilon^{(1)}_i & \tilde{\varepsilon}^{(1)}_i & i=i' & i \leftrightarrow i'
\\\hline\hline
{\bf AAA} &  - \eta_{(1)} \, \varepsilon^{(1)}_{i'} & \eta_{(1)} \left( \tilde{\varepsilon}^{(1)}_{i'} -\varepsilon^{(1)}_{i'} \right) & 
1,2,3 & 4,5
\\\hline 
{\bf AAB}  & - \eta_{(1)} \, \varepsilon^{(1)}_{i'} & \eta_{(1)} \left( \tilde{\varepsilon}^{(1)}_{i'} -\varepsilon^{(1)}_{i'} \right) & 1,5 & 3,4
\\
& \eta_{(1)} \left( \varepsilon^{(1)}_{2} -  \tilde{\varepsilon}^{(1)}_{2} \right) & - \eta_{(1)} \, \tilde{\varepsilon}^{(1)}_{2} & 2 &  - 
\\\hline 
{\bf ABB} & \eta_{(1)} \left( \varepsilon^{(1)}_{i'} -  \tilde{\varepsilon}^{(1)}_{i'}\right) & - \eta_{(1)} \, \tilde{\varepsilon}^{(1)}_{i'} & 1,2,3 & 4,5
\\\hline 
{\bf BBB} & \eta_{(1)} \, \varepsilon^{(1)}_{i'} & \eta_{(1)} \left( \varepsilon^{(1)}_{i'} -  \tilde{\varepsilon}^{(1)}_{i'}\right)
& 1,2,3 & 4,5
\\\hline
\end{array}
\end{equation*}
\end{center}
\caption{Orientifold projection of the exceptional three-cycles from the $\Z_2^{(1)}$ twisted sector on the 
$T^6/(\Z_2 \times \Z_6' \times \OR)$ background. The sign factor is again $\eta_{(1)} \equiv \eta_{\OR} \eta_{\OR\Z_2^{(1)}}$.  }
\label{Tab:Z2Z6p-OR-ex-1}
\end{table}

\begin{table}[h!]
\renewcommand{\arraystretch}{1.3}
  \begin{center}
\begin{equation*}
\begin{array}{|c||c|c||c|c|}\hline
\multicolumn{5}{|c|}{\OR \; \text{\bf  on exceptional 3-cycles on $T^6/\Z_2 \times \Z_6'$, Part II}}
\\\hline\hline
\text{3-cycle} & \varepsilon^{(2)}_i & \tilde{\varepsilon}^{(2)}_i & i=i' & i \leftrightarrow i'
\\\hline\hline
{\bf AAA} &  - \eta_{(2)} \, \varepsilon^{(2)}_{i'} & \eta_{(2)} \left( \tilde{\varepsilon}^{(2)}_{i'} -\varepsilon^{(2)}_{i'} \right) & 
1,2,3 & 4,5
\\\hline 
{\bf AAB} & -\eta_{(2)} \, \varepsilon^{(2)}_{i'} & \eta_{(2)} \left( \tilde{\varepsilon}^{(2)}_{i'} -\varepsilon^{(2)}_{i'} \right) & 1,5 & 3,4
\\
& \eta_{(2)} \left( \varepsilon^{(2)}_{2} -  \tilde{\varepsilon}^{(2)}_{2}\right) & -\eta_{(2)} \,  \tilde{\varepsilon}^{(2)}_{2} & 2 & - 
\\\hline 
{\bf ABB} & -\eta_{(2)} \,  \tilde{\varepsilon}^{(2)}_{i'} & - \eta_{(2)} \, \varepsilon^{(2)}_{i'} & 1,5 & 3,4 
\\ & \eta_{(2)} \, \varepsilon^{(2)}_{2} & \eta_{(2)} \left( \varepsilon^{(2)}_{2} -  \tilde{\varepsilon}^{(3)}_{2}\right) & 2 & -
\\\hline 
{\bf BBB} & \eta_{(2)} \, \varepsilon^{(2)}_{i'} & \eta_{(2)} \left( \varepsilon^{(2)}_{i'} -  \tilde{\varepsilon}^{(2)}_{i'}\right)
& 1,2,3 & 4,5
\\\hline
\end{array}
\end{equation*}
\end{center}
\caption{Orientifold projection of the exceptional three-cycles from the $\Z_2^{(2)}$ twisted sector on the 
$T^6/(\Z_2 \times \Z_6' \times \OR)$ background.}
\label{Tab:Z2Z6p-OR-ex-2}
\end{table}

\begin{table}[h!]
\renewcommand{\arraystretch}{1.3}
  \begin{center}\begin{equation*}
\begin{array}{|c||c|c||c|c|}\hline
\multicolumn{5}{|c|}{\OR \,  \text{\bf  on exceptional 3-cycles on $T^6/\Z_2 \times \Z_6'$, Part III}}
\\\hline\hline
\text{3-cycle} & \varepsilon^{(3)}_i & \tilde{\varepsilon}^{(3)}_i & i=i' & i \leftrightarrow i'
\\\hline\hline
{\bf AAA} &  - \eta_{(3)} \, \varepsilon^{(3)}_{i'} & \eta_{(3)} \left( \tilde{\varepsilon}^{(3)}_{i'} -\varepsilon^{(3)}_{i'} \right) & 
1,2,3 & 4,5
\\\hline 
{\bf AAB} & -\eta_{(3)} \,  \tilde{\varepsilon}^{(3)}_{i'} &  - \eta_{(3)} \, \varepsilon^{(3)}_{i'} & 1,2,3 & 4,5
\\\hline 
{\bf ABB} & -\eta_{(3)} \,  \tilde{\varepsilon}^{(3)}_{i'} & - \eta_{(3)} \, \varepsilon^{(3)}_{i'} & 1,5 & 3,4 
\\
& \eta_{(3)} \, \varepsilon^{(3)}_{2} & \eta_{(3)} \left( \varepsilon^{(3)}_{2} -  \tilde{\varepsilon}^{(3)}_{2}\right) & 2 & -
\\\hline 
{\bf BBB} & \eta_{(3)} \, \varepsilon^{(3)}_{i'} & \eta_{(3)} \left( \varepsilon^{(3)}_{i'} -  \tilde{\varepsilon}^{(3)}_{i'}\right)
& 1,2,3 & 4,5
\\\hline
\end{array}
\end{equation*}
\end{center}
\caption{Orientifold projection of the exceptional three-cycles from the $\Z_2^{(3)}$ twisted sector on the 
$T^6/(\Z_2 \times \Z_6' \times \OR)$ background.}
\label{Tab:Z2Z6p-OR-ex-3}
\end{table}
If we write a fractional three-cycle in the form 
\begin{equation}\label{Eq:Z2Z6p-frac-expansion}
\begin{aligned}
\Pi^{\rm frac} =&  \frac{1}{4} \Pi^{\rm bulk}  + \frac{1}{4} \sum_{\alpha=1}^3 \Pi^{\Z_2^{(\alpha)}}
\\
=& \frac{1}{4} \left( X \rho_1 + Y \rho_2  \right)
+ \frac{1}{4} \sum_{\alpha=1}^3 \sum_{i=1}^5 \left( x^{(\alpha)}_i \, \varepsilon_i^{(\alpha)} + y ^{(\alpha)}_i \, \tilde{\varepsilon}_i^{(\alpha)} \right)
,
\end{aligned}
\end{equation}
the twisted RR tadpole cancellation conditions take the following form for the four inequivalent lattices,
{\allowdisplaybreaks[4]
\begin{equation*}
\begin{aligned}
& {\bf AAA:}\\
& \sum_{\alpha=1}^3 \sum_{i=1}^3 \left(\sum_a N_a 
[  (1 -\eta_{(\alpha)}) \, x^{(\alpha)}_{i,a} - \eta_{(\alpha)}\,  y^{(\alpha)}_{i,a}  ] \right) \varepsilon_i^{(\alpha)}
+ \left(\sum_a N_a (1 +\eta_{(\alpha)}) \,  y^{(\alpha)}_{i,a}
\right) \, \tilde{\varepsilon}_i^{(\alpha)} =
\\
& - \sum_{\alpha=1}^3 \left(\sum_a N_a [ x^{(\alpha)}_{4,a}-\eta_{(\alpha)} x^{(\alpha)}_{5,a} + \frac{1}{2}(y^{(\alpha)}_{4,a} - \eta_{(\alpha)} y^{(\alpha)}_{5,a}) ] \right) \, (\varepsilon_4^{(\alpha)} - \eta_{(\alpha)}\varepsilon_5^{(\alpha)} )
\\
& - \sum_{\alpha=1}^3\left(\sum_a N_a (y^{(\alpha)}_{4,a} + \eta_{(\alpha)}y^{(\alpha)}_{5,a}  ) \right)\, [\tilde{\varepsilon}_4^{(\alpha)} + \eta_{(\alpha)}\, \tilde{\varepsilon}_5^{(\alpha)}-\frac{1}{2} (\varepsilon_4^{(\alpha)} + \eta_{(1)}\varepsilon_5^{(\alpha)} ) ]
,
\end{aligned}
\end{equation*}
%
\begin{align*}
&{\bf AAB:}\\
& \sum_{\alpha=1}^2 \sum_{i=1,5}\!\! \left\{ \left(\sum_a N_a [
    (1 -\eta_{(\alpha)}) \, x^{(\alpha)}_{i,a} - \eta_{(\alpha)}\,
    y^{(\alpha)}_{i,a} ] \right) \varepsilon_i^{(\alpha)} 
\!+\! \left(\!\sum_a N_a\! (1 +\eta_{(\alpha)}) \,  y^{(\alpha)}_{i,a}
\right)  \tilde{\varepsilon}_i^{(\alpha)} 
\right\} =
\\
& - \sum_{\alpha=1}^2 \left(\sum_a N_a [ x^{(\alpha)}_{3,a}-\eta_{(\alpha)} x^{(\alpha)}_{4,a} + \frac{1}{2}(y^{(\alpha)}_{3,a} - \eta_{(\alpha)} y^{(\alpha)}_{4,a}) ] \right)\, (\varepsilon_3^{(\alpha)} - \eta_{(\alpha)}\varepsilon_4^{(\alpha)} )
\\
& - \sum_{\alpha=1}^2 \left(\sum_a N_a (y^{(\alpha)}_{3,a} + \eta_{(\alpha)}y^{(\alpha)}_{4,a}  ) \right)
\, [\tilde{\varepsilon}_3^{(\alpha)} + \eta_{(\alpha)}\, \tilde{\varepsilon}_4^{(\alpha)}-\frac{1}{2} (\varepsilon_3^{(\alpha)} + \eta_{(\alpha)}\varepsilon_4^{(\alpha)} ) ]
 \\
& - \sum_{\alpha=1}^2\! \left\{ \!\left(\!\sum_a\! N_a (1
    +\eta_{(\alpha)}) x^{(\alpha)}_{2,a}\right) 
  \varepsilon_{2}^{(\alpha)}  
-  \left(\sum_a N_a [-\eta_{(1)} \, x^{(\alpha)}_{2,a} + (1
  -\eta_{(\alpha)})  y^{(\alpha)}_{2,a} ]\right) \tilde{
  \varepsilon}_{2}^{(\alpha)} \right\} ,
\\
& \sum_{i=1}^3 \left( \sum_a N_a (x^{(3)}_{i,a} -\eta_{(3)}
  y^{(3)}_{i,a}) \right) \, (\varepsilon_{i}^{(3)}  -\eta_{(3)}
\tilde{\varepsilon}_{i}^{(3)} ) =
\\
& - \left( \sum_a N_a (x^{(3)}_{4,a}\! -\!\eta_{(3)}
  y^{(3)}_{5,a})\right)( \varepsilon_{4}^{(3)} - \eta_{(3)}
\tilde{\varepsilon}_{5}^{(3)} ) \\ 
& - \left( \sum_a N_a  (x^{(3)}_{5,a} -\eta_{(3)} y^{(3)}_{4,a} )
\right)(\varepsilon_{5}^{(3)}- \eta_{(3)}
\tilde{\varepsilon}_{4}^{(3)} ) 
,
\end{align*}
\begin{align*}
&{\bf ABB:} \\ & \!\!\sum_{i=1}^3\!\! \left\{\left(\sum_a N_a  (1
    +\eta_{(1)}) \, x^{(1)}_{i,a} \right) \, \varepsilon_{i}^{(1)}\!  
+\! \left(\!\sum_a\! N_a [-\eta_{(1)} \, x^{(1)}_{i,a}\! +\! (1 -\eta_{(1)}) \,
  y^{(1)}_{i,a} ] \right) \tilde{ \varepsilon}_{i}^{(1)} \right\} =
\\
& - \left( \sum_a N_a   (x^{(1)}_{4,a}+ \eta_{(1)} x^{(1)}_{5,a}) \right)
\, [\varepsilon_4^{(1)} +\eta_{(1)}  \varepsilon_5^{(1)} -\frac{1}{2}( \tilde{\varepsilon}_4^{(1)} + \eta_{(1)}  \tilde{\varepsilon}_5^{(1)}) ]
\\
& -  \left( \sum_a N_a  [y^{(1)}_{4,a} - \eta_{(1)} y^{(1)}_{5,a} +\frac{1}{2}(x^{(1)}_{4,a}- \eta_{(1)} x^{(1)}_{5,a})  ] \right)
 \,( \tilde{\varepsilon}_4^{(1)} - \eta_{(1)}  \tilde{\varepsilon}_5^{(1)}),
\\
& \sum_{\alpha=2}^3 \sum_{i=1,5} \left( \sum_a N_a
  (x^{(\alpha)}_{i,a} -\eta_{(\alpha)}  y^{(\alpha)}_{i,a}) \right) \,
(\varepsilon_{i}^{(\alpha)}  -\eta_{(\alpha)}
\tilde{\varepsilon}_{i}^{(\alpha)} ) =
\\
& - \sum_{\alpha=2}^3 \left(\sum_a N_a (x^{(\alpha)}_{3,a} -\eta_{(\alpha)} y^{(\alpha)}_{4,a}) \right)
( \varepsilon_{3}^{(\alpha)} - \eta_{(\alpha)} \tilde{\varepsilon}_{4}^{(\alpha)} )
\\ 
& -  \sum_{\alpha=2}^3 \left(\sum_a N_a  (x^{(\alpha)}_{4,a} -\eta_{(\alpha)} y^{(\alpha)}_{3,a} )\right)(\varepsilon_{4}^{(\alpha)}- \eta_{(\alpha)} \tilde{\varepsilon}_{3}^{(\alpha)} )
\\
& - \sum_{\alpha=2}^3 \left\{\left( \sum_a N_a  [  (1 +\eta_{(\alpha)}) \, x^{(\alpha)}_{2,a} + \eta_{(\alpha)}\,  y^{(\alpha)}_{2,a} ] \right)
\varepsilon_2^{(\alpha)}- \left( \sum_a N_a (1 -\eta_{(\alpha)}) \,  y^{(\alpha)}_{2,a} \right) \, \tilde{\varepsilon}_2^{(\alpha)}
\right\}
,
\end{align*}
\begin{align*}
& {\bf BBB:} \\& \sum_{\alpha=1}^3 \sum_{i=1}^3 \left(\sum_a N_a 
[  (1 + \eta_{(\alpha)}) \, x^{(\alpha)}_{i,a} + \eta_{(\alpha)}\,  y^{(\alpha)}_{i,a}  ] \right) \varepsilon_i^{(\alpha)}
+ \left(\sum_a N_a (1 - \eta_{(\alpha)}) \,  y^{(\alpha)}_{i,a}
\right) \, \tilde{\varepsilon}_i^{(\alpha)} =
\\
& - \sum_{\alpha=1}^3 \left(\sum_a N_a [ x^{(\alpha)}_{4,a}+\eta_{(\alpha)} x^{(\alpha)}_{5,a} + \frac{1}{2}(y^{(\alpha)}_{4,a} + \eta_{(\alpha)} y^{(\alpha)}_{5,a}) ] \right) \, (\varepsilon_4^{(\alpha)} + \eta_{(\alpha)}\varepsilon_5^{(\alpha)} )
\\
& - \sum_{\alpha=1}^3\left(\sum_a N_a (y^{(\alpha)}_{4,a} - \eta_{(\alpha)}y^{(\alpha)}_{5,a}  ) \right)\, [\tilde{\varepsilon}_4^{(\alpha)} - \eta_{(\alpha)}\, \tilde{\varepsilon}_5^{(\alpha)}-\frac{1}{2} (\varepsilon_4^{(\alpha)} - \eta_{(1)}\varepsilon_5^{(\alpha)} ) ]
.
\end{align*}
}
Notice that the twisted RR tadpole conditions on {\bf AAA} and {\bf BBB} are related by replacing $\eta_{(\alpha)} \rightarrow - \eta_{(\alpha)}$.

The integer coefficients $( x^{(\alpha)}_{i},y^{(\alpha)}_{i})$ can be read off from tables~\ref{Tab:Z2Z6p-Part1} and~\ref{Tab:Z2Z6p-Part2}
in appendix~\ref{App:Tables-Ex-Sectors}.
Per twist sector and D6-brane, exactly three of the five pairs are non-vanishing. Each of the pairs originates from either
a single fixed point (I) or two fixed point contributions (II). Their shape is restricted to the following six options,
\begin{equation}\nonumber
\begin{array}{|c|c|}\hline
\multicolumn{2}{|c|}{(x,y)}
\\\hline
{\rm I.} & {\rm II.}
\\\hline\hline
\pm (n,m) & \pm(-n+[z-1] \, m,-z \, m+[1-z] \, n)
\\
\pm (-n-m,n) & \pm (n+z \, m,-z \, n +[1-z] \, m)
\\
\pm (m,-n-m) &  \pm ([1-z] \, n-z \, m , z \, n + m)
\\\hline
\end{array}
\end{equation}
with $z \in \{(-1)^{\tau_j},(-1)^{\tau_k},(-1)^{\tau_j+ \tau_k}\}$ in the $\Z_2^{(i)}$ twisted sector.
The global signs depend also on these Wilson lines, as well as the $\Z_2^{(i)}$ eigenvalue $\tau_0^{(i)}$.
Note that this shape is very similar to the one for the $\Z_2^{(2)}$ and $\Z_2^{(3)}$ sectors on $T^6/\Z_2 \times \Z_6$ with discrete torsion
discussed in section~\ref{Ss:Z2Z6-twisted-part}.

The knowledge of the general form of the coefficients in the twisted sector contributions is expected to be useful when simplifying the 
K-theory constraints in the following section.

\subsection{K-theory constraint}\label{Ss:Z2Z6p-Ktheory}

The classification of $\OR$ invariant D6-branes follows again closely the one for $T^6/\Z_2 \times \Z_2$ in section~\ref{Ss:Z2Z2Ktheory}.
Possible probe D6-branes are of the form~(\ref{Eq:define-probes}), where the bulk contributions can be read off from 
table~\ref{Tab:Z2Z6p-ORonBulkO6plane-bulk}, 
and the exceptional contributions are listed in tables~\ref{Tab:Z2Z6p-Ktheory-AAA+BBB} to~\ref{Tab:Z2Z6p-Ktheory-ABB} for the four inequivalent lattice
orientations.
The existence of $\OR$ invariant D6-branes boils again down to the relations in table~\ref{Tab:Conditions-on_b+t+s-SOSp} among the sets of signs in the
orientifold projection $\{\eta_{(i)}\}$ and discrete displacements $\vec{\sigma}$ and Wilson lines $\vec{\tau}$, where 
$b_1=b_2=b_3=\frac{1}{2}$ due to the $SU(3)^3$ compactifaction lattice.
\begin{table}[h!]
\renewcommand{\arraystretch}{1.3}
  \begin{center}
\begin{equation*}
\begin{array}{|c|c|}\hline
\multicolumn{2}{|c|}{\text{\bf Exceptional contributions to $\OR$ invariant D6-branes on $T^6/\Z_2 \times \Z_6'$, Part I}}
\\\hline\hline
\multicolumn{2}{|c|}{\bf AAA}
\\\hline\hline
\Pi^{\Z_2^{(i)}}_{h,(0,0)} & (-1)^{\tau_j} \varepsilon^{(i)}_1 + (-1)^{\tau_k} \varepsilon^{(i)}_2  +(-1)^{\tau_j+\tau_k} \varepsilon^{(i)}_3 
\\\hline
\Pi^{\Z_2^{(i)}}_{h,(1,0)} & - \varepsilon^{(i)}_1 +[1-(-1)^{\tau_j}]\tilde{\varepsilon}^{(i)}_1
-(-1)^{\tau_j+\tau_k}\tilde{\varepsilon}^{(i)}_4
 +(-1)^{\tau_k}[- \varepsilon^{(i)}_5 + \tilde{\varepsilon}^{(i)}_5]
\\\hline
\Pi^{\Z_2^{(i)}}_{h,(0,1)} &-\varepsilon^{(i)}_2 + (1-(-1)^{\tau_k})\tilde{\varepsilon}^{(i)}_2
+ (-1)^{\tau_j} \varepsilon^{(i)}_4 + (-1)^{\tau_j+\tau_k} \varepsilon^{(i)}_5 
\\\hline
\Pi^{\Z_2^{(i)}}_{h,(1,1)} &
-\varepsilon^{(i)}_3 + (1-(-1)^{\tau_j+\tau_k})\tilde{\varepsilon}^{(i)}_3+(-1)^{\tau_k}[- \varepsilon^{(i)}_4 + \tilde{\varepsilon}^{(i)}_4]
 -(-1)^{\tau_j}  \tilde{\varepsilon}^{(i)}_5 
\\\hline\hline
\Pi^{\Z_2^{(i)}}_{v,(0,0)} &
(-1)^{\tau_j}[- \varepsilon^{(i)}_1 + 2 \tilde{\varepsilon}^{(i)}_1] + (-1)^{\tau_k} [- \varepsilon^{(i)}_2 + 2 \tilde{\varepsilon}^{(i)}_2]
 +(-1)^{\tau_j+\tau_k}[- \varepsilon^{(i)}_3 + 2 \tilde{\varepsilon}^{(i)}_3]
\\\hline
\Pi^{\Z_2^{(i)}}_{v,(1,0)} &
[2(-1)^{\tau_j}-1)]\varepsilon^{(i)}_1 +[-1-(-1)^{\tau_j}]\tilde{\varepsilon}^{(i)}_1
+(-1)^{\tau_j+\tau_k}[2 \varepsilon^{(i)}_4 - \tilde{\varepsilon}^{(i)}_4] +(-1)^{\tau_k}[- \varepsilon^{(i)}_5 - \tilde{\varepsilon}^{(i)}_5] 
\\\hline
\Pi^{\Z_2^{(i)}}_{v,(0,1)} & [-1+2(-1)^{\tau_k}]\varepsilon^{(i)}_2
-[1+(-1)^{\tau_k}]\tilde{\varepsilon}^{(i)}_2
+ (-1)^{\tau_j}[ - \varepsilon^{(i)}_4 + 2\tilde{\varepsilon}^{(i)}_4] 
+ (-1)^{\tau_j+\tau_k}[ - \varepsilon^{(i)}_5 + 2 \tilde{\varepsilon}^{(i)}_5]
\\\hline
\Pi^{\Z_2^{(i)}}_{v,(1,1)} &
[-1+2 (-1)^{\tau_j+\tau_k}]\varepsilon^{(i)}_3 - [1+(-1)^{\tau_j+\tau_k}]\tilde{\varepsilon}^{(i)}_3
- (-1)^{\tau_k}[ \varepsilon^{(i)}_4 +  \tilde{\varepsilon}^{(i)}_4]
 +(-1)^{\tau_j}[2 \varepsilon^{(i)}_5 - \tilde{\varepsilon}^{(i)}_5]
\\\hline\hline\hline
\multicolumn{2}{|c|}{\bf BBB}
\\\hline\hline
\Pi^{\Z_2^{(i)}}_{h,(0,0)} & (-1)^{\tau_j}\left[\varepsilon^{(i)}_1 - 2 \tilde{\varepsilon}^{(i)}_1   \right]
+ (-1)^{\tau_k}\left[\varepsilon^{(i)}_2 - 2 \tilde{\varepsilon}^{(i)}_2   \right]
+ (-1)^{\tau_j+\tau_k}\left[\varepsilon^{(i)}_3 - 2 \tilde{\varepsilon}^{(i)}_3   \right]
\\\hline
\Pi^{\Z_2^{(i)}}_{h,(1,0)} &(1-2 (-1)^{\tau_j}) \varepsilon^{(i)}_1  + (1 +(-1)^{\tau_j}) \tilde{\varepsilon}^{(i)}_1  
+(-1)^{\tau_j + \tau_k}[-2  \varepsilon^{(i)}_4 + \tilde{\varepsilon}^{(i)}_4 ] + (-1)^{\tau_k} [\varepsilon^{(i)}_5 + \tilde{\varepsilon}^{(i)}_5]
\\\hline
\Pi^{\Z_2^{(i)}}_{h,(0,1)} &(1 -2(-1)^{\tau_k}) \varepsilon^{(i)}_2
 + (1 +(-1)^{\tau_k}) \tilde{\varepsilon}^{(i)}_2  
+(-1)^{\tau_j} [ \varepsilon^{(i)}_4 - 2  \tilde{\varepsilon}^{(i)}_4]
+(-1)^{\tau_j+\tau_k} [ \varepsilon^{(i)}_5 - 2  \tilde{\varepsilon}^{(i)}_5]
\\\hline
\Pi^{\Z_2^{(i)}}_{h,(1,1)} &(1-2(-1)^{\tau_j+\tau_k}) \varepsilon^{(i)}_3
+ (1 +(-1)^{\tau_j+\tau_k}) \tilde{\varepsilon}^{(i)}_3  
+(-1)^{\tau_k}[  \varepsilon^{(i)}_4 +  \tilde{\varepsilon}^{(i)}_4]
+(-1)^{\tau_j}[ -2 \varepsilon^{(i)}_5 +  \tilde{\varepsilon}^{(i)}_5]
\\\hline\hline
\Pi^{\Z_2^{(i)}}_{v,(0,0)} & -(-1)^{\tau_j}\varepsilon^{(i)}_1- (-1)^{\tau_k}\varepsilon^{(i)}_2 - (-1)^{\tau_j+\tau_k} \varepsilon^{(i)}_3 
\\\hline
\Pi^{\Z_2^{(i)}}_{v,(1,0)} &  \varepsilon^{(i)}_1 + (-1 +(-1)^{\tau_j}) \tilde{\varepsilon}^{(i)}_1  
+(-1)^{\tau_j + \tau_k}  \tilde{\varepsilon}^{(i)}_4  + (-1)^{\tau_k} [ \varepsilon^{(i)}_5 - \tilde{\varepsilon}^{(i)}_5]
\\\hline
\Pi^{\Z_2^{(i)}}_{v,(0,1)} & \varepsilon^{(i)}_2 + (-1 +(-1)^{\tau_k}) \tilde{\varepsilon}^{(i)}_2  
-(-1)^{\tau_j}  \varepsilon^{(i)}_4
-(-1)^{\tau_j+\tau_k}  \varepsilon^{(i)}_5 
\\\hline
\Pi^{\Z_2^{(i)}}_{v,(1,1)} & \varepsilon^{(i)}_3
+ (-1 +(-1)^{\tau_j+\tau_k}) \tilde{\varepsilon}^{(i)}_3  
+(-1)^{\tau_k}[\varepsilon^{(i)}_4 - \tilde{\varepsilon}^{(i)}_4]
+(-1)^{\tau_j} \tilde{\varepsilon}^{(i)}_5
\\\hline
\end{array}
\end{equation*}
\end{center}
\caption{Exceptional three-cycles for $\OR$ invariant D6-branes, which contribute to the K-theory constraint on 
the {\bf AAA} and {\bf BBB} lattices on $T^6/(\Z_2 \times \Z_6' \times \OR)$ with discrete torsion.
All cycles habe to be multiplied by the $\Z_2^{(i)}$ eigenvalues $(-1)^{\tau_0^{(i)}}$.
The subscript $h$ labels 1-cycles parallel to the $\OR$ plane, cycles with subscript $v$ have the 1-cycle contribution 
perpendicular to the $\OR$ plane.
The discrete displacements on the remaining four-torus are given in parenthesis. 
Notice the symmetry
$ {\bf AAA:} ( \Pi^{\Z_2^{(i)}}_{h,(\sigma_j,\sigma_k)},\Pi^{\Z_2^{(i)}}_{v,(\sigma_j,\sigma_k)}) \Leftrightarrow
 {\bf BBB:} ( -\Pi^{\Z_2^{(i)}}_{v,(\sigma_j,\sigma_k)} , - \Pi^{\Z_2^{(i)}}_{h,(\sigma_j,\sigma_k)})$
for the same choice of discrete displacements $(\sigma_j,\sigma_k)$.}
\label{Tab:Z2Z6p-Ktheory-AAA+BBB}
\end{table}
\begin{table}[h!]
\renewcommand{\arraystretch}{1.3}
  \begin{center}
\begin{equation*}
\begin{array}{|c|c|}\hline
\multicolumn{2}{|c|}{\text{\bf Exceptional contributions to $\OR$ invariant D6-branes on $T^6/\Z_2 \times \Z_6'$, Part II}}
\\\hline\hline
\multicolumn{2}{|c|}{{\bf AAB}: \Z_2^{(i)} \text{ with } i=1,2}
\\\hline\hline
\Pi^{\Z_2^{(i)}}_{h,(0,0)} &(-1)^{\tau_j}  \varepsilon^{(i)}_1 
- (-1)^{\tau_3}  \tilde{\varepsilon}^{(i)}_2   + (-1)^{\tau_j+\tau_3}  \varepsilon^{(i)}_5
\\\hline
\Pi^{\Z_2^{(i)}}_{h,(1,0)} & - \varepsilon^{(i)}_1
+ [1 - (-1)^{\tau_j}]\tilde{\varepsilon}^{(i)}_1
- (-1)^{\tau_j+\tau_3}\tilde{\varepsilon}^{(i)}_3 
+(-1)^{\tau_3}[ - \varepsilon^{(i)}_4 +  \tilde{\varepsilon}^{(i)}_4]
\\\hline
\Pi^{\Z_2^{(i)}}_{h,(0,1)} & (1-(-1)^{\tau_3}) \varepsilon^{(i)}_2 +(-1)^{\tau_3} \tilde{\varepsilon}^{(i)}_2  
 +(-1)^{\tau_j}\varepsilon^{(i)}_3+ (-1)^{\tau_j+\tau_3}\varepsilon^{(i)}_4 
\\\hline
\Pi^{\Z_2^{(i)}}_{h,(1,1)} & (-1)^{\tau_3}[ - \varepsilon^{(i)}_3 +  \tilde{\varepsilon}^{(i)}_3]
-(-1)^{\tau_j} \tilde{\varepsilon}^{(i)}_4   
- \varepsilon^{(i)}_5 + [1 - (-1)^{\tau_j+\tau_3}]\tilde{\varepsilon}^{(i)}_5
\\\hline\hline
\Pi^{\Z_2^{(i)}}_{v,(0,0)} & (-1)^{\tau_j} [- \varepsilon^{(i)}_1 +2 \tilde{\varepsilon}^{(i)}_1]  
+ (-1)^{\tau_3} [2 \varepsilon^{(i)}_2 - \tilde{\varepsilon}^{(i)}_2 ]  
+ (-1)^{\tau_j+\tau_3} [-\varepsilon^{(i)}_5 +2 \tilde{\varepsilon}^{(i)}_5] 
\\\hline
\Pi^{\Z_2^{(i)}}_{v,(1,0)} & [2(-1)^{\tau_j}-1]\varepsilon^{(i)}_1
- [ (-1)^{\tau_j} +1]\tilde{\varepsilon}^{(i)}_1
+ (-1)^{\tau_j+\tau_3}[2\varepsilon^{(i)}_3 - \tilde{\varepsilon}^{(i)}_3 ]
+(-1)^{\tau_3}[ - \varepsilon^{(i)}_4 - \tilde{\varepsilon}^{(i)}_4]
\\\hline
\Pi^{\Z_2^{(i)}}_{v,(0,1)} & -(1+(-1)^{\tau_3}) \varepsilon^{(i)}_2
 + (2 -(-1)^{\tau_3}) \tilde{\varepsilon}^{(i)}_2  
 +(-1)^{\tau_j}[-\varepsilon^{(i)}_3 +2 \tilde{\varepsilon}^{(i)}_3 ]
 + (-1)^{\tau_j+\tau_3}[-\varepsilon^{(i)}_4 +2 \tilde{\varepsilon}^{(i)}_4 ]
\\\hline
\Pi^{\Z_2^{(i)}}_{v,(1,1)} & - (-1)^{\tau_3}[  \varepsilon^{(i)}_3 + \tilde{\varepsilon}^{(i)}_3]
+(-1)^{\tau_j}[2\varepsilon^{(i)}_4 - \tilde{\varepsilon}^{(i)}_4 ]
[2(-1)^{\tau_j+\tau_3}-1]\varepsilon^{(i)}_5
-[1 + (-1)^{\tau_j+\tau_3}]\tilde{\varepsilon}^{(i)}_5
\\\hline\hline\hline
\multicolumn{2}{|c|}{{\bf AAB}: \Z_2^{(3)}}
\\\hline\hline
\Pi^{\Z_2^{(3)}}_{h,(0,0)} & (-1)^{\tau_1}[ \varepsilon^{(3)}_1 +  \tilde{\varepsilon}^{(3)}_1]
 + (-1)^{\tau_2} [ \varepsilon^{(3)}_2 +  \tilde{\varepsilon}^{(3)}_2]
 +(-1)^{\tau_1+\tau_2}[\varepsilon^{(3)}_3 + \tilde{\varepsilon}^{(3)}_3]
\\\hline
\Pi^{\Z_2^{(3)}}_{h,(1,0)} & [(-1)^{\tau_1}-2]\varepsilon^{(3)}_1
+[1-2(-1)^{\tau_1}]\tilde{\varepsilon}^{(3)}_1
 +(-1)^{\tau_1+\tau_2}[ \varepsilon^{(3)}_4 - 2 \tilde{\varepsilon}^{(3)}_4]
 +(-1)^{\tau_2}[-2 \varepsilon^{(3)}_5 + \tilde{\varepsilon}^{(3)}_5]
\\\hline
\Pi^{\Z_2^{(3)}}_{h,(0,1)} & [(-1)^{\tau_2}-2]\varepsilon^{(3)}_2
+ [1-2(-1)^{\tau_2}]\tilde{\varepsilon}^{(3)}_2
+ (-1)^{\tau_1}[ \varepsilon^{(3)}_4 +  \tilde{\varepsilon}^{(3)}_4]
+ (-1)^{\tau_1+\tau_2}[ \varepsilon^{(i)}_5 +  \tilde{\varepsilon}^{(3)}_5]
\\\hline
\Pi^{\Z_2^{(3)}}_{h,(1,1)} &[(-1)^{\tau_1+\tau_2}-2]\varepsilon^{(3)}_3
 + [1-2(-1)^{\tau_1+\tau_2}]\tilde{\varepsilon}^{(3)}_3
 +(-1)^{\tau_2}[- 2 \varepsilon^{(3)}_4 + \tilde{\varepsilon}^{(3)}_4]
 +(-1)^{\tau_1}[ \varepsilon^{(3)}_5 - 2 \tilde{\varepsilon}^{(3)}_5]
\\\hline\hline
\Pi^{\Z_2^{(3)}}_{v,(0,0)} & (-1)^{\tau_1}[ \varepsilon^{(3)}_1 -  \tilde{\varepsilon}^{(3)}_1]
 + (-1)^{\tau_2} [ \varepsilon^{(3)}_2 -  \tilde{\varepsilon}^{(3)}_2]
 +(-1)^{\tau_1+\tau_2}[\varepsilon^{(3)}_3 - \tilde{\varepsilon}^{(3)}_3]
\\\hline
\Pi^{\Z_2^{(3)}}_{v,(1,0)} & -(-1)^{\tau_1} \varepsilon^{(3)}_1 +\tilde{\varepsilon}^{(3)}_1
 -(-1)^{\tau_1+\tau_2} \varepsilon^{(3)}_4  +(-1)^{\tau_2} \tilde{\varepsilon}^{(3)}_5
\\\hline
\Pi^{\Z_2^{(3)}}_{v,(0,1)} & -(-1)^{\tau_2} \varepsilon^{(3)}_2+\tilde{\varepsilon}^{(3)}_2
 + (-1)^{\tau_1}[ \varepsilon^{(3)}_4 - \tilde{\varepsilon}^{(3)}_4]
 + (-1)^{\tau_1+\tau_2}[ \varepsilon^{(i)}_5 - \tilde{\varepsilon}^{(3)}_5]
\\\hline
\Pi^{\Z_2^{(3)}}_{v,(1,1)} & -(-1)^{\tau_1+\tau_2} \varepsilon^{(3)}_3  +\tilde{\varepsilon}^{(3)}_3
 +(-1)^{\tau_2} \tilde{\varepsilon}^{(3)}_4 -(-1)^{\tau_1}\varepsilon^{(3)}_5
\\\hline
\end{array}
\end{equation*}
\end{center}
\caption{Exceptional three-cycles for $\OR$ invariant D6-branes, which contribute to the K-theory constraint on 
the {\bf AAB} lattice on $T^6/(\Z_2 \times \Z_6' \times \OR)$ with discrete torsion.
All cycles habe to be multiplied by the $\Z_2^{(i)}$ eigenvalues $(-1)^{\tau_0^{(i)}}$.
For more details on the notation see  table~\protect\ref{Tab:Z2Z6p-Ktheory-AAA+BBB}.}
\label{Tab:Z2Z6p-Ktheory-AAB}
\end{table}
\begin{table}[h!]
\renewcommand{\arraystretch}{1.3}
  \begin{center}
\begin{equation*}
\begin{array}{|c|c|}\hline
\multicolumn{2}{|c|}{\text{\bf Exceptional contributions to $\OR$ invariant D6-branes on $T^6/\Z_2 \times \Z_6'$, Part III}}
\\\hline\hline
\multicolumn{2}{|c|}{{\bf ABB}: \Z_2^{(1)}}
\\\hline\hline
\Pi^{\Z_2^{(1)}}_{h,(0,0)} & -(-1)^{\tau_2}\tilde{\varepsilon}^{(1)}_1  
-(-1)^{\tau_3}\tilde{\varepsilon}^{(1)}_2  
- (-1)^{\tau_2+\tau_3}\tilde{\varepsilon}^{(1)}_3  
\\\hline
\Pi^{\Z_2^{(1)}}_{h,(1,0)} & (1-(-1)^{\tau_2}) \varepsilon^{(1)}_1 + (-1)^{\tau_2} \tilde{\varepsilon}^{(1)}_1  
+(-1)^{\tau_2 + \tau_3}[- \varepsilon^{(1)}_4 +  \tilde{\varepsilon}^{(1)}_4 ]+ (-1)^{\tau_3}  \varepsilon^{(1)}_5 
\\\hline
\Pi^{\Z_2^{(1)}}_{h,(0,1)} & (1-(-1)^{\tau_3}) \varepsilon^{(1)}_2 + (-1)^{\tau_3} \tilde{\varepsilon}^{(1)}_2  
-(-1)^{\tau_2}  \tilde{\varepsilon}^{(1)}_4 -(-1)^{\tau_2+\tau_3}  \tilde{\varepsilon}^{(1)}_5
\\\hline
\Pi^{\Z_2^{(1)}}_{h,(1,1)} & (1-(-1)^{\tau_2+\tau_3})\varepsilon^{(1)}_3 + (-1)^{\tau_2+\tau_3} \tilde{\varepsilon}^{(1)}_3  
+(-1)^{\tau_3} \varepsilon^{(1)}_4 +(-1)^{\tau_2}[ -  \varepsilon^{(1)}_5 +  \tilde{\varepsilon}^{(1)}_5]
\\\hline\hline
\Pi^{\Z_2^{(1)}}_{v,(0,0)} & (-1)^{\tau_2}\left[2 \varepsilon^{(1)}_1 - \tilde{\varepsilon}^{(1)}_1   \right]
+ (-1)^{\tau_3}\left[2 \varepsilon^{(1)}_2 -\tilde{\varepsilon}^{(1)}_2   \right]
+ (-1)^{\tau_2+\tau_3}\left[2 \varepsilon^{(1)}_3 - \tilde{\varepsilon}^{(1)}_3   \right]
\\\hline
\Pi^{\Z_2^{(1)}}_{v,(1,0)} & -[1+(-1)^{\tau_2}] \varepsilon^{(1)}_1 
+ (2 -(-1)^{\tau_2}) \tilde{\varepsilon}^{(1)}_1  
-(-1)^{\tau_2 + \tau_3}[ \varepsilon^{(1)}_4 + \tilde{\varepsilon}^{(1)}_4 ] 
+ (-1)^{\tau_3} [- \varepsilon^{(1)}_5 + 2 \tilde{\varepsilon}^{(1)}_5]
\\\hline
\Pi^{\Z_2^{(1)}}_{v,(0,1)} & -[1+(-1)^{\tau_3}] \varepsilon^{(1)}_2  
 + (2-(-1)^{\tau_3}) \tilde{\varepsilon}^{(1)}_2  
+(-1)^{\tau_2} [2 \varepsilon^{(1)}_4 -  \tilde{\varepsilon}^{(1)}_4]
+(-1)^{\tau_2+\tau_3} [2 \varepsilon^{(1)}_5 -   \tilde{\varepsilon}^{(1)}_5]
\\\hline
\Pi^{\Z_2^{(1)}}_{v,(1,1)} & -[1+(-1)^{\tau_2+\tau_3}] \varepsilon^{(1)}_3 
 + (2-(-1)^{\tau_2+\tau_3}) \tilde{\varepsilon}^{(1)}_3  
+(-1)^{\tau_3}[  - \varepsilon^{(1)}_4 + 2 \tilde{\varepsilon}^{(1)}_4]
-(-1)^{\tau_2}[  \varepsilon^{(1)}_5 + \tilde{\varepsilon}^{(1)}_5]
\\\hline\hline\hline
\multicolumn{2}{|c|}{{\bf ABB}: \Z_2^{(i)} \text{ with } i=2,3}
\\\hline\hline
\Pi^{\Z_2^{(i)}}_{h,(0,0)} & (-1)^{\tau_1} [ \varepsilon^{(i)}_1 + \tilde{\varepsilon}^{(i)}_1]  
+ (-1)^{\tau_k} [ \varepsilon^{(i)}_2 -2 \tilde{\varepsilon}^{(i)}_2 ]  
+ (-1)^{\tau_1+\tau_k} [ \varepsilon^{(i)}_5 + \tilde{\varepsilon}^{(i)}_5] 
\\\hline
\Pi^{\Z_2^{(i)}}_{h,(1,0)} &
[(-1)^{\tau_1}-2]\varepsilon^{(i)}_1
 + [1 - 2 (-1)^{\tau_1}]\tilde{\varepsilon}^{(i)}_1
 + (-1)^{\tau_1+\tau_k}[\varepsilon^{(i)}_3 - 2 \tilde{\varepsilon}^{(i)}_3 ]
 +(-1)^{\tau_k}[ -2 \varepsilon^{(i)}_4 + \tilde{\varepsilon}^{(i)}_4]
\\\hline
\Pi^{\Z_2^{(i)}}_{h,(0,1)} & (1-2(-1)^{\tau_k}) \varepsilon^{(i)}_2
 + (1+(-1)^{\tau_k}) \tilde{\varepsilon}^{(i)}_2  
 +(-1)^{\tau_1}[\varepsilon^{(i)}_3 + \tilde{\varepsilon}^{(i)}_3 ]
 + (-1)^{\tau_1+\tau_k}[\varepsilon^{(i)}_4 + \tilde{\varepsilon}^{(i)}_4 ]
\\\hline
\Pi^{\Z_2^{(i)}}_{h,(1,1)} &  (-1)^{\tau_k}[ - 2 \varepsilon^{(i)}_3 + \tilde{\varepsilon}^{(i)}_3]
 +(-1)^{\tau_1}[\varepsilon^{(i)}_4 - 2 \tilde{\varepsilon}^{(i)}_4 ]
 + [(-1)^{\tau_1+\tau_k}-2]\varepsilon^{(i)}_5
 + [(1 - 2 (-1)^{\tau_1+\tau_k}]\tilde{\varepsilon}^{(i)}_5
\\\hline\hline
\Pi^{\Z_2^{(i)}}_{v,(0,0)} & (-1)^{\tau_1} [ \varepsilon^{(i)}_1 - \tilde{\varepsilon}^{(i)}_1]  
- (-1)^{\tau_k}  \varepsilon^{(i)}_2   
+ (-1)^{\tau_1+\tau_k} [ \varepsilon^{(i)}_5 - \tilde{\varepsilon}^{(i)}_5] 
\\\hline
\Pi^{\Z_2^{(i)}}_{v,(1,0)} & -(-1)^{\tau_1} \varepsilon^{(i)}_1+\tilde{\varepsilon}^{(i)}_1
 - (-1)^{\tau_1+\tau_k}\varepsilon^{(i)}_3 +(-1)^{\tau_k}\tilde{\varepsilon}^{(i)}_4
\\\hline
\Pi^{\Z_2^{(i)}}_{v,(0,1)} & \varepsilon^{(i)}_2+ ((-1)^{\tau_k}-1) \tilde{\varepsilon}^{(i)}_2  
 +(-1)^{\tau_1}[\varepsilon^{(i)}_3 - \tilde{\varepsilon}^{(i)}_3 ]
 + (-1)^{\tau_1+\tau_k}[\varepsilon^{(i)}_4 -  \tilde{\varepsilon}^{(i)}_4 ]
\\\hline
\Pi^{\Z_2^{(i)}}_{v,(1,1)} & (-1)^{\tau_k} \tilde{\varepsilon}^{(i)}_3
-(-1)^{\tau_1}\varepsilon^{(i)}_4  
- (-1)^{\tau_1+\tau_k} \varepsilon^{(i)}_5 +\tilde{\varepsilon}^{(i)}_5
\\\hline
\end{array}
\end{equation*}
\end{center}
\caption{Exceptional three-cycles for $\OR$ invariant D6-branes, which contribute to the K-theory constraint on 
the {\bf ABB} lattice on $T^6/(\Z_2 \times \Z_6' \times \OR)$ with discrete torsion.
All cycles habe to be multiplied by the $\Z_2^{(i)}$ eigenvalues $(-1)^{\tau_0^{(i)}}$.
For more details on the notation see  table~\protect\ref{Tab:Z2Z6p-Ktheory-AAA+BBB}.}
\label{Tab:Z2Z6p-Ktheory-ABB}
\end{table}
The K-theory constraints can be explicitly derived by using the
intersection numbers, 
\begin{equation}
\begin{aligned}
\Pi^{\rm frac}_a \circ \Pi^{\rm frac}_b = & \frac{1}{4} \left(
X_a Y_b - Y_a X_b
-  \sum_{\alpha=1}^3 \left(\vec{x}^{(\alpha)}_a \cdot \vec{y}^{(\alpha)}_b - \vec{x}^{(\alpha)}_b \cdot \vec{y}^{(\alpha)}_a \right)
\right)\\
& \text{with}
\quad
\vec{x}^{(\alpha)}_a \cdot \vec{y}^{(\alpha)}_b \equiv \sum_{m=1}^5 x^{(\alpha)}_{m,a} y^{(\alpha)}_{m,b}
, \end{aligned}  
\end{equation}
when the expansion~(\ref{Eq:Z2Z6p-frac-expansion}) of a fractional three-cycle on $T^6/\Z_2 \times \Z_6'$ is used.

The bulk contributions to the K-theory constraints are given in table~\ref{Tab:Z2Z6p-Ktheory-bulk-v1},
which after RR tadpole cancellation simplify to the expression in table~\ref{Tab:Z2Z6p-Ktheory-bulk-v2}.
\begin{table}[h]
  \renewcommand{\arraystretch}{1.3}
  \begin{center}
\begin{equation*}\!\!\!\!\!\!\!\!\!\!\!\!\!\!\!\!\!\!\!\!\!\!\!\!
\begin{array}{|c||c||c|c|c|c|}\hline
\multicolumn{6}{|c|}{\text{\bf Bulk part of the K-theory constraints on } T^6/\Z_2 \times \Z_6'}
\\\hline\hline
& & {\bf AAA} & {\bf AAB} & {\bf ABB} & {\bf BBB}
\\\hline\hline
\OR &\! \frac{1}{2^{3- \eta}} \sum_a N_a \Pi^{\rm bulk}_a \circ \Pi^{\rm bulk}_{Sp(2)_0} \!\!
& \frac{-1}{2^{1- \eta}} \sum_a N_a Y_a\!\! & \frac{1}{2^{1- \eta}} \sum_a N_a \left(X_a - Y_a  \right)\!\!
& \frac{3}{2^{1- \eta}} \sum_a N_a X_a\!\! & \frac{3}{2^{1- \eta}} \sum_a N_a \left( 2X_a + Y_a   \right)\!\!
\\\hline
\!\!\!\OR\Z_2^{(1)}\!\!\! &\! \frac{1}{2^{3- \eta}} \sum_a N_a \Pi^{\rm bulk}_a \circ \Pi^{\rm bulk}_{Sp(2)_1} \!\!
& \frac{-3}{2^{1- \eta}} \sum_a N_a Y_a\!\! & \frac{1}{2^{1- \eta}} \sum_a N_a \left(X_a - Y_a  \right) \!\!
& \frac{1}{2^{1- \eta}} \sum_a N_a X_a\!\! & \frac{1}{2^{1- \eta}} \sum_a N_a \left( 2X_a + Y_a \right) \!\!
\\\hline
\!\!\!\OR\Z_2^{(2)}\!\!\! &\!\frac{1}{2^{3- \eta}} \sum_a N_a \Pi^{\rm bulk}_a \circ \Pi^{\rm bulk}_{Sp(2)_2}\!\! 
& \frac{-3}{2^{1- \eta}} \sum_a N_a Y_a\!\! & \frac{3}{2^{1- \eta}} \sum_a N_a \left(X_a - Y_a  \right) \!\!
& \frac{3}{2^{1- \eta}} \sum_a N_a X_a\!\! & \frac{1}{2^{1- \eta}} \sum_a N_a \left( 2X_a + Y_a   \right)\!\!
\\\hline
\!\!\!\OR\Z_2^{(3)}\!\!\! &\!\frac{1}{2^{3- \eta}} \sum_a N_a \Pi^{\rm bulk}_a \circ \Pi^{\rm bulk}_{Sp(2)_3} \!\!
&  \frac{-3}{2^{1- \eta}} \sum_a N_a Y_a\!\! & \frac{1}{2^{1- \eta}} \sum_a N_a \left( X_a - Y_a \right) \!\!
& \frac{3}{2^{1- \eta}} \sum_a N_a X_a\!\! & \frac{1}{2^{1- \eta}} \sum_a N_a \left( 2X_a + Y_a   \right)\!\!
\\\hline
\end{array}
\end{equation*}
\end{center}
\caption{Bulk contributions to the K-theory constraints on $T^6/(\Z_2 \times \Z_6' \times \OR)$ without ($\eta=1$) and 
with  ($\eta=-1$) discrete torsion. The sums can be simplified using the bulk RR tadpole cancellation 
conditions~(\protect\ref{Eq:Z2Z6p-bulkRRtcc}).
The result is displayed in table~\protect\ref{Tab:Z2Z6p-Ktheory-bulk-v2}. }
\label{Tab:Z2Z6p-Ktheory-bulk-v1}
\end{table}
\begin{table}[h]
  \renewcommand{\arraystretch}{1.3}
  \begin{center}
\begin{equation*}
\begin{array}{|c|c||c|c||c|c||c|c|}\hline
\multicolumn{8}{|c|}{\text{\bf Simplified contributions to the K-theory constraints on } T^6/\Z_2 \times \Z_6'}
\\\hline\hline
{\bf AAA} &\!\!\!\begin{array}{c} \text{after RR tcc} \\ \text{mod  2} \end{array}\!\!\!\!& {\bf AAB}  &\!\!\!\begin{array}{c}\text{after RR tcc} \\ \text{mod  2} \end{array}\!\!\!\!& {\bf ABB}  
 &\!\!\!\begin{array}{c} \text{after RR tcc} \\ \text{mod  2} \end{array}\!\!\!\! & {\bf BBB} &\!\!\!\begin{array}{c} \text{after RR tcc} \\ \text{mod  2} \end{array}\!\!\!\!
 \\\hline\hline
\OR & -2^{ \eta} \sum_a N_a X_a 
& \OR & 2^{ \eta} \sum_a N_a Y_a
&\OR\Z_2^{(1)} & -2^{ \eta} \sum_a N_a Y_a
&\OR\Z_2^{(1)} & 2^{ \eta} \sum_a N_a X_a 
\\\hline
\end{array}
\end{equation*}
\end{center}
\caption{Simplified bulk contributions to the K-theory constraints upon RR tadpole cancellation~(\protect\ref{Eq:Z2Z6p-bulkRRtcc}).
The other contributions are given by integer multiples of the listed ones, as shown in table~\protect\ref{Tab:Z2Z6p-Ktheory-bulk-v1}.}
\label{Tab:Z2Z6p-Ktheory-bulk-v2}
\end{table}
The K-theory constraints on $ T^6/(\Z_2 \times \Z_6' \times \OR)$ without discrete torsion are thus trivially fulfilled, if RR tadpoles are cancelled.
In the case with discrete torsion, the exceptional contributions have to be evaluated explicitly.
Due to the large number of combinatorial possibilities displayed in tables~\ref{Tab:Z2Z6p-Ktheory-AAA+BBB} to~\ref{Tab:Z2Z6p-Ktheory-ABB},
we do not write out all constraints, but we expect that also the twisted contributions to the K-theory constraints can be simplified due to the 
shape of the coefficients $(x,y)$ presented at the end of section~\ref{Ss:Z2Z6p-twisted}.
The explicit examples below fulfill the K-theory constraint trivially by only having gauge groups of even rank.

The discussion in appendix~\ref{sec:thres} confirms again that all D6-branes in the classification of $\OR$ cycles 
carry indeed $Sp(2M)$ gauge groups.

{\boldmath
\subsection{A $T^6/(\Z_2 \times \Z_6' \times \OR)$ example without discrete torsion}\label{S:Z2Z6p-example-no-torsion}
}

In~\cite{Honecker:2004kb}, a supersymmetric model with fractional D6-branes on the {\bf AAB} lattice 
on the orbifold $T^6/\Z_6$ was presented. After the choice of a representant per orbifold and orientifold orbit as
 discussed at the end of section~\ref{Ss:Z2Z6p-bulk}, the torus cycles are 
those listed in table~\ref{Tab:Z2Z6p-example-bulk}.
\begin{table}[h]
  \renewcommand{\arraystretch}{1.3}
  \begin{center}
  \begin{equation*}
    \begin{array}{|c|c|c||c|c|}   \hline
\multicolumn{5}{|c|}{\text{\bf Supersymmetric bulk 3-cycles on the AAB lattice on } T^6/\Z_2 \times \Z_6'}
\\\hline\hline
\text{brane} & \frac{\text{angle}}{\pi} & (n^1,m^1;n^2,m^2;n^3,m^3) & X & Y
\\\hline\hline
a,b,d & (\frac{1}{3},-\frac{1}{3},0) & (0,1;1,-1;1,1)  &  1 & 1
\\\hline
c,e/\OR & (0,0,0) & (1,0;1,0;1,1)  & 1 & 1  
\\\hline
\OR\Z_2^{(3)} & (\frac{1}{2},-\frac{1}{2},0) & (-1,2;1,-2;1,1) & 3 & 3 
\\
\OR\Z_2^{(1)} & (0,\frac{1}{2},-\frac{1}{2}) & (1,0;-1,2;1,-1) & 1 & 1  
\\
\OR\Z_2^{(2)} & (\frac{1}{2},0,-\frac{1}{2}) & (-1,2;1,0;1,-1) & 1 & 1  
\\\hline
   \end{array}
   \end{equation*}
    \caption{Some supersymmetric bulk cycles on the {\bf AAB} on the $T^6/(\Z_2 \times \Z_6' \times \OR)$ background.
The names $a \ldots e$ are taken from~\cite{Honecker:2004kb}. $\OR$ labels a cycle which is parallel to the first O6-plane orbit.}
\label{Tab:Z2Z6p-example-bulk}
  \end{center}
\end{table}
Combining the untwisted RR tadpole cancellation condition~(\ref{Eq:Z2Z6p-bulkRRtcc}) with the supersymmetry constraint~(\ref{Eq:Z2Z6p-SUSY}) on the {\bf AAB} lattice leads to 
\begin{equation}
\sum_a N_a X_a = 6
\qquad\qquad
\end{equation}
without discrete torsion.
This equation can be solved for brane $a$ in table~\ref{Tab:Z2Z6p-example-bulk} with $N_a=6$, i.e.
\begin{equation}
\Pi_a = \frac{1}{2} \left( \rho_1 + \rho_2 \right)
.
\end{equation}
The gauge group is thus $U(6)$, where in contrast to the earlier examples on 
$T^6/\Z_2 \times \Z_4$ and $T^6/\Z_2 \times \Z_6'$, the Abelian subgroup is anomaly-free and stays massless.
The massless matter spectrum is completely non-chiral and can be computed along the lines described in appendix~\ref{sec:thres}
by using the torus intersection numbers $I_{a(\omega^k a)}$, $I_{a(\omega^k a')}$ and $I_{a,\OR\theta^n\omega^m}$.
This leads to
\begin{itemize}
\item
three multiplets in the adjoint representation of $U(6)$ from the $aa$ sector,
\item
 three adjoints from the $a(\omega^k a)_{k=1,2}$ sectors,
\item 
one non-chiral pair of  antisymmetric representations from the $aa'$ sector,
\item
five non-chiral pairs of antisymmetrics from the  $a(\omega a') + a(\omega^2 a')$ sectors,
\item
one non-chiral pair of symmetrics from the $a(\omega a') + a(\omega^2 a')$ sectors.
\end{itemize}
The total massless spectrum is thus completely non-chiral and consists thus of six multiplets in the adjoint representation of $U(6)$, 
six non-chiral pairs of antisymmetric representations and one non-chiral pair of symmetric representations.

The K-theory constraints are trivivally fulfilled due to the even rank of the single gauge factor $U(6)$.

{\boldmath
\subsection{A $T^6/(\Z_2 \times \Z_6' \times \OR)$ example with discrete torsion}\label{S:Z2Z6p-example-torsion}
}

We choose again the {\bf AAB} lattice with $\eta_{\OR\Z_2^{(3)}}=1$, and the exotic O6-plane
is within one of the orbits $\{\OR,\OR\Z_2^{(1)},\OR\Z_2^{(2)} \}$. Imposing the supersymmetry condition~(\ref{Eq:Z2Z6p-SUSY})
leads to the untwisted RR tadpole cancellation condition
\begin{equation}
 \sum_a N_a X_a = 8
\end{equation}
on the {\bf AAB} background with discrete torsion.
The twisted RR tadpole conditions can be solved by adding four D6-branes $a_m$ with different $\Z_2^{(i)}$ eigenvalues
such that the sum wraps the bulk cycle $a$ in table~\ref{Tab:Z2Z6p-example-bulk}, 
\begin{equation}
\begin{aligned}
\Pi_{a_m} =& \frac{1}{4} \left(\rho_1 +\rho_2   \right)
+  \frac{(-1)^{\tau_{0,m}^{(1)}}}{4} \left( \varepsilon_1^{(1)} - \tilde{\varepsilon}_1^{(1)}  
+\varepsilon_2^{(1)} - \tilde{\varepsilon}_2^{(1)}
+\varepsilon_3^{(1)} - \tilde{\varepsilon}_3^{(1)}
\right)
\\
& +  \frac{(-1)^{\tau_{0,m}^{(2)}}}{4}
\left(\tilde{\varepsilon}_1^{(2)} -\varepsilon_2^{(2)} +
  \tilde{\varepsilon}_4^{(2)}  \right) \\
& +  \frac{(-1)^{\tau_{0,m}^{(3)}}}{4} \left(-2 \, \varepsilon_1^{(3)}
  + \tilde{\varepsilon}_1^{(3)}  
+\varepsilon_2^{(3)} -2 \,  \tilde{\varepsilon}_2^{(3)} 
 -2 \, \varepsilon_4^{(3)} + \tilde{\varepsilon}_4^{(3)}
   \right)
\end{aligned}
\end{equation}
where all discrete displacements and Wilson lines have been set to zero, $\vec{\sigma}=\vec{\tau}=0$,
and the exceptional contributions are read off from table~\ref{Tab:Z2Z6p-Part1}.
Since $\sum_m N_{a_m} \Pi_{a_m} = 2 \left(\rho_1 +\rho_2 \right)$, the K-theory contraint is trivially fulfilled.

We use the following assignment of $\Z_2^{(i)}$ eigenvalues to the four fractional D6-branes.
\begin{equation*}
\begin{array}{|c|ccc|}\hline
\text{brane} & \tau_{0,m}^{(1)} & \tau_{0,m}^{(2)} & \tau_{0,m}^{(3)}
\\\hline
a_0 & 0 & 0 & 0
\\
a_1 & 0 & 1 & 1 
\\
a_2 & 1 & 0 & 1
\\
a_3 & 1 & 1 & 0
\\\hline
\end{array}
\end{equation*}

The orientifold image branes are computed using tables~\ref{Tab:Z2Z6p-OR-ex-1} to~\ref{Tab:Z2Z6p-OR-ex-3} with the result
\begin{equation}
\begin{aligned}
\Pi_{a_m}' =& \frac{1}{4} \left(\rho_1 +\rho_2   \right)
+  \eta_{(1)} \, \frac{(-1)^{\tau_{0,m}^{(1)}}}{4} \left( - \tilde{\varepsilon}_1^{(1)}  + \varepsilon_2^{(1)} - \tilde{\varepsilon}_4^{(1)}
\right)
\\
& + \eta_{(2)} \, \frac{(-1)^{\tau_{0,m}^{(2)}}}{4}
 \left(-\varepsilon_1^{(2)} + \tilde{\varepsilon}_1^{(2)} -\varepsilon_2^{(2)} + \tilde{\varepsilon}_2^{(2)}-\varepsilon_3^{(2)} + \tilde{\varepsilon}_3^{(2)}  \right)\\
& + \eta_{(3)} \, \frac{(-1)^{\tau_{0,m}^{(3)}}}{4} 
\left(- \, \varepsilon_1^{(3)} + 2\, \tilde{\varepsilon}_1^{(3)} 
+ 2 \, \varepsilon_2^{(3)} - \tilde{\varepsilon}_2^{(3)}
 -  \varepsilon_5^{(3)} + 2 \, \tilde{\varepsilon}_5^{(3)} \right),
\end{aligned}
\end{equation}
which leads to vanishing intersection numbers
\begin{equation}
\Pi_{a_m} \circ \Pi_{a_n} =\Pi_{a_m} \circ \Pi_{a_n}' =0. 
\end{equation}
One can check explicitly that the terms vanish for each untwisted and twisted sector separately. 
The massless spectrum is thus completely non-chiral.

The gauge group is $U(2)^4$, and the massless matter spectrum splits into a part which is independent of 
the choice of the exotic O6-plane
\begin{itemize}
\item
bifundamental representations on parallel D6-branes $a_ma_n$ with $m \neq n$
\begin{equation}\nonumber
\bigl[(\2,\ov{\2},\1,\1) + (\1,\1,\2,\ov{\2}) 
+ (\2,\1,\ov{\2},\1) + (\1,\2,\1,\ov{\2}) 
+ (\2,\1,\1,\ov{\2}) + (\1,\2,\ov{\2},\1) 
+c.c. \bigr]
\end{equation}
\item
bifundamental representations at intersections $a_m(\omega^k a_n)_{k=1,2}$,
\begin{equation*}
2 \times \bigl[(\2,\ov{\2},\1,\1) + (\1,\1,\2,\ov{\2}) +c.c. \bigr]
+ \bigl[ (\2,\1,\ov{\2},\1) + (\1,\2,\1,\ov{\2}) +c.c. \bigr],
\end{equation*}
which stem from one intersection point that is $\Z_2^{(1)}$ and $\Z_2^{(3)}$ invariant plus a pair
of $\Z_2^{(3)}$ invariant intersections, which are exchanged under $\Z_2^{(1)}$,
\end{itemize}
and a part which depends on the choice of the exotic O6-plane and contains:
\begin{itemize}
\item
bifundamental representations at intersections  $a_m(\theta^k a_m')_{k=0,1,2}$
\begin{equation}\nonumber
\begin{aligned}
\frac{7 + \eta_{(1)} - \eta_{(2)} - \eta_{(3)} }{4} & [ (\2,\2,\1,\1) + (\1,\1,\2,\2) 
+ c.c.],\\
+\frac{7 - \eta_{(1)} + \eta_{(2)} - \eta_{(3)} }{4} & [(\2,\1,\2,\1)+(\1,\2,\1,\2) 
+ c.c.],\\
+\frac{7 - \eta_{(1)} - \eta_{(2)} + \eta_{(3)} }{4} & [(\2,\1,\1,\2) + (\1,\2,\2,\1)
+ c.c.],
\end{aligned}
\end{equation}
\item
two non-chiral pairs of antisymmetric representations per $U(2)$ gauge factor if $\eta_{\OR} = \eta_{\OR\Z_2^{(3)}} =1$ 
or one pair of symmetric representations if $\eta_{\OR} =-1$.
\end{itemize}
The diagonal Abelian factor $U(1)=\sum_{m=0}^3 U(1)_m$ stays massless, while the other three
linear combinations are anomalous and receive a mass by the generalised Green-Schwarz mechanism.
The K-theory constraint is trivially fulfilled for $N_{a_m}=2$ for all stacks.

Note again, that there are no multiplets in the adjoint representation. The D6-branes are thus again completely rigid.

\section{Conclusions and Outlook}\label{S:Conclusions}

In this work, we gave the complete list of supersymmetric D6-brane model building rules on 
$T^6/(\Z_2 \times \Z_{2M} \times \OR)$ orientifolds with discrete torsion and factorisable tori.

For $2M=2$, in section~\ref{S:Z2Z2orient} we completed the classification of orientifold invariant D6-branes in the presence of discrete torsion, 
which can lead to new K-theory constraints for backgrounds with at least one tilted torus, which has to our knowledge not been noted before.
We also found that the computation of intersection numbers is even on this orbifold
not sufficient in order to derive the full matter spectrum, if D6-branes are parallel on some two-torus.

For $2M=4$, we found that for both cases without and with discrete torsion, D6-branes wrap the same half-bulk three-cycles, and 
in section~\ref{Ss:Z2Z4-3generations}  we showed that three generation left-right symmetric models or Standard Model vacua with all quarks realised as 
bifundamental representations are excluded, and $SU(5)$ GUT vacua are by the same simple arguments excluded on four of the six
inequivalent background lattices. We expect that a more detailed case-by-case study, which goes beyond the scope of this paper,
rules also out the remaining model building possibilities of some right-handed quarks realised as antisymmetric representations of the QCD stack.

For $2M=6,6'$, we constructed the complete lattice of untwisted and $\Z_2$ twisted three-cycles, on which D6-brane model building can be performed.
We gave a chiral example on $T^6/\Z_2 \times \Z_6$ without and with discrete torsion in section~\ref{Ss:Z2Z6-example-without-torsion} 
and~\ref{Ss:Z2Z6-example-with-torsion}, respectively. The D6-branes in the latter case were completely rigid.
For  $T^6/\Z_2 \times \Z_6'$ without discrete torsion, supersymmetry excludes chirality due to $b_3^{\text{no torsion}}=2$, but 
with discrete torsion, there is ample possiblity for chiral spectra. In section~\ref{S:Z2Z6p-example-torsion}, we showed for an example
that completely rigid D6-branes also occur on this orbifold background.

The  $T^6/\Z_2 \times \Z_6$  and  $T^6/\Z_2 \times \Z_6'$ orientifolds with discrete torsion
look very promising in view of obtaining three Standard Model generations.
However, rigid D6-branes without any adjoint representation arising at intersections of orbifold image branes only occur under special circumstances,
namely on the former orbifold rigidity constrains the wrapping numbers on $T_{(2)}^2 \times T_{(3)}^2$ 
to just $3 \times 3$ possibilities, while also requiring the relation $\sigma_2 \tau_2=\sigma_3\tau_3$ for the 
discrete displacements and Wilson lines, 
as detailed in appendix~\ref{App:Sss-no_adjoints}. For the latter
orbifold, the condition on rigid D6-branes is not as straightforwardly
classified, but still very restrictive. 
This considerably narrows the search for phenomenologically interesting models,
since the cases of completely rigid D6-branes are of paramount interest as candidates 
for the QCD and electro-weak stacks.
We will come back to analysing model building on these two orbifold backgrounds in future work~\cite{Forste:201xxx}.
If Standard Model vacua on rigid D6-branes are found, we will be in the unique position of being able to
compute the  low-energy effective field theory exactly by means of conformal field theory -
in contrast to the present focus on F-theory models on smooth Calabi-Yau manifolds, for which there is at present 
no such tool available. 
For the expected Standard Model vacua on rigid D6-branes, it will also be of great interest to 
investigate the blow-up of fixed points
 and if twisted moduli are stabilised at the orbifold fixed point,
but also if de Sitter vacua are feasible, see e.g. the analogous discussion on heterotic 
orbifolds~\cite{Lowen:2008fm,Parameswaran:2010ec}.
The blown-up models will also provide a more geometric understanding of our
constructions as discussed e.g. in~\cite{Sharpe:2009hr}.
By using M/F-theory duality, the blown-up IIA orbifold results will help to derive the missing low-energy 
properties of F-theory models. 

As a next step it will also be interesting to introduce closed string background fluxes in the set-up, 
in particular as means to stabilise the untwisted K\"ahler moduli and the dilaton.

Finally, Standard Models on rigid D6-branes in orbifold backgrounds with torsion
might provide explicit examples for the proposed phenomenon of 
low-scale string signatures at the LHC, see e.g.~\cite{Lust:2008qc,Lust:2009pz,Feng:2010yx,Blumenhagen:2010dt}.

\subsection*{Acknowledgements}

We thank Guhan Sukumaran and Cristina Timirgaziu for collaboration at
early stages of this project. 
G.H. thanks the KITP in Santa Barbara for kind hospitality during the
workshop ``Strings at the LHC and in the Early Universe''. 

The work of G.\ H.\ was supported in part by the FWO - Vlaanderen,
project G.0235.05 and in part by the Federal Office for Scientific,  
Technical and Cultural Affairs through the ÔInteruniversity Attraction
Poles Programme Ð Belgian Science PolicyÕ P6/11-P. 
 This research was supported in part by the National Science
 Foundation under Grant No. NSF PHY05-51164. S.\ F.\ is supported by
 the SFB--Transregio ``The Dark Universe'' and the European Union
 7$^{\mbox{\scriptsize th}}$ network programme ``Unification in the LHC
 era'' (PITN-GA-2009-237920).

The work of G.\ H. \ is partially supported by the 
``Research Center Elementary Forces and Mathematical Foundations'' (EMG)
at the Johannes Gutenberg-Universit\"at Mainz.


\clearpage
\begin{appendix}

{\boldmath
\section{The IIA closed string spectrum on $T^6/\Z_2 \times \Z_{2M}$ 
without and with discrete torsion}\label{App:A}

\subsection{The massless closed IIA string spectrum on Calabi-Yau manifolds and their orientifolds}
}

The massless closed string spectrum of  IIA string theory on smooth Calabi-Yau-threefolds
and its orientifold by $\OR$ has been discussed in detail in~\cite{Grimm:2004ua}.
The multiplicities of multiplets in terms of Hodge numbers and the bosonic content of each multiplet 
are summarised in table~\ref{Tab:N2-N1-IIAspectra}, 
\begin{table}[h!]
\renewcommand{\arraystretch}{1.3}
  \begin{center}
\begin{equation*}
\begin{array}{|c|c|c||c|c|c|}\hline
\multicolumn{6}{|c|}{\text{\bf Four dimensionsal closed IIA spectra on Calabi-Yaus and their orientifolds}}
\\\hline\hline
{\cal N}=2 \text{ multiplet} & \text{mult.} & \text{bosons} & {\cal N}=1 \text{ multiplet} & \text{mult.} & \text{bosons}
\\\hline\hline
\text{gravity} & 1 & (G_{\mu\nu},A^0) & \text{gravity} & 1 & (G_{\mu\nu})
\\\hline
\text{tensor} & 1 & (B_{\mu\nu},\phi,\zeta^0,\tilde{\zeta}_0) & \text{linear (dilaton-axion)} & 1 & (\phi,\xi^0)
\\\hline
\text{hyper} & h_{21} & (z^k,\bar{z}_k,\zeta^k,\tilde{\zeta}_k) & \text{chiral (complex structures)} & h_{21} & (c^k,\xi^k)
\\\hline
\text{vector} & h_{11} & (A^i,v^i,b^i) & \text{chiral (K\"ahler moduli)} & h_{11}^- & (v^i,b^i)
\\
& & & \text{vector} & h_{11}^+ & (A^i)
\\\hline
\end{array}
\end{equation*}
\end{center}
\caption{The massless ${\cal N}=2$ and ${\cal N}=1$ four dimensional multiplets for type IIA string theory on 
a Calabi-Yau manifold (left) and its orientifold (right). The vector, hyper and chiral multiplets are counted by the Hodge numbers 
of the manifold and their transformation under $\OR$. The bosonic field content is listed explicitly.}
\label{Tab:N2-N1-IIAspectra}
\end{table}
where the four dimensional bosonic degrees of freedom are obtained from the ten dimensional 
RR-fields $C_{\rm odd}$, the NSNS-two form $B_2$ and the K\"ahler form $J_{\text{K\"ahler}}$ 
(which is a reparameterisation of the metric $G$) of the 
Calabi-Yau manifold via integrals over two- and three-cycles as follows,  
\begin{equation}
\begin{aligned}
A^0 &=C_1,
\\
A^i &= \int_{i^{th} \, (1,1)-\text{cycle}} C_3, 
\quad
v^i = \int_{i^{th} \, (1,1)-\text{cycle}} J_{\text{K\"ahler}},
\quad
b^i = \int_{i^{th} \, (1,1)-\text{cycle}} B_2,
\quad
i=1 \ldots h_{11},
\\
\zeta^K &= \int_{K^{th} \text{ 3-cycle}} C_3,
\qquad
\tilde{\zeta}_K = \int_{K^{th} \text{ dual 3-cycle}} C_3,
\qquad
k =0 \ldots h_{21},
\\
z^k,\bar{z}_k & \quad \text{complex structure deformations}
\qquad\qquad
k=1 \ldots h_{21}.
\end{aligned}
\end{equation}
Making use of the worldsheet parity of the ten dimensional massless IIA fields,
\begin{equation}
\Omega(G)=G,
\qquad
\Omega(\phi)=\phi,
\qquad
\Omega(B_2) = - B_2,
\qquad
\Omega(C_1) = - C_1,
\qquad
\Omega(C_3)=C_3,
\end{equation}
leads to the ${\cal N}=1$ fields in four dimensions,
\begin{equation}
\begin{aligned}
A^i &= \int_{i^{th} \, {\cal R} \text{even } (1,1)-\text{cycle}} C_3, 
\qquad
v^i = \int_{i^{th} \,{\cal R} \text{odd }  (1,1)-\text{cycle}} J_{\text{K\"ahler}},
\qquad
b^i = \int_{i^{th} \,{\cal R} \text{odd }  (1,1)-\text{cycle}} B_2,
\\
\xi^K &= \int_{K^{th} {\cal R} \text{even}  \text{ 3-cycle}} C_3,
\qquad
k =1 \ldots h_{21},
\\
c^k & \sim z^k+ \bar{z}_k  
\qquad \text{complex structure deformations}
\qquad
k=1 \ldots h_{21}.
\end{aligned}
\end{equation}

In the orbifold limit $T^6/\Z_2 \times \Z_{2M}$, both without and with discrete torsion, the closed string spectrum
can be either computed by the same means using all two- and three-cycles from the bulk and twisted sectors.
Alternatively, the closed string spectrum can be obtained at the orbifold point in a separate computation
(see the following section), and the decomposition of the Hodge number $h_{11}$ into ${\cal R}$ even and odd cycles per twist 
sector can be read off from the number of chiral multiplets. This information is important for the issue of moduli stabilisation, which goes beyond the scope of the present work.

\subsection{The massless closed IIA string spectrum on orbifolds}

The untwisted closed string spectrum can be computed as follows:
any left-moving string is given by the four-vector $|s_0,\vec{s}\rangle$, and 
right moving strings are parameterised by $|s_0',\vec{s}'\rangle$, where the first entry encodes the spin along the four non-compact dimensions and the remaining three entries spins per complex two-torus.

 In the NS-sectors, all entries $s_i, s_i' \in \Z$ with $|0,\vec{0}\rangle_{\rm NS}$ the tachyonic vacuum state. The GSO projection selects physical states with both $ \sum_i s_i$ and $ \sum_i s_i'$  odd.
The massless untwisted left-moving NS-sector states are $|\pm 1,\vec{0}\rangle \equiv \psi^{\mu}_{-1/2} |0 \rangle_{\rm NS}$ and $|0,\underline{\pm 1_i,0_j,0_k}\rangle \equiv \psi^{i}_{-1/2} |0 \rangle_{\rm NS},\psi^{\ov{i}}_{-1/2} |0 \rangle_{\rm NS} $ and similarly for the right-moving NS-sector with oscillators $\tilde{\psi}$.

The R sectors have half-integer entries $s_i, s_i'$ with the GSO projection
enforcing $\sum_i s_i$ odd and $ \sum_i s_i'$  even for the IIA string theory. The massless R 
sector states in the left moving sector are $|\frac{-1}{2},\frac{1}{2},\frac{1}{2},\frac{1}{2}\rangle_{\rm R} \equiv|-+++\rangle_{\rm R}$ or any state with an even number of signs  
flip, whereas the massless right-moving R states are of the form $|\frac{1}{2},\frac{1}{2},\frac{1}{2},\frac{1}{2}\rangle_{\rm R} \equiv|++++\rangle_{\rm R}$ or an even number of signs
flipped.

The orbifold projectors act identically on the left- and right-moving sectors by a phase,
\begin{equation}
\theta: |s_0,\vec{s}\rangle \longrightarrow e^{2 \pi i \vec{v} \cdot \vec{s}} |s_0,\vec{s}\rangle
\qquad \text{ and } \qquad
\omega: |s_0,\vec{s}\rangle \longrightarrow e^{2 \pi i \vec{w} \cdot \vec{s}} |s_0,\vec{s}\rangle,
\end{equation}
whereas the orientifold projection flips the spins along the compact directions while exchanging 
left- and right-moving sectors,
\begin{equation}
\begin{aligned}
\OR: |s_0,\vec{s}\rangle_{\rm NS}|s_0',\vec{s}'\rangle_{\rm NS} \longrightarrow  
|s_0',-\vec{s}'\rangle_{\rm NS} |s_0,-\vec{s}\rangle_{\rm NS} ,
\\
\OR: |s_0,\vec{s}\rangle_{\rm R}|s_0',\vec{s}'\rangle_{\rm R} \longrightarrow  
- |s_0',-\vec{s}'\rangle_{\rm R} |s_0,-\vec{s}\rangle_{\rm R} .
\end{aligned}
\end{equation}
In the RR-sector, the exchange of the left- and right-moving sector is accompanied by a minus sign. 
As a result, the untwisted IIA closed string spectra before and after the orientifold projection are given in table~\ref{Tab:N2-N1-IIAuntwistedOrb}.
\begin{table}[h!]
\renewcommand{\arraystretch}{1.5}
  \begin{center}
{\small 
\begin{equation*}\!\!\!\!\!\!\!\!\!\!\!
{\tiny
\begin{array}{|c|c||c|c|}\hline
\multicolumn{4}{|c|}{\text{\bf Bosonic part of the untwisted massless closed string spectrum on $T^6/\Z_2 \times \Z_{2M}$ orbifolds and orientifolds}}
\\\hline\hline
\multicolumn{2}{|c|}{T^6/\Z_2 \times \Z_{2M}} & \multicolumn{2}{|c|}{T^6/(\Z_2 \times \Z_{2M} \times \OR)} 
\\\hline\hline
\text{particle} & \text{massless state} & \text{particle} & \text{massless state} 
\\\hline\hline
\multicolumn{4}{|c|}{\text{NS-NS sector} \qquad (i) \quad \text{universal}}
\\\hline\hline
\!\!\!G_{\mu\nu} + B_{\mu\nu}+\phi \!\!\! & \psi^{\mu}_{-1/2}\tilde{\psi}^{\nu}_{-1/2} |0\rangle_{\rm NSNS}
& G_{\mu\nu} +\phi & \left( \psi^{\mu}_{-1/2}\tilde{\psi}^{\nu}_{-1/2} + \psi^{\nu}_{-1/2}\tilde{\psi}^{\mu}_{-1/2} \right) |0\rangle_{\rm NSNS}
\\\hline
(v^i,b^i)_{i=1,2,3} 
& \psi^{i}_{-1/2}\tilde{\psi}^{\ov{i}}_{-1/2}|0\rangle_{\rm NSNS}, \quad \psi^{\ov{i}}_{-1/2}\tilde{\psi}^{i}_{-1/2}|0\rangle_{\rm NSNS}
& \!\!\!(v^i,b^i)_{i=1,2,3} \!\!\!
& \psi^{i}_{-1/2}\tilde{\psi}^{\ov{i}}_{-1/2}|0\rangle_{\rm NSNS}, \quad \psi^{\ov{i}}_{-1/2}\tilde{\psi}^{i}_{-1/2}|0\rangle_{\rm NSNS}
\\\hline\hline
\multicolumn{4}{|c|}{\text{NS-NS sector} \qquad (ii) \quad \Z_2 \times \Z_2, \Z_2 \times \Z_4, \Z_2 \times \Z_6 \text{ only}}
\\\hline\hline
z^1, \bar{z}_1 &  \psi^{1}_{-1/2}\tilde{\psi}^{1}_{-1/2} |0\rangle_{\rm NSNS}, \quad
 \psi^{\ov{1}}_{-1/2}\tilde{\psi}^{\ov{1}}_{-1/2}  |0\rangle_{\rm NSNS}
& 
c^1 &  \left( \psi^{1}_{-1/2}\tilde{\psi}^{1}_{-1/2}+  \psi^{\ov{1}}_{-1/2}\tilde{\psi}^{\ov{1}}_{-1/2} \right)  |0\rangle_{\rm NSNS}
\\\hline\hline
\multicolumn{4}{|c|}{\text{NS-NS sector} \qquad (iii) \quad  \Z_2 \times \Z_2 \text{ only}}
\\\hline\hline
(z^i,\bar{z}_i)_{i=2,3} &  \psi^{i}_{-1/2}\tilde{\psi}^{i}_{-1/2} |0\rangle_{\rm NSNS}, \quad \psi^{\ov{i}}_{-1/2}\tilde{\psi}^{\ov{i}}_{-1/2}  |0\rangle_{\rm NSNS}
& 
(c^i)_{i=2,3}  & \left( \psi^{i}_{-1/2}\tilde{\psi}^{i}_{-1/2} + \psi^{\ov{i}}_{-1/2}\tilde{\psi}^{\ov{i}}_{-1/2} \right) |0\rangle_{\rm NSNS}
\\\hline\hline\hline
\multicolumn{4}{|c|}{\text{RR sector} \qquad (i) \quad \text{universal}}
\\\hline\hline
(\zeta^0,\tilde{\zeta}_0) & \left\{\begin{array}{c}
|-+++\rangle|++++\rangle_{\rm RR} \\  |+---\rangle|----\rangle_{\rm RR}
\end{array}\right.
& 
\xi^0  & {\footnotesize |-+++\rangle|++++\rangle_{\rm RR} -  |+---\rangle|----\rangle_{\rm RR} }
\\\hline
\!\!(A^0,A^i)_{i=1,2,3}\! & \!\!\!{\tiny \left\{\!\!\begin{array}{c}
  |-+++\rangle|----\rangle_{\rm RR}
\\|-+--\rangle|--++\rangle_{\rm RR}
\\|---+\rangle|-++-\rangle_{\rm RR}
\\|--+-\rangle|-+-+\rangle_{\rm RR}
\end{array}\right.\!\!
,
\left\{\!\!\begin{array}{c}
  |+-++\rangle|++--\rangle_{\rm RR}
\\|++-+\rangle|+-+-\rangle_{\rm RR}
\\|+++-\rangle|+--+\rangle_{\rm RR}
\\|+---\rangle|++++\rangle_{\rm RR}
\end{array}\right.
}\!\!\!\!\!
& \multicolumn{2}{|c|}{\emptyset} 
\\\hline\hline
\multicolumn{4}{|c|}{\text{RR sector} \qquad (ii) \quad \Z_2 \times \Z_2, \Z_2 \times \Z_4, \Z_2 \times \Z_6 \text{ only}}
\\\hline\hline
(\zeta^1,\tilde{\zeta}_1) & \left\{\begin{array}{c}
|+-++\rangle|--++\rangle_{\rm RR} \\  |-+--\rangle|++--\rangle_{\rm RR}
\end{array}\right.
& \xi^1 &  {\tiny |+-++\rangle|--++\rangle_{\rm RR} -  |-+--\rangle|++--\rangle_{\rm RR} }
\\\hline\hline
\multicolumn{4}{|c|}{\text{RR sector} \qquad (iii) \quad \Z_2 \times \Z_2 \text{ only}}
\\\hline\hline
(\zeta^i,\tilde{\zeta}_i)_{i=2,3} & {\tiny \left\{\begin{array}{c}
|---+\rangle|+--+\rangle_{\rm RR}\\
|--+-\rangle|+-+-\rangle_{\rm RR}\\
|++-+\rangle|-+-+\rangle_{\rm RR}\\
|+++-\rangle|-++-\rangle_{\rm RR}
\end{array}\right. }
&  (\xi^i)_{i=2,3} & {\tiny \left\{\begin{array}{c}
|---+\rangle|+--+\rangle_{\rm RR} -|+++-\rangle|-++-\rangle_{\rm RR} \\
|--+-\rangle|+-+-\rangle_{\rm RR} - |++-+\rangle|-+-+\rangle_{\rm RR}
\end{array}\right. }
\\\hline\hline\hline
\multicolumn{4}{|c|}{ (h_{11}^+,h_{11}^-)^U=(0,3) \qquad\qquad
h_{21}^U= 3 (\Z_2 \times \Z_2), \;
1 (\Z_2 \times \Z_4 \text{ and  } \Z_2 \times \Z_6), \;
0 (\Z_2 \times \Z_6')}
\\\hline
\end{array}
}
\end{equation*}
}
\end{center}
\caption{The massless untwisted ${\cal N}=2$ and ${\cal N}=1$ four dimensional multiplets for type IIA string theory on $T^6/\Z_2 \times \Z_{2M}$ 
(left) and its orientifold by $\OR$ (right). Only bosonic states are listed; the fermionic superpartners arise in the NS-R and R-NS sectors.
The untwisted spectrum is independent of the choice of discrete torsion and the exotic O6-plane.}
\label{Tab:N2-N1-IIAuntwistedOrb}
\end{table}

The untwisted closed string spectrum is independent of the choice of discrete torsion.

The twisted ${\cal N}=2$ spectrum depends on the choice of discrete torsion, and moreover
the orientifold projected twisted ${\cal N}=1$ states depend on the choice of the exotic O6-plane as follows.
Closed string states in the $n\vec{v} + m\vec{w}$ twisted sector are obtained from the states 
\begin{equation}\label{Eq:twisted-closed-states}
|p_0, \vec{p} \rangle \equiv |s_0,\vec{s}+( n\vec{v} + m\vec{w})\rangle,
\qquad
|p_0', \vec{p}' \rangle \equiv |s_0',\vec{s}' -( n\vec{v} + m\vec{w})\rangle,
\end{equation}
which shows that the $\OR$ projection preserves each twist sector. The GSO projection is the same as in the untwisted sector, 
but the orbifold action in the case of discrete torsion is modified by the signs discussed in section~\ref{Sss:DT+EO}.

The masses of the twisted states are computed from 
\begin{equation}
\frac{\alpha'}{4} m^2 = \frac{1}{2} p^2 + E_T - \frac{1}{2}
\qquad
\text{with}
\qquad
E_T = \frac{1}{2} \sum_{i=1}^3 |nv_i + mw_i| \left(1 -  |nv_i + mw_i|\right)
\end{equation}
where $|nv_i + mw_i| \in [0,1)$ (up to an integer shift which might need to be performed).

The complete twisted closed string spectrum for the $T^6/\Z_2 \times \Z_{2M}$ orbifolds without and with discrete torsion is 
displayed in table~\ref{tab:Closed-string-T6ZNxZM-torsion}.
\mathsidetabfix{
\renewcommand{\arraystretch}{1.3}
      \begin{array}{|c||c|c|c|c|c|c|c|} \hline
        \multicolumn{8}{|c|}{
\text{\bf Twisted massless closed strings on $T^6/\Z_2 \times \Z_{2M}$ without and with torsion}
}\\ \hline\hline
\begin{array}{c} T^6/ \\\hline V/H \\ {\rm f.p.} \\\hline (\# V,\# H) \end{array} &  \vec{w} &  2\vec{w}  & 3\vec{w}
& \vec{v}   & (\vec{v}+\vec{w}) & (\vec{v}+2\vec{w}) &  (\vec{v}+3\vec{w})
\\\hline\hline
\!\!\!\Z_2 \times \Z_2\!\!\!& {\color{blue} (0,\frac{1}{2},-\frac{1}{2})} & \multicolumn{2}{|c|}{} &  {\color{blue} (\frac{1}{2},-\frac{1}{2},0)} &  {\color{blue} (\frac{1}{2},0,-\frac{1}{2})}
 & \multicolumn{2}{|c|}{} 
\\\hline
V & {\footnotesize \frac{1+\eta}{2}\left\{\!\!\!\begin{array}{c}|00++\rangle00--\rangle_{\rm NSNS} \\ |00--\rangle|00++\rangle_{\rm NSNS} \\ 
|++00\rangle|+-00\rangle_{\rm RR} \\ |--00\rangle|-+00\rangle_{\rm RR}  \end{array}\!\!\!\right\}^{(ij)} }
& \multicolumn{2}{|c|}{} 
& {\footnotesize \frac{1+\eta}{2}\left\{\!\!\!\begin{array}{c}|0++0\rangle0--0\rangle_{\rm NSNS} \\ |0--0\rangle|0++0\rangle_{\rm NSNS} \\ 
|+00+\rangle|+00-\rangle_{\rm RR} \\ |-00-\rangle|-00+\rangle_{\rm RR}  \end{array}\!\!\!\right\}^{(ij)} }
& {\footnotesize \frac{1+\eta}{2}\left\{\!\!\!\begin{array}{c}|0+0+\rangle0-0-\rangle_{\rm NSNS} \\ |0-0-\rangle|0+0+\rangle_{\rm NSNS} \\ 
|+0+0\rangle|+0-0\rangle_{\rm RR} \\ |-0-0\rangle|-0+0\rangle_{\rm RR}  \end{array}\!\!\!\right\}^{(ij)} }
&  \multicolumn{2}{|c|}{} 
\\
H & {\footnotesize \frac{1-\eta}{2}\left\{\!\!\!\begin{array}{c}  |00++\rangle|00++\rangle_{\rm NSNS}\\|00--\rangle|00--\rangle_{\rm NSNS}
\\  |++00\rangle|-+00\rangle_{\rm RR} \\ |--00\rangle|+-00\rangle_{\rm RR}  \end{array}\!\!\!\right\}^{(ij)} }
&\multicolumn{2}{|c|}{} 
& {\footnotesize \frac{1-\eta}{2}\left\{\!\!\!\begin{array}{c}  |0++0\rangle|0++0\rangle_{\rm NSNS}\\|0--0\rangle|0--0\rangle_{\rm NSNS}
\\  |+00+\rangle|-00+\rangle_{\rm RR} \\ |-00-\rangle|+00-\rangle_{\rm RR}  \end{array}\!\!\!\right\}^{(ij)} }
& {\footnotesize \frac{1-\eta}{2}\left\{\!\!\!\begin{array}{c}  |0+0+\rangle|0+0+\rangle_{\rm NSNS}\\|0-0-\rangle|0-0-\rangle_{\rm NSNS}
\\  |+0+0\rangle|-0+0\rangle_{\rm RR} \\ |-0-0\rangle|+0-0\rangle_{\rm RR}  \end{array}\!\!\!\right\}^{(ij)} }
&  \multicolumn{2}{|c|}{} 
\\
{\rm f.p.} & {\footnotesize i,j \in \{1 \ldots 4\} \text{ on } T_2 \times T_3}  & \multicolumn{2}{|c|}{} 
& {\footnotesize i,j \in \{1 \ldots 4\} \text{ on } T_1 \times T_2} 
& {\footnotesize i,j \in \{1 \ldots 4\} \text{ on } T_1 \times T_3} & \multicolumn{2}{|c|}{} 
\\\hline
(\# V,\# H) & (\frac{1+\eta}{2} \cdot 16, \frac{1-\eta}{2} \cdot 16) &  \multicolumn{2}{|c|}{} 
 & (\frac{1+\eta}{2} \cdot 16, \frac{1-\eta}{2} \cdot 16) 
& (\frac{1+\eta}{2} \cdot 16, \frac{1-\eta}{2} \cdot 16)&  \multicolumn{2}{|c|}{} 
\\ \hline\hline\hline
\!\!\!\Z_2 \times \Z_4\!\!\!& (0,\frac{1}{4},-\frac{1}{4}) & {\color{blue} (0,\frac{1}{2},-\frac{1}{2})} & & {\color{blue} (\frac{1}{2},-\frac{1}{2},0)} & 
(\frac{1}{2},-\frac{1}{4},-\frac{1}{4}) & {\color{blue} (\frac{1}{2},0,-\frac{1}{2})} &  
\\\hline
V &  {\footnotesize \frac{1+\eta}{2}\left\{\!\!\!\begin{array}{c}|00\frac{1}{4}\frac{3}{4}\rangle|00\frac{-1}{4}\frac{-3}{4}\rangle_{\rm NSNS}
\\ |00\frac{-3}{4}\frac{-1}{4}\rangle|00\frac{3}{4}\frac{1}{4}\rangle_{\rm NSNS}
\\|\frac{-1}{2} \frac{-1}{2}\frac{-1}{4}\frac{1}{4}\rangle|\frac{-1}{2} \frac{1}{2}\frac{1}{4}\frac{-1}{4}\rangle_{\rm RR}
\\ |\frac{1}{2} \frac{1}{2}\frac{-1}{4}\frac{1}{4}\rangle|\frac{1}{2}\frac{-1}{2}\frac{1}{4}\frac{-1}{4}\rangle_{\rm RR}
 \end{array}\!\!\!\right\}^{(ij)} }
& {\footnotesize  \left\{\!\!\!\begin{array}{c}|00++\rangle00--\rangle_{\rm NSNS} \\ |00--\rangle|00++\rangle_{\rm NSNS} \\ 
|++00\rangle|+-00\rangle_{\rm RR} \\ |--00\rangle|-+00\rangle_{\rm RR}  \end{array}\!\!\!\right\}^{(ij)} }
& 
& {\footnotesize \left\{\!\!\!\begin{array}{c}|0++0\rangle0--0\rangle_{\rm NSNS} \\ |0--0\rangle|0++0\rangle_{\rm NSNS} \\ 
|+00+\rangle|+00-\rangle_{\rm RR} \\ |-00-\rangle|-00+\rangle_{\rm RR}  \end{array}\!\!\!\right\}^{(ij)} }
&{\footnotesize \frac{1+\eta}{2}\left\{\!\!\!\begin{array}{c}
|0\frac{-1}{2}\frac{-1}{4}\frac{-1}{4}\rangle|0\frac{1}{2}\frac{1}{4}\frac{1}{4}\rangle_{\rm NSNS}
\\ |\frac{1}{2}0\frac{1}{4}\frac{1}{4}\rangle|\frac{1}{2}0\frac{-1}{4}\frac{-1}{4}\rangle_{\rm RR}
 \end{array}\!\!\!\right\}^{(ijk)} }
&{\footnotesize \left\{\!\!\!\begin{array}{c}|0+0+\rangle0-0-\rangle_{\rm NSNS} \\ |0-0-\rangle|0+0+\rangle_{\rm NSNS} \\ 
|+0+0\rangle|+0-0\rangle_{\rm RR} \\ |-0-0\rangle|-0+0\rangle_{\rm RR}  \end{array}\!\!\!\right\}^{(ij)} }
& 
\\
H &  {\footnotesize \frac{1-\eta}{2}\left\{\!\!\!\begin{array}{c} |00\frac{1}{4}\frac{3}{4}\rangle|00\frac{3}{4}\frac{1}{4}\rangle_{\rm NSNS}
\\|00\frac{-3}{4}\frac{-1}{4}\rangle|00\frac{-1}{4}\frac{-3}{4}\rangle_{\rm NSNS}
\\ |\frac{-1}{2} \frac{-1}{2}\frac{-1}{4}\frac{1}{4}\rangle|\frac{1}{2}\frac{-1}{2}\frac{1}{4}\frac{-1}{4}\rangle_{\rm RR}
\\ |\frac{1}{2} \frac{1}{2}\frac{-1}{4}\frac{1}{4}\rangle|\frac{-1}{2} \frac{1}{2}\frac{1}{4}\frac{-1}{4}\rangle_{\rm RR}
 \end{array}\!\!\!\right\}^{(ij)} }
& \emptyset 
&
& \emptyset 
&  \emptyset
& \emptyset
&
\\
{\rm f.p.} & {\footnotesize i,j \in \{1,2\} \text{ on } T_2 \times T_3} 
& {\footnotesize \begin{array}{c} (ij) \in \{(11),(12),(21),(22), \\ (13+14),(31+41),(23+24),(32+42), \\ (33+44),(34+43)\}
\text{ on } T_2 \times T_3  \end{array}} & 
& {\footnotesize \begin{array}{c} (ij) \in \{(i1\pm i1),(i2 \pm i2),(i3 \pm i4)\}
\\ \text{ on } T_1 \times T_2 \text{ with } i=1\ldots 4 
\end{array}} 
&  {\footnotesize \begin{array}{c} (ijk) \text{ on } T_1 \times T_2 \times T_3 
\\ \text{ with } i=1\ldots 4; j,k=1,2 \end{array}}
& {\footnotesize \begin{array}{c} (ij) \in \{(i1\pm i1),(i2 \pm i2),(i3 \pm i4)\}
\\ \text{ on } T_1 \times T_3 \text{ with } i=1\ldots 4  \end{array}} 
&
\\\hline
(\# V,\# H) & (\frac{1+\eta}{2} \cdot 8, \frac{1-\eta}{2} \cdot 8) & (10,0) & 
& (4 + \frac{1+\eta}{2} \cdot 8,0) & (\frac{1+\eta}{2} \cdot 16,0) & (4 + \frac{1+\eta}{2} \cdot 8,0)  & 
\\ \hline\hline\hline
\!\!\!\Z_2 \times \Z_6\!\!\!& (0,\frac{1}{6},-\frac{1}{6}) & (0,\frac{1}{3},-\frac{1}{3}) & {\color{blue} (0,\frac{1}{2},-\frac{1}{2})} & {\color{blue} (\frac{1}{2},-\frac{1}{2},0)} & (\frac{1}{2},-\frac{1}{3},-\frac{1}{6}) & (\frac{1}{2},-\frac{1}{6},-\frac{1}{3}) &  {\color{blue} (\frac{1}{2},0,-\frac{1}{2})}  
\\\hline
V &   {\footnotesize \frac{1+\eta}{2}\left\{\!\!\!\begin{array}{c}|00\frac{1}{6}\frac{5}{6}\rangle|00\frac{-1}{6} \frac{-5}{6}\rangle_{\rm NSNS}
\\ |00\frac{-5}{6} \frac{-1}{6}\rangle |00 \frac{5}{6}\frac{1}{6} \rangle_{\rm NSNS}
\\|\frac{1}{2},\frac{1}{2},\frac{-1}{3},\frac{1}{3}\rangle|\frac{1}{2},\frac{-1}{2},\frac{1}{3},\frac{-1}{3}\rangle_{\rm RR}
\\
|\frac{-1}{2},\frac{-1}{2},\frac{-1}{3},\frac{1}{3}\rangle|\frac{-1}{2},\frac{1}{2} ,\frac{1}{3},\frac{-1}{3}\rangle_{\rm RR}
 \end{array}\!\!\!\right\}^{(11)} }
& {\footnotesize \left\{\!\!\!\begin{array}{c}|00\frac{1}{3}\frac{2}{3}\rangle|00\frac{-1}{3} \frac{-2}{3}\rangle_{\rm NSNS}
\\ |00\frac{-2}{3} \frac{-1}{3}\rangle |00 \frac{2}{3}\frac{1}{3} \rangle_{\rm NSNS}
\\|\frac{1}{2},\frac{1}{2},\frac{-1}{6},\frac{1}{6}\rangle|\frac{1}{2},\frac{-1}{2},\frac{1}{6},\frac{-1}{6}\rangle_{\rm RR}
\\|\frac{-1}{2},\frac{-1}{2},\frac{-1}{6},\frac{1}{6}\rangle|\frac{-1}{2},\frac{1}{2},\frac{1}{6},\frac{-1}{6}\rangle_{\rm RR}
 \end{array}\!\!\!\right\}^{(ij)} }
& {\footnotesize \frac{1+\eta}{2}\left\{\!\!\!\begin{array}{c}|00++\rangle00--\rangle_{\rm NSNS} \\ |00--\rangle|00++\rangle_{\rm NSNS} \\ 
|++00\rangle|+-00\rangle_{\rm RR} \\ |--00\rangle|-+00\rangle_{\rm RR}  \end{array}\!\!\!\right\}^{(ij)} }
&  {\footnotesize \frac{1+\eta}{2}\left\{\!\!\!\begin{array}{c}|0++0\rangle0--0\rangle_{\rm NSNS} \\ |0--0\rangle|0++0\rangle_{\rm NSNS} \\ 
|+00+\rangle|+00-\rangle_{\rm RR} \\ |-00-\rangle|-00+\rangle_{\rm RR}  \end{array}\!\!\!\right\}^{(ij)} }
&{\footnotesize \left\{\!\!\!\begin{array}{c} |0,\frac{-1}{2},\frac{-1}{3},\frac{-1}{6}\rangle|0,\frac{1}{2},\frac{1}{3},\frac{1}{6}\rangle_{\rm NSNS}\\ 
|\frac{1}{2},0,\frac{1}{6},\frac{1}{3}\rangle|\frac{1}{2},0,\frac{-1}{6},\frac{-1}{3}\rangle_{\rm RR}  \end{array}\!\!\!\right\}^{(ijk)} }
&{\footnotesize \left\{\!\!\!\begin{array}{c}  |0,\frac{-1}{2},\frac{-1}{6},\frac{-1}{3}\rangle|0,\frac{1}{2},\frac{1}{6},\frac{1}{3}\rangle_{\rm NSNS} 
\\ |\frac{1}{2},0,\frac{1}{3},\frac{1}{6}\rangle |\frac{1}{2},0,\frac{-1}{3},\frac{-1}{6}\rangle_{\rm RR}
\end{array}\!\!\!\right\}^{(ijk)} }
&{\footnotesize \frac{1+\eta}{2}\left\{\!\!\!\begin{array}{c}|0+0+\rangle0-0-\rangle_{\rm NSNS} \\ |0-0-\rangle|0+0+\rangle_{\rm NSNS} \\ 
|+0+0\rangle|+0-0\rangle_{\rm RR} \\ |-0-0\rangle|-0+0\rangle_{\rm RR}  \end{array}\!\!\!\right\}^{(ij)} }
\\
H &  {\footnotesize \frac{1-\eta}{2}\left\{\!\!\!\begin{array}{c}
|00\frac{1}{6}\frac{5}{6}\rangle|00\frac{5}{6} \frac{1}{6}\rangle_{\rm NSNS}
\\ |00\frac{-5}{6} \frac{-1}{6}\rangle |00 \frac{-1}{6}\frac{-5}{6} \rangle_{\rm NSNS}
\\ |\frac{1}{2},\frac{1}{2},\frac{-1}{3},\frac{1}{3}\rangle|\frac{-1}{2},\frac{1}{2},\frac{1}{3},\frac{-1}{3}\rangle_{\rm RR} 
\\|\frac{-1}{2},\frac{-1}{2},\frac{-1}{3},\frac{1}{3}\rangle|\frac{1}{2}.\frac{-1}{2},\frac{1}{3},\frac{-1}{3}\rangle_{\rm RR}
 \end{array}\!\!\!\right\}^{(11)} }
& {\footnotesize\left\{\!\!\!\begin{array}{c} |00\frac{1}{3}\frac{2}{3}\rangle|00\frac{2}{3} \frac{1}{3}\rangle_{\rm NSNS}
\\ |00\frac{-2}{3} \frac{-1}{3}\rangle |00 \frac{-1}{3}\frac{-2}{3} \rangle_{\rm NSNS}
\\|\frac{1}{2},\frac{1}{2},\frac{-1}{6},\frac{1}{6}\rangle|\frac{-1}{2},\frac{1}{2},\frac{1}{6},\frac{-1}{6}\rangle_{\rm RR}
\\|\frac{-1}{2},\frac{-1}{2},\frac{-1}{6},\frac{1}{6}\rangle|\frac{1}{2},\frac{-1}{2},\frac{1}{6},\frac{-1}{6}\rangle_{\rm RR}
 \end{array}\!\!\!\right\}^{(kl)} }
&{\footnotesize \frac{1-\eta}{2}\left\{\!\!\!\begin{array}{c}  |00++\rangle|00++\rangle_{\rm NSNS}\\|00--\rangle|00--\rangle_{\rm NSNS}
\\  |++00\rangle|-+00\rangle_{\rm RR} \\ |--00\rangle|+-00\rangle_{\rm RR}  \end{array}\!\!\!\right\}^{(ij)} }
& {\footnotesize \frac{1-\eta}{2}\left\{\!\!\!\begin{array}{c}  |0++0\rangle|0++0\rangle_{\rm NSNS}\\|0--0\rangle|0--0\rangle_{\rm NSNS}
\\  |+00+\rangle|-00+\rangle_{\rm RR} \\ |-00-\rangle|+00-\rangle_{\rm RR}  \end{array}\!\!\!\right\}^{(ij)} }
& \emptyset
& \emptyset
&{\footnotesize \frac{1-\eta}{2}\left\{\!\!\!\begin{array}{c}  |0+0+\rangle|0+0+\rangle_{\rm NSNS}\\|0-0-\rangle|0-0-\rangle_{\rm NSNS}
\\  |+0+0\rangle|-0+0\rangle_{\rm RR} \\ |-0-0\rangle|+0-0\rangle_{\rm RR}  \end{array}\!\!\!\right\}^{(ij)} }
\\
{\rm f.p.} & (11) \text{ on } T_2 \times T_3
& {\footnotesize \begin{array}{c} (ij) \in \{(11),(12+13),(21+31),\\ 
(22+33+23+32)\} \text{ on } T_2 \times T_3 \\
(kl)=(22+33+23+32)
\end{array}}
& {\footnotesize \begin{array}{c} (ij) \in \{(11),(14+15+16),\\ (41+51+61),(44+56+65), \\(45+54+66),(46+55+64)\}
\text{ on } T_2 \times T_3 \end{array}}
& {\footnotesize \begin{array}{c} (ij) \in \{(i1 \pm i1), (i4+i5+i6)\} \\
\text{ on } T_1 \times T_2 \text{ with } i=1,2,3,4 \end{array}}
&{\footnotesize \begin{array}{c} (ijk) \in \{(i11\pm i11),(i21\pm i31)\}
\\
\text{ on } T_1 \times T_2 \times T_3 \text{ with } i=1,2,3,4
\end{array} }
&{\footnotesize \begin{array}{c} (ijk) \in \{(i11\pm i11),(i12\pm i13)\}
\\
\text{ on } T_1 \times T_2 \times T_3 \text{ with } i=1,2,3,4 \end{array} }
& {\footnotesize \begin{array}{c} (ij) \in \{(i1 \pm i1), (i4+i5+i6)\} \\
\text{ on } T_1 \times T_3 \end{array}}
\\\hline
(\# V,\# H) & (\frac{1+\eta}{2} \cdot 2, \frac{1-\eta}{2} \cdot 2) & (8,2) & (\frac{1+\eta}{2} \cdot 6, \frac{1-\eta}{2} \cdot 6) & 
 (\frac{1+\eta}{2} \cdot 8, \frac{1-\eta}{2} \cdot 4) & (4 + \frac{1+\eta}{2} \cdot 4,0) & (4 + \frac{1+\eta}{2} \cdot 4,0) 
&  (\frac{1+\eta}{2} \cdot 8, \frac{1-\eta}{2} \cdot 4)
\\ \hline\hline\hline
\!\!\!\Z_2 \times \Z_6'\!\!\!&  (-\frac{1}{3},\frac{1}{6},\frac{1}{6}) & (-\frac{2}{3},\frac{1}{3},\frac{1}{3}) & {\color{blue} (0,\frac{1}{2},-\frac{1}{2})} & {\color{blue} (\frac{1}{2},-\frac{1}{2},0)} & (\frac{1}{6},-\frac{1}{3},\frac{1}{6}) & (-\frac{1}{6},-\frac{1}{6},\frac{1}{3})  &  {\color{blue} (\frac{1}{2},0,-\frac{1}{2})}  
\\\hline
V &  {\footnotesize \left\{\!\!\!\begin{array}{c} |0,\frac{2}{3},\frac{1}{6},\frac{1}{6}\rangle |0,\frac{-2}{3},\frac{-1}{6},\frac{-1}{6}\rangle_{\rm NSNS}
\\ |\frac{-1}{2},\frac{1}{6},\frac{-1}{3},\frac{-1}{3}\rangle |\frac{-1}{2},\frac{-1}{6},\frac{1}{3},\frac{1}{3}\rangle_{\rm RR}
\end{array}\!\!\!\right\}^{(ijk)} }
& {\footnotesize \left\{\!\!\!\begin{array}{c} |0,\frac{1}{3},\frac{1}{3},\frac{1}{3}\rangle |0,\frac{-1}{3},\frac{-1}{3},\frac{-1}{3}\rangle_{\rm NSNS}
\\ |\frac{-1}{2},\frac{-1}{6},\frac{-1}{6},\frac{-1}{6}\rangle |\frac{-1}{2},\frac{1}{6},\frac{1}{6},\frac{1}{6}\rangle_{\rm RR}
\end{array}\!\!\!\right\}^{(ijk)} }
& {\footnotesize \frac{1+\eta}{2}\left\{\!\!\!\begin{array}{c}|00++\rangle00--\rangle_{\rm NSNS} \\ |00--\rangle|00++\rangle_{\rm NSNS} \\ 
|++00\rangle|+-00\rangle_{\rm RR} \\ |--00\rangle|-+00\rangle_{\rm RR}  \end{array}\!\!\!\right\}^{(ij)} }
& {\footnotesize \frac{1+\eta}{2}\left\{\!\!\!\begin{array}{c}|0++0\rangle0--0\rangle_{\rm NSNS} \\ |0--0\rangle|0++0\rangle_{\rm NSNS} \\ 
|+00+\rangle|+00-\rangle_{\rm RR} \\ |-00-\rangle|-00+\rangle_{\rm RR}  \end{array}\!\!\!\right\}^{(ij)} }
& {\footnotesize \left\{\!\!\!\begin{array}{c} |0,\frac{1}{6},\frac{2}{3},\frac{1}{6}\rangle |0,\frac{-1}{6},\frac{-2}{3},\frac{-1}{6}\rangle_{\rm NSNS}
\\ |\frac{-1}{2},\frac{-1}{3},\frac{1}{6},\frac{-1}{3}\rangle |\frac{-1}{2},\frac{1}{3},\frac{-1}{6},\frac{1}{3}\rangle_{\rm RR}
\end{array}\!\!\!\right\}^{(ijk)} }
& {\footnotesize \left\{\!\!\!\begin{array}{c} |0,-\frac{1}{6},-\frac{1}{6},-\frac{2}{3}\rangle|0,\frac{1}{6},\frac{1}{6},\frac{2}{3}\rangle_{\rm NSNS}
\\ |\frac{1}{2},\frac{1}{3},\frac{1}{3},\frac{-1}{6}\rangle |\frac{1}{2},\frac{-1}{3},\frac{-1}{3},\frac{1}{6}\rangle_{\rm RR}
\end{array}\!\!\!\right\}^{(ijk)} }
& {\footnotesize \frac{1+\eta}{2}\left\{\!\!\!\begin{array}{c}|0+0+\rangle0-0-\rangle_{\rm NSNS} \\ |0-0-\rangle|0+0+\rangle_{\rm NSNS} \\ 
|+0+0\rangle|+0-0\rangle_{\rm RR} \\ |-0-0\rangle|-0+0\rangle_{\rm RR}  \end{array}\!\!\!\right\}^{(ij)} }
\\
H & \emptyset
& \emptyset
&{\footnotesize \frac{1-\eta}{2}\left\{\!\!\!\begin{array}{c}  |00++\rangle|00++\rangle_{\rm NSNS}\\|00--\rangle|00--\rangle_{\rm NSNS}
\\  |++00\rangle|-+00\rangle_{\rm RR} \\ |--00\rangle|+-00\rangle_{\rm RR}  \end{array}\!\!\!\right\}^{(ij)} }
& {\footnotesize \frac{1-\eta}{2}\left\{\!\!\!\begin{array}{c}  |0++0\rangle|0++0\rangle_{\rm NSNS}\\|0--0\rangle|0--0\rangle_{\rm NSNS}
\\  |+00+\rangle|-00+\rangle_{\rm RR} \\ |-00-\rangle|+00-\rangle_{\rm RR}  \end{array}\!\!\!\right\}^{(ij)} }
& \emptyset
& \emptyset
&{\footnotesize \frac{1-\eta}{2}\left\{\!\!\!\begin{array}{c}  |0+0+\rangle|0+0+\rangle_{\rm NSNS}\\|0-0-\rangle|0-0-\rangle_{\rm NSNS}
\\  |+0+0\rangle|-0+0\rangle_{\rm RR} \\ |-0-0\rangle|+0-0\rangle_{\rm RR}  \end{array}\!\!\!\right\}^{(ij)} }
\\
{\rm f.p.} & {\footnotesize \begin{array}{c} (ijk) \in \{(111 \pm 111), (211 \pm 311)\} 
\\ \text{ on } T_1 \times T_2 \times T_3 \end{array} }
& {\footnotesize \begin{array}{c} (ijk) \in \{(111), (211+311), (121+131),(112+113),\\
(221+231+321+331), (212+213+312+313),\\
 (122+133+132+123), (222+233+332+323),\\
(232+223+322+333)  \}
 \text{ on } T_1 \times T_2 \times T_3 \end{array} }
& {\footnotesize \begin{array}{c} (ij) \in \{(11\pm 11),(14+15+16), \\(41+51+61),(44+55+66),(45+56+64), \\ (46+54+65)\}
\text{ on } T_2 \times T_3 \end{array}}
& {\footnotesize \begin{array}{c} (ij) \in \{(11\pm 11),(14+15+16), \\(41+51+61),(44+55+66),(45+56+64), \\ (46+54+65)\}
\text{ on } T_1 \times T_2 \end{array}}
&{\footnotesize \begin{array}{c} (ijk) \in \{(111 \pm 111), (121 \pm 131)\} \\ \text{ on } T_1 \times T_2 \times T_3 \end{array} }
&{\footnotesize \begin{array}{c} (ijk) \in \{(111 \pm 111), (112 \pm 113)\} \\ \text{ on } T_1 \times T_2 \times T_3 \end{array} }
& {\footnotesize \begin{array}{c} (ij) \in \{(11\pm 11),(14+15+16), \\(41+51+61),(44+55+66),(45+56+64), \\ (46+54+65)\}
\text{ on } T_1 \times T_3 \end{array}}
\\\hline
(\# V,\# H) & (1+\frac{1+\eta}{2},0) & (9,0) & (\frac{1+\eta}{2} \cdot
6, \frac{1-\eta}{2} \cdot 5)  
& (\frac{1+\eta}{2} \cdot 6, \frac{1-\eta}{2} \cdot 5) &
(1+\frac{1+\eta}{2},0) & (1+\frac{1+\eta}{2},0)  
& (\frac{1+\eta}{2} \cdot 6, \frac{1-\eta}{2} \cdot 5) 
\\\hline
     \end{array}
}{Closed-string-T6ZNxZM-torsion}{Massless closed string spectrum
  from the twisted sectors in $T^6/\Z_2 \times \Z_{2M}$ orbifolds  
without ($\eta=1$) and with ($\eta=-1$) discrete torsion. Only the NS-NS and RR sectors, which provide for spacetime bosons, are 
listed explicitly. The R-NS and NS-R sectors, which are spacetime fermions, can be infered by ${\cal N}=2$ supersymmetry in four dimensions or
constructed explicitly in an analogous manner. All twisted sectors $n\vec{v} + m\vec{w}$ other than $\Z_2$ have inverse sectors
$(N-n)\vec{v}+(M-m)\vec{w}$ which are not listed explicitly. 
The counting $(\#V,\#H)$ of ${\cal N}=2$ vector and hyper multiplets per sector
includes these inverse sectors.
The decomposition under the orientifold projection of the ${\cal N}=2$ vector multiplets into ${\cal N}=1$ 
vector and chiral multiplets
is given in table~\protect\ref{tab:DecompositionHodge-T6ZNxZM-torsion}. }  

Under the orientifold projection, the $\theta^k\omega^l$ twisted closed string sector picks up a sign $\eta_{\OR} \eta_{\OR\theta^k\omega^l}$
as explained in section~\ref{Sss:DT+EO}. Since $\eta_{\OR\theta^k\omega^l} = \eta_{\OR\theta^k\omega^{l+2}} = \eta_{\OR\theta^{k+2}\omega^l} 
= \eta_{\OR\theta^{k+2}\omega^{l+2}}$, only three different non-trivial prefactors arise which we labelled for $\Z_2 \times \Z_2$, $\Z_2 \times \Z_6$ and $\Z_2 \times \Z_6'$ by $\eta_{(i)} \equiv \eta_{\OR} \eta_{\OR\Z_2^{(i)}}$. As a result, the number of K\"ahler moduli and Abelian vectors
does not only depend on the choice of orbifold and discrete torsion, but also on the choice of the exotic O6-plane, as can be seen in the complete 
list of the orientifolded twisted closed string spectra in
table~\ref{tab:DecompositionHodge-T6ZNxZM-torsion}. 
\mathsidetabfix{
\renewcommand{\arraystretch}{1.3}
      \begin{array}{|c|c||c||c|c|c|c|c|c|c||c|} \hline
        \multicolumn{11}{|c|}{\rule[-3mm]{0mm}{8mm}
\text{\bf Hodge numbers $(h_{11}^+, h_{11}^{-})$ per twist sector on $T^6/(\Z_2 \times \Z_{2M} \times \OR)$ without and with discrete torsion}
}\\ \hline\hline
\begin{array}{c} T^6/ \\
{\rm torsion} \end{array}& & \begin{sideways}\!\!\!\!\!\!\textcolor{blue}{Untwisted}\; \end{sideways}&  \vec{w} 
&  2\vec{w}  & 3\vec{w}
& \vec{v}   & (\vec{v}+\vec{w})
& (\vec{v}+2\vec{w}) &  (\vec{v}+3\vec{w})
& \text{total}
\\\hline\hline
\Z_2 \times \Z_2 &  & & {\color{blue} (0,\frac{1}{2},-\frac{1}{2})} & \multicolumn{2}{|c|}{} &  {\color{blue} (\frac{1}{2},-\frac{1}{2},0)} &  {\color{blue} (\frac{1}{2},0,-\frac{1}{2})} & \multicolumn{2}{|c||}{} & 
\\\hline
\eta =1&\!\!\!\!\begin{array}{c} h_{11}^+ \\ h_{11}^- \end{array}\!\!\!\!& \begin{array}{c} 0 \\ 3   \end{array} 
& \!\!\!\!{\footnotesize  \begin{array}{c} 8(b_2+b_3- b_2b_3) \\  8(2-b_2-b_3+b_2b_3)  \end{array}  }\!\!\!\!
& \multicolumn{2}{|c|}{} 
&   \!\!\!\!{\footnotesize  \begin{array}{c} 8(b_1+b_2- b_1b_2) \\  8(2-b_1-b_2+b_1b_2)  \end{array}  }\!\!\!\!
&  \!\!\!\!{\footnotesize  \begin{array}{c} 8(b_1+b_3- b_1b_3) \\  8(2-b_1-b_3+b_1b_3)  \end{array}  }\!\!\!\!
& \multicolumn{2}{|c||}{} & {\footnotesize \!\!\!\!\begin{array}{c} 16 \sum_{i=1}^3b_i -8\, \sum_{i<j} b_ib_j
\\  51-16 \sum_{i=1}^3b_i + 8\, \sum_{i<j} b_ib_j  \end{array}\!\!\!\!}
\\\hline
\eta =-1 &\!\!\!\!\begin{array}{c} h_{11}^+ \\ h_{11}^- \end{array}\!\!\!\!& \begin{array}{c} 0 \\ 3   \end{array}  & 0 &  \multicolumn{2}{|c|}{} & 0 & 0 & \multicolumn{2}{|c||}{}  & \begin{array}{c} 0 \\ 3   \end{array} 
\\\hline\hline
\Z_2 \times \Z_4 && & (0,\frac{1}{4},-\frac{1}{4}) & {\color{blue} (0,\frac{1}{2},-\frac{1}{2})} & & {\color{blue} (\frac{1}{2},-\frac{1}{2},0)} & 
(\frac{1}{2},-\frac{1}{4},-\frac{1}{4}) & {\color{blue} (\frac{1}{2},0,-\frac{1}{2})} &  &
\\\hline
\eta =1 &\!\!\!\!\begin{array}{c} h_{11}^+ \\ h_{11}^- \end{array}\!\!\!\!& \begin{array}{c} 0 \\ 3   \end{array} 
&  \begin{array}{c} 0 \\  8   \end{array} &  \begin{array}{c} 0 \\ 10  \end{array}
& & \begin{array}{c} 6 \, b \\6(2-b)  \end{array} &  \begin{array}{c}  8 \, b \\ 8 (2-b) \end{array}
& \begin{array}{c} 6 \, b \\ 6(2-b) \end{array}  &  & \begin{array}{c} 20\, b \\  61 -20 \, b \end{array} 
\\\hline
\eta =-1 &\!\!\!\!\begin{array}{c} h_{11}^+ \\ h_{11}^- \end{array}\!\!\!\!& \begin{array}{c} 0 \\ 3   \end{array} 
& 0 &  \begin{array}{c} 0 \\ 10  \end{array} &  & 
\!\!\!\!\!{\footnotesize \begin{array}{cc} {\bf a/bAA}, {\bf a/bAB} & {\bf a/bBB} \\
2\left[1-(1-b)\eta_{(3)}\right] &  2\left[1+(1-b)\eta_{(3)}\right] 
\\  2\left[1+(1-b)\eta_{(3)}\right] &  2\left[1-(1-b)\eta_{(3)}\right]   \end{array}} \!\!\!\!
& 0 
& \!\!\!\!\!{\footnotesize \begin{array}{cc} 
 {\bf a/bAA} &  {\bf a/bAB}, {\bf a/bBB} \\ 2\left[1-(1-b)\eta_{(3)}\right] & 2\left[1+(1-b)\eta_{(3)}\right] 
\\   2\left[1+(1-b)\eta_{(3)}\right] &   2\left[1-(1-b)\eta_{(3)}\right]   \end{array}} \!\!\!\!
& & 
 \!\!\!\!\!{\footnotesize \begin{array}{ccc} {\bf a/bAA} & {\bf a/bAB} & {\bf a/bBB}  \\ 4\left[1-(1-b)\eta_{(3)}\right]   &    4  & 4\left[1+(1-b)\eta_{(3)}\right] 
\\ 
   17+4(1-b)\eta_{(3)}  &   17 &  17-4(1-b)\eta_{(3)}
 \end{array}} \!\!\!\!
 \\\hline\hline
\Z_2 \times \Z_6 & & & (0,\frac{1}{6},-\frac{1}{6}) & (0,\frac{1}{3},-\frac{1}{3}) & {\color{blue} (0,\frac{1}{2},-\frac{1}{2})} & {\color{blue} (\frac{1}{2},-\frac{1}{2},0)} & (\frac{1}{2},-\frac{1}{3},-\frac{1}{6}) & (\frac{1}{2},-\frac{1}{6},-\frac{1}{3}) &  {\color{blue} (\frac{1}{2},0,-\frac{1}{2})} &  
\\\hline
\eta =1 &\!\!\!\!\begin{array}{c} h_{11}^+ \\ h_{11}^- \end{array}\!\!\!\!& \begin{array}{c} 0 \\ 3   \end{array}  
&  \begin{array}{c} 0 \\ 2   \end{array}
& \begin{array}{c} 0 \\ 8  \end{array}& \begin{array}{c} 1 \\ 5 \end{array}& \begin{array}{c} 4 \, b \\ 4(2-b) \end{array} 
& \begin{array}{c} 4 \, b \\ 4(2-b) \end{array} & \begin{array}{c} 4 \, b \\ 4(2-b) \end{array}
& \begin{array}{c} 4 \, b \\ 4(2-b) \end{array} & \begin{array}{c} 1 + 16 \, b \\ 50 - 16 \, b\end{array}
\\\hline
\eta =-1 &\!\!\!\!\begin{array}{c} h_{11}^+ \\ h_{11}^- \end{array}\!\!\!\!& \begin{array}{c} 0 \\ 3   \end{array}  
& 0 & \begin{array}{c} 0 \\ 8 \end{array} & 0 & 0 
&  \!\!\!\!\!{\footnotesize \begin{array}{cc}
{\bf a/bAA}, {\bf a/bAB} & {\bf a/bBB} \\ 2\left[1+(1-b)\eta_{(2)}\right] &  2\left[1-(1-b)\eta_{(2)}\right] \\
 2\left[1-(1-b)\eta_{(2)}\right] &  2\left[1+(1-b)\eta_{(2)}\right]   \end{array}} \!\!\!\!
& \!\!\!\!\!{\footnotesize \begin{array}{cc} 
 {\bf a/bAA} & {\bf a/bAB}, {\bf a/bBB} \\ 2\left[1+(1-b)\eta_{(3)}\right] & 2\left[1-(1-b)\eta_{(3)}\right] 
\\   2\left[1-(1-b)\eta_{(3)}\right] &   2\left[1+(1-b)\eta_{(3)}\right]   \end{array}} \!\!\!\! 
& 0 
& \!\!\!\!\!{\footnotesize \begin{array}{ccc}  {\bf a/bAA} & {\bf a/bAB} & {\bf a/bBB}  
\\ 4+ 2(1-b)(\eta_{(2)}+\eta_{(3)}) &  4+ 2(1-b)(\eta_{(2)}-\eta_{(3)})  &   4- 2(1-b)(\eta_{(2)}+\eta_{(3)}) 
\\ 15 - 2(1-b)(\eta_{(2)}+\eta_{(3)})  &  15 - 2(1-b)(\eta_{(2)}-\eta_{(3)}) &     15 + 2(1-b)(\eta_{(2)}+\eta_{(3)})
 \end{array}} \!\!\!\! 
\\ \hline\hline
\Z_2 \times \Z_6' & & &  (-\frac{1}{3},\frac{1}{6},\frac{1}{6}) & (-\frac{2}{3},\frac{1}{3},\frac{1}{3}) & {\color{blue} (0,\frac{1}{2},-\frac{1}{2})} & {\color{blue} (\frac{1}{2},-\frac{1}{2},0)} & (\frac{1}{6},-\frac{1}{3},\frac{1}{6}) & (-\frac{1}{6},-\frac{1}{6},\frac{1}{3})  &  {\color{blue} (\frac{1}{2},0,-\frac{1}{2})} &  
\\\hline
\eta =1 &\!\!\!\!\begin{array}{c} h_{11}^+ \\ h_{11}^- \end{array}\!\!\!\!& \begin{array}{c} 0 \\ 3   \end{array}  
& \begin{array}{c} 0 \\ 2 \end{array} & {\footnotesize \begin{array}{cc} {\bf AAA,ABB}& {\bf AAB,BBB} \\1 & 0 \\    8 &  9 \end{array} }
& \begin{array}{c} 1 \\ 5 \end{array} &   \begin{array}{c} 1 \\ 5 \end{array}
&  \begin{array}{c} 0 \\ 2 \end{array}  &  \begin{array}{c} 0 \\ 2 \end{array}  &   \begin{array}{c} 1 \\ 5 \end{array}
& {\footnotesize \begin{array}{cc} {\bf AAA,ABB}& {\bf AAB,BBB}\\ 4 & 3 \\    32 &  33 \end{array} }
\\\hline
\eta =-1 &\!\!\!\!\begin{array}{c} h_{11}^+ \\ h_{11}^- \end{array}\!\!\!\!& \begin{array}{c} 0 \\ 3   \end{array} 
&  \!\!\!\!{\footnotesize \begin{array}{cc} {\bf AAA}, {\bf AAB}, {\bf ABB}  & {\bf BBB} \\ \frac{1+\eta_{(1)}}{2} &   \frac{1-\eta_{(1)}}{2}  \\
\frac{1-\eta_{(1)}}{2} & 
\frac{1+\eta_{(1)}}{2} \end{array}}\!\!\!\! 
& {\footnotesize \begin{array}{cc} {\bf AAA,ABB} & {\bf AAB,BBB} \\ 1 & 0 \\    8 &  9  \end{array}} & 0 & 0 
&  {\footnotesize \begin{array}{cc}  {\bf AAA}, {\bf AAB} & {\bf ABB}, {\bf BBB} \\ \frac{1+\eta_{(2)}}{2} &  \frac{1-\eta_{(2)}}{2}  \\
  \frac{1-\eta_{(2)}}{2} &   \frac{1+\eta_{(2)}}{2}\end{array}} 
&  {\footnotesize \begin{array}{cc}  {\bf AAA}  & {\bf AAB}, {\bf ABB}, {\bf BBB}
\\ \frac{1+\eta_{(3)}}{2} &    \frac{1-\eta_{(3)}}{2}  \\
  \frac{1-\eta_{(3)}}{2} &   \frac{1+\eta_{(3)}}{2}\end{array}} & 0 
&  {\footnotesize \begin{array}{cccc} {\bf AAA} & {\bf AAB} & {\bf ABB} &  {\bf BBB}
\\ \frac{5 + \sum_{i=1}^3 \eta_{(i)} }{2}   
&  \frac{3 + \eta_{(1)} + \eta_{(2)} - \eta_{(3)} }{2} 
&  \frac{5 + \eta_{(1)} -\eta_{(2)} - \eta_{(3)} }{2}  
&   \frac{3- \sum_{i=1}^3 \eta_{(i)} }{2}
\\ \frac{25 - \sum_{i=1}^3 \eta_{(i)} }{2}   
&  \frac{27 - \eta_{(1)} - \eta_{(2)} + \eta_{(3)} }{2} 
&    \frac{25 - \eta_{(1)} +\eta_{(2)} + \eta_{(3)} }{2} 
&  \frac{27 + \sum_{i=1}^3 \eta_{(i)} }{2}
\end{array}}
\\ \hline
     \end{array}}{DecompositionHodge-T6ZNxZM-torsion}{Decomposition of Hodge number $h_{11}= h_{11}^+ + h_{11}^-$ under the orientifold projection per twist-sector for orbifolds 
     \mbox{$T^6/(\Z_2 \times \Z_{2M}\times \OR)$}  without and with torsion. $h_{11}^-$ counts the K\"ahler moduli, whereas $h_{11}^+$ labels the number of Abelian vectors.}

Massless open string states can be counted by e.g. taking the expressions~(\ref{Eq:twisted-closed-states}) 
for the right-moving NS and R sectors and replacing the twist sector $n\vec{v} + m \vec{w}$ by the relative angles $\vec{\varphi}$ of the D6-branes
on which the open string under question ends.

{\boldmath
\section{The open string spectrum on $T^6/\Z_2 \times \Z_{2M}$ via Chan-Paton labels and gauge threshold amplitudes}\label{S:App-OpenStrings}
}

\subsection{Chan-Paton labels}\label{S:App-ChanPaton}

The Chan-Paton label $\lambda$ associated to some open string state transforms as follows under 
the $\Z_2$ orbifold projections,
\begin{equation}
\Z_2^{(i)}: \quad \lambda |{\rm state}\rangle \longrightarrow c^{\Z_2^{(i)} }_{\rm state}
\left( \gamma_{\Z_2^{(i)} } \lambda\gamma_{\Z_2^{(i)} }^{-1} \right)  |{\rm state}\rangle.
\end{equation}
 In the absence of discrete Wilson lines, $c^{\Z_2^{(i)} }_{\rm state}= \pm 1$ is simply the $\Z_2^{(i)}$ 
eigenvalue of the massless state.
Orbifold generators other than $\Z_2$ change the positions of the D6-branes where the open string
ends on, thereby identifying orbifold images.

In order to evaluate the Chan-Paton labels, representations of the $\Z_2$ gamma matrices are needed. We start with
\begin{equation}
\gamma_{\Z_2^{(1)} }=\left(\begin{array}{cccc}  1 & 0 & 0 & 0 \\ 0 & 1 & 0 & 0 \\ 0 & 0 & -1 & 0 
\\ 0 & 0 & 0 & -1
\end{array}\right)
\quad
\gamma_{\Z_2^{(2)}} =\left(\begin{array}{cccc}  1 & 0 & 0 & 0 \\ 0 & -1 & 0 & 0 \\ 0 & 0 & 1 & 0 \\
0 & 0 & 0 & -1
\end{array}\right)
\quad
\gamma_{\Z_2^{(3)} }=\left(\begin{array}{cccc}  1 & 0 & 0 & 0 \\ 0 & -1 & 0 & 0 \\ 0 & 0 & -1 & 0 \\
0 & 0 & 0 & 1
\end{array}\right)
\end{equation}
and check the consistency conditions: $\gamma_{\Z_2^{(i)}}^2 = \unity$ for all $i$ and 
$\gamma_{\Z_2^{(1)} }\cdot \gamma_{\Z_2^{(2)}} = \gamma_{\Z_2^{(3)} }$.
A Chan-Paton label contains the representations 
\begin{equation}
\lambda=
\left(\begin{array}{ccccc}
(\N_a^{1},\ov{\N}_b^{1}) & (\N_a^{1},\ov{\N}_b^{2}) & (\N_a^{1},\ov{\N}_b^{3}) & (\N_a^{1},\ov{\N}_b^{4}) \\
(\N_a^{2},\ov{\N}_b^{1}) & (\N_a^{2},\ov{\N}_b^{2}) & (\N_a^{2},\ov{\N}_b^{3}) & (\N_a^{2},\ov{\N}_b^{4}) \\
(\N_a^{3},\ov{\N}_b^{1}) & (\N_a^{3},\ov{\N}_b^{2}) & (\N_a^{3},\ov{\N}_b^{3}) & (\N_a^{3},\ov{\N}_b^{4}) \\
(\N_a^{4},\ov{\N}_b^{1}) & (\N_a^{4},\ov{\N}_b^{2}) & (\N_a^{4},\ov{\N}_b^{3}) & (\N_a^{4},\ov{\N}_b^{4}) 
\end{array}\right)
\end{equation}
of $\prod_{i=1}^4 U(N^i_a) \times U(N^i_b)$.
A single $\Z_2^{(k)}$ projection acts as follows.
{\tiny
\begin{equation*}
\begin{array}{|c|c||c|c|}\hline
(\Z_2^{(1)},\Z_2^{(2)},\Z_2^{(3)}) & \lambda & (\Z_2^{(1)},\Z_2^{(2)},\Z_2^{(3)}) & \lambda
\\\hline\hline
(+,*,*) &\left(\begin{array}{ccccc}
(\N_a^{1},\ov{\N}_b^{1}) & (\N_a^{1},\ov{\N}_b^{2}) & 0 & 0  \\
(\N_a^{2},\ov{\N}_b^{1}) & (\N_a^{2},\ov{\N}_b^{2}) & 0 & 0 \\
0 & 0 & (\N_a^{3},\ov{\N}_b^{3}) & (\N_a^{3},\ov{\N}_b^{4}) \\
0 & 0  & (\N_a^{4},\ov{\N}_b^{3}) & (\N_a^{4},\ov{\N}_b^{4}) 
\end{array}\right)
&
(-,*,*) & \left(\begin{array}{ccccc}
0 & 0  & (\N_a^{1},\ov{\N}_b^{3}) & (\N_a^{1},\ov{\N}_b^{4}) \\
0 & 0  & (\N_a^{2},\ov{\N}_b^{3}) & (\N_a^{2},\ov{\N}_b^{4}) \\
(\N_a^{3},\ov{\N}_b^{1}) & (\N_a^{3},\ov{\N}_b^{2}) & 0 & \\
(\N_a^{4},\ov{\N}_b^{1}) & (\N_a^{4},\ov{\N}_b^{2}) & 0 & 0
\end{array}\right)
\\\hline
(*,+,*) &  \left(\begin{array}{ccccc}
(\N_a^{1},\ov{\N}_b^{1}) & 0 & (\N_a^{1},\ov{\N}_b^{3}) & 0 \\
0 & (\N_a^{2},\ov{\N}_b^{2}) & 0 & (\N_a^{2},\ov{\N}_b^{4})  \\
(\N_a^{3},\ov{\N}_b^{1}) & 0 & (\N_a^{3},\ov{\N}_b^{3}) & 0\\
0 & (\N_a^{4},\ov{\N}_b^{2}) & 0 & (\N_a^{4},\ov{\N}_b^{4}) 
\end{array}\right)
&
(*,-,*) &  \left(\begin{array}{ccccc}
0 & (\N_a^{1},\ov{\N}_b^{2}) & 0 & (\N_a^{1},\ov{\N}_b^{4}) \\
(\N_a^{2},\ov{\N}_b^{1}) & 0 & (\N_a^{2},\ov{\N}_b^{3}) & 0 \\
0 & (\N_a^{3},\ov{\N}_b^{2}) & 0 & (\N_a^{3},\ov{\N}_b^{4}) \\
(\N_a^{4},\ov{\N}_b^{1}) & 0 & (\N_a^{4},\ov{\N}_b^{3}) & 0
\end{array}\right)
\\\hline
(*,*,+) & \left(\begin{array}{ccccc}
(\N_a^{1},\ov{\N}_b^{1}) & 0 & 0 & (\N_a^{1},\ov{\N}_b^{4}) \\
0 & (\N_a^{2},\ov{\N}_b^{2}) & (\N_a^{2},\ov{\N}_b^{3}) & 0 \\
0 & (\N_a^{3},\ov{\N}_b^{2})  & (\N_a^{3},\ov{\N}_b^{3})  & 0\\
(\N_a^{4},\ov{\N}_b^{1}) & 0& 0 & (\N_a^{4},\ov{\N}_b^{4}) 
\end{array}\right)
&
(*,*,-) &  \left(\begin{array}{ccccc}
0 & (\N_a^{1},\ov{\N}_b^{2}) & (\N_a^{1},\ov{\N}_b^{3}) & 0 \\
(\N_a^{2},\ov{\N}_b^{1}) & 0 & 0 & (\N_a^{2},\ov{\N}_b^{4}) \\
(\N_a^{3},\ov{\N}_b^{1}) & 0 & 0 & (\N_a^{3},\ov{\N}_b^{4}) \\
0 & (\N_a^{4},\ov{\N}_b^{2}) & (\N_a^{4},\ov{\N}_b^{3})  & 0
\end{array}\right)
\\\hline
\end{array}
\end{equation*}
}
If some $\lambda$ is subject to two $\Z_2^{(i)}$ projections, the third one is automatically fulfilled.

The open string states can be computed analogously to the right-moving sector of the closed string states.
This leads to the massless NS and R states with given $\Z_2^{(i)}$ eigenvalues in table~\ref{Tab:Open-String-States-CP}.
\begin{table}[h!]
\renewcommand{\arraystretch}{1.3}
  \begin{center}
{\footnotesize
\begin{equation*}
\begin{array}{|c|c|c|c|c|}\hline
\multicolumn{5}{|c|}{\text{\bf Massless open string states}}
\\\hline\hline
\frac{\text{angle}}{\pi} & \text{NS-sector} & \text{R-sector} & (c^{\Z_2^{(1)}},c^{\Z_2^{(2)}},c^{\Z_2^{(3)}}) & \text{representation}
\\\hline\hline
(0,0,0) & \psi^{\mu}_{-1/2}|0\rangle_{\rm NS} & |0 \rangle_{\rm R}, \psi^{\mu}_0 \prod_{i=1}^3 \psi^i_0 |0 \rangle_{\rm R} & (+,+,+) & \prod_{k=1}^4 U(N^k)
\\
& \psi^{1}_{-1/2}|0\rangle_{\rm NS}, \psi^{\ov{1}}_{-1/2}|0\rangle_{\rm NS}& \psi^2_0\psi^3_0 |0 \rangle_{\rm R},\psi^{\mu}_0\psi^1_0 |0 \rangle_{\rm R}
 & (+,-,-) & (\N^1,\ov{\N}^2) + (\N^3,\ov{\N}^4) + c.c.
\\
& \psi^{2}_{-1/2}|0\rangle_{\rm NS},\psi^{\ov{2}}_{-1/2}|0\rangle_{\rm NS} & \psi^1_0\psi^3_0 |0 \rangle_{\rm R},\psi^{\mu}_0\psi^2_0 |0 \rangle_{\rm R}& (-,+,-) & (\N^1,\ov{\N}^3) + (\N^2,\ov{\N}^4) + c.c.
\\
& \psi^{3}_{-1/2}|0\rangle_{\rm NS},\psi^{\ov{3}}_{-1/2}|0\rangle_{\rm NS} &\psi^1_0\psi^2_0 |0 \rangle_{\rm R},\psi^{\mu}_0\psi^3_0 |0 \rangle_{\rm R} & (-,-,+) & (\N^1,\ov{\N}^4) + (\N^2,\ov{\N}^3) + c.c.
\\\hline
(0,\phi,-\phi) & 
\psi^3_{-1/2+\phi} |0 \rangle_{\rm NS}^{({\rm tw})} & |\tilde{0}\rangle_{\rm R}^{({\rm tw}, 1)}
& (-,-,+) & \!\!\left\{\!\!\begin{array}{c} (\N^{1},\ov{\N}^{4}) +  (\N^{2},\ov{\N}^{3}) \\+ (\N^{3},\ov{\N}^{2}) + (\N^{4},\ov{\N}^{1}) \end{array}\right.
\\
& \psi^{\ov{2}}_{-1/2+\phi} |0 \rangle_{\rm NS}^{({\rm tw})} & \psi^{\mu}_0 \psi^1_0|\tilde{0}\rangle_{\rm R}^{({\rm tw}, 1)} 
& (-,+,-) &  \!\!\left\{\!\! \begin{array}{c}(\ov{\N}^{1},\N^{3}) + (\ov{\N}^{2},\N^{4})\\ + (\ov{\N}^{3},\N^{1}) + (\ov{\N}^{4},\N^{2}) \end{array}\right.
\\\hline
\!\!\!\!
\begin{array}{c} (\phi_1,\phi_2,\phi_3)_{\sum_i \phi_i=0} \\ \phi_2,\phi_3 > 0, \phi_1 < 0 \\
|\phi_1 | > |\phi_2|,|\phi_3| \end{array}\!\!\!\!
& \psi^1_{-1/2+\phi_1} |0 \rangle_{\rm NS}^{({\rm tw})} &  |\tilde{0}\rangle_{\rm R}^{({\rm tw})}
& (+,-,-) & (\N^1,\ov{\N}^2) + (\N^3,\ov{\N}^4)
\\\hline
\end{array}
\end{equation*}
}
\end{center}
\caption{Counting of massless open string states, their $\Z_2^{(i)}$ eigenvalues on $T^6/\Z_2 \times \Z_{2M}$ with discrete torsion and representations computed from Chan-Paton labels. The multiplets from D6-branes at angles have chiralities depending on the values of the angles which in turn determine
the massless R-sector states.
The representations in the last column correspond to full ${\cal N}=1$ multiplets, where for some non-vanishing angle, the explicitly listed NS and R states 
have been paired up with those from the inverse angles, e.g. $\pm (0, \phi, - \phi)$.
 }
\label{Tab:Open-String-States-CP}
\end{table}
In the absence of discrete Wilson lines, parallel D6-branes thus provide three non-chiral ${\cal N}=2$ hyper multiplets with different $\Z_2^{(i)}$ eigenvalues.
D6-branes at angle $\pi(0,\phi,-\phi)$ provide two chiral multiplets of opposite chirality and with different $\Z_2^{(2)}$ and  $\Z_2^{(3)}$  eigenvalues.
If the D6-brane intersection point does not coincide with a $\Z_2^{(1)}$ fixed point, a non-chiral ${\cal N}=2$ hyper multiplet arises, which is identified
 with another hyper multiplet at the $\Z_2^{(1)}$ image of the intersection point.

This means that not even for the so far existing literature on $T^6/\Z_2 \times \Z_2$~\cite{Blumenhagen:2005tn,Blumenhagen:2007ip}, 
the open string spectrum is fully determined by intersection numbers! The running of gauge couplings can only be computed if these non-chiral states
are included in the beta function coefficient.

\subsection{The gauge thresholds and beta functions}\label{sec:thres}

The complete massless charged matter content can be determined if the corresponding beta function coefficient is known.
The beta function coefficient arises in the computation of threshold corrections to the gauge kinetic function as follows:
these threshold corrections can be derived by using a gauged partition function to describe the Annulus and M\"obius strip amplitudes and expanding in powers of the newly introduced non-compact magnetic field,
see e.g.~\cite{Blumenhagen:2006ci} and references therein. 
As a result, one obtains a sum of three terms:
(1) a tadpole, which cancels among all possible contributions from various D-branes in a RR tadpole free configuration;
(2) a term proportional to $1/\varepsilon$, where $\varepsilon$ is the power of the dimensional regularisation and $1/\varepsilon$ is identified with $\ln \frac{M_{\rm string}^2}{\mu^2}$, whose numerical prefactor is the beta function contribution from the D6-brane and O6-plane configuration under consideration;
(3) a finite term, the actual gauge threshold due to massive strings. 

We can thus use the extensive inspection of gauge thresholds on $T^6/\Z_{2N}$ performed in~\cite{Gmeiner:2009fb} to find beta function
coefficients on $T^6/\Z_2 \times \Z_{2M}$ without and with discrete torsion and compare with the field theoretic expression for
the beta function coefficients
\begin{equation}\label{Eq:beta-fctn}
\begin{aligned}
b_{SU(N_a)} =&  -3 \, N_a  +\sum_{b\neq a}
  \frac{N_b}{2} \left( \varphi^{ab} + \varphi^{ab'}\right)  + N_a \,
  \varphi^{\Adj_a} 
+ \frac{N_a}{2} \, \left(\varphi^{\Sym_a} +  \varphi^{\Anti_a}
\right)
\\ &+ \left(\varphi^{\Sym_a} -  \varphi^{\Anti_a}
\right) ,
\\
\left.\begin{array}{c}
b_{Sp(2M_x)} \\ b_{SO(2M_x)} 
\end{array}\!\!\right\}
=& -3 \, (M_x \pm 1) + \sum_{a \neq x} \frac{N_a}{2}
  \varphi^{ax} + M_x \, \left( \varphi^{\Sym_x} +  \varphi^{\Anti_x}
  \right) 
\\
& + \left( \varphi^{\Sym_x} -  \varphi^{\Anti_x} \right) .
\end{aligned}
\end{equation}

\begin{table}[h!]
\renewcommand{\arraystretch}{1.3}
  \begin{center}
\begin{equation*}
\begin{array}{|c|c|c|}\hline
\multicolumn{3}{|c|}{b_{SU(N_a)} \text{ \bf  for bifundamental and adjoints: } T^6/\Z_2 \times \Z_{2M} \text{ \bf  without and with discrete torsion}}
\\\hline\hline
\frac{\text{Angle}}{\pi} & b_{SU(N_a)}^{\text{no torsion}} & b_{SU(N_a)}^{\text{with torsion}} 
\\\hline\hline
(0,0,0) &  -
& - \,  N_b \, \left(\prod_{n=1}^3 \delta_{\sigma_n^{ab},0} \delta_{\tau_n^{ab},0}\right) \sum_{i=1}^3  (-1)^{\tau^{\Z_2^{(i)}}_{ab}} 
\\
(0,\phi,-\phi) &   N_b \,\delta_{\sigma_1^{ab},0} \, \delta_{\tau_1^{ab},0} \, |I_{ab}^{(2 \cdot 3)} | 
&  \frac{N_b}{4} \,\delta_{\sigma_1^{ab},0} \, \delta_{\tau_1^{ab},0} \, \left( |I_{ab}^{(2 \cdot 3)} | - I_{ab}^{\Z_2^{(1)},(2 \cdot 3)} \right)
\\
(\phi^{(1)},\phi^{(2)},\phi^{(3)})_{\sum_{n=1}^3 \phi^{(n)}=0}  & \frac{N_b}{2} \, | I_{ab} |
&\frac{N_b}{8} \, \left( | I_{ab} | + \sgn(I_{ab}) \, \sum_{i=1}^3 I_{ab}^{\Z_2^{(i)}}  \right)
\\\hline
\end{array}
\end{equation*}
\end{center}
\caption{Counting bifundamental and adjoint representations using their beta function coefficients. By comparison with~(\protect\ref{Eq:beta-fctn}),
in the case {\it without discrete torsion} one obtains three multiplets in the adjoint representation from the $aa$ sector and for 
$T^6/\Z_2 \times \Z_4$ and $T^6/\Z_2 \times \Z_6$ without torsion additionally $k \times |I_{a(\theta^k a)}^{(2 \cdot 3)} |$  adjoint multiplets,
where $k=1$ for the former and $k=2$ for the latter.
For $T^6/\Z_2 \times \Z_6'$ without discrete torsion, $|I_{a(\theta^k a)}|$ multiplets in the adjoint representation parameterise the freedom for brane
recombination of orbifold images.
In the presence of {\it discrete torsion}, as anticipated in table~\protect\ref{Tab:Open-String-States-CP}, the $aa$ sector only contains the 
vector of $U(N_a)$, and in the absence of relative discrete Wilson lines and displacements, parallel branes $ab$ with different $\Z_2^{(i)}$ eigenvalues 
support one non-chiral pair of multiplets in the bifundamental representation.  
The symbols $\sigma^{ab}_i$, $\tau^{ab}_i$, $\tau_{ab}^{{\mathbb
    Z}_2^{i}}$ are related to relative
quantities between brane $a$ and $b$: 
relative displacement, relative Wilson line along the
$i^{\mbox{\scriptsize th}}$
$T^2$ and the relative eigenvalue under ${\mathbb Z}_2^{(i)}$. (`Related
to' means that they, as always, appear as powers of $(-1)$.) 
\label{Tab:BetaFct-Bifund}}
\end{table}

\begin{table}[ht!]
\renewcommand{\arraystretch}{1.3}
  \begin{center}
{\small
\begin{equation*}\!\!\!\!\!\!\!\!\!\!\!\!\!\!\!\!
\begin{array}{|c|c|c|}\hline
\multicolumn{3}{|c|}{b_{SU(N_a)} \text{ \bf  for (anti)symmetrics: } T^6/\Z_2 \times \Z_{2M} \text{ \bf  without and with discrete torsion}}
\\\hline\hline
\frac{\text {Angle}}{\pi} & b_{SU(N_a)}^{\text{no torsion}}  & b_{SU(N_a)}^{\text{with torsion}} 
\\\hline\hline
\begin{array}{c} (0,0,0) \\ \pp \OR \end{array}  
&  \sum_{i=1}^3 \delta_{\sigma_i^{aa'},0} \delta_{\tau_i^{aa'},0}\,   \tilde{I}_a^{\OR\Z_2^{(i)},(j \cdot k)} 
&\begin{array}{c} 
-  \frac{N_a}{4}  \,  \sum_{i=1}^3  I_{aa'}^{\Z_2^{(i)},(j \cdot k)} \\  + \frac{1}{2} \sum_{i=1}^3 \eta_{\OR\Z_2^{(i)}}  \tilde{I}_a^{\OR\Z_2^{(i)},(j \cdot k)} 
\end{array} 
\\\hline
\begin{array}{c} (0,0,0) \\ \pp \OR \Z_2^{(i)} \end{array}  &  
\begin{array}{c} \delta_{\sigma_i^{aa'},0} \delta_{\tau_i^{aa'},0}\,   \tilde{I}_a^{\OR,(j \cdot k)} \\ 
+  \sum_{j \neq i}   \delta_{\sigma_k^{aa'},0} \delta_{\tau_k^{aa'},0}\, \tilde{I}_a^{\OR\Z_2^{(j)} ,(i \cdot j)} \end{array} 
 & \begin{array}{c} -  N_a  \,  \sum_{i=1}^3  (-1)^{\tau^{\Z_2^{(i)}}_{aa'}} \\
  + \frac{1}{2} \left( \eta_{\OR}  \,   \tilde{I}_a^{\OR,(j \cdot k)} 
+ \sum_{j \neq i} \eta_{\OR\Z_2^{(j)}}  \,   \tilde{I}_a^{\OR\Z_2^{(j)} ,(i \cdot j)}
\right)
\end{array} \!\!\!
\\\hline 
\!\!\!
\begin{array}{c}(0_i,\phi_j,\phi_k)_{\phi_k = -\phi_j \neq \pm \frac{1}{2}} \\ \pp \left( \OR + \OR\Z_2^{(i)} \right)\end{array}\!\!\!
&  \begin{array}{c}\delta_{\sigma_i^{aa'}}\delta_{\tau_i^{aa'},0} \, \Bigl\{  N_a \,   |I_{aa'}^{(j \cdot k)} |  \\
+     \tilde{I}_a^{\OR,(j \cdot k)} +   \tilde{I}_a^{\OR\Z_2^{(i)} ,(j \cdot k)}  \Bigr\}
\end{array}
 &  \begin{array}{c} \frac{N_a}{4} \, \left( |I_{aa'}^{(j \cdot k)} |  
- I_{aa'}^{\Z_2^{(i)},(j \cdot k)} \right)\\
+\frac{1}{2}  \,   \left(\eta_{\OR}  \, \tilde{I}_a^{\OR,(j \cdot k)} +  \eta_{\OR\Z_2^{(i)}} \, \tilde{I}_a^{\OR\Z_2^{(i)} ,(j \cdot k)}
\right)  
\end{array}
\\\hline
\!\!\!\!\!
\begin{array}{c}(0_i,\phi_j,\phi_k)_{\phi_k = -\phi_j \neq \pm \frac{1}{2}} \\ \pp \left( \OR\Z_2^{(j)} + \OR\Z_2^{(k)} \right)\end{array}\!\!\!\!\!
& \begin{array}{c} \delta_{\sigma_i^{aa'},0}\delta_{\tau_i^{aa'},0}\, \Bigl\{ N_a \, |I_{aa'}^{(j \cdot k)} |  \\ 
 +    \tilde{I}_a^{\OR\Z_2^{(j)} ,(j \cdot k)}   +  \tilde{I}_a^{\OR\Z_2^{(k)} ,(j \cdot k)}      \Bigr\} 
\end{array}
&  \begin{array}{c}  \frac{N_a}{4} \, \left( |I_{aa'}^{(j \cdot k)} |  - I_{aa'}^{\Z_2^{(i)},(j \cdot k)} \right)\\
 + \frac{1}{2} \,\left(   \eta_{\OR\Z_2^{(j)}}\, \tilde{I}_a^{\OR\Z_2^{(j)} ,(j \cdot k)}   +   \eta_{\OR\Z_2^{(k)}}\, \tilde{I}_a^{\OR\Z_2^{(k)} ,(j \cdot k)}     \right) 
\end{array}\!\!\!\!
\\\hline
\!\!\!\!\!
\begin{array}{c}
(\phi^{(1)},\phi^{(2)},\phi^{(3)}) \\ {\sum_{n=1}^3 \phi^{(n)}=0} 
\end{array}
& \begin{array}{c}  \frac{N_a}{2} \, | I_{aa'} | \\ 
+\frac{c^{\OR}_a1}{2}  \tilde{I}_a^{\OR} + \sum_{i=1}^3  \frac{c^{\OR\Z_2^{(i)}}_a}{2}  \tilde{I}_a^{\OR\Z_2^{(i)}}
\end{array}
&\!\!\!\!\! \begin{array}{c}\frac{N_a}{8} \, \left( | I_{aa'} |
+ \sgn(I_{aa'} ) \sum_{i=1}^3 I_{aa'}^{\Z_2^{(i)}}  \right)
\\
+ \frac{1}{4} \left( c^{\OR}_a\,  \eta_{\OR}  \, \tilde{I}_a^{\OR} 
+ \sum_{i=1}^3  c^{\OR\Z_2^{(i)}}_a \, \eta_{\OR\Z_2^{(i)}}\,  \tilde{I}_a^{\OR\Z_2^{(i)}} \right)
\end{array}\!\!\!\!\!\!
\\\hline
\end{array}
\end{equation*}
}
\end{center}
\caption{Counting symmetric and antisymmetric matter states. 
The constants \mbox{$c^{\OR\Z_2^{(i)}}_a=\sgn(\tilde{\phi}^{(k)})\left[2H(|\tilde{\phi}^{(k)}|-\frac{1}{2}) -1\right]\in \{-1,0,1\}$}
arise from M\"obius strip contributions when the D$6_a$-branes are at three non-vanishing angles with the $\OR\Z_2^{(i)}$ planes 
preserving the $aa'$ open string sector. $0<|\tilde{\phi}^{(k)}|<1$ is the largest absolute value of the three angles between the D$6_a$-brane
and the $\OR\Z_2^{(i)}$ plane, and $H(x)=0,\frac{1}{2},1$ for $x<0,x=0,x>0$, respectively, is the Heavyside step function.
Note that the Kronecker deltas $\delta_{\sigma_i^{aa'},0}\delta_{\tau_i^{aa'},0}$ can be dropped in the 
expressions with discrete torsion due to the discrete nature of the displacements and Wilson lines.
For the M\"obius strip amplitude on a tilted torus, however, we run into an inconsistency in the classification
of orientifold invariant D$6_c$-branes, unless we replace e.g. $\eta_{\OR\Z_2^{(i)}} \rightarrow 
(-1)^{2 b_i \sigma_i^c \tau_i^c}\eta_{\OR\Z_2^{(i)}}$ in the formula for the beta function coefficient of
a D$6_c$-brane parallel to the $\OR$ invariant plane. We expect that a thorough re-examination of 
the lattice contributions, which goes beyond the scope of the present work, will provide the  missing sign factor. 
 }
\label{Tab:BetaFct-AntiSym}
\end{table}

The advantage of this method of counting the massless spectrum is its completeness, whereas it is blind to chiralities. The derivation of the beta function coefficients can therefore not replace the computation of the chiral spectrum via three-cycle intersection numbers.

\subsubsection{Rigid D6-branes without adjoint matter}\label{App:Sss-no_adjoints}

The contributions to beta function coefficients from bifundamental and adjoint representations of $SU(N_a)$ are displayed in table~\ref{Tab:BetaFct-Bifund}.
Using the expressions~(\ref{Eq:Z4-1cycle-Orb}),~(\ref{Eq:Z6-1cycle-Orb}) and~(\ref{Eq:Z6p-1cycle-Orb}) for the torus wrapping numbers of orbifold images on $T^6/\Z_2 \times \Z_4$, $T^6/\Z_2 \times \Z_6$ and $T^6/\Z_2 \times \Z_6'$, respectively, one obtains
\begin{equation}\nonumber
\begin{array}{ll}
T^6/\Z_2 \times \Z_4: \quad & I_{a(\theta a)}^{(2 \cdot 3)} = - \prod_{i=2}^3 \left( (n_a^i)^2 + (m_a^i)^2 \right) ,
\\
T^6/\Z_2 \times \Z_6: \quad &  I_{a(\theta a)}^{(2 \cdot 3)} = I_{a(\theta^2 a)}^{(2 \cdot 3)} =- \prod_{i=2}^3 \left( (n_a^i)^2 + n_a^i m_a^i
+ (m_a^i)^2 \right) ,
\\
T^6/\Z_2 \times \Z_6': \quad &  I_{a(\theta a)} = - I_{a(\theta^2 a)} =- \prod_{i=1}^3 \left( (n_a^i)^2 + n_a^i m_a^i + (m_a^i)^2 \right)
.
\end{array}
\end{equation}
Completely rigid D6-branes with no adjoint representation arising at orbifold images is only possible 
if $|I_{(\theta a)}^{(2 \cdot 3)}| = I_{(\theta a)}^{\Z_2^{(1)},(2 \cdot 3)}$ for $T^6/\Z_2 \times \Z_6$
with discrete torsion, whereas it is impossible for $T^6/\Z_2 \times \Z_4$.
For $T^6/\Z_2 \times \Z_6$ one has $n^2 + nm + m^2 = \frac{3}{4}(n+m)^2 + \frac{1}{4}(n-m)^2$, which only has solutions to no adjoint representation for $(n,m) \in \{(1,0),(0,1),(1,-1)\}$, all with $n^2+nm+m^2=1$. 
The necessary conditions on the absence of adjoint representations needs to be supplemented by the condition that $I_{(\theta a)}^{\Z_2^{(1)},(2 \cdot 3)} >0$, which restricts the combinations of discrete Wilson lines and displacements on $T^2_{(2)} \times T^2_{(3)}$ due to the following relations
\begin{equation}\nonumber
\begin{array}{ccc}
i \in \{2,3\} & (n^i_a,m^i_a) & I_{(\theta a)}^{\Z_2^{(1)},(i)}
\\
\Z_2 \times \Z_6 & (1,0),(0,1),(1,-1) & (-1)^{\sigma_i^a \tau_i^a}
.
\end{array}
\end{equation}
This has been derived by inspection of the intersection points along the lines of appendix A.1 in~\cite{Gmeiner:2009fb}.
For $T^6/\Z_2 \times \Z_6$, supersymmetric solutions for $SU(N_a)$ gauge factors without any adjoint representation
are found by choosing an appropriate one-cycle on $T^2_{(1)}$.
For $T^6/\Z_2 \times \Z_6'$, the situation is different since the torus intersection number can be cancelled by the sum of all three $\Z_2^{(i)}$ invariant intersection 
numbers, see the last row in table~\ref{Tab:BetaFct-Bifund}. In the example in section~\ref{S:Z2Z6p-example-torsion}, the cancellation occured due to
$I^{\Z_2^{(1)}}_{a_m(\omega^k \, a_m)} = - I^{\Z_2^{(2)}}_{a_m(\omega^k \, a_m)}$ and $I^{\Z_2^{(3)}}_{a_m(\omega^k \, a_m)}= - I_{a_m(\omega^k \, a_m)}$,
which shows the existence of completely rigid D6-branes on this orbifold background.

The condition of completely rigid D6-branes severly resctrict the search for fully-fledged Standard Model vacua, which we will address in future work~\cite{Forste:201xxx}.

{\boldmath
\subsubsection{$Sp(2M)$ and $SO(2M)$ gauge factors}\label{App:Sss-SO+Sp}
}

The $\OR$ invariant three-cycles can be easily classified, as done in sections~\ref{Ss:Z2Z2Ktheory},
\ref{Ss:Z2Z4-Ktheory},~\ref{Ss:Z2Z6-Ktheory} and~\ref{Ss:Z2Z6p-Ktheory}
for the $T^6/\Z_2 \times \Z_{2M}$ backgrounds without and with discrete torsion for $2M=2,4,6$ and $6'$, respectively.
A priori, the gauge group which D6-branes wrapped on these cycles support, can be either of $SO(2N)$ or $Sp(2N)$ 
type.

One way of determining the correct group assignment consists of extracting the $\OR$ eigenvalue of the massless
open string state from the M\"obius strip amplitude and finding a viable $\OR$ projection matrix for the 
Chan-Paton label of a given D6-brane configuration.

However, the method of extracting the beta function coefficient from the gauge threshold computation turns out to be more economic. As discussed in detail in~\cite{Gmeiner:2009fb}, the expressions for $b_{SO(2N)}$ and $b_{Sp(2N)}$
in terms of intersection numbers are (up to a global factor of $\frac{1}{2}$) the same as for $b_{SU(N)}$ in 
tables~\ref{Tab:BetaFct-Bifund} and~\ref{Tab:BetaFct-AntiSym}.
From the first two rows of the latter, one can extract the type of gauge group supported by orientifold invariant 
D6-branes:
\begin{itemize}
\item
since $\tilde{I}_x^{\OR\Z_2^{(i)},(j \cdot k)} = -4$ for any of the orientifold invariant three-cycles $x$ 
on all $T^6/\Z_2 \times \Z_{2M}$ orbifolds and all lattice orientations discussed in this article, 
the $xx$ sector contribute $b_{SO/Sp(2N_x)}=-6$ in the case {\it without discrete torsion}, and therefore the
gauge group is $Sp(2N_x)$ with three chiral multiplets in the antisymmetric representation;
\item
on $T^6/\Z_2 \times \Z_4$ {\it with discrete torsion}, the formulas for $b_{SU(N)}^{\text{no torsion}}$ in 
table~\ref{Tab:BetaFct-AntiSym} dressed with the charges $\eta_{\OR}$ and $\eta_{\OR\Z_2^{(i)}}$, 
lead to  $SO(2N_x)$ gauge factors, if the stack of D6-branes is perpendicular to the orbit of exotic O6-planes, or
 $Sp(2N_x)$ otherwise. 
\item
for $T^6/\Z_2 \times \Z_{2M}$ {\it with discrete torsion} and $2M \in \{2,6,6'\}$, the gauge group is $Sp(2N_x)$.
For $b_i\sigma_i \tau_i=0$, one can verify explicitly that this is due to the fact that orientifold invariant 
fractional 
D6-branes are parallel to the exotic O6-plane, cf. table~\ref{Tab:Class-OR-inv-cycles}.
\end{itemize}
Gauge groups of $SO(2N_x)$ type  occur for $T^6/(\Z_2 \times \Z_{2M} \times \OR)$ with discrete torsion and $2M \in \{2,6,6'\}$ 
only in the presence of three exotic O6-planes, which 
is incompatible with supersymmetric D6-branes cancelling  the untwisted RR tadpoles.

In general, there are more antisymmetric or symmetric representations of $Sp(2N_x)$ or $SO(2N_x)$ supported at intersections 
$x(\omega^k x)$, which need to be computed on a case by case basis.

{\boldmath
\section{Tables for exceptional sectors in $T^6/\Z_2 \times \Z_{2M}$ orbifolds with discrete torsion}\label{App:Tables-Ex-Sectors}
}

In this appendix, we collect tables for the exceptional sectors of $T^6/\Z_2 \times \Z_{2M}$ with 
$2M \in \{2,6,6'\}$.

In table~\ref{tab:Z2fixed_Z2Z2}, the transformations of fixed points on $T^6/\Z_2 \times \Z_2$ for bulk parts 
parallel to some O6-plane are evaluated, which lead to the classification of orientifold invariant fractional
D6-branes in table~\ref{Tab:Class-OR-inv-cycles}. The latter also holds for $T^6/\Z_2 \times \Z_6$ and 
 $T^6/\Z_2 \times \Z_6'$.

In tables~\ref{Tab:Z2Z6-Z1sector}  to~\ref{Tab:Z2Z6-Z3sector}, the assignment of exceptional three-cycles on
$T^6/\Z_2 \times \Z_6$ for a given choice of torus wrapping numbers $(n^i,m^i)$, discrete displacements 
$\vec{\sigma}$ and Wilson lines $\tau$ is presented. In tables~\ref{Tab:Z2Z6p-Part1} and~\ref{Tab:Z2Z6p-Part2}, 
the same relations for $T^6/\Z_2 \times \Z_6'$ are given.

\mathsidetabfix{
\renewcommand{\arraystretch}{1.1}
\begin{array}{|cc|c|c|c|c|c|}\hline
\multicolumn{7}{|c|}{\OR \; \text{\bf images of $\Z_2^{(k)}$ fixed points on $T^6/\Z_2 \times \Z_2$ and classification of $\OR$  invariant three-cycles}}
\\\hline\hline
T_i & T_j & T_k  & \multicolumn{4}{|c|}{\Z_2^{(k)} \text{ fixed points on } T_i \times T_j: (x_1y_1|x_2y_1|x_1y_2|x_2y_2)}
\\
\multicolumn{3}{|c|}{(\sigma_i,\sigma_j)}   & (0,0) & (1,0) & (0,1) & (1,1)
\\\hline\hline
\rightarrow & \rightarrow & \rightarrow   
& \!\!\!(1,1|\frac{2}{1-b_i},1|1,\frac{2}{1-b_j}|\frac{2}{1-b_i},\frac{2}{1-b_j})\!\!\!
& \!\!\!(4(1-b_i),1|3,1|4(1-b_i),\frac{2}{1-b_j}|3,\frac{2}{1-b_j} )\!\!\!
& \!\!\!(1,4(1-b_j) |\frac{2}{1-b_i}, 4(1-b_j)|1, 3|\frac{2}{1-b_i}, 3)\!\!\!
& \!\!\!(4(1-b_i),4(1-b_j)|3,4(1-b_j)|4(1-b_i),3|3,3 )\!\!\!
\\
\multicolumn{3}{|c|}{\OR} & (1,1|\frac{2}{1-b_i},1|1,\frac{2}{1-b_j}|\frac{2}{1-b_i},\frac{2}{1-b_j})\!\!\!
& \!\!\!(4-2b_i,1|2+2b_i,1|4-2b_i,\frac{2}{1-b_j}|3-2b_i,\frac{2}{1-b_j})\!\!\!
& \!\!\!(1,4-2b_j|\frac{2}{1-b_i},3-2b_j|1,4-2b_j|\frac{2}{1-b_i},3-2b_j)
& \!\!\!(4-2b_i,4-2b_j|3-2b_i,4-2b_j|4-2b_i,3-2b_j|3-2b_i,3-2b_j)
\\\hline
\multicolumn{3}{|c|}{\bf aa} & \multicolumn{4}{|c|}{\text{all sets pointwise invariant}}
\\
\multicolumn{3}{|c|}{\text{3-cycle invariant for}} &  \multicolumn{4}{|c|}{\eta_{(k)} = -1}
\\\hline
\multicolumn{3}{|c|}{\bf ab} & \multicolumn{2}{|c|}{\text{sets pointwise invariant}} &  \multicolumn{2}{|c|}{\text{permututation within sets } (x_1y_2|x_2y_2|x_1y_1|x_2y_1)} 
\\
\multicolumn{3}{|c|}{\text{3-cycle invariant for}} &  \multicolumn{2}{|c|}{\eta_{(k)} = -1} &   \multicolumn{2}{|c|}{\eta_{(k)} = -(-1)^{\tau_j} }
\\\hline
\multicolumn{3}{|c|}{\bf ba} &  \text{pointwise invariant} & (x_2y_1|x_1y_1|x_2y_2|x_1y_2)    &   \text{pointwise invariant} & (x_2y_1|x_1y_1|x_2y_2|x_1y_2)
\\
\multicolumn{3}{|c|}{\text{3-cycle invariant for}} &  \eta_{(k)} = -1 &   \eta_{(k)} = -(-1)^{\tau_i}   &  \eta_{(k)} = -1 & \eta_{(k)} = -(-1)^{\tau_i} 
\\\hline
\multicolumn{3}{|c|}{\bf bb} &  \text{pointwise invariant} &  (x_2y_1|x_1y_1|x_2y_2|x_1y_2)  & (x_1y_2|x_2y_2|x_1y_1|x_2y_1) & (x_2y_2|x_1y_2|x_2y_1|x_1y_1)
\\
\multicolumn{3}{|c|}{\text{3-cycle invariant for}} &  \eta_{(k)} = -1 & \eta_{(k)} = -(-1)^{\tau_i} & \eta_{(k)} = -(-1)^{\tau_j} & \eta_{(k)} = -(-1)^{\tau_i+ \tau_j}
\\\hline\hline
\multicolumn{3}{|c|}{\text{3-cycle invariant for}} &    \multicolumn{4}{|c|}{\eta_{(k)} = -(-1)^{2(b_i\sigma_i\tau_i+b_j\sigma_j\tau_j)} }
\\\hline\hline\hline
\uparrow & \downarrow & \rightarrow 
& (1,1|4,1|1,4|4,4) 
& (2,1|3,1|2,4|3,4)
& (1,2|4,2|1,3|4,3)
& (2,2|3,2|2,3|3,3)
\\
\multicolumn{3}{|c|}{\OR} & (1,1|4,1|1,4|4,4) 
& (2+2b_i,1|3-2b_i,1|2+2b_i,4|3-2b_i,4)
& (1,2+2b_j|4,2+2b_j|1,3-2b_j|4,3-2b_j)
& (2+2b_i,2+2b_j|3-2b_i,2+2b_j|2+2b_i,3-2b_j|3-2b_i,3-2b_j)
\\\hline\hline
\multicolumn{3}{|c|}{\bf aa} & \multicolumn{4}{|c|}{\text{all sets pointwise invariant}}
\\
\multicolumn{3}{|c|}{\text{3-cycle invariant for}} &  \multicolumn{4}{|c|}{\eta_{(k)} = -1}
\\\hline
\multicolumn{3}{|c|}{\bf ab} & \multicolumn{2}{|c|}{\text{sets pointwise invariant}} &  \multicolumn{2}{|c|}{\text{permututation within sets }  (x_1y_2|x_2y_2|x_1y_1|x_2y_1)}  
\\
\multicolumn{3}{|c|}{\text{3-cycle invariant for}} &   \multicolumn{2}{|c|}{\eta_{(k)} = -1}  &   \multicolumn{2}{|c|}{\eta_{(k)} = -(-1)^{\tau_j} }
\\\hline
\multicolumn{3}{|c|}{\bf ba} &  \text{pointwise invariant} &  (x_2y_1|x_1y_1|x_2y_2|x_1y_2) &   \text{pointwise invariant} & (x_2y_1|x_1y_1|x_2y_2|x_1y_2)  
\\
\multicolumn{3}{|c|}{\text{3-cycle invariant for}} & \eta_{(k)} = -1 & \eta_{(k)} = -(-1)^{\tau_i} & \eta_{(k)} = -1 & \eta_{(k)} = -(-1)^{\tau_i}
\\\hline
\multicolumn{3}{|c|}{\bf bb} &  \text{pointwise invariant} &  (x_2y_1|x_1y_1|x_2y_2|x_1y_2) &(x_1y_2|x_2y_2|x_1y_1|x_2y_1) & (x_2y_2|x_1y_2|x_2y_1|x_1y_1)
\\
\multicolumn{3}{|c|}{\text{3-cycle invariant for}} &  \eta_{(k)} = -1 & \eta_{(k)} = -(-1)^{\tau_i} & \eta_{(k)} = -(-1)^{\tau_j} &  \eta_{(k)} = -(-1)^{\tau_i+ \tau_j}
\\\hline\hline
\multicolumn{3}{|c|}{\text{3-cycle invariant for}} &    \multicolumn{4}{|c|}{\eta_{(k)} = -(-1)^{2(b_i\sigma_i\tau_i+b_j\sigma_j\tau_j)}
}
\\\hline\hline\hline
\uparrow & \rightarrow & \downarrow 
& (1,1|4,1|1,\frac{2}{1-b_j}|4,\frac{2}{1-b_j})
& (2,1|3,1|2,\frac{2}{1-b_j}|3,\frac{2}{1-b_j})
& (1, 4(1-b_j) |4, 4(1-b_j) |1,3|4,3)
& (2, 4(1-b_j) |3,4(1-b_j)  |2,3|3,3)
\\
\multicolumn{3}{|c|}{\OR} & (1,1|4,1|1,\frac{2}{1-b_j}|4,\frac{2}{1-b_j})
& (2+2b_i,1|3-2b_i,1|2+2b_i,\frac{2}{1-b_j}|3-2b_i,\frac{2}{1-b_j})
& (1,4-2b_j|4,4-2b_j|1,3-2b_j|4,3-2b_j)
& (2+2b_i,4-2b_j|3-2b_i,4-2b_j|2+2b_i,3-2b_j|3-2b_i,3-2b_j)
\\\hline\hline
\multicolumn{3}{|c|}{\bf aa} & \multicolumn{4}{|c|}{\text{all sets pointwise invariant}}
\\
\multicolumn{3}{|c|}{\text{3-cycle invariant for}} &  \multicolumn{4}{|c|}{\eta_{(k)} = 1}
\\\hline
\multicolumn{3}{|c|}{\bf ab} & \multicolumn{2}{|c|}{\text{sets pointwise invariant}} &   \multicolumn{2}{|c|}{\text{permututation within sets }  (x_1y_2|x_2y_2|x_1y_1|x_2y_1)  }
\\
\multicolumn{3}{|c|}{\text{3-cycle invariant for}} &    \multicolumn{2}{|c|}{\eta_{(k)} = 1} & \multicolumn{2}{|c|}{\eta_{(k)} = (-1)^{\tau_j}}
\\\hline
\multicolumn{3}{|c|}{\bf ba} &  \text{pointwise invariant} &   (x_2y_1|x_1y_1|x_2y_2|x_1y_2)  &   \text{pointwise invariant} &  (x_2y_1|x_1y_1|x_2y_2|x_1y_2)
\\
\multicolumn{3}{|c|}{\text{3-cycle invariant for}} &  \eta_{(k)} = 1 & \eta_{(k)} = (-1)^{\tau_i}   & \eta_{(k)} = 1 & \eta_{(k)} = (-1)^{\tau_i} 
\\\hline
\multicolumn{3}{|c|}{\bf bb} &  \text{pointwise invariant} &  (x_2y_1|x_1y_1|x_2y_2|x_1y_2) &  (x_1y_2|x_2y_2|x_1y_1|x_2y_1) & (x_2y_2|x_1y_2|x_2y_1|x_1y_1)
\\
\multicolumn{3}{|c|}{\text{3-cycle invariant for}} & \eta_{(k)} = 1 & \eta_{(k)} = (-1)^{\tau_i}   & \eta_{(k)} = (-1)^{\tau_j} & \eta_{(k)} = (-1)^{\tau_i+ \tau_j}
\\\hline\hline
\multicolumn{3}{|c|}{\text{3-cycle invariant for}} &    \multicolumn{4}{|c|}{\eta_{(k)} = (-1)^{2(b_i\sigma_i\tau_i+b_j\sigma_j\tau_j)}
}
\\\hline\hline\hline
\rightarrow & \uparrow & \downarrow 
& (1,1|\frac{2}{1-b_i},1|1,4|\frac{2}{1-b_i},4)
& (4(1-b_i),1|3,1|4(1-b_i),4|3,4)
& (1,2|\frac{2}{1-b_i},2|1,3|\frac{2}{1-b_i},3)
& (4(1-b_i),2|3,2|4(1-b_i),3|3,3)
\\
\multicolumn{3}{|c|}{\OR} & (1,1|\frac{2}{1-b_i},1|1,4|\frac{2}{1-b_i},4)
& (4-2b_i,1|3-2b_i,1|4-2b_i,4|3-2b_i,4)
& (1,2+2b_i|\frac{2}{1-b_i},2+2b_i|1,3-2b_i|\frac{2}{1-b_i},3-2b_i)
& (4-2b_i,2+2b_j|3-2b_i,2+2b_j|4-2b_i,3-2b_j|3-2b_i,3-2b_j)
\\\hline\hline
\multicolumn{3}{|c|}{\bf aa} & \multicolumn{4}{|c|}{\text{all sets pointwise invariant}}
\\
\multicolumn{3}{|c|}{\text{3-cycle invariant for}} &  \multicolumn{4}{|c|}{\eta_{(k)} = 1}
\\\hline
\multicolumn{3}{|c|}{\bf ab} & \multicolumn{2}{|c|}{\text{sets pointwise invariant}} &   \multicolumn{2}{|c|}{\text{permututation within sets }   (x_1y_2|x_2y_2|x_1y_1|x_2y_1)  }
\\
\multicolumn{3}{|c|}{\text{3-cycle invariant for}} &    \multicolumn{2}{|c|}{\eta_{(k)} = 1} & \multicolumn{2}{|c|}{\eta_{(k)} =(-1)^{\tau_j} } 
\\\hline
\multicolumn{3}{|c|}{\bf ba} &  \text{pointwise invariant} &  (x_2y_1|x_1y_1|x_2y_2|x_1y_2)  &   \text{pointwise invariant} &  (x_2y_1|x_1y_1|x_2y_2|x_1y_2)
\\
\multicolumn{3}{|c|}{\text{3-cycle invariant for}} &  \eta_{(k)} = 1 & \eta_{(k)} = (-1)^{\tau_i}   & \eta_{(k)} = 1 & \eta_{(k)} = (-1)^{\tau_i} 
\\\hline
\multicolumn{3}{|c|}{\bf bb} &  \text{pointwise invariant} & (x_2y_1|x_1y_1|x_2y_2|x_1y_2)  &  (x_1y_2|x_2y_2|x_1y_1|x_2y_1) & (x_2y_2|x_1y_2|x_2y_1|x_1y_1)
\\
\multicolumn{3}{|c|}{\text{3-cycle invariant for}} &  \eta_{(k)} = 1 & \eta_{(k)} = (-1)^{\tau_i}   & \eta_{(k)} = (-1)^{\tau_j} &  \eta_{(k)} = (-1)^{\tau_i+ \tau_j}
\\\hline\hline
\multicolumn{3}{|c|}{\text{3-cycle invariant for}} &    \multicolumn{4}{|c|}{\eta_{(k)} = (-1)^{2(b_i\sigma_i\tau_i+b_j\sigma_j\tau_j)}
}
\\\hline
\end{array} 
}{Z2fixed_Z2Z2}{Transformation of $T^6/\Z_2 \times \Z_2$ fixed points in the $\Z_2^{(k)}$ twisted sector and conditions for $\OR$ invariance of the three-cycle
in dependence of the choice of the exotic O6-plane.}

\begin{table}[h!]
\renewcommand{\arraystretch}{1.3}
  \begin{center}
{\tiny
\begin{equation*}
\begin{array}{|c||c|c|c|c|}\hline
\multicolumn{5}{|c|}{\text{Existence of $Sp(2M)$ or $SO(2M)$ gauge factors on the $T^6/\Z_2 \times \Z_{2M}$ orientifolds with discrete torsion}}
\\\hline\hline
\!\!\!\begin{array}{c} \text{3-cycle} \pp  \\ \text{to O6-plane} \end{array}\!\!\!  & \multicolumn{4}{|c|}{(\eta_{(1)},\eta_{(2)},\eta_{(3)})}
\\\hline
 (\sigma_1,\sigma_2,\sigma_3)  & (-1,-1,-1) & (-1,1,1) & (1,-1,1) & (1,1,-1)
\\\hline\hline
\OR   & b_1\sigma_1\tau_1=b_2\sigma_2\tau_2=b_3\sigma_3\tau_3  
& \begin{array}{c}
 b_1\sigma_1\tau_1\neq \\ b_2\sigma_2\tau_2=b_3\sigma_3\tau_3  
\end{array} 
& \begin{array}{c}
 b_2\sigma_2\tau_2\neq \\ b_1\sigma_1\tau_1=b_3\sigma_3\tau_3  
\end{array} 
& \begin{array}{c}
 b_3\sigma_3\tau_3\neq \\ b_1\sigma_1\tau_1=b_2\sigma_2\tau_2  
\end{array} 
\\\hline
 (0,0,0) & \text{any choice of } b_i\tau_i & \times & \times & \times 
\\
 (1,0,0) & b_1\tau_1=0 &  2b_1\tau_1=1 & \times & \times 
\\
 (0,1,0) &  b_2\tau_2=0 & \times & 2b_2\tau_2=1 & \times 
\\
 (0,0,1) &  b_3\tau_3=0 & \times & \times & 2b_3\tau_3=1
\\
 (1,1,0) &  b_1\tau_1= b_2\tau_2=0 & 2b_1\tau_1=1; b_2\tau_2=0 & b_1\tau_1=0;2b_2\tau_2=1 &  2b_1\tau_1=2b_2\tau_2=1
\\
 (1,0,1) &   b_1\tau_1= b_3\tau_3=0 &  2b_1\tau_1=1; b_3\tau_3=0 & 2b_1\tau_1= 2b_3\tau_3=1 & b_1\tau_1=0; 2b_3\tau_3=1
\\
 (0,1,1) &  b_2\tau_2= b_3\tau_3=0 &  2b_2\tau_2= 2b_3\tau_3=1 &  2b_2\tau_2=1;  b_3\tau_3=0 &  b_2\tau_2=0;  2b_3\tau_3=1  
\\
 (1,1,1) &   b_1\tau_1= b_2\tau_2= b_3\tau_3 &  b_1\tau_1\neq b_2\tau_2= b_3\tau_3 & b_1\tau_1= b_3\tau_3\neq b_2\tau_2 & b_1\tau_1= b_2\tau_2 \neq b_3\tau_3 
\\\hline\hline
\OR\Z_2^{(1)}   
& \begin{array}{c}
 b_1\sigma_1\tau_1\neq \\ b_2\sigma_2\tau_2=b_3\sigma_3\tau_3  
\end{array} 
&  b_1\sigma_1\tau_1=b_2\sigma_2\tau_2=b_3\sigma_3\tau_3  
& \begin{array}{c}
 b_3\sigma_3\tau_3\neq \\ b_1\sigma_1\tau_1=b_2\sigma_2\tau_2  
\end{array} 
& \begin{array}{c}
 b_2\sigma_2\tau_2\neq \\ b_1\sigma_1\tau_1=b_3\sigma_3\tau_3  
\end{array} 
 \\\hline
(0,0,0) &  \times &\text{any choice of } b_i\tau_i & \times & \times 
\\
(1,0,0) &  2b_1\tau_1=1 &  b_1\tau_1=0 & \times & \times 
\\
(0,1,0)  & \times &  b_2\tau_2=0 & \times & 2b_2\tau_2=1
\\
(0,0,1) & \times &  b_3\tau_3=0  & 2b_3\tau_3=1 & \times
\\
(1,1,0) & 2b_1\tau_1=1; b_2\tau_2=0 & b_1\tau_1= b_2\tau_2=0 & 2b_1\tau_1=2b_2\tau_2=1 & b_1\tau_1=0;2b_2\tau_2=1
\\
(1,0,1) &  2b_1\tau_1=1; b_3\tau_3=0&   b_1\tau_1= b_3\tau_3=0 & b_1\tau_1=0; 2b_3\tau_3=1 & 2b_1\tau_1= 2b_3\tau_3=1 
\\
(0,1,1)  &  2b_2\tau_2= 2b_3\tau_3=1 &  b_2\tau_2= b_3\tau_3=0 &  b_2\tau_2=0;  2b_3\tau_3=1  &  2b_2\tau_2=1;  b_3\tau_3=0 
\\
(1,1,1) &  b_1\tau_1\neq b_2\tau_2= b_3\tau_3 & b_1\tau_1= b_2\tau_2= b_3\tau_3 & b_1\tau_1= b_2\tau_2 \neq b_3\tau_3 & b_1\tau_1= b_3\tau_3\neq b_2\tau_2 
\\\hline\hline
\OR\Z_2^{(2)}   
& \begin{array}{c}
 b_2\sigma_2\tau_2\neq \\ b_1\sigma_1\tau_1=b_3\sigma_3\tau_3  
\end{array} 
&  \begin{array}{c}
 b_3\sigma_3\tau_3\neq \\ b_1\sigma_1\tau_1=b_2\sigma_2\tau_2  
\end{array} 
&  b_1\sigma_1\tau_1=b_2\sigma_2\tau_2=b_3\sigma_3\tau_3  
& \begin{array}{c}
 b_1\sigma_1\tau_1\neq \\ b_2\sigma_2\tau_2=b_3\sigma_3\tau_3  
\end{array} 
\\\hline
(0,0,0) & \times & \times &\text{any choice of } b_i\tau_i & \times 
\\
(1,0,0)  & \times & \times &  b_1\tau_1=0 &  2b_1\tau_1=1
\\
(0,1,0) & 2b_2\tau_2=1 & \times &  b_2\tau_2=0 & \times 
\\
(0,0,1)  & \times & 2b_3\tau_3=1 &  b_3\tau_3=0 & \times
\\
(1,1,0) & b_1\tau_1=0;2b_2\tau_2=1 &  2b_1\tau_1=2b_2\tau_2=1 &  b_1\tau_1= b_2\tau_2=0 & 2b_1\tau_1=1; b_2\tau_2=0 
\\
(1,0,1)  & 2b_1\tau_1= 2b_3\tau_3=1 & b_1\tau_1=0; 2b_3\tau_3=1 &  b_1\tau_1= b_3\tau_3=0 &  2b_1\tau_1=1; b_3\tau_3=0
\\
(0,1,1)  &  2b_2\tau_2=1;  b_3\tau_3=0 &  b_2\tau_2=0;  2b_3\tau_3=1  & b_2\tau_2= b_3\tau_3=0 &  2b_2\tau_2= 2b_3\tau_3=1
\\
(1,1,1) & b_1\tau_1= b_3\tau_3\neq b_2\tau_2 & b_1\tau_1= b_2\tau_2 \neq b_3\tau_3 &  b_1\tau_1= b_2\tau_2= b_3\tau_3 &  b_1\tau_1\neq b_2\tau_2= b_3\tau_3 
\\\hline\hline
\OR\Z_2^{(3)}   
& \begin{array}{c}
 b_3\sigma_3\tau_3\neq \\ b_1\sigma_1\tau_1=b_2\sigma_2\tau_2  
\end{array} 
&  \begin{array}{c}
 b_2\sigma_2\tau_2\neq \\ b_1\sigma_1\tau_1=b_3\sigma_3\tau_3  
\end{array} 
& \begin{array}{c}
 b_1\sigma_1\tau_1\neq \\ b_2\sigma_2\tau_2=b_3\sigma_3\tau_3  
\end{array} 
&  b_1\sigma_1\tau_1=b_2\sigma_2\tau_2=b_3\sigma_3\tau_3  
\\\hline
(0,0,0)  & \times & \times & \times &\text{any choice of } b_i\tau_i
\\
(1,0,0)  & \times & \times &  2b_1\tau_1=1 &  b_1\tau_1=0 
\\
(0,1,0) & \times & 2b_2\tau_2=1  & \times &  b_2\tau_2=0 
\\
(0,0,1) &  2b_3\tau_3=1 & \times & \times & b_3\tau_3=0
\\
(1,1,0)  & 2b_1\tau_1=2b_2\tau_2=1 & b_1\tau_1=0;2b_2\tau_2=1&  2b_1\tau_1=1; b_2\tau_2=0 & b_1\tau_1= b_2\tau_2=0
\\
(1,0,1) & b_1\tau_1=0; 2b_3\tau_3=1 & 2b_1\tau_1= 2b_3\tau_3=1 &  2b_1\tau_1=1; b_3\tau_3=0&   b_1\tau_1= b_3\tau_3=0 
\\
(0,1,1)  &  b_2\tau_2=0;  2b_3\tau_3=1  &  2b_2\tau_2=1;  b_3\tau_3=0 &  2b_2\tau_2= 2b_3\tau_3=1 &  b_2\tau_2= b_3\tau_3=0
\\
(1,1,1)  & b_1\tau_1= b_2\tau_2 \neq b_3\tau_3 & b_1\tau_1= b_3\tau_3\neq b_2\tau_2 &  b_1\tau_1\neq b_2\tau_2= b_3\tau_3 & b_1\tau_1= b_2\tau_2= b_3\tau_3
\\\hline
\end{array}
\end{equation*}
}
\end{center}
\caption{Existence of $\OR$ invariant three-cycles on $T^6/\Z_2 \times \Z_{2M}$ orientifolds with discrete torsion. For $T^6/\Z_2 \times \Z_2$,
all choices of $b_i$ correspond to untilted or tilted tori. For $T^6/\Z_2 \times \Z_6$, the second and third two torus are tilted, $b_2=b_3=1/2$, 
but the shape of the first two torus remains a free parameter, $b_1 \equiv b$. For  $T^6/\Z_2 \times \Z_6'$, all tori are tilted, i.e. 
$b_1=b_2=b_3=1/2$.}
\label{Tab:Class-OR-inv-cycles}
\end{table}

\begin{sidewaystable}%
\parbox{\textheight}{%
{\footnotesize
\begin{equation*}\!\!\!\!\!\!\!\!\!\!\!\!\!\!\!\!\!\!\!\!\!\!\!\!
\begin{array}{|c|c||c|c||c|c||c|c|}\hline
\multicolumn{8}{|c|}{\text{\bf Exceptional three-cycles from the $\Z_2^{(1)}$ sector of $T^6/\Z_2 \times \Z_6$
 in dependence of wrapping numbers, Wilson lines and displacements}} 
\\\hline\hline
\!\!{\rm f.p.}^{(1)}\!\! & {\rm orbit} &\!\!{\rm f.p.}^{(1)}\!\! & {\rm orbit} & \!\!{\rm f.p.}^{(1)}\!\! & {\rm orbit} & \!\!{\rm f.p.}^{(1)}\!\! & {\rm orbit}
\\
\multicolumn{8}{|c|}{(n^2,m^2;n^3,m^3)=\text{(odd,odd;odd,odd)}}
\\
 \multicolumn{2}{|c|}{(\sigma_2;\sigma_3)=(0;0)} & \multicolumn{2}{|c|}{(1;0)} & \multicolumn{2}{|c|}{(0;1)} & \multicolumn{2}{|c|}{(1;1)}
\\\hline\hline
\begin{array}{c} 11 \\ 61 \\ 16 \\ 66
\end{array} &
\!\!\!
\begin{array}{c} 
[n^1 \varepsilon^{(1)}_0 + m^1 \tilde{\varepsilon}^{(1)}_0]
\\
+ (-1)^{\tau_2}[n^1 \varepsilon^{(1)}_1 + m^1 \tilde{\varepsilon}^{(1)}_1]
\\
+ (-1)^{\tau_3}[ n^1 \varepsilon^{(1)}_2 + m^1 \tilde{\varepsilon}^{(1)}_2]
\\
+ (-1)^{\tau_2 + \tau_3} [ n^1 \varepsilon^{(1)}_4 + m^1 \tilde{\varepsilon}^{(1)}_4]
\end{array}\!\!\! &  
\begin{array}{c}  41 \\ 51 \\ 46 \\ 56 
\end{array} & \!\!\! 
\begin{array}{c} [1+(-1)^{\tau_2}][n^1 \varepsilon^{(1)}_1 + m^1 \tilde{\varepsilon}^{(1)}_1]
\\
+(-1)^{\tau_3}[n^1 \varepsilon^{(1)}_5 + m^1 \tilde{\varepsilon}^{(1)}_5]
\\
+(-1)^{\tau_2 + \tau_3}[n^1 \varepsilon^{(1)}_3 + m^1 \tilde{\varepsilon}^{(1)}_3]
\end{array}\!\!\! &  
\begin{array}{c} 14 \\ 64 \\ 15 \\ 65
\end{array} &  \!\!\!
\begin{array}{c}
[1+(-1)^{\tau_3}][n^1 \varepsilon^{(1)}_2 + m^1 \tilde{\varepsilon}^{(1)}_2]
\\
+(-1)^{\tau_2}[n^1 \varepsilon^{(1)}_5 + m^1 \tilde{\varepsilon}^{(1)}_5]
\\
+(-1)^{\tau_2 + \tau_3}[n^1 \varepsilon^{(1)}_3 + m^1 \tilde{\varepsilon}^{(1)}_3]
\end{array}\!\!\! &  
\begin{array}{c} 44 \\ 54 \\ 45 \\ 55
\end{array} &\!\!\!\!\!
\begin{array}{c}
 n^1 \varepsilon^{(1)}_3 + m^1 \tilde{\varepsilon}^{(1)}_3
 \\
 +[(-1)^{\tau_2} + (-1)^{\tau_3}][n^1 \varepsilon^{(1)}_4 + m^1 \tilde{\varepsilon}^{(1)}_4]
 \\
 + (-1)^{\tau_2 + \tau_3}[n^1 \varepsilon^{(1)}_5 + m^1 \tilde{\varepsilon}^{(1)}_5]
\end{array}\!\!\!\!\!
\\\hline\hline\hline
\multicolumn{8}{|c|}{(n^2,m^2;n^3,m^3)=\text{(odd,even;odd,odd)}}
\\
 \multicolumn{2}{|c|}{(\sigma_2;\sigma_3)=(0;0)} & \multicolumn{2}{|c|}{(1;0)} & \multicolumn{2}{|c|}{(0;1)} & \multicolumn{2}{|c|}{(1;1)}
\\\hline\hline
\begin{array}{c} 11 \\ 41 \\ 16 \\ 46
\end{array} & \!\!\! 
\begin{array}{c} 
[n^1 \varepsilon^{(1)}_0 + m^1 \tilde{\varepsilon}^{(1)}_0]
\\
+ (-1)^{\tau_2}[n^1 \varepsilon^{(1)}_1 + m^1 \tilde{\varepsilon}^{(1)}_1]
\\
+ (-1)^{\tau_3}[ n^1 \varepsilon^{(1)}_2 + m^1 \tilde{\varepsilon}^{(1)}_2]
\\
+ (-1)^{\tau_2 + \tau_3} [ n^1 \varepsilon^{(1)}_5 + m^1 \tilde{\varepsilon}^{(1)}_5]
\end{array}\!\!\! &  
\begin{array}{c} 51 \\ 61 \\ 56 \\ 66 
\end{array} &  \!\!\!
\begin{array}{c} [1+(-1)^{\tau_2}][n^1 \varepsilon^{(1)}_1 + m^1 \tilde{\varepsilon}^{(1)}_1]
\\ +(-1)^{\tau_3} [n^1 \varepsilon^{(1)}_3 + m^1 \tilde{\varepsilon}^{(1)}_3]
\\ +(-1)^{\tau_2 + \tau_3}[n^1 \varepsilon^{(1)}_4 + m^1 \tilde{\varepsilon}^{(1)}_4]
\end{array} \!\!\! &  
\begin{array}{c} 14 \\ 44 \\ 15 \\ 45 
\end{array} & \!\!\!  
\begin{array}{c} [1+(-1)^{\tau_3}][n^1 \varepsilon^{(1)}_2 + m^1 \tilde{\varepsilon}^{(1)}_2]
\\ +(-1)^{\tau_2} [n^1 \varepsilon^{(1)}_3 + m^1 \tilde{\varepsilon}^{(1)}_3]
\\ +(-1)^{\tau_2 + \tau_3}[n^1 \varepsilon^{(1)}_4 + m^1 \tilde{\varepsilon}^{(1)}_4]
\end{array} \!\!\! & 
\begin{array}{c} 54 \\ 64 \\ 55 \\ 65  
\end{array} & \!\!\!\!\! 
\begin{array}{c} [n^1 \varepsilon^{(1)}_4 + m^1 \tilde{\varepsilon}^{(1)}_4]
\\ + [(-1)^{\tau_2} + (-1)^{\tau_3}][n^1 \varepsilon^{(1)}_5 + m^1 \tilde{\varepsilon}^{(1)}_5]
\\ + (-1)^{\tau_2 + \tau_3}[n^1 \varepsilon^{(1)}_3 + m^1 \tilde{\varepsilon}^{(1)}_3]
\end{array} \!\!\!\!\!
\\\hline\hline\hline
\multicolumn{8}{|c|}{(n^2,m^2;n^3,m^3)=\text{(even,odd;odd,odd)}}
\\
 \multicolumn{2}{|c|}{(\sigma_2;\sigma_3)=(0;0)} & \multicolumn{2}{|c|}{(1;0)} & \multicolumn{2}{|c|}{(0;1)} & \multicolumn{2}{|c|}{(1;1)}
\\\hline\hline
\begin{array}{c} 11 \\ 51 \\ 16 \\ 56
\end{array} &\!\!\!
\begin{array}{c} 
[n^1 \varepsilon^{(1)}_0 + m^1 \tilde{\varepsilon}^{(1)}_0]
\\
+ (-1)^{\tau_2}[n^1 \varepsilon^{(1)}_1 + m^1 \tilde{\varepsilon}^{(1)}_1]
\\
+ (-1)^{\tau_3}[ n^1 \varepsilon^{(1)}_2 + m^1 \tilde{\varepsilon}^{(1)}_2]
\\
+ (-1)^{\tau_2 + \tau_3} [ n^1 \varepsilon^{(1)}_3 + m^1 \tilde{\varepsilon}^{(1)}_3]
\end{array}\!\!\! &  
\begin{array}{c} 41 \\ 61 \\ 46 \\ 66
\end{array} &   \!\!\!
\begin{array}{c} [1+(-1)^{\tau_2}][n^1 \varepsilon^{(1)}_1 + m^1 \tilde{\varepsilon}^{(1)}_1]
\\ +(-1)^{\tau_3}[n^1 \varepsilon^{(1)}_5 + m^1 \tilde{\varepsilon}^{(1)}_5]
\\ +(-1)^{\tau_2+ \tau_3}[n^1 \varepsilon^{(1)}_4 + m^1 \tilde{\varepsilon}^{(1)}_4]
\end{array} \!\!\! &  
\begin{array}{c} 14 \\ 54 \\ 15 \\ 55 
\end{array} & \!\!\!  
\begin{array}{c} [1+(-1)^{\tau_3}][n^1 \varepsilon^{(1)}_2 + m^1 \tilde{\varepsilon}^{(1)}_2]
\\ + (-1)^{\tau_2}[n^1 \varepsilon^{(1)}_4 + m^1 \tilde{\varepsilon}^{(1)}_4]
\\ + (-1)^{\tau_2+ \tau_3}[n^1 \varepsilon^{(1)}_5 + m^1 \tilde{\varepsilon}^{(1)}_5]
\end{array} \!\!\! &  
\begin{array}{c} 44 \\ 64 \\ 45 \\ 65
\end{array} &  \!\!\! 
\begin{array}{c} [1 + (-1)^{\tau_2+ \tau_3}][n^1 \varepsilon^{(1)}_3 + m^1 \tilde{\varepsilon}^{(1)}_3]
\\ + (-1)^{\tau_2}[n^1 \varepsilon^{(1)}_5 + m^1 \tilde{\varepsilon}^{(1)}_5]
\\ +(-1)^{\tau_3}[n^1 \varepsilon^{(1)}_4 + m^1 \tilde{\varepsilon}^{(1)}_4]
\end{array}  \!\!\!
\\\hline
\end{array}
\end{equation*}
}
}
\caption{
Relation among wrapping numbers $(n^i,m^i)$,  discrete displacements $\vec{\sigma}$ and Wilson lines $\vec{\tau}$, 
 $\Z_2^{(1)}$ fixed points and exceptional three-cycles on $T^6/\Z_2 \times \Z_6$. The exceptional contribution to 
a fractional cycle will be multiplied by an over-all factor $(-1)^{\tau^{(1)}_0}/4$, where $\tau_0^{(1)} \in \{0,1\}$ 
parameterises the  $\Z_2^{(1)}$ eigenvalue.}
\label{Tab:Z2Z6-Z1sector}
\end{sidewaystable}

\begin{table}[ht]
\renewcommand{\arraystretch}{1.3}
  \begin{center}
\begin{equation*}
\begin{array}{|c|c||c|c|}\hline
\multicolumn{4}{|c|}{\text{\bf Exceptional three-cycles from the $\Z_2^{(2)}$ sector of } T^6/\Z_2 \times \Z_6} 
\\\hline\hline
\!\!{\rm f.p.}^{(2)}\!\! & {\rm orbit} & \!\!{\rm f.p.}^{(2)}\!\! & {\rm orbit} 
\\
\multicolumn{4}{|c|}{(n^3,m^3)=\text{(odd,odd)}}
\\
 \multicolumn{2}{|c|}{\sigma_3=0} & \multicolumn{2}{|c|}{1} 
\\\hline\hline
\begin{array}{c} k1 \\ k6
\end{array} & (-1)^{\tau_3}[-(n^2 +m^2) \varepsilon^{(2)}_k + n^2 \tilde{\varepsilon}^{(2)}_k]
& \begin{array}{c} k4 \\ k5
\end{array} & [ n^2 +  (-1)^{\tau_3} m^2 ] \varepsilon^{(2)}_k +[(1-(-1)^{\tau_3})m^2 -(-1)^{\tau_3} n^2] \tilde{\varepsilon}^{(2)}_k
\\\hline
\end{array}
\end{equation*}
\end{center}
\caption{Relation among torus wrapping numbers $(n^i,m^i)$, discrete displacements $\vec{\sigma}$ and Wilson lines $\vec{\tau}$, $\Z_2^{(2)}$ fixed points and exceptional three-cycles on $T^6/\Z_2 \times \Z_6$.
For brevity of the table, we keep $k \in \{1 \ldots 4\}$ as a free label on the first two-torus $T^2_{(1)}$.
The contribution to a fractional cycle is multiplied by the $\Z_2^{(2)}$ eigenvalue times a normalisation factor, $(-1)^{\tau^{(2)}_0}/4$.}
\label{Tab:Z2Z6-Z2sector}
\end{table}

\begin{table}[ht]
\renewcommand{\arraystretch}{1.3}
  \begin{center}
\begin{equation*}
\begin{array}{|c|c||c|c|}\hline\multicolumn{4}{|c|}{\text{\bf Exceptional three-cycles from the $\Z_2^{(3)}$ sector of } T^6/\Z_2 \times \Z_6} 
\\\hline\hline
\!\!{\rm f.p.}^{(3)}\!\! & {\rm orbit} & \!\!{\rm f.p.}^{(3)}\!\! & {\rm orbit} 
\\
\multicolumn{4}{|c|}{(n^2,m^2)=\text{(odd,odd)}}
\\
 \multicolumn{2}{|c|}{\sigma_2=0} & \multicolumn{2}{|c|}{1} 
\\\hline\hline
\begin{array}{c} k1 \\ k6
\end{array} & (-1)^{\tau_2} [  -(n^3 + m^3)  \varepsilon^{(3)}_k + n^3 \tilde{\varepsilon}^{(3)}_k]
& \begin{array}{c} k4 \\ k5
\end{array} & [n^3 +  (-1)^{\tau_2} m^3]   \varepsilon^{(3)}_k +[(1-(-1)^{\tau_2})m^3 -(-1)^{\tau_2} n^3 ]\tilde{\varepsilon}^{(3)}_k
\\\hline\hline\hline
\multicolumn{4}{|c|}{(n^2,m^2)=\text{(odd,even)}}
\\
 \multicolumn{2}{|c|}{\sigma_2=0} & \multicolumn{2}{|c|}{1} 
\\\hline\hline
\begin{array}{c} k1 \\ k4
\end{array} & (-1)^{\tau_2} [ n^3  \varepsilon^{(3)}_k + m^3 \tilde{\varepsilon}^{(3)}_k]
& \begin{array}{c} k5 \\ k6
\end{array} &  
 \begin{array}{c}
[m^3 -(-1)^{\tau_2} (n^3 + m^3)]  \varepsilon^{(3)}_k 
\\ + [ -(n^3 + m^3) + (-1)^{\tau_2}n^3] \tilde{\varepsilon}^{(3)}_k
\end{array}
\\\hline\hline\hline
\multicolumn{4}{|c|}{(n^2,m^2)=\text{(even,odd)}}
\\
 \multicolumn{2}{|c|}{\sigma_2=0} & \multicolumn{2}{|c|}{1} 
\\\hline\hline
\begin{array}{c} k1 \\ k5
\end{array} & (-1)^{\tau_2} [ m^3  \varepsilon^{(3)}_k -(n^3 + m^3) \tilde{\varepsilon}^{(3)}_k]
& \begin{array}{c} k4 \\ k6
\end{array} & [n^3  -(-1)^{\tau_2} (n^3 + m^3)]  \varepsilon^{(3)}_k + [ m^3 + (-1)^{\tau_2}n^3] \tilde{\varepsilon}^{(3)}_k
\\\hline
\end{array}
\end{equation*}
 \end{center}
\caption{Relation among wrapping numbers $(n^i,m^i)$, discrete displacements $\vec{\sigma}$ and Wilson lines $\vec{\tau}$, 
$\Z_2^{(3)}$ fixed points and exceptional three-cycles on $T^6/\Z_2 \times \Z_6$.
For brevity of the table, we keep $k \in \{1 \ldots 4\}$ as a free label on the first two-torus $T^2_{(1)}$.
The contribution to a fractional cycle is multiplied by the $\Z_2^{(3)}$ eigenvalue times a normalisation factor, $(-1)^{\tau^{(3)}_0}/4$.}
\label{Tab:Z2Z6-Z3sector}
\end{table}

\begin{sidewaystable}[ht]
\renewcommand{\arraystretch}{1.3}
  \begin{center}
\begin{equation*}\!\!\!\!\!\!\!\!\!\!\!\!\!\!\!\!\!\!\!\!\!
{\tiny
\begin{array}{|c|c||c|c||c|c||c|c|}\hline
\multicolumn{8}{|c|}{\text{\bf Exceptional three-cycles, wrapping numbers, Wilson lines and displacements on }
 T^6/\Z_2 \times \Z_6' \;  \text{\bf with discrete torsion, Part I}}
\\\hline\hline
 \!\!{\rm f.p.}^{(i)}\!\! & {\rm orbit} &  \!\!{\rm f.p.}^{(i)}\!\! & {\rm orbit} &  \!\!{\rm f.p.}^{(i)}\!\! & {\rm orbit} &  \!\!{\rm f.p.}^{(i)}\!\! & {\rm orbit}
\\
\multicolumn{8}{|c|}{(n^j,m^j;n^k,m^k)=\text{(odd,odd;odd,odd)}}
\\
 \multicolumn{2}{|c|}{(\sigma_j;\sigma_k)=(0;0)} & \multicolumn{2}{|c|}{(1;0)} & \multicolumn{2}{|c|}{(0;1)} & \multicolumn{2}{|c|}{(1;1)}
\\\hline\hline
\begin{array}{c} 11 \\ 61 \\ 16 \\ 66
\end{array} &
 \!\!\!\!\!\!\begin{array}{c} (-1)^{\tau_j}\left[m^i \varepsilon^{(i)}_1 - (n^i+m^i)\tilde{\varepsilon}^{(i)}_1   \right]\\
+ (-1)^{\tau_k}\left[m^i\varepsilon^{(i)}_2 - (n^i+m^i)\tilde{\varepsilon}^{(i)}_2   \right]\\
+ (-1)^{\tau_j+\tau_k}\left[m^i\varepsilon^{(i)}_3 - (n^i+m^i)\tilde{\varepsilon}^{(i)}_3   \right]\\
\end{array}\!\!\!\!\!\!
& \begin{array}{c} 41 \\ 51 \\ 46 \\ 56
\end{array} &
\!\!\!\!\!\!\begin{array}{c} ((1-(-1)^{\tau_j})n^i -(-1)^{\tau_j}m^i) \varepsilon^{(i)}_1\\ + (m^i +(-1)^{\tau_j}n^i) \tilde{\varepsilon}^{(i)}_1  \\
+(-1)^{\tau_j + \tau_k}[-(n^i+m^i)  \varepsilon^{(i)}_4 + n^i \tilde{\varepsilon}^{(i)}_4 ]\\ + (-1)^{\tau_k} [n^i \varepsilon^{(i)}_5 + m^i \tilde{\varepsilon}^{(i)}_5]
\end{array}\!\!\!\!\!\!
& \begin{array}{c} 14 \\ 64 \\ 15 \\ 65
\end{array} &
 \!\!\!\!\!\!\begin{array}{c} ((1-(-1)^{\tau_k})n^i -(-1)^{\tau_k}m^i) \varepsilon^{(i)}_2\\ + (m^i +(-1)^{\tau_k}n^i) \tilde{\varepsilon}^{(i)}_2  \\
+(-1)^{\tau_j} [m^i \varepsilon^{(i)}_4 - (n^i + m^i)  \tilde{\varepsilon}^{(i)}_4]\\
+(-1)^{\tau_j+\tau_k} [m^i \varepsilon^{(i)}_5 - (n^i + m^i)  \tilde{\varepsilon}^{(i)}_5]
\end{array}\!\!\!\!\!\!
& \begin{array}{c} 44 \\ 54 \\ 45 \\ 55 
\end{array} &
 \!\!\!\!\!\!\begin{array}{c} ((1-(-1)^{\tau_j+\tau_k})n^i -(-1)^{\tau_j+\tau_k}m^i) \varepsilon^{(i)}_3\\ + (m^i +(-1)^{\tau_j+\tau_k}n^i) \tilde{\varepsilon}^{(i)}_3  \\
+(-1)^{\tau_k}[  n^i \varepsilon^{(i)}_4 + m^i \tilde{\varepsilon}^{(i)}_4]
\\
+(-1)^{\tau_j}[ -(n^i + m^i)  \varepsilon^{(i)}_5 + n^i \tilde{\varepsilon}^{(i)}_5]
\end{array}\!\!\!\!\!\!
\\\hline\hline\hline
\multicolumn{8}{|c|}{(n^j,m^j;n^k,m^k)=\text{(odd,even;odd,odd)}}
\\
 \multicolumn{2}{|c|}{(\sigma_j;\sigma_k)=(0;0)} & \multicolumn{2}{|c|}{(1;0)} & \multicolumn{2}{|c|}{(0;1)} & \multicolumn{2}{|c|}{(1;1)}
\\\hline\hline
\begin{array}{c} 11 \\ 41 \\ 16 \\ 46
\end{array} & 
 \!\!\!\!\!\!\begin{array}{c} (-1)^{\tau_j} [n^i \varepsilon^{(i)}_1 +m^i \tilde{\varepsilon}^{(i)}_1]  \\
+ (-1)^{\tau_k} [m^i \varepsilon^{(i)}_2 -(n^i+m^i) \tilde{\varepsilon}^{(i)}_2 ]  \\
+ (-1)^{\tau_j+\tau_k} [n^i \varepsilon^{(i)}_5 +m^i \tilde{\varepsilon}^{(i)}_5] 
\end{array}\!\!\!\!\!\!
& \begin{array}{c} 51 \\ 61 \\ 56 \\ 66
\end{array} &
 \!\!\!\!\!\!\begin{array}{c} [-n^i +((-1)^{\tau_j}-1)m^i]\varepsilon^{(i)}_1
\\
+ [(1 - (-1)^{\tau_j}) n^i - (-1)^{\tau_j} m^i]\tilde{\varepsilon}^{(i)}_1
\\
+(-1)^{\tau_k}[ -(n^i+m^i) \varepsilon^{(i)}_4 + n^i \tilde{\varepsilon}^{(i)}_4]
\\
+ (-1)^{\tau_j+\tau_k}[m^i\varepsilon^{(i)}_3 - (n^i+m^i)\tilde{\varepsilon}^{(i)}_3 ]
\end{array}\!\!\!\!\!\!
& \begin{array}{c} 14 \\ 44 \\ 15 \\ 45 
\end{array} &
 \!\!\!\!\!\!\begin{array}{c}  ((1-(-1)^{\tau_k})n^i -(-1)^{\tau_k}m^i) \varepsilon^{(i)}_2
\\ + (m^i +(-1)^{\tau_k}n^i) \tilde{\varepsilon}^{(i)}_2  
\\ +(-1)^{\tau_j}[n^i\varepsilon^{(i)}_3 +m^i \tilde{\varepsilon}^{(i)}_3 ]
\\ + (-1)^{\tau_j + \tau_k}[n^i\varepsilon^{(i)}_4 +m^i \tilde{\varepsilon}^{(i)}_4 ]
\end{array}\!\!\!\!\!\!
& \begin{array}{c} 54 \\ 64 \\ 55 \\ 65
\end{array} &
 \!\!\!\!\!\!\begin{array}{c} [-n^i +((-1)^{\tau_j+\tau_k}-1)m^i]\varepsilon^{(i)}_5
\\
+ [(1 - (-1)^{\tau_j+\tau_k}) n^i - (-1)^{\tau_j+\tau_k} m^i]\tilde{\varepsilon}^{(i)}_5
\\
+(-1)^{\tau_j}[m^i\varepsilon^{(i)}_4 - (n^i+m^i)\tilde{\varepsilon}^{(i)}_4 ]
\\
+ (-1)^{\tau_k}[ -(n^i+m^i) \varepsilon^{(i)}_3 + n^i \tilde{\varepsilon}^{(i)}_3]
\end{array}\!\!\!\!\!\!
%
\\\hline\hline\hline
\multicolumn{8}{|c|}{(n^j,m^j;n^k,m^k)=\text{(even,odd;odd,odd)}}
\\
 \multicolumn{2}{|c|}{(\sigma_j;\sigma_k)=(0;0)} & \multicolumn{2}{|c|}{(1;0)} & \multicolumn{2}{|c|}{(0;1)} & \multicolumn{2}{|c|}{(1;1)}
\\\hline\hline
\begin{array}{c} 11 \\ 51 \\ 16 \\ 56
\end{array} & 
 \!\!\!\!\!\!\begin{array}{c} 
(-1)^{\tau_j}[-(n^i+m^i) \varepsilon^{(i)}_1 +n^i \tilde{\varepsilon}^{(i)}_1]
\\
+(-1)^{\tau_k}[m^i \varepsilon^{(i)}_2 -(n^i+m^i) \tilde{\varepsilon}^{(i)}_2 ] 
\\
+(-1)^{\tau_j+\tau_k}[-(n^i+m^i) \varepsilon^{(i)}_4 + n^i \tilde{\varepsilon}^{(i)}_4]
\end{array}\!\!\!\!\!\!
& \begin{array}{c} 41 \\ 61 \\ 46 \\ 66
\end{array} &
 \!\!\!\!\!\!\begin{array}{c} 
[n^i+(-1)^{\tau_j}m^i] \varepsilon^{(i)}_1 
\\
+ [ -(-1)^{\tau_j}n^i + (1-(-1)^{\tau_j})n^i] \tilde{\varepsilon}^{(i)}_1
\\
+(-1)^{\tau_k} [n^i \varepsilon^{(i)}_5 + m^i \tilde{\varepsilon}^{(i)}_5]
\\
+ (-1)^{\tau_j+\tau_k}[m^i\varepsilon^{(i)}_3 - (n^i+m^i)\tilde{\varepsilon}^{(i)}_3 ]
\end{array}\!\!\!\!\!\!
& \begin{array}{c} 14 \\ 54 \\ 15 \\ 55
\end{array} &
 \!\!\!\!\!\!\begin{array}{c} ((1-(-1)^{\tau_k})n^i -(-1)^{\tau_k}m^i) \varepsilon^{(i)}_2
\\ + (m^i +(-1)^{\tau_k}n^i) \tilde{\varepsilon}^{(i)}_2  
\\ +(-1)^{\tau_j}[-(n^i+m^i) \varepsilon^{(i)}_5 +n^i \tilde{\varepsilon}^{(i)}_5]
\\ + (-1)^{\tau_j+\tau_k}[-(n^i+m^i) \varepsilon^{(i)}_3 +n^i \tilde{\varepsilon}^{(i)}_3]
\end{array}\!\!\!\!\!\!
& \begin{array}{c} 44 \\ 64 \\ 45 \\ 65
\end{array} &
 \!\!\!\!\!\!\begin{array}{c} 
n^i\varepsilon^{(i)}_3 +m^i \tilde{\varepsilon}^{(i)}_3 
\\
+ ((-1)^{\tau_j} m^i+(-1)^{\tau_k}n^i)\varepsilon^{(i)}_4
\\
+(-(-1)^{\tau_j} n^i+ ((-1)^{\tau_k}-(-1)^{\tau_j})m^i) \tilde{\varepsilon}^{(i)}_4
\\
+(-1)^{\tau_j+\tau_k}[m^i \varepsilon^{(i)}_5 -(n^i+m^i) \tilde{\varepsilon}^{(i)}_5 ] 
\end{array}\!\!\!\!\!\!
%
\\\hline\hline\hline
\multicolumn{8}{|c|}{(n^j,m^j;n^k,m^k)=\text{(odd,odd;odd,even)}}
\\
 \multicolumn{2}{|c|}{(\sigma_j;\sigma_k)=(0;0)} & \multicolumn{2}{|c|}{(1;0)} & \multicolumn{2}{|c|}{(0;1)} & \multicolumn{2}{|c|}{(1;1)}
\\\hline\hline
\begin{array}{c} 11 \\ 61 \\ 14 \\ 64
\end{array} & 
 \!\!\!\!\!\!\begin{array}{c} 
(-1)^{\tau_j}[m^i \varepsilon^{(i)}_1 -(n^i+m^i) \tilde{\varepsilon}^{(i)}_1]
\\
+(-1)^{\tau_k}[n^i \varepsilon^{(i)}_2 +m^i \tilde{\varepsilon}^{(i)}_2 ] 
\\
+(-1)^{\tau_j+\tau_k}[m^i \varepsilon^{(i)}_4 -( n^i+m^i) \tilde{\varepsilon}^{(i)}_4]
\end{array}\!\!\!\!\!\!
& \begin{array}{c} 41 \\ 51 \\ 44 \\ 54 
\end{array} &
\begin{array}{c}  ((1-(-1)^{\tau_j})n^i -(-1)^{\tau_j}m^i) \varepsilon^{(i)}_1
\\ 
+ (m^i +(-1)^{\tau_j}n^i) \tilde{\varepsilon}^{(i)}_1  
\\
+(-1)^{\tau_k}[n^i \varepsilon^{(i)}_3 + m^i \tilde{\varepsilon}^{(i)}_3]
\\
+ (-1)^{\tau_j+\tau_k}[-(n^i+m^i) \varepsilon^{(i)}_5 +n^i \tilde{\varepsilon}^{(i)}_5]
\end{array} 
& \begin{array}{c} 15 \\ 65 \\ 16 \\ 66
\end{array} 
& \!\!\!\!\!\!\begin{array}{c} 
[-n^i + ((-1)^{\tau_k}-1) m^i] \varepsilon^{(i)}_2
\\
+ [(1-(-1)^{\tau_k})n^i -(-1)^{\tau_k} m^i]\tilde{\varepsilon}^{(i)}_2 
\\
+(-1)^{\tau_j}[m^i \varepsilon^{(i)}_5 -(n^i+m^i) \tilde{\varepsilon}^{(i)}_5]
\\
+ (-1)^{\tau_j+\tau_k}[m^i \varepsilon^{(i)}_3 -(n^i+m^i) \tilde{\varepsilon}^{(i)}_3]
\end{array}\!\!\!\!\!\!
& \begin{array}{c} 45 \\ 55 \\ 46 \\ 56 
\end{array} 
&
 \!\!\!\!\!\!\begin{array}{c} 
[(1-(-1)^{\tau_j+\tau_k})n^i - (-1)^{\tau_j+\tau_k} m^i]\varepsilon^{(i)}_4 
\\
+ [(-1)^{\tau_j+\tau_k}n^i + m^i ] \tilde{\varepsilon}^{(i)}_4
\\
+(-1)^{\tau_j}[-(n^i+m^i) \varepsilon^{(i)}_3 +n^i \tilde{\varepsilon}^{(i)}_3]
\\
+(-1)^{\tau_k}[n^i \varepsilon^{(i)}_5 + m^i \tilde{\varepsilon}^{(i)}_5]
\end{array}\!\!\!\!\!\!
\\\hline\hline\hline
\multicolumn{8}{|c|}{(n^j,m^j;n^k,m^k)=\text{(odd,odd;even,odd)}}
\\
 \multicolumn{2}{|c|}{(\sigma_j;\sigma_k)=(0;0)} & \multicolumn{2}{|c|}{(1;0)} & \multicolumn{2}{|c|}{(0;1)} & \multicolumn{2}{|c|}{(1;1)}
\\\hline\hline
\begin{array}{c} 11 \\  61  \\ 15  \\ 65
\end{array} &
 \!\!\!\!\!\!\begin{array}{c} 
(-1)^{\tau_j}[m^i \varepsilon^{(i)}_1 -(n^i+m^i) \tilde{\varepsilon}^{(i)}_1]
\\
+(-1)^{\tau_k}[-(n^i+m^i) \varepsilon^{(i)}_2 +n^i \tilde{\varepsilon}^{(i)}_2 ] 
\\
+(-1)^{\tau_j+\tau_k}[m^i \varepsilon^{(i)}_5 -( n^i+m^i) \tilde{\varepsilon}^{(i)}_5]
\end{array}\!\!\!\!\!\!
& \begin{array}{c} 41 \\ 51 \\ 45 \\ 55
\end{array} 
&  \begin{array}{c} ((1-(-1)^{\tau_j})n^i -(-1)^{\tau_j}m^i) \varepsilon^{(i)}_1
\\ 
+ (m^i +(-1)^{\tau_j}n^i) \tilde{\varepsilon}^{(i)}_1  
\\
+(-1)^{\tau_k}[n^i \varepsilon^{(i)}_4 + m^i \tilde{\varepsilon}^{(i)}_4]
\\
+ (-1)^{\tau_j+\tau_k}[-(n^i+m^i) \varepsilon^{(i)}_3 +n^i \tilde{\varepsilon}^{(i)}_3]
\end{array} 
& \begin{array}{c} 14 \\ 64 \\ 16 \\ 66
\end{array} 
& \!\!\!\!\!\!\begin{array}{c} [n^i + (-1)^{\tau_k}m^i]\varepsilon^{(i)}_2
\\
+[-(-1)^{\tau_k}n^i + (1 -(-1)^{\tau_k}) m^i]  \tilde{\varepsilon}^{(i)}_2 
\\
+(-1)^{\tau_j}[m^i \varepsilon^{(i)}_4 -(n^i+m^i) \tilde{\varepsilon}^{(i)}_4]
\\
+(-1)^{\tau_j+\tau_k}[m^i \varepsilon^{(i)}_3 -(n^i+m^i) \tilde{\varepsilon}^{(i)}_3]
\end{array}\!\!\!\!\!\!
& \begin{array}{c} 44 \\ 54 \\ 46 \\ 56 
\end{array} 
& \begin{array}{c}
 n^i\varepsilon^{(i)}_3 +m^i \tilde{\varepsilon}^{(i)}_3 
\\
+ ([(-1)^{\tau_k}-(-1)^{\tau_j}] n^i-(-1)^{\tau_j}m^i)\varepsilon^{(i)}_5
\\
+ ((-1)^{\tau_j} n^i + (-1)^{\tau_k} m^i) \tilde{\varepsilon}^{(i)}_5
\\
+(-1)^{\tau_j+\tau_k}[-(n^i+m^i) \varepsilon^{(i)}_4 +n^i \tilde{\varepsilon}^{(i)}_4]
\end{array}
\\\hline
\end{array}
}
\end{equation*}
 \end{center}
\caption{Relation among wrapping numbers, $\Z_2^{(i)}$ fixed points and exceptional three-cycles on $T^6/\Z_2 \times \Z_6'$ with discrete torsion, Part I.
The contribution to a fractional cycle is multiplied by the $\Z_2^{(i)}$ eigenvalue times a normalisation factor, $(-1)^{\tau^{(i)}_0}/4$.}
\label{Tab:Z2Z6p-Part1}
\end{sidewaystable}

\begin{sidewaystable}[ht]
\renewcommand{\arraystretch}{1.3}
  \begin{center}
\begin{equation*}\!\!\!\!\!\!\!\!\!\!\!\!\!\!\!\!\!\!\!\!\!\!\!\!
{\tiny
\begin{array}{|c|c||c|c||c|c||c|c|}\hline
\multicolumn{8}{|c|}{\text{\bf Exceptional three-cycles, wrapping numbers, Wilson lines and displacements on } T^6/\Z_2 \times \Z_6' \;  \text{\bf with discrete torsion, Part II}}
\\\hline\hline  {\rm f.p.} & {\rm orbit} & {\rm f.p.} & {\rm orbit} & {\rm f.p.} & {\rm orbit} & {\rm f.p.} & {\rm orbit}
\\
\multicolumn{8}{|c|}{(n^j,m^j;n^k,m^k)=\text{(odd,even;odd,even)}}
\\
 \multicolumn{2}{|c|}{(\sigma_j;\sigma_k)=(0;0)} & \multicolumn{2}{|c|}{(1;0)} & \multicolumn{2}{|c|}{(0;1)} & \multicolumn{2}{|c|}{(1;1)}
\\\hline\hline
\begin{array}{c} 11 \\ 41  \\ 14 \\ 44
\end{array} & 
 \!\!\!\!\!\!\begin{array}{c} 
 (-1)^{\tau_j}[n^i \varepsilon^{(i)}_1 + m^i \tilde{\varepsilon}^{(i)}_1]
\\ + (-1)^{\tau_k} [n^i \varepsilon^{(i)}_2 + m^i \tilde{\varepsilon}^{(i)}_2]
\\ +(-1)^{\tau_j+\tau_k}[n^i \varepsilon^{(i)}_3 + m^i \tilde{\varepsilon}^{(i)}_3]
\end{array}\!\!\!\!\!\!
& \begin{array}{c} 51 \\ 61 \\ 54 \\ 64 
\end{array} &
\!\!\!\!\!\!\begin{array}{c} 
[-n^i+((-1)^{\tau_j}-1)m^i]\varepsilon^{(i)}_1
\\ +[(1-(-1)^{\tau_j})n^i-(-1)^{\tau_j}m^i]\tilde{\varepsilon}^{(i)}_1
\\ +(-1)^{\tau_k}[-(n^i+m^i) \varepsilon^{(i)}_5 +n^i \tilde{\varepsilon}^{(i)}_5]
\\ +(-1)^{\tau_j+\tau_k}[m^i \varepsilon^{(i)}_4 -(n^i+m^i) \tilde{\varepsilon}^{(i)}_4]
 \end{array}\!\!\!\!\!\!
& \begin{array}{c} 15 \\ 45 \\ 16 \\ 46 
\end{array} &
\!\!\!\!\!\!\begin{array}{c} 
[-n^i+((-1)^{\tau_k}-1)m^i]\varepsilon^{(i)}_2
\\+ [(1-(-1)^{\tau_k})n^i-(-1)^{\tau_k}m^i]\tilde{\varepsilon}^{(i)}_2
\\+ (-1)^{\tau_j}[n^i \varepsilon^{(i)}_4 + m^i \tilde{\varepsilon}^{(i)}_4]
\\+ (-1)^{\tau_j+\tau_k}[n^i \varepsilon^{(i)}_5 + m^i \tilde{\varepsilon}^{(i)}_5]
 \end{array}\!\!\!\!\!\!
& \begin{array}{c} 55 \\ 65 \\ 56 \\ 66
\end{array} &
\!\!\!\!\!\!\begin{array}{c} 
[-n^i+((-1)^{\tau_j+\tau_k}-1)m^i]\varepsilon^{(i)}_3
\\ + [(1-(-1)^{\tau_j+\tau_k})n^i-(-1)^{\tau_j+\tau_k}m^i]\tilde{\varepsilon}^{(i)}_3
\\ +(-1)^{\tau_j}[m^i \varepsilon^{(i)}_5 -(n^i+m^i) \tilde{\varepsilon}^{(i)}_5]
\\ +(-1)^{\tau_k}[-(n^i+m^i) \varepsilon^{(i)}_4 +n^i \tilde{\varepsilon}^{(i)}_4]
 \end{array}\!\!\!\!\!\!
%
\\\hline\hline\hline
\multicolumn{8}{|c|}{(n^j,m^j;n^k,m^k)=\text{(odd,even;even,odd)}}
\\
 \multicolumn{2}{|c|}{(\sigma_j;\sigma_k)=(0;0)} & \multicolumn{2}{|c|}{(1;0)} & \multicolumn{2}{|c|}{(0;1)} & \multicolumn{2}{|c|}{(1;1)}
\\\hline\hline
\begin{array}{c} 11 \\ 41 \\ 15 \\ 45
\end{array} & 
\!\!\!\!\!\!\begin{array}{c} 
 (-1)^{\tau_j}[n^i \varepsilon^{(i)}_1 + m^i \tilde{\varepsilon}^{(i)}_1]
\\ + (-1)^{\tau_k} [-(n^i + m^i) \varepsilon^{(i)}_2 + n^i \tilde{\varepsilon}^{(i)}_2]
\\ +(-1)^{\tau_j+\tau_k}[n^i \varepsilon^{(i)}_4 + m^i \tilde{\varepsilon}^{(i)}_4]
\end{array}\!\!\!\!\!\!
& \begin{array}{c} 51 \\ 61 \\ 55 \\ 65 
\end{array} &
\!\!\!\!\!\!\begin{array}{c} 
[-n^i+((-1)^{\tau_j}-1)m^i]\varepsilon^{(i)}_1
\\ +[(1-(-1)^{\tau_j})n^i-(-1)^{\tau_j}m^i]\tilde{\varepsilon}^{(i)}_1
\\ +(-1)^{\tau_k}[-(n^i + m^i) \varepsilon^{(i)}_3 + n^i \tilde{\varepsilon}^{(i)}_3]
\\ +(-1)^{\tau_j+\tau_k}[m^i \varepsilon^{(i)}_5 -(n^i+m^i) \tilde{\varepsilon}^{(i)}_5]
 \end{array}\!\!\!\!\!\!
& \begin{array}{c} 14 \\ 44 \\ 16 \\ 46 
\end{array} &
\!\!\!\!\!\!\begin{array}{c} 
[n^i + (-1)^{\tau_k} m^i ] \varepsilon^{(i)}_2
\\ +[-(-1)^{\tau_k} n^i + (1-(-1)^{\tau_k} )m^i]\tilde{\varepsilon}^{(i)}_2
\\ + (-1)^{\tau_j}[n^i \varepsilon^{(i)}_3 + m^i \tilde{\varepsilon}^{(i)}_3]
\\ +(-1)^{\tau_j+\tau_k}[n^i \varepsilon^{(i)}_5 + m^i \tilde{\varepsilon}^{(i)}_5]
 \end{array}\!\!\!\!\!\!
& \begin{array}{c} 54 \\ 64 \\ 56 \\ 66
\end{array} &
\!\!\!\!\!\!\begin{array}{c} 
 [-(n^i + m^i) \varepsilon^{(i)}_5 + n^i \tilde{\varepsilon}^{(i)}_5]
\\+[- (-1)^{\tau_k} n^i+((-1)^{\tau_j}-(-1)^{\tau_k})m^i] \varepsilon^{(i)}_4 
\\+[((-1)^{\tau_k}-(-1)^{\tau_j})n^i -(-1)^{\tau_j}m^i]\tilde{\varepsilon}^{(i)}_4
\\+(-1)^{\tau_j+\tau_k}[m^i \varepsilon^{(i)}_3 -(n^i+m^i) \tilde{\varepsilon}^{(i)}_3]
 \end{array}\!\!\!\!\!\!
%
\\\hline\hline\hline
\multicolumn{8}{|c|}{(n^j,m^j;n^k,m^k)=\text{(even,odd;odd,even)}}
\\
 \multicolumn{2}{|c|}{(\sigma_j;\sigma_k)=(0;0)} & \multicolumn{2}{|c|}{(1;0)} & \multicolumn{2}{|c|}{(0;1)} & \multicolumn{2}{|c|}{(1;1)}
\\\hline\hline
\begin{array}{c} 11 \\ 51 \\ 14 \\ 54
\end{array} & 
\!\!\!\!\!\!\begin{array}{c} 
(-1)^{\tau_j} [-(n^i + m^i) \varepsilon^{(i)}_1 + n^i \tilde{\varepsilon}^{(i)}_1]
\\ + (-1)^{\tau_k}  [n^i \varepsilon^{(i)}_2 + m^i \tilde{\varepsilon}^{(i)}_2]
\\ +(-1)^{\tau_j+\tau_k} [-(n^i + m^i) \varepsilon^{(i)}_5 + n^i \tilde{\varepsilon}^{(i)}_5]
 \end{array}\!\!\!\!\!\!
& \begin{array}{c} 41 \\ 61 \\ 44 \\ 64 
\end{array} &
\!\!\!\!\!\!\begin{array}{c}   [n^i + (-1)^{\tau_j} m^i ] \varepsilon^{(i)}_1
\\ +[-(-1)^{\tau_j} n^i + (1-(-1)^{\tau_j} )m^i]\tilde{\varepsilon}^{(i)}_1
\\ +(-1)^{\tau_k}[n^i \varepsilon^{(i)}_3 + m^i \tilde{\varepsilon}^{(i)}_3]
\\ +(-1)^{\tau_j+\tau_k}[m^i \varepsilon^{(i)}_4 -(n^i+m^i) \tilde{\varepsilon}^{(i)}_4]
 \end{array}\!\!\!\!\!\!
& \begin{array}{c} 15 \\ 55 \\ 16 \\ 56 
\end{array} &
\!\!\!\!\!\!\begin{array}{c} 
[-n^i+((-1)^{\tau_k}-1)m^i]\varepsilon^{(i)}_2
\\+ [(1-(-1)^{\tau_k})n^i-(-1)^{\tau_k}m^i]\tilde{\varepsilon}^{(i)}_2
\\+ (-1)^{\tau_j} [-(n^i + m^i) \varepsilon^{(i)}_3 + n^i \tilde{\varepsilon}^{(i)}_3]
\\+ (-1)^{\tau_j+\tau_k} [-(n^i + m^i) \varepsilon^{(i)}_4 + n^i \tilde{\varepsilon}^{(i)}_4]
 \end{array}\!\!\!\!\!\!
& \begin{array}{c} 45 \\ 65 \\ 46 \\ 66
\end{array} &
\!\!\!\!\!\!\begin{array}{c}
[n^i \varepsilon^{(i)}_4 + m^i \tilde{\varepsilon}^{(i)}_4]
\\+ [(-1)^{\tau_k}n^i +(-1)^{\tau_j}m^i]  \varepsilon^{(i)}_5 
\\+[-(-1)^{\tau_j}n^i + ((-1)^{\tau_k}-(-1)^{\tau_j})m^i]\tilde{\varepsilon}^{(i)}_5
\\+(-1)^{\tau_j+\tau_k} [m^i \varepsilon^{(i)}_3 -(n^i+m^i) \tilde{\varepsilon}^{(i)}_3]
 \end{array}\!\!\!\!\!\!
%
\\\hline\hline\hline
\multicolumn{8}{|c|}{(n^j,m^j;n^k,m^k)=\text{(even,odd;even,odd)}}
\\
 \multicolumn{2}{|c|}{(\sigma_j;\sigma_k)=(0;0)} & \multicolumn{2}{|c|}{(1;0)} & \multicolumn{2}{|c|}{(0;1)} & \multicolumn{2}{|c|}{(1;1)}
\\\hline\hline
\begin{array}{c} 11 \\ 51 \\ 15 \\ 55
\end{array} & 
\!\!\!\!\!\!\begin{array}{c} 
(-1)^{\tau_j} [-(n^i + m^i) \varepsilon^{(i)}_1 + n^i \tilde{\varepsilon}^{(i)}_1]
\\ + (-1)^{\tau_k} [-(n^i + m^i) \varepsilon^{(i)}_2 + n^i \tilde{\varepsilon}^{(i)}_2]
\\ +(-1)^{\tau_j+\tau_k} [-(n^i + m^i) \varepsilon^{(i)}_3 + n^i \tilde{\varepsilon}^{(i)}_3]
 \end{array}\!\!\!\!\!\!
& \begin{array}{c} 41 \\ 61 \\ 45 \\ 65
\end{array} &
\!\!\!\!\!\!\begin{array}{c}  [n^i + (-1)^{\tau_j} m^i ] \varepsilon^{(i)}_1
\\ +[-(-1)^{\tau_j} n^i + (1-(-1)^{\tau_j} )m^i]\tilde{\varepsilon}^{(i)}_1
\\ +(-1)^{\tau_k}  [n^i \varepsilon^{(i)}_4 + m^i \tilde{\varepsilon}^{(i)}_4]
\\ +(-1)^{\tau_j+\tau_k}[m^i \varepsilon^{(i)}_5 -(n^i+m^i) \tilde{\varepsilon}^{(i)}_5]
 \end{array}\!\!\!\!\!\!
& \begin{array}{c} 14 \\ 54 \\ 16 \\ 56 
\end{array} &
\!\!\!\!\!\!\begin{array}{c}  [n^i + (-1)^{\tau_k} m^i ] \varepsilon^{(i)}_2
\\ +[-(-1)^{\tau_k} n^i + (1-(-1)^{\tau_k} )m^i]\tilde{\varepsilon}^{(i)}_2
\\ + (-1)^{\tau_j}[-(n^i + m^i) \varepsilon^{(i)}_5 + n^i \tilde{\varepsilon}^{(i)}_5]
\\ +(-1)^{\tau_j+\tau_k}[-(n^i + m^i) \varepsilon^{(i)}_4 + n^i \tilde{\varepsilon}^{(i)}_4]
 \end{array}\!\!\!\!\!\!
& \begin{array}{c} 44 \\ 64 \\ 46 \\ 66
\end{array} &
\!\!\!\!\!\!\begin{array}{c} 
[n^i+(-1)^{\tau_j+\tau_k}m^i]\varepsilon^{(i)}_3 
\\ +[-(-1)^{\tau_j+\tau_k}n^i + (1-(-1)^{\tau_j+\tau_k})m^i]\tilde{\varepsilon}^{(i)}_3
\\ +(-1)^{\tau_j}[m^i \varepsilon^{(i)}_4 -(n^i+m^i) \tilde{\varepsilon}^{(i)}_4]
\\ +(-1)^{\tau_k}[n^i \varepsilon^{(i)}_5 + m^i \tilde{\varepsilon}^{(i)}_5]
 \end{array}\!\!\!\!\!\!
\\\hline
\end{array}
}
\end{equation*}
 \end{center}
\caption{Relation among wrapping numbers, $\Z_2^{(i)}$ fixed points and exceptional three-cycles on $T^6/\Z_2 \times \Z_6'$ with discrete torsion, Part II.}
\label{Tab:Z2Z6p-Part2}
\end{sidewaystable}

{\boldmath
\section{From $T^4/\Z_N$ and $T^6/\Z_N$ orbifolds to $T^6/\Z_N \times \Z_M$ without and with discrete torsion}\label{App:ZN}
}

The twisted sectors of $T^6/\Z_2 \times \Z_{2M}$ orbifolds without and with discrete torsion are inherited from various
$T^4/\Z_N$ and $T^6/\Z_N$ sub-sectors, but the number of multiplets can be reduced by new identifications of fixed points.
For example the sector twisted by $(0,\frac{1}{3},-\frac{1}{3})$ contributes $h_{11}=18$ on $T^4/\Z_3$, which splits 
into $(h_{11},h_{21})=(12,6)$ on $T^6/\Z_6'$ which in turn is reduced to $(h_{11},h_{21})=(8,2)$ on $T^6/\Z_2 \times \Z_6$
due to new $\Z_2$ identifications of the $\Z_3$ fixed points.

The Hodge numbers for factorisable $T^4/\Z_N$ and $T^6/\Z_N$ orbifolds are tabulated in table~\ref{Tab:Hodge-T6ZN}.
\begin{table}[ht]
\renewcommand{\arraystretch}{1.3}
  \begin{minipage}[b]{0.5\linewidth}\centering
   \begin{equation*}
{\tiny
\begin{array}{|c||c||c||c|c|c||c|} \hline
        \multicolumn{7}{|c|}{\rule[-3mm]{0mm}{8mm}
\text{\bf Hodge numbers per twist sector on $T^4/\Z_N$}
}\\ \hline\hline
\!\!\!T^4/\!\!\!& \begin{array}{c} {\rm lattice} \\ \text{Hodge nr.} \end{array} & \begin{sideways}\!\!\!\!\!\!\!\textcolor{blue}{\rm Untwisted} \end{sideways}&  \vec{w} 
&2\vec{w}  &  3\vec{w}  & \text{total}
\\\hline\hline
\Z_2 & SU(2)^4 &  & \color{blue}{(\frac{1}{2},-\frac{1}{2})} & \multicolumn{2}{|c||}{} & 
\\\hline
& h_{11} & \color{blue}{4} & \color{blue}{16} & \multicolumn{2}{|c||}{} & \color{blue}{20}
\\\hline\hline
\Z_3 & SU(3)^2  & &  (\frac{1}{3},-\frac{1}{3}) & \multicolumn{2}{|c||}{} & 
\\\hline
& h_{11} & \color{blue}{2} & 18 & \multicolumn{2}{|c||}{} & {\color{blue} 2} + 18
\\\hline\hline
\Z_4 & SO(5)^2 & &  (\frac{1}{4},-\frac{1}{4}) &  \color{blue}{(\frac{1}{2},-\frac{1}{2})} & &
\\\hline
& h_{11} & \color{blue}{2} & 8 & \color{blue}{10} &  & {\color{blue} 12} + 8 
\\\hline\hline
\Z_6 & SU(3)^2 & &  (\frac{1}{6},-\frac{1}{6}) &  (\frac{1}{3},-\frac{1}{3}) &  \color{blue}{(\frac{1}{2},-\frac{1}{2})} &
\\\hline
& h_{11} & \color{blue}{2} & 2 & 10 & \color{blue}{6} & {\color{blue} 8} + 12
\\ \hline
     \end{array}
}
    \end{equation*}
\end{minipage}
\hspace{0.5cm}
\begin{minipage}[b]{0.5\linewidth}
\centering
%
%
    \begin{equation*}\!\!\!\!\!\!\!
{\tiny
      \begin{array}{|c|c||c||c|c|c||c|} \hline
        \multicolumn{7}{|c|}{\rule[-3mm]{0mm}{8mm}
\text{\bf Hodge numbers per twist sector on $T^6/\Z_N$}
}\\ \hline\hline
\!\!\!T^6/\!\!\! &\begin{array}{c} {\rm lattice} \\ \text{Hodge numbers} \end{array} & \begin{sideways}\!\!\!\!\!\!\!\textcolor{blue}{\rm Untwisted} \end{sideways} & \vec{w} 
&  2\vec{w}  &  3\vec{w} & \text{total}
\\\hline\hline
\Z_3 & SU(3)^3  & & (-\frac{2}{3},\frac{1}{3},\frac{1}{3}) & \multicolumn{2}{|c||}{} & 
\\\hline
& h_{11} & 9 & 27 & \multicolumn{2}{|c||}{} & 36
\\
& h_{21} & \color{blue}{0} & \color{blue}{0} & \multicolumn{2}{|c||}{} & \color{blue}{0}
\\\hline\hline
\Z_4 & SU(2)^2 \times SO(5)^2 & & (-\frac{1}{2},\frac{1}{4},\frac{1}{4}) &  \color{blue}{(0,\frac{1}{2},-\frac{1}{2})} & &
\\\hline
& h_{11} & 5 & 16 & 10 &  & 31 
 \\
& h_{21} & \color{blue}{1} & 0 & \color{blue}{6} &  & \color{blue}{7}
\\\hline\hline
\Z_6 & SU(3)^3 & &  (-\frac{1}{3},\frac{1}{6},\frac{1}{6}) & (-\frac{2}{3},\frac{1}{3},\frac{1}{3}) &  \color{blue}{(0,\frac{1}{2},-\frac{1}{2})} &
\\\hline
& h_{11} & 5 & 3 & 15 & 6 & 29
\\
& h_{21} & \color{blue}{0} & 0 & 0 & \color{blue}{5} & \color{blue}{5} 
\\ \hline\hline
\Z_6' & SU(2)^2 \times SU(3)^2 & & (-\frac{1}{2},\frac{1}{3},\frac{1}{6}) & (0,-\frac{1}{3},\frac{1}{3}) & \color{blue}{(\frac{1}{2},0,-\frac{1}{2})} &
\\\hline
& h_{11} & 3 & 12 & 12 & 8 & 35   
\\
& h_{21} & \color{blue}{1} & 0 & 6 & \color{blue}{4} & {\color{blue} 5} + 6
\\ \hline
     \end{array}
}
    \end{equation*}
 \end{minipage}
\caption{Distribution of the Hodge numbers of toroidal orbifold limits of K3 (left) and of
 Calabi-Yau three-folds (right) per twist sector. Intersecting D6-branes (D7-branes) in Type IIA (IIB)
orientifolds can wrap three-cycles (two-cycles) from the untwisted and $\Z_2$ twisted sectors only
whose counting is highlighted in blue. Since $b_3= 2 h_{21} + 2$ for Calabi-Yau three-folds ($b_2=h_{11} +2$ for K3), 
there are two more three-cycles (two-cycles) from the untwisted sector on which D6-branes (D7-branes) can be wrapped.}
\label{Tab:Hodge-T6ZN}
\end{table}
Since the splitting of $h_{11}=h_{11}^+ + h_{11}^-$ has to our knowledge not been performed systematically before for orientifolds of $T^6/\Z_N$,
we list the complete result in table~\ref{Tab:Hodge-even+odd-T6ZN}.  
\begin{table}[ht]
\renewcommand{\arraystretch}{1.3}
  \begin{center}
    \begin{equation*}
{\tiny
      \begin{array}{|c||c||c|c|c||c|} \hline
        \multicolumn{6}{|c|}{\rule[-3mm]{0mm}{8mm}
\text{\bf Hodge numbers $(h_{11}^+,h_{11}^-)$ per twist sector for IIA on $T^6/(\Z_N \times \OR)$}}
\\ \hline\hline
\begin{array}{c} T^6/ \\ \text{Hodge nr.}\end{array}  & {\rm Untw.} & \vec{w} 
&  2\vec{w}  &  3\vec{w} & \text{total}
\\\hline\hline
\Z_3  &   & (-\frac{2}{3},\frac{1}{3},\frac{1}{3}) & \multicolumn{2}{|c||}{} & 
\\\hline
 (h_{11}^+,h_{11}^-) & (3,6) & \begin{array}{cc} {\bf AAA} & (13,14)\\ {\bf AAB}& (12,15)\\ {\bf ABB} & (9,18)\\ {\bf BBB} & (0,27)\end{array} & \multicolumn{2}{|c||}{} &  \begin{array}{cc} {\bf AAA} & (16,20)\\ {\bf AAB}& (15,21)\\ {\bf ABB} & (12,24)\\ {\bf BBB} & (3,33)\end{array}  
\\\hline\hline
\Z_4  & & (-\frac{1}{2},\frac{1}{4},\frac{1}{4}) &  (0,\frac{1}{2},-\frac{1}{2}) & &
\\\hline
 (h_{11}^+,h_{11}^-) & (1,4) & (8 \, b, 16-8b) & \begin{array}{cc} \left.\begin{array}{c} {\bf a/bAA}\\{\bf a/bBB}\end{array}\right\}  & (0,10) \\ {\bf a/bAB} & (1,9) \end{array} &  & \!\!\!\!\!\!\!\!
 \begin{array}{cc} \left.\begin{array}{c} {\bf a/bAA}\\{\bf a/bBB}\end{array}\right\}\!\!\!\!  & (1+8 \, b,30 - 8 \, b  ) \\ {\bf a/bAB} & (2 + 8\, b, 29 - 8\, b ) \end{array}
\\\hline\hline
\Z_6   & &  (-\frac{1}{3},\frac{1}{6},\frac{1}{6}) & (-\frac{2}{3},\frac{1}{3},\frac{1}{3}) & (0,\frac{1}{2},-\frac{1}{2}) &
\\\hline
 (h_{11}^+,h_{11}^-) & (1,4) & \begin{array}{cc} \left.\begin{array}{c}{\bf AAA}\\{\bf AAB}\\{\bf ABB}\end{array}\right\} & (1,2) \\ {\bf BBB} & (0,3) \end{array} 
& \begin{array}{cc} \left.\begin{array}{c} {\bf AAA}\\{\bf ABB}\end{array}\right\} & (5,10) \\ {\bf AAB} & (6,9) \\ {\bf BBB} & (0,15) \end{array} & (1,5) 
& \begin{array}{cc} \left.\begin{array}{c} {\bf AAA}\\{\bf ABB}\end{array}\right\} & (8,21) \\ {\bf AAB} & (9,20) \\ {\bf BBB} & (2,27) \end{array}
\\ \hline\hline
\Z_6'   & & (-\frac{1}{2},\frac{1}{3},\frac{1}{6}) & (0,-\frac{1}{3},\frac{1}{3}) & (\frac{1}{2},0,-\frac{1}{2}) &
\\\hline
 (h_{11}^+,h_{11}^-) & (0,3) & \!\!\!\!\!\!\!\!\begin{array}{cc} \left.\begin{array}{c} {\bf a/bAA}\\{\bf a/bAB}\end{array}\right\}\!\!\!\! 
& (4+2b,8-2b) \\ \left.\begin{array}{c}{\bf a/bBA} \\ {\bf a/bBB} \end{array}\right\}\!\!\!\! 
& (6b,12-6b) \end{array} 
& \begin{array}{cc} \left.\begin{array}{c} {\bf a/bAA}\\{\bf a/bAB}\end{array}\right\}\!\!\!\! & (4,8) 
\\ \left.\begin{array}{c}{\bf a/bBA} \\ {\bf a/bBB} \end{array}\right\}\!\!\!\! & (0,12) \end{array}  
& (4b,8-4b) &  \!\!\!\!\!\!\!\!  \begin{array}{cc} \left.\begin{array}{c} {\bf a/bAA}\\{\bf a/bAB}\end{array}\right\}\!\!\!\! & (8+6b,27-6b ) 
\\ \left.\begin{array}{c}{\bf a/bBA} \\ {\bf a/bBB} \end{array}\right\}\!\!\!\! & (10b,35-10b ) \end{array}   
\\ \hline
     \end{array}
}
    \end{equation*}
  \end{center}
\caption{The number of K\"ahler moduli and Abelian vectors in $T^6/\Z_N$ orientifolds depends on the orientation of the factorisable lattices.}
\label{Tab:Hodge-even+odd-T6ZN}
\end{table}

In tables~\ref{Tab:T6Z2Z3}, \ref{Tab:T6Z2Z2}, \ref{Tab:T6Z2Z4}, \ref{Tab:T6Z2Z6}, \ref{Tab:T6Z2Z6p}, we review how untwisted two- and three-cycles (denoted, e.g., by $\pi_{35}$ and $\pi_{135}$) 
and twisted two- and three-cycles (two-cycles denoted by $d^{\mathbb{C}^2/\Z_N}$ for $N \neq 2$ and $e$ for $\Z_2$ singularities, 
three-cycles denoted by $\varepsilon$ or  $\tilde{\varepsilon}$ if they stem from a $\Z_2$ twisted sector and $\delta$ or $\tilde{\delta}$ otherwise)
of the $T^6/\Z_N \times \Z_M$ orbifolds arise from $T^4/\Z_M$ and $T^6/\Z_N$ sectors. For completeness, we also include the case 
$T^6/\Z_6' = T^6/\Z_2 \times \Z_3$ which does not admit discrete torsion.

\begin{table}[ht]
\begin{minipage}[b]{0.65\linewidth}\centering
\begin{equation*}\!\!\!\!\!\!\!\!\!\!\!\!\!\!\!\!\!\!\!\!\!\!\!\!\!\!\!\!\!\!
\mbox{\resizebox{0.8\textwidth}{!}{%
$\begin{array}{|c||c|c|}\hline
\multicolumn{3}{|c|}{\text{\bf Two-cycles on } T^6/\Z_2 \times \Z_3}
\\\hline\hline
T^4/\Z_3  & T^6/\Z_6' & \text{fixed point counting}
\\\hline\hline
\sum_{k=0}^2 \omega (\pi_{35})  & \emptyset & 
\\
\sum_{k=0}^2 \omega (\pi_{36})  & \emptyset & 
\\
3 \, \pi_{34} & 6 \, \pi_{34} & 
\\
3 \, \pi_{56} & 6 \, \pi_{56} & 
\\
\emptyset &  6 \, \pi_{12} & 
\\\hline
d_{l,l\in\{1\ldots 9\}}^{\bC^2/\Z_3(k), k \in\{1,2\}}  & 2 \, d_{l, l\in\{1,2,3\}}^{\bC^2/\Z_3^{(1)} (k), k \in\{1,2\}} 
& l \in \{1 \ldots 3\} \Leftrightarrow \{(1i)\}
\\
& \begin{array}{l}2 \, \bigl( d_{l=3+i}^{\bC^2/\Z_3^{(1)} (k)} \\ \qquad + d_{l=6+i}^{\bC^2/\Z_3^{(1)} (k)} \bigr)_{ i\in\{1,2,3\}}^{k \in\{1,2\} }\end{array}
 & \begin{array}{c} l \in \{4 \ldots 6\} \Leftrightarrow \{(2i)\}
\\
l \in \{7 \ldots 9\} \Leftrightarrow \{(3i)\}
\\
\text{ on } T_2 \times T_3 \text{ with } i=1,2,3
\end{array}
\\\hline
\emptyset & d^{\bC^3/\Z_6'}_{l,l\in\{1 \ldots 12\}} & \begin{array}{c} l \in \{1 \ldots 12 \} \Leftrightarrow \\
\{(i1k)\} \text{ on } T_1 \times T_2 \times T_3 \\
\text{with } i=1,2,3,4 \text{ and } k=1,2,3  \end{array}
\\\hline
\emptyset & 2 \, e_{l,l\in\{1 \ldots 8 \}}^{(2)} &  \begin{array}{c} l \in \{1 \ldots 4 \} \Leftrightarrow \{(i1)\}
\\
l \in \{5 \ldots 8 \} \Leftrightarrow \{(i4+i5+i6)\}
\\
\text{ on } T_1 \times T_3 \text{ with } i=1,2,3,4
\end{array}
\\\hline
\end{array}$}}
\end{equation*}
\end{minipage}
\begin{minipage}[b]{0.65\linewidth}
\centering
\hspace{-30mm}
\begin{equation*}\hspace{-58mm}
\mbox{\resizebox{0.8\textwidth}{!}{%
$\begin{array}{|c|c|c|c|}\hline
\multicolumn{4}{|c|}{\text{\bf Three-cycles on } T^6/\Z_2 \times \Z_3}
\\\hline\hline
T^4/\Z_3 & & T^6/\Z_6'& \text{fixed point counting}
\\\hline\hline
\sum_{k=0}^2 \omega (\pi_{35})  & \stackrel{2\pi_1 \otimes}{\longrightarrow} & \hat{\rho}_1 & 
\\
& \stackrel{2\pi_2 \otimes}{\longrightarrow} &  \hat{\rho}_3 & 
\\
\sum_{k=0}^2 \omega (\pi_{35})  & \stackrel{2\pi_1 \otimes}{\longrightarrow} &  \hat{\rho}_2 & 
\\
& \stackrel{2\pi_2 \otimes}{\longrightarrow} &  \hat{\rho}_4 & 
\\
3 \, \pi_{34} &  & \emptyset  & 
\\
3 \, \pi_{56} & & \emptyset & 
\\\hline
d_{l,l\in\{1\ldots 9\}}^{\bC^2/\Z_3 (k), k \in\{1,2\}}    & \stackrel{2\pi_1 \otimes}{\longrightarrow} 
& \begin{array}{l}  2 \, \pi_1 \otimes \bigl( d_{l=3+i}^{\bC^2/\Z_3^{(1)} (k)}\\
\qquad  - d_{l=6+i}^{\bC^2/\Z_3^{(1)} (k)} \bigr)_{ i\in\{1,2,3\}}^{k \in\{1,2\}}\end{array}
& \begin{array}{c} l \in \{4 \ldots 6\} \Leftrightarrow \{(2i)\}
\\
l \in \{7 \ldots 9\} \Leftrightarrow \{(3i)\}
\end{array}
\\
& \stackrel{2\pi_2 \otimes}{\longrightarrow} 
& \begin{array}{l}  2 \, \pi_2 \otimes ( d_{l=3+i}^{\bC^2/\Z_3^{(1)} (k)}
\\ \qquad - d_{l=6+i}^{\bC^2/\Z_3^{(1)} (k)} \bigr)_{ i\in\{1,2,3\}}^{k \in\{1,2\}}
\end{array}
& \text{ on } T_2 \times T_3 \text{ with } i=1,2,3
\\\hline
\emptyset & & \hat{\varepsilon}_{k,k\in\{1 \ldots 4\}} &  \begin{array}{c} k \in\{1 \ldots 4\} \Leftrightarrow
\\
\text{Orbit of } \pi_1 \otimes \{(i4)\}
\\
\text{ with } (i4) \text{ on } T_1 \times T_3
\end{array}
\\\hline
\emptyset & & \hat{\tilde{\varepsilon}}_{k,k\in\{1 \ldots 4\}} &  \begin{array}{c} k \in\{1 \ldots 4\} \Leftrightarrow
\\
\text{Orbit of } \pi_2 \otimes \{(i4)\}
\\
\text{ with } (i4) \text{ on } T_1 \times T_3
\end{array}
\\\hline
\end{array}$}}
\end{equation*}
\end{minipage}
\caption{Two and three-cycles on $T^6/\Z_2 \times \Z_3$ and their origin from the $T^4/\Z_3$ and 
$T^6/\Z_6'$ sub-sectors of the product orbifold, where $T^4 =T^2_{(2)} \times T^2_{(3)}$.}
\label{Tab:T6Z2Z3}
\end{table}

\begin{table}[ht]
\begin{minipage}[b]{0.5\linewidth}\centering
\begin{equation*}
\begin{array}{|c||c|c|}\hline
\multicolumn{3}{|c|}{\text{\bf Two-cycles on } T^6/\Z_2 \times \Z_2}
\\\hline\hline
T^4/\Z_2  & \multicolumn{2}{|c|}{T^6/\Z_2 \times \Z_2}
\\
&  \eta=1 & \eta=-1 
\\\hline\hline
2 \, \pi_{35} & \multicolumn{2}{|c|}{\emptyset}
\\
2 \, \pi_{46} &\multicolumn{2}{|c|}{\emptyset}
\\
2 \, \pi_{36} &\multicolumn{2}{|c|}{\emptyset}
\\
2 \, \pi_{45} &\multicolumn{2}{|c|}{\emptyset}
\\
2 \, \pi_{34} & \multicolumn{2}{|c|}{4 \, \pi_{34}}
\\
2 \, \pi_{56} & \multicolumn{2}{|c|}{4 \,\pi_{56}}
\\ 
\emptyset & \multicolumn{2}{|c|}{4 \, \pi_{12} }
\\\hline\hline
e_{l,l\in\{1\ldots 16\}} & 2 \, e_{l,l\in\{1\ldots 16\}}^{(1)} & \emptyset
\\\hline
\emptyset &  2 \, e_{l,l\in\{1\ldots 16\}}^{(2)} & \emptyset
\\\hline
\emptyset &  2 \, e_{l,l\in\{1\ldots 16\}}^{(3)} & \emptyset
\\\hline
\end{array}
\end{equation*}
\end{minipage}
\hspace{0.5cm}
\begin{minipage}[b]{0.5\linewidth}
\centering
\begin{equation*}
\begin{array}{|c|c|c|c|}\hline
\multicolumn{4}{|c|}{\text{\bf Three-cycles on } T^6/\Z_2 \times \Z_2}
\\\hline\hline
T^4/\Z_2 & & \multicolumn{2}{|c|}{T^6/\Z_2 \times \Z_2}
\\
& & \eta=1 & \eta=-1 
\\\hline\hline
2 \, \pi_{35} & \stackrel{2\pi_1 \otimes}{\longrightarrow} & \multicolumn{2}{|c|}{4 \, \pi_{135}}
\\
& \stackrel{2\pi_2 \otimes}{\longrightarrow} & \multicolumn{2}{|c|}{4 \, \pi_{235}}
\\
2 \, \pi_{46} & \stackrel{2\pi_1 \otimes}{\longrightarrow} & \multicolumn{2}{|c|}{4 \, \pi_{146}}
\\
& \stackrel{2\pi_2 \otimes}{\longrightarrow} & \multicolumn{2}{|c|}{4 \, \pi_{246}}
\\
2 \, \pi_{36} & \stackrel{2\pi_1 \otimes}{\longrightarrow} & \multicolumn{2}{|c|}{4 \, \pi_{136}}
\\
& \stackrel{2\pi_2 \otimes}{\longrightarrow} & \multicolumn{2}{|c|}{4 \, \pi_{236}}
\\
2 \, \pi_{45} & \stackrel{2\pi_1 \otimes}{\longrightarrow} & \multicolumn{2}{|c|}{4 \, \pi_{145}}
\\
& \stackrel{2\pi_2 \otimes}{\longrightarrow} & \multicolumn{2}{|c|}{4 \, \pi_{245}}
\\
2 \, \pi_{34} & \stackrel{2\pi_1 \otimes}{\longrightarrow} & \multicolumn{2}{|c|}{\emptyset}
\\
& \stackrel{2\pi_2 \otimes}{\longrightarrow} & \multicolumn{2}{|c|}{\emptyset}
\\
2 \, \pi_{56} & \stackrel{2\pi_1 \otimes}{\longrightarrow} & \multicolumn{2}{|c|}{\emptyset}
\\
& \stackrel{2\pi_2 \otimes}{\longrightarrow} & \multicolumn{2}{|c|}{\emptyset}
\\\hline
e_l & \stackrel{2\pi_1 \otimes}{\longrightarrow} & \emptyset & 2 \, e_{l,l\in\{1\ldots 16\}}^{(1)} \otimes \pi_1
\\
& \stackrel{2\pi_2 \otimes}{\longrightarrow} & \emptyset &  2 \, e_{l,l\in\{1\ldots 16\}}^{(1)} \otimes \pi_2
\\\hline
\emptyset  & & \emptyset & 2 \, e_{l,l\in\{1\ldots 16\}}^{(2)} \otimes \pi_3
\\
& & \emptyset &  2 \, e_{l,l\in\{1\ldots 16\}}^{(2)} \otimes \pi_4
\\\hline
\emptyset &  & \emptyset & 2 \, e_{l,l\in\{1\ldots 16\}}^{(3)} \otimes \pi_5
\\
& & \emptyset &  2 \, e_{l,l\in\{1\ldots 16\}}^{(3)} \otimes \pi_6
\\\hline
\end{array}
\end{equation*}
\end{minipage}
\caption{Two and three-cycles on $T^6/\Z_2 \times \Z_2$ and their origin from the $T^4/\Z_2$ sub-sector with
$T^4 =T^2_{(2)} \times T^2_{(3)}$ and $\Z_2=\Z_2^{(1)}$ for the case without ($\eta=1$) and with
 ($\eta=-1$) discrete torsion.}
\label{Tab:T6Z2Z2}
\end{table}

\begin{sidewaystable}[ht]
\begin{minipage}[b]{0.65\linewidth}\centering
\begin{equation*}\hspace{-28mm}\mbox{
\resizebox{0.8\textwidth}{!}{%
$\begin{array}{|c|c||c|c|c|}\hline
\multicolumn{5}{|c|}{\text{\bf Two-cycles on } T^6/\Z_2 \times \Z_4}
\\\hline\hline
T^4/\Z_4 & T^6/\Z_4  & \multicolumn{2}{|c|}{T^6/\Z_2 \times \Z_4} &
\\
& &  \eta=1 & \eta=-1 & \text{fixed point counting}
\\\hline\hline
\emptyset & 4 \, \pi_{12} \stackrel{ \times 2 }{\longrightarrow}   & \multicolumn{2}{|c|}{8 \, \pi_{12}} & 
\\
\multicolumn{2}{|c||}{ 4  \, \pi_{34} \stackrel{ \times 2 }{\longrightarrow}}  & \multicolumn{2}{|c|}{8 \, \pi_{34}} & 
\\
\multicolumn{2}{|c||}{  4 \, \pi_{56} \stackrel{ \times 2 }{\longrightarrow}}   & \multicolumn{2}{|c|}{8 \, \pi_{56}} & 
\\
  2(\pi_{35} - \pi_{46}) & 2(\pi_{35} + \pi_{46}) &  \multicolumn{2}{|c|}{\emptyset} & 
\\
 2 (\pi_{36}+\pi_{45})  &  2 (\pi_{36}-\pi_{45})  &  \multicolumn{2}{|c|}{\emptyset} &
\\\hline
d_{l,l\in\{1\ldots 4\}}^{\bC^2/\Z_4(k),k\in\{0,1,2\}} \stackrel{ \times 2 }{\longrightarrow}  &  e_{l,l\in\{1 \ldots 4\}}^{(1)} \stackrel{ \times 2 }{\longrightarrow} 
 &  2\, d_{l,l\in\{1\ldots 4\}}^{\bC^4/\Z_4^{(1)}(k),k\in\{0,1,2\}}
& \left\{\begin{array}{c}  2 \, e_{l,l\in\{1 \ldots 4\}}^{(1)} \\ \equiv   2 \, d_{l,l\in\{1\ldots 4\}}^{\bC^4/\Z_4^{(1)}(0)}
\end{array}\right.
&
\begin{array}{c} l \in \{1\ldots 4\} \Leftrightarrow
\\
\{(11),(13),(31),(33)\} \\
\text{ on } T_2 \times T_3
\end{array}
\\\hline
\multicolumn{2}{|c||}{e_{l,l\in\{5 \ldots 10\}} \stackrel{ \times 2 }{\longrightarrow} } & \multicolumn{2}{|c|}{2 \, e_{l,l\in\{5 \ldots 10\}}^{(1)} }
&
\begin{array}{c} l \in \{5\ldots 10\} \Leftrightarrow
\\
\{(12+14),(32+34),(21+41),\\(23+43),(22+44),(24+42)\} \\
\text{ on } T_2 \times T_3
\end{array}
\\\hline
\emptyset & d_{l,l \in \{1 \ldots 16\}}^{\bC^6/\Z_4}\stackrel{ \times 2 }{\longrightarrow} &  2 \, d_{l,l \in \{1 \ldots 16\}}^{\bC^6/\Z_4}  & \emptyset
&
\begin{array}{c} l \in \{1 \ldots  16\} \Leftrightarrow
\\
\{(i11),(i13),(i31),(i33)\}\\
\text{ with } i=1,2,3,4
\end{array}
\\\hline
\multicolumn{2}{|c||}{\emptyset } 
&  2 \, e_{l,l \in\{1\ldots 12\}}^{(2)}
& 2 \, e_{l,l \in\{9\ldots 12\}}^{(2)} 
&
\begin{array}{c} l \in \{1 \ldots 8\} \Leftrightarrow
\{(i1),(i3)\},\\
 l \in \{9 \ldots 12\} \Leftrightarrow
\{(i2 \pm i4)\},\\ 
\text{ on } T_1 \times T_3 \text{ with } i=1,2,3,4 
\end{array}
\\\hline
\multicolumn{2}{|c||}{\emptyset } 
&  2 \, e_{l,l \in\{1\ldots 12\}}^{(3)}
& 2 \, e_{l,l \in\{9\ldots 12\}}^{(3)} 
&
\begin{array}{c} l \in \{1 \ldots 8\} \Leftrightarrow
\{(i1),(i3)\},\\
 l \in \{9 \ldots 12\} \Leftrightarrow
\{(i2 \pm i4)\}, 
\\
\text{ on } T_1 \times T_2 \text{ with } i=1,2,3,4 
\end{array} 
\\\hline
\end{array}$}}
\end{equation*} 
\end{minipage}
\hspace{-20mm}
\begin{minipage}[b]{0.65\linewidth}
\centering
\begin{equation*}\hspace{-44mm}
\mbox{\resizebox{0.8\textwidth}{!}{%
$\begin{array}{|c|c||c|c|c|}\hline
\multicolumn{5}{|c|}{\text{\bf Three-cycles on } T^6/\Z_2 \times \Z_4}
\\\hline\hline
T^4/\Z_4 & T^6/\Z_4  & \multicolumn{2}{|c|}{T^6/\Z_2 \times \Z_4} & 
\\
& &  \eta=1 & \eta=-1 & \text{fixed point counting}
\\\hline\hline
 2(\pi_{35} - \pi_{46})   \stackrel{\otimes 2\pi_1 }{\longrightarrow} &  \hat{\rho}_1 \stackrel{ \times 2 }{\longrightarrow} &  \multicolumn{2}{|c|}{\rho_1} &
\\
 2 (\pi_{36}+\pi_{45})  \stackrel{\otimes 2\pi_2 }{\longrightarrow}  &   \hat{\rho}_2  \stackrel{ \times 2 }{\longrightarrow} &  \multicolumn{2}{|c|}{\rho_2} & 
 \\
-  2(\pi_{35} - \pi_{46})  \stackrel{\otimes 2 \pi_2 }{\longrightarrow}  &  \hat{\rho}_3  \stackrel{ \times 2 }{\longrightarrow} &  \multicolumn{2}{|c|}{\rho_3} & 
\\
 2 (\pi_{36}+\pi_{45})  \stackrel{\otimes 2\pi_1 }{\longrightarrow}  &   \hat{\rho}_4  \stackrel{ \times 2 }{\longrightarrow} &  \multicolumn{2}{|c|}{\rho_4} & 
\\
4 \, \pi_{34} & \emptyset & \multicolumn{2}{|c|}{\emptyset} & 
\\
4 \, \pi_{56} & \emptyset & \multicolumn{2}{|c|}{\emptyset} & 
 \\\hline
d_{l,l\in\{0\ldots 4\}}^{\bC^2/\Z_4(k),k\in\{0,1,2\}} \stackrel{\otimes 2\pi_1 }{\longrightarrow}   & \emptyset & \emptyset & \delta^{(k), k \in \{1,2\}}_{l,l \in \{1 \ldots 4\}}
&\begin{array}{c} l \in \{1\ldots 4\} \Leftrightarrow
\\
\{(11),(13),(31),(33)\}
\end{array}
 \\
d_{l,l\in\{0\ldots 4\}}^{\bC^2/\Z_4(k),k\in\{0,1,2\}}  \stackrel{\otimes 2\pi_2 }{\longrightarrow}    & \emptyset & \emptyset & \tilde{\delta}^{(k), k \in \{1,2\}}_{l,l \in \{1 \ldots 4\}}
& \text{ on } T_2 \times T_3
\\\hline
e_{l,l\in\{5 \ldots 10\}} \stackrel{\otimes 2\pi_1 }{\longrightarrow} &  \varepsilon^{(1)}_{l,l\in\{5 \ldots 10\}}
& \multicolumn{2}{|c|}{ \emptyset }
&\begin{array}{c} l \in \{5\ldots 10\} \Leftrightarrow
\\\{(12 \pm 14),(32 \pm 34),(21 \pm 41), \end{array}
\\
\qquad\qquad\quad \stackrel{\otimes 2\pi_2 }{\longrightarrow}    & \tilde{\varepsilon}^{(1)}_{l,l\in\{5 \ldots 10\}}
& \multicolumn{2}{|c|}{ \emptyset }&\begin{array}{c} 
(23 \pm 43),(22 \pm 44),(24 \pm 42)\} \\
\text{ on } T_2 \times T_3
\end{array}
\\\hline
\end{array}$}}
\end{equation*} 
\end{minipage}
\caption{Two and three-cycles on $T^6/\Z_2 \times \Z_4$ and their origin from $T^4/\Z_4$ with 
$T^4 =T^2_{(2)} \times T^2_{(3)}$ and $\Z_4=\Z_4^{(1)}$ and from $T^6/\Z_4$
for the case without ($\eta=1$) and with ($\eta=-1$) discrete torsion. .}
\label{Tab:T6Z2Z4}
\end{sidewaystable}

\begin{sidewaystable}[ht]
\begin{minipage}[b]{0.65\linewidth}\centering
\begin{equation*}\hspace{-30mm}
\mbox{\resizebox{0.8\textwidth}{!}{%
$\begin{array}{|c|c||c|c|c|}\hline
\multicolumn{5}{|c|}{\text{\bf Two-cycles on } T^6/\Z_2 \times \Z_6}
\\\hline\hline
T^4/\Z_6 & T^6/\Z_6'  & \multicolumn{2}{|c|}{T^6/\Z_2 \times \Z_6} & 
\\
& &  \eta=1 & \eta=-1  & \text{fixed point counting}
\\\hline\hline
\emptyset & 6 \, \pi_{12}  \stackrel{\times 2}{\longrightarrow} &  \multicolumn{2}{|c|}{12 \, \pi_{12}} & 
\\
\multicolumn{2}{|c||}{6 \, \pi_{34}  \stackrel{\times 2}{\longrightarrow}} &  \multicolumn{2}{|c|}{12 \, \pi_{34}} & 
\\
\multicolumn{2}{|c||}{6 \, \pi_{56} \stackrel{\times 2}{\longrightarrow}} &  \multicolumn{2}{|c|}{12 \, \pi_{56}} &
\\
2 \, \sum_{k=0}^2 \omega(\pi_{35}) & \emptyset &  \multicolumn{2}{|c|}{\emptyset} & 
 \\
2 \, \sum_{k=0}^2 \omega(\pi_{36}) & \emptyset &  \multicolumn{2}{|c|}{\emptyset} & 
\\\hline
\!\!d^{\bC^2/\Z_6 (k),k\in\{0\ldots 4\}}_l \!\! & \!\! d^{\bC^2/\Z_3^{(1)} (k), k \in \{1,2\}}_l \!\!
&\!\! 2 d^{\bC^2/\Z_6^{(1)} (k),k\in\{0\ldots 4\}}_l \!\! & \!\! 2  d^{\bC^2/\Z_6^{(1)} (k),k\in\{1,2\}}_l \!\! & l=(11) \text{ on } T_2 \times T_3 
\\\hline
\!\! d_{l,l\in\{1 \ldots 4\}}^{\bC^2/\Z_3 (k),k\in \{1,2\}} \!\! & \!\! d_{l',l'\in\{1 \ldots 5\}}^{\bC^2/\Z_3^{(1)} (k),k\in \{1,2\}} \!\!
& \multicolumn{2}{|c|}{2 \, d_{l',l'\in\{1 \ldots 3\}}^{\bC^2/\Z_3^{(1)} (k),k\in \{1,2\}} } & 
\begin{array}{c}
l \in \{1 \ldots 4 \} \Leftrightarrow \\
 \{(12+13),(21+31),(22+33),(23+32)\},
\\
l'\in \{1 \ldots 5\} \Leftrightarrow \\
\{(21),(31),(12+13),(22+23),(32+33)\},
\\
l'' \in \{1 \ldots 3 \} \Leftrightarrow \\
 \{(12+13),(21+31),(22+33 + 23+32)\}
\\
\text{on } T_2 \times T_3
\end{array}
\\\hline
e_{l,l\in\{1 \ldots 5\}} \stackrel{\times 2}{\longrightarrow} & \emptyset & 2 \, e_{l,l\in\{1 \ldots 5\}}^{(1)} & \emptyset
&
\begin{array}{c}
l \in \{1 \ldots 5 \} \Leftrightarrow
\\
 \{(14 + 15+16),(41+51+61),(44+56+65),\\
 (45+54+66), (46+55+64)\}
\text{ on } T_2 \times T_3
\end{array}
\\\hline
\emptyset & e_{l,l\in\{1 \ldots 8\}}^{(2)}\stackrel{\times 2}{\longrightarrow} & 2 \, e_{l,l\in\{1 \ldots 8\}}^{(2)} & \emptyset
&
\begin{array}{c}
l \in \{1 \ldots 8 \} \Leftrightarrow
 \{(i1),(i4+i5+i6)\}
\\
\text{on } T_1 \times T_3
\text{ with } i=1,2,3,4
\end{array}
\\\hline
\emptyset & \emptyset &  2 \, e_{l,l\in\{1 \ldots 8\}}^{(3)} & \emptyset
&
\begin{array}{c}
l \in \{1 \ldots 8 \} \Leftrightarrow
 \{(i1),(i4+i5+i6)\}
\\
\text{on } T_1 \times T_2
\text{ with } i=1,2,3,4
\end{array}
\\\hline
\emptyset & d_{l,l\in\{1\ldots 12\}}^{\bC^3/\Z_6'}  \stackrel{\times 2}{\longrightarrow}  & 2 \, d_{l',l'\in\{1\ldots 8\}}^{\bC^3/\Z_6'}  & 2 \, d_{l'',l''\in\{1\ldots 4\}}^{(\bC^3/\Z_6')} 
& \begin{array}{c} l \in \{1 \ldots 12\} \Leftrightarrow \{(i11),(i21),(i31)\}
\\
l' \in \{1 \ldots 8\} \Leftrightarrow \{(i11),(i21 +i31)\}
\\
l'' \in \{1 \ldots 4\} \Leftrightarrow \{(i21 - i31)\}
\\
\text{ on } T_1 \times T_2 \times T_3
\text{ with } i=1 \ldots 4
\end{array}
\\\hline
\multicolumn{2}{|c||}{\emptyset} & 2 \, d_{l',l'\in\{1\ldots 8\}}^{\bC^3/\tilde{\Z}_6'}  & 2 \, d_{l'',l''\in\{1\ldots 4\}}^{\bC^3/\tilde{\Z}_6'} 
& \begin{array}{c}
l' \in \{1 \ldots 8\} \Leftrightarrow \{(i11),(i12 +i13)\}
\\
l'' \in \{1 \ldots 4\} \Leftrightarrow \{(i12 - i13)\}
\\
\text{ on } T_1 \times T_2 \times T_3
\text{ with } i=1 \ldots 4
\end{array}
\\\hline
\end{array}$}}
\end{equation*}
\end{minipage}
\hspace{-20mm}
\begin{minipage}[b]{0.65\linewidth}
\centering
\begin{equation*}\hspace{-50mm}
\mbox{\resizebox{0.8\textwidth}{!}{%
$\begin{array}{|c|c||c|c|c|}\hline
\multicolumn{5}{|c|}{\text{\bf Three-cycles on } T^6/\Z_2 \times \Z_6}
\\\hline\hline
T^4/\Z_6 & T^6/\Z_6' & \multicolumn{2}{|c|}{T^6/\Z_2 \times \Z_6} & 
\\
& &  \eta=1 & \eta=-1  & \text{fixed point counting}
\\\hline\hline
2 \sum_{k=0}^2 \omega(\pi_{35})  \stackrel{2 \pi_1 \otimes }{\longrightarrow}  & \hat{\rho}_1  \stackrel{\times 2}{\longrightarrow} &  \multicolumn{2}{|c|}{ \rho_1 } & 
\\
2 \sum_{k=0}^2 \omega(\pi_{36})  \stackrel{2 \pi_1 \otimes }{\longrightarrow}  & \hat{\rho}_2  \stackrel{\times 2}{\longrightarrow} &  \multicolumn{2}{|c|}{ \rho_2 } & 
\\
2 \sum_{k=0}^2 \omega(\pi_{35})  \stackrel{2 \pi_2 \otimes }{\longrightarrow}  & \hat{\rho}_3  \stackrel{\times 2}{\longrightarrow} &  \multicolumn{2}{|c|}{ \rho_3 } & 
\\
2 \sum_{k=0}^2 \omega(\pi_{36})  \stackrel{2 \pi_2 \otimes }{\longrightarrow}  & \hat{\rho}_4  \stackrel{\times 2}{\longrightarrow} &  \multicolumn{2}{|c|}{ \rho_4 } & 
\\
6 \, \pi_{34} & \emptyset & \multicolumn{2}{|c|}{ \emptyset } & 
\\
6 \, \pi_{56} & \emptyset  & \multicolumn{2}{|c|}{ \emptyset } &  
\\\hline
e_{l,l\in\{1\ldots 5\}}  \stackrel{2 \pi_1 \otimes }{\longrightarrow}  & \emptyset & \emptyset & \varepsilon_{l,l\in\{1\ldots 5\}}^{(1)}
& \begin{array}{c}
l \in \{1 \ldots 5 \} \Leftrightarrow \{(14+15+16),(41+51+61),\\
(44+56+65),(45+54+66),(46+55+64)\}
\end{array}
\\
\qquad\qquad\quad \stackrel{2 \pi_2 \otimes }{\longrightarrow} & \emptyset & \emptyset & \tilde{\varepsilon}_{l,l\in\{1\ldots 5\}}^{(1)}
&\text{ on } T_2 \times T_3 
\\\hline
d^{\bC^2/\Z_6(k),k\in\{0\ldots 4\}}  \stackrel{2\pi_1 \otimes}{\longrightarrow} & \emptyset & \emptyset
& \varepsilon_0^{(1)}, \delta^{(1), k\in \{3,4\}}
& (11) \text{ on } T_2 \times T_3 
\\
\qquad\qquad\qquad\quad \stackrel{2\pi_2 \otimes}{\longrightarrow} & \emptyset  & \emptyset 
& \tilde{\varepsilon}_0^{(1)}, \tilde{\delta}^{(1), k\in \{3,4\}}
&
\\\hline
d^{\bC^2/\Z_3 (k),k\in\{1,2\}}_{l,l\in\{1\ldots 4\}}
 \stackrel{2 \pi_1 \otimes }{\longrightarrow}
& \hat{\delta}_{l',l'\in\{1\ldots 3\}}^{k,k\in\{1,2\}}  \stackrel{\times 2}{\longrightarrow}  & \emptyset & \delta_{l''}^{(1), k\in\{1,2\}}
&\!\!\!\! \begin{array}{c}
 l \in \{1 \ldots 4\} \Leftrightarrow \{(12+13),(21+31),(22+33),(23+32)\}
\\
 l' \in \{1 \ldots 3\} \Leftrightarrow \{(i2+i3)\} \text{ with } i=1,2,3
\end{array}\!\!\!\!
\\
\qquad\qquad\qquad\quad \stackrel{2 \pi_2 \otimes }{\longrightarrow} &  \hat{\tilde{\delta}}_{l',l'\in\{1\ldots 3\}}^{k,k\in\{1,2\}}  \stackrel{\times 2}{\longrightarrow}  & \emptyset & \tilde{\delta}_{l''}^{(1),k\in\{1,2\}}
& l''=(22+23+32+33)
\qquad\qquad \text{ on } T_2 \times T_3 
\\\hline
\emptyset & \hat{\varepsilon}_{l,l\in\{1\ldots 4\}}  \stackrel{\times 2}{\longrightarrow}  & \emptyset & \varepsilon_{l,l\in\{1\ldots 4\}}^{(2)}
& l \in \{1 \ldots 4 \} \Leftrightarrow \{(i4+i5+i6)\}
\\
\emptyset &  \hat{\tilde{\varepsilon}}_{l,l\in\{1\ldots 4\}}  \stackrel{\times 2}{\longrightarrow}  & \emptyset & \tilde{\varepsilon}_{l,l\in\{1\ldots 4\}}^{(2)}
& \text{ on } T_1 \times T_3 \text{ with } i=1 \ldots 4
\\\hline
\multicolumn{2}{|c||}{\emptyset} & \emptyset & \varepsilon_{l,l\in\{1\ldots 4\}}^{(3)}
& l \in \{1 \ldots 4 \} \Leftrightarrow \{(i4+i5+i6)\}
\\
\multicolumn{2}{|c||}{\emptyset} & \emptyset & \tilde{\varepsilon}_{l,l\in\{1\ldots 4\}}^{(3)}
& \text{ on } T_1 \times T_2 \text{ with } i=1 \ldots 4
\\\hline
\end{array}$}}
\end{equation*}
\end{minipage}
\caption{Two and three-cycles on $T^6/\Z_2 \times \Z_6$ and their origin from $T^4/\Z_6$ with 
$T^4 =T^2_{(2)} \times T^2_{(3)}$ and $\Z_6=\Z_6^{(1)}$ and from $T^6/\Z_6'$ for the case without $(\eta=1)$
and with  $(\eta=-1)$ discrete torsion.}
\label{Tab:T6Z2Z6}
\end{sidewaystable}

\clearpage

\begin{sidewaystable}[ht]
\begin{minipage}[b]{0.65\linewidth}\centering
\begin{equation*}\hspace{-30mm}
\mbox{\resizebox{0.8\textwidth}{!}{%
$\begin{array}{|c||c|c|c|}\hline
\multicolumn{4}{|c|}{\text{\bf Two-cycles on } T^6/\Z_2 \times \Z_6'}
\\\hline\hline
T^6/\Z_6  & \multicolumn{2}{|c|}{T^6/\Z_2 \times \Z_6'} &
\\
&  \eta=1 & \eta=-1  & \text{fixed point counting}
\\\hline\hline
6 \, \pi_{12} \stackrel{\times 2}{\longrightarrow} &  \multicolumn{2}{|c|}{12 \, \pi_{12}} &
\\
6 \, \pi_{34} \stackrel{\times 2}{\longrightarrow}  &   \multicolumn{2}{|c|}{12 \, \pi_{34}} &
\\
6 \pi_{56} \stackrel{\times 2}{\longrightarrow}  &   \multicolumn{2}{|c|}{12 \, \pi_{56}} &
\\
\left.\begin{array}{c} \sum_{k=0}^5 \theta^k(\pi_{35}) = \\ 4 \pi_{35} + 4 \pi_{46} - 2 \pi_{45} - 2 \pi_{36} \end{array}\right\} &  \multicolumn{2}{|c|}{\emptyset } &
\\
\left.\begin{array}{c} \sum_{k=0}^5 \theta^k(\pi_{36}) = \\ 2 \pi_{35} + 2 \pi_{36} +2 \pi_{46} - 4 \pi_{45} \end{array}\right\} &  \multicolumn{2}{|c|}{\emptyset} &
\\\hline
d^{\bC^3/\Z_6^{(1)}}_{l,l\in\{1 \ldots 3\}} & 2 \, d^{\bC^3/\Z_6^{(1)}}_{l',l' \in\{1,2\}} & 2 \, d^{\bC^3/\Z_6^{(1)}}_{l''} & \begin{array}{c}
l\in\{1 \ldots 3\} \Leftrightarrow \{(111),(211),(311)\}
\\
l' \in \{1,2\}  \Leftrightarrow \{(111),(211+311)\}
\\
l'' = (211-311)
\\
\text{ on } T_1 \times T_2 \times T_3
\end{array}
\\\hline
d^{\bC^3/\Z_3}_{l,l\in\{1\ldots 15\}} & \multicolumn{2}{|c|}{d^{\bC^3/\Z_3}_{l',l'\in\{1\ldots 9\}}} & \begin{array}{c}
l,l' \in \{1,2,3\} \Leftrightarrow \{(111),(121+131),(112+113)\}
\\
l \in \{4 \ldots 9\} \Leftrightarrow \{(211), (221+231),\\(212+213),(122+133),(222+233),(232+223)\}
\\
l \in \{10 \ldots 15\} \Leftrightarrow \{(311),(321+331),\\(312+313),(132+123),(332+323),(322+333)\}
\\
l' \in \{4 \ldots 9\} \Leftrightarrow \{ (l + [l+6]) \text{ with } l \in 4 \ldots 9\} ,
\end{array}
\\\hline
e_{l,l\in\{1\ldots 6\}} \stackrel{\times 2}{\longrightarrow}  & 2 \, e^{(1)}_{l,l\in\{1\ldots 6\}} & \emptyset & \begin{array}{c} l \in \{1 \ldots 5\} \Leftrightarrow \{(11),(14+15+16),(41+51+61),\\ (44+55+66),(45+56+64),(46+54+65)\} \\
\text{ on } T_2 \times T_3
\end{array}
\\\hline
\emptyset &   2 \, e^{(2)}_{l,l\in\{1\ldots 6\}} & \emptyset & \text{see } 2 \, e^{(1)}_l \text{ and permute } T_2 \leftrightarrow T_1
\\
\emptyset &   2 \, e^{(3)}_{l,l\in\{1\ldots 6\}} & \emptyset &  \ldots \text{ permute } T_3 \leftrightarrow T_1
\\\hline
\emptyset & 2 \,  d^{\bC^3/\Z_6^{(2)}}_{l',l'\in \{1,2\}} & 2 \, d^{\bC^3/\Z_6^{(2)}}_{l''} & \text{see } 2 \, d^{\bC^3/\Z_6^{(1)}} \text{ and permute } T_2 \leftrightarrow T_1
\\
\emptyset & 2 \,  d^{\bC^3/\Z_6^{(3)}}_{l',l'\in \{1,2\}} & 2 \, d^{\bC^3/\Z_6^{(3)}}_{l''} & \ldots  \text{ permute } T_3 \leftrightarrow T_1
\\\hline
\end{array}$}}
\end{equation*}
 \end{minipage}
\hspace{-20mm}
\begin{minipage}[b]{0.65\linewidth}
\centering
\begin{equation*}\hspace{-48mm}
\mbox{\resizebox{0.8\textwidth}{!}{%
$\begin{array}{|c||c|c|c|}\hline
\multicolumn{4}{|c|}{\text{\bf Three-cycles on } T^6/\Z_2 \times \Z_6'}
\\\hline\hline
T^6/\Z_6 & \multicolumn{2}{|c|}{T^6/\Z_2 \times \Z_6'} &
\\
&  \eta=1 & \eta=-1  & \text{fixed point counting}
\\\hline\hline
\hat{\rho}_1  \stackrel{\times 2}{\longrightarrow}  &  \multicolumn{2}{|c|}{\rho_1} &
\\
\hat{\rho}_2  \stackrel{\times 2}{\longrightarrow}  &  \multicolumn{2}{|c|}{\rho_2 } &
\\\hline
\hat{\varepsilon}_{l,l\in\{1 \ldots 5\}}  \stackrel{\times 2}{\longrightarrow}  & \emptyset & \varepsilon_{l,l\in\{1 \ldots 5\}} ^{(1)}
& \begin{array}{c} l \in \{1 \ldots 5\} \Leftrightarrow \{(14+15+16),
\\ (41+51+61),(44+55+66),\\
(45+56+64),(46+54+65)\}
\end{array}
\\
\hat{\tilde{\varepsilon}}_{l,l\in\{1 \ldots 5\}} ^{(1)} \stackrel{\times 2}{\longrightarrow}  & \emptyset &  \tilde{\varepsilon}_{l,l\in\{1 \ldots 5\}} ^{(1)}
& \text{ on } T_2 \times T_3
\\\hline
\emptyset  & \emptyset &  \varepsilon_{l,l\in\{1 \ldots 5\}} ^{(2)} &
\\
\emptyset &  \emptyset &   \tilde{\varepsilon}_{l,l\in\{1 \ldots 5\}} ^{(2)} &  \ldots \text{ on } T_1 \times T_3
\\\hline
\emptyset & \emptyset &  \varepsilon_{l,l\in\{1 \ldots 5\}} ^{(3)} &
\\
\emptyset &  \emptyset &   \tilde{\varepsilon}_{l,l\in\{1 \ldots 5\}} ^{(3)} &   \ldots \text{ on } T_1 \times T_2
\\\hline
\end{array}$}}
\end{equation*}
\end{minipage}
\caption{Two and three cycles on $T^6/\Z_2 \times \Z_6'$ and their origin from $T^6/\Z_6$
for the case without $(\eta=1)$ and with  $(\eta=-1)$ discrete torsion.}
\label{Tab:T6Z2Z6p}
\end{sidewaystable}

\end{appendix}

\clearpage	 
\addcontentsline{toc}{section}{References}	 
\bibliographystyle{ieeetr}	 
\bibliography{refs_rigid}	 

\end{document}

%% file: Z2-lattices.pdf_t
\begin{picture}(0,0)%
\includegraphics{Z2-lattices.pdf}%
\end{picture}%
\setlength{\unitlength}{2565sp}%
\begingroup\makeatletter\ifx\SetFigFontNFSS\undefined%
\gdef\SetFigFontNFSS#1#2#3#4#5{%
  \reset@font\fontsize{#1}{#2pt}%
  \fontfamily{#3}\fontseries{#4}\fontshape{#5}%
  \selectfont}%
\fi\endgroup%
\begin{picture}(7380,3129)(-4439,-8068)
\put(-4424,-5986){\makebox(0,0)[lb]{\smash{{\SetFigFontNFSS{14}{16.8}{\familydefault}{\mddefault}{\updefault}$\pi_{2i}$}}}}
\put(-1649,-7936){\makebox(0,0)[lb]{\smash{{\SetFigFontNFSS{14}{16.8}{\familydefault}{\mddefault}{\updefault}$\pi_{2i-1}$}}}}
\put(2326,-7186){\makebox(0,0)[lb]{\smash{{\SetFigFontNFSS{14}{16.8}{\familydefault}{\mddefault}{\updefault}$\pi_{2i-1}$}}}}
\put(-149,-6061){\makebox(0,0)[lb]{\smash{{\SetFigFontNFSS{14}{16.8}{\familydefault}{\mddefault}{\updefault}$\pi_{2i}$}}}}
\put(-1124,-6661){\makebox(0,0)[lb]{\smash{{\SetFigFontNFSS{12}{14.4}{\rmdefault}{\mddefault}{\updefault}$R_2$}}}}
\put(2926,-5986){\makebox(0,0)[lb]{\smash{{\SetFigFontNFSS{12}{14.4}{\rmdefault}{\mddefault}{\updefault}$R_2$}}}}
\put(1651,-7861){\makebox(0,0)[lb]{\smash{{\SetFigFontNFSS{12}{14.4}{\rmdefault}{\mddefault}{\updefault}$R_1$}}}}
\put(-2549,-5611){\makebox(0,0)[lb]{\smash{{\SetFigFontNFSS{12}{14.4}{\rmdefault}{\mddefault}{\updefault}$R_1$}}}}
\end{picture}%

%% file: Z4-Z6-lattices.pdf_t
\begin{picture}(0,0)%
\includegraphics{Z4-Z6-lattices.pdf}%
\end{picture}%
\setlength{\unitlength}{2565sp}%
\begingroup\makeatletter\ifx\SetFigFontNFSS\undefined%
\gdef\SetFigFontNFSS#1#2#3#4#5{%
  \reset@font\fontsize{#1}{#2pt}%
  \fontfamily{#3}\fontseries{#4}\fontshape{#5}%
  \selectfont}%
\fi\endgroup%
\begin{picture}(6702,2682)(-2864,-8068)
\put(1126,-6211){\makebox(0,0)[lb]{\smash{{\SetFigFontNFSS{14}{16.8}{\familydefault}{\mddefault}{\updefault}$\pi_{2i}$}}}}
\put(2551,-7936){\makebox(0,0)[lb]{\smash{{\SetFigFontNFSS{14}{16.8}{\familydefault}{\mddefault}{\updefault}$\pi_{2i-1}$}}}}
\put(-524,-7936){\makebox(0,0)[lb]{\smash{{\SetFigFontNFSS{14}{16.8}{\familydefault}{\mddefault}{\updefault}$\pi_{2i-1}$}}}}
\put(-2849,-5986){\makebox(0,0)[lb]{\smash{{\SetFigFontNFSS{14}{16.8}{\familydefault}{\mddefault}{\updefault}$\pi_{2i}$}}}}
\end{picture}%

%% file: kbtree.pdf_t
\begin{picture}(0,0)%
\includegraphics{kbtree.pdf}%
\end{picture}%
\setlength{\unitlength}{1243sp}%
\begingroup\makeatletter\ifx\SetFigFont\undefined%
\gdef\SetFigFont#1#2#3#4#5{%
  \reset@font\fontsize{#1}{#2pt}%
  \fontfamily{#3}\fontseries{#4}\fontshape{#5}%
  \selectfont}%
\fi\endgroup%
\begin{picture}(14170,5485)(386,-5931)
\end{picture}%

%% file: kbloop.pdf_t
\begin{picture}(0,0)%
\includegraphics{kbloop.pdf}%
\end{picture}%
\setlength{\unitlength}{1243sp}%
\begingroup\makeatletter\ifx\SetFigFont\undefined%
\gdef\SetFigFont#1#2#3#4#5{%
  \reset@font\fontsize{#1}{#2pt}%
  \fontfamily{#3}\fontseries{#4}\fontshape{#5}%
  \selectfont}%
\fi\endgroup%
\begin{picture}(13450,17140)(386,-17676)
\end{picture}%